\documentclass[a4paper,11pt,twoside]{ThesisStyle}
\usepackage{pdfpages}

\usepackage{amsmath,amssymb}
\usepackage[T1]{fontenc}
\usepackage[utf8]{inputenc}
\usepackage[english]{babel}

\usepackage[left=1.0in,right=1.0in,top=1.1in,bottom=1.1in,includefoot,includehead,headheight=13.6pt]{geometry}

\usepackage{silence}

\usepackage{aecompl}
\usepackage{url}

\usepackage[printonlyused,withpage]{acronym}

\usepackage{ifpdf}

\graphicspath{{.}{images/}}

\usepackage{color}
\definecolor{linkcol}{rgb}{0,0,0.4}
\definecolor{citecol}{rgb}{0.5,0,0}

\usepackage[nottoc, notlof, notlot]{tocbibind}
\usepackage{minitoc}
\usepackage{rotating}                    
\usepackage{fancyhdr}                    

\let\headruleORIG\headrule
\renewcommand{\headrule}{\color{black} \headruleORIG}

\usepackage{colortbl}
\arrayrulecolor{black}

\fancypagestyle{plain}{
  \fancyhead{}
  \fancyfoot{}
  
}

\usepackage{algorithm}
\usepackage[noend]{algorithmic}
\usepackage{scrextend}

\makeatletter

\def\cleardoublepage{\clearpage\if@twoside \ifodd\c@page\else%
  \hbox{}%
 \thispagestyle{empty}
  \newpage%
  \if@twocolumn\hbox{}\newpage\fi\fi\fi}

\makeatother

{%

\hrulefill
\vspace*{0.5cm}%
\end{minipage}
}

\let\minitocORIG\minitoc
\renewcommand{\minitoc}{\minitocORIG \vspace{1.5em}}

\usepackage{subfigure}
\usepackage{multirow}
{ \begin{list}%
	{$\bullet$}%
	{\setlength{\labelwidth}{25pt}%
	 \setlength{\leftmargin}{30pt}%
	 \setlength{\itemsep}{\parsep}}}%
{ \end{list} }

\renewcommand{\epsilon}{\varepsilon}

\ifpdf
  \usepackage[pagebackref,hyperindex=true]{hyperref}
\else
  \usepackage[dvipdfm,pagebackref,hyperindex=true]{hyperref}
\fi

\renewcommand*{\backref}[1]{}
\renewcommand*{\backrefalt}[4]{%
\ifcase #1 %
(Not cited.)%
\or
(Cited on page~#2.)%
\else
(Cited on pages~#2.)%
\fi}

\hypersetup
{
bookmarksopen=true,
pdftitle="Manuscript title",
pdfauthor="Your name",
pdfsubject="Manuscript topic in a few words", 
pdfmenubar=true, 
pdfhighlight=/O, 
colorlinks=true, 
pdfpagemode=UseNone, 
pdfpagelayout=SinglePage, 
pdffitwindow=true, 
linkcolor=linkcol, 
citecolor=citecol, 
urlcolor=linkcol 
}

\begin{document}

\pagenumbering{Alph}

\begin{titlepage}
\begin{center}
\vspace*{0.3cm}
\noindent {\Huge \textbf{DHOST theories as disformal gravity: \smallskip \\ From black holes to radiative \smallskip \\ \vspace*{0.5cm} spacetimes}} \\
\vspace*{0.8cm}
\noindent \LARGE Jibril \textsc{Ben Achour} \\
\vspace*{0.8cm}
\vspace*{0.8cm}
\noindent \Large Arnold Sommerfeld Center for Theoretical Physics, Munich, Germany \\
Laboratoire de physique, \'Ecole Normale Sup\'erieure de Lyon, France \\
IJCLab, Paris-Saclay University, Orsay, France \\
\vspace*{0.2cm}
\begin{figure}[t]
\includegraphics[width=3cm]{LMU.pdf} \qquad 
\includegraphics[width=3cm]{ORSAY.pdf}
\centering
\end{figure}
\noindent \Large Habilitation Thesis - HDR\\ 15th November 2024
\vspace*{0.8cm} 
\begin{figure}[t]
\includegraphics[width=5cm]{ENS1.pdf} 
\centering
\end{figure}
\end{center}
     
      				
                                          
\end{titlepage}

\pagenumbering{arabic}
\sloppy

\titlepage

\pagenumbering{roman}

\setcounter{page}{0}
\cleardoublepage

\section*{}

\textit{\`A toi qui m'as transmis la flamme,}\\
\textit{Merci d'être \`a mes côtés, tout simplement, et de voir plus loin que moi ...}

\newpage
\section*{Acknowledgments}

Let me first thank the members of the jury: my referees \'Eric Gourgoulhon, Mokhtar Hassaine, Christos Charmousis, and the rest of the jury, David Langlois, Alexander Vikman, Misao Sasaki, Danièle Steer and Eugeny Babichev.  I am excited to explore further  the open questions of the field with some of you in the near future.

It is a pleasure to thank all the people I had the chance to meet and collaborate with during these last ten years. It has been a true journey from Orsay to China, Japan, Germany and back to France. The list of friends and colleagues who had an impact in one way or another on my trajectory is long. A special thanks to my long-term collaborators and friends: Jean-Philippe Uzan for his enthusiasm to explore new territories and for his constant support, Etera R. Livine for his insatiable curiosity, for enlightening  ydiscussions and for accepting to go back to 1d physics and finally to Karim Noui who has been there from the start. Thanks to you, I have witnessed over the last ten years your complementary styles and approaches to study theoretical physics, providing a quite rare but definitely enjoyable cocktail. I would also like to warmly thank my postdoc friends who have made this scientific (but not only) journey possible through a close collaboration from the first postdoc up to now: Suddhasattwa Brahma, Hongguang Liu, Mohammad Ali Gorji and Hugo Roussille. 

A first special thanks goes to Shinji Mukohyama for the opportunity to join the Yukawa Institute and live in the magnificent city of Kyoto, probably the best spot in the world to think peacefully about gravity.  A second goes to Daniele Oriti for the last four years, for the opportunity to join the Arnold Sommerfeld Center for Theoretical Physics, for the new exciting research direction initiated during our numerous discussions, but above all for his understanding during these last three years. At this stage, I would also like to thank Jean-Christophe Geminard and with him the whole theory group and administrative staff at \'Ecole Normale Sup\'erieure (ENS) de Lyon for welcoming me as a long term visitor. 

These acknowledgements would be incomplete without having a final thanks for all the old and new friends with who I have learned and experienced the up and downs of the researcher life. A big thanks to Marc Geiller of course. To the trio Ileyk El Mellah, Lea Jouvin and Thibaut Houdy. To Alexis Helou, Vivien Scottez, Julien Peloton, Julien Grain, Vincent Vennin, Julian Bautista, Ibrahim Akal, Francesco Di Filippo, C\'eline Zwikel, Daniele Pranzetti, Sylvain Carrozza, Danilo Artigas, Andreas Pithis, Christophe Goeller, Alexender Jercher, George Zahariade, Julien Larena and ones of the latest newcomers in the game, Luca Lionni and Ady.

A last thank you to those people who have been around me during part of this journey and who feel anxious when looking at the last released images from the James Webb Space Telescope. You don't need to see your name written explicitly here to know the importance you had and still have in my life.

\newpage

\section*{Abstract}

This manuscript gathers and reviews part of our work focusing on the exploration of modified theories of gravity known as degenerate higher order scalar-tensor (DHOST) theories. It focuses on the construction of exact solutions describing both black holes and radiative spacetimes. After motivating the need for alternatives theories of gravity beyond general relativity, we discuss in more details the long terms objectives of this research program. 
The first one is to characterize both the theory and some sectors of the solution space of DHOST gravity. The second one is to provide concrete and exact solutions of the DHOST field equations describing compact objects, in particular black holes, that can be used to confront DHOST theories to current and future observations in the strong field regime. A key tool towards these two objectives is the concept of disformal field redefinition (DFR) which plays a central role in this exploration. We start be reviewing the structure of the DHOST theory space, the notion of degeneracy conditions and the stability of these degeneracy classes under DFR. Then we review several key notions related to stationary and axi-symmetric black holes, and in particular the no-hair theorems derived in GR and in its scalar-tensor extensions. The rest of the chapters are devoted to a review of the disformal solution generating map, the subtle role of matter coupling and how it can be used to construct new hairy black holes solutions. The case of spherically symmetric solutions, axi-symmetric but non-rotating solutions, and finally rotating solutions are discussed, underlining the advantages and the limitations of this approach. A brief review of the rotating black holes solutions found so far in this context is followed by the detailed description of the disformed Kerr black hole. We further comment on on-going efforts to construct rotating black hole solutions mimicking the closest the Kerr geometry. Then, we discuss how DFR affects the algebraic properties of a gravitational field and in particular its Petrov type. This provides a first systematic characterization of this effect, paving the road for constructing new solutions with a fixed Petrov type.
Finally, we review more recent works aiming at characterizing the effect of a DFR on non-linear radiative geometries. We derive the general conditions for the generation of disformal tensorial gravitational wave and we study in detail a concrete example in DHOST gravity.
While most of the material presented here is a re-organized and augmented version of our published works, we have included new results and also new proposals to construct phenomenologically interesting solutions.

\newpage

\section*{Publications}

The results discussed in this manuscript are based on one part of our research works presented in the following articles:
\begin{itemize}
\item "\textit{Degenerate higher order scalar-tensor theories beyond Horndeski and disformal transformations}"\\ \textbf{J. BA}, David Langlois, Karim Noui, \\ Phs. Rev. D 93 (2016), \cite{BenAchour:2016cay}
\item "\textit{Degenerate higher order scalar-tensor theories beyond Horndeski up to cubic order}"\\ \textbf{J. BA}, Marc Crisostomi, Kazuya Koyama, David Langlois, Karim Noui, and G. Tasinato, \\ JHEP. 12 (2016) 100, \cite{BenAchour:2016fzp}
\item "\textit{Hairy Schwarzschild-(A)dS black hole solutions in degenerate higher order scalar-tensor theories beyond shift symmetry}"\\ \textbf{J. BA}, Hongguang Liu, \\ Phs. Rev. D 99 (2019) 6, 064042, \cite{BenAchour:2018dap}
\item "\textit{Hairy black holes in DHOST theories: Exploring disformal transformation as a solution-generating method}"\\ \textbf{J. BA}, Hongguang Liu, Shinji Mukohyama,\\ JCAP 02 (2020) 023, \cite{BenAchour:2020wiw}
\item "\textit{On rotating black holes in DHOST theories}"\\ \textbf{J. BA}, Hongguang Liu, Hayato Motohashi, Shinji Mukohyama, Karim Noui, \\ JCAP 11 (2020) 001, \cite{BenAchour:2020fgy}
\item "\textit{Disformal map and Petrov classification in modified gravity}"\\ \textbf{J. BA}, Antonio De Felice, Mohammad Ali Gorji, Shinji Mukohyama, Masroor C. Pookkillath, \\ JCAP 10 (2021) 067, \cite{BenAchour:2021pla}
\item "\textit{Nonlinear gravitational waves in Horndeski gravity: scalar pulse and memories}"\\ \textbf{J. BA}, Mohammad Ali Gorji, Hugo Roussille, \\JCAP 05 (2024) 026, \cite{BenAchour:2024zzk}
\item "\textit{Disformal gravitational waves}"\\ \textbf{J. BA}, Mohammad Ali Gorji, Hugo Roussille, \\ JCAP (2024), \cite{BenAchour:2024tqt}
\end{itemize}



\dominitoc
\tableofcontents

\mainmatter


\chapter{Introduction}
\label{Chapter1}

\textit{"[L'homme est la seule créature\footnote{Je sais que ce mot te chiffonne, mais tu leur pardonneras!} qui refuse d'être ce qu'elle est.] Et ces insoumissions déroulent, sans fin, le fil de nos rêveries contre le silence de nos étoiles."}\\ \\
\bigskip
\textit{J. Monod, "Le hasard et la nécéssité"} \\ \vspace*{0.3cm}
 \textit{L. Slimak, "Homo Sapiens nu: Le premier âge du rêve"}

\bigskip

\minitoc
\def\rd{\mathrm{d}}

\section{Why modify Einstein gravity ?}

Since this work focuses on characterizing alternatives theories of gravity beyond general relativity, we first discuss the key motivations behind this research program.  
We start with the observational motivations and then address the theoretical ones. 

\subsection{Observational motivations} 

\textit{Motivations from the observations of the gravitational field at the cosmological scales:} \\Over the last decades, early cosmology has entered in a new era of high precision tests. This unprecedented flow of data offers for the first time the possibility to shed light on the fine structure of the early universe and question the mechanisms at play which have led to its present state.
The standard model of cosmology, i.e. the $\Lambda$-CDM model, which combines general relativity, quantum field theory and nuclear physics, has succeeded to fit the data with a remarkably small numbers of parameters. Nevertheless, exploring the degeneracy of scenarii associated to the measured values of the cosmological parameters is a challenging task.

The picture which emerges from this standard model of cosmology is that at early time, the universe can be accurately described by a very dense and hot plasma which cools down due to cosmic expansion. During this cooling process, the universe undergoes  successive transitions during which specific fields (photons, baryons, electrons, neutrinos, etc ...) decouple from the rest of the plasma. The high energy phase transitions are still largely speculative. However, the observational consequences of the transitions taking place at lower energy scales and predicted by the standard model have been successfully confirmed and nowadays represent the pillars of the predictive power of the standard model of cosmology. Typically, at an energy scale between 10 to 0.1 MeV, the Big-Bang nucleosynthesis (BBN) takes place and the first light elements are formed. At an energy scale of the order of the eV, the electrons and protons recombine allowing the photons to decouple from the plasma and travel freely resulting today in the cosmic microwave background (CMB). At this stage, the universe becomes neutral allowing gravitational structures to form and grow. The existence of the BBN and of the CMB stand as some of the corner stones predictions of the standard model of cosmology, these predictions being themselves consequences of the cosmic expansion predicted first by GR\footnote{The first theory of gravity showing that the geometry of the universe filled with matter has to be dynamical, expanding or contracting, and exhibiting a dynamical horizon was indeed GR. Yet, in the first model, Einstein used an ad hoc assumption, demanding that the energy density compensate the effect of the cosmological constant such that the geometry has to remain static. Despite this akward assumption, the theory indeed predicted that the geometry of the universe is dynamical, the fine-tuned system designed by Einstein being itself unstable against perturbations. See \cite{Bianchi:2010uw} for an account on this point.}. 

However, in order to understand the observed statistical properties of the CMB and in particular the nearly scale invariant distribution of the temperature anisotropies, one has to introduce new physics. The leading paradigm to account for this property of the CMB assumes an early phase of cosmic inflation driven by a yet unknown matter content (generally taken as a self-interacting scalar field absent from the standard model of particle physics), a mechanism introduced in the early eighties \cite{Starobinsky:1979ty, Starobinsky:1980te, Guth:1980zm, Linde:1981mu, Mukhanov:1981xt, Starobinsky:1982ee, Hawking:1982cz, Bardeen:1983qw}. This powerful mechanism based on this phase of cosmic acceleration allows one to solve several issues of the previous hot big bang model and successfully
predicts the statistical properties of the anisotropies of the CMB measured first by the COBE and WMAP missions \cite{1992ApJ...396L...1S, WMAP:2003elm} and later with great accuracy by the Planck mission \cite{Planck:2013oqw} \\ \cite{Planck:2018jri}, in particular the tiny deviation of the spectral index from unity \cite{Sasaki:1995aw}. Moreover, it also provides a mechanism to seed the large scales structures observed currently in the
universe from the initial quantum fluctuations of the gravitational and matter fields in the early universe. Nevertheless, the price to pay is the introduction of new physics which does not fit in the standard picture. In particular, the way the inflaton couples to the standard model of particles is crucial to describe how the universe transits from this inflationary phase to the radiation era, a process dubbed the reheating.

Alongside this paradigm aiming at providing a coherent scenario of the early phases of the universe, the observations of the gravitational dynamics at both the cosmological and galactic scales also reveal the need for new physics. Indeed one also has to assume the existence of yet unknown contributions to the energy budget of the universe, the dark energy and the dark matter, commonly referred as the dark sector. The first one is needed to explain the late acceleration of the cosmic expansion first identified in the late nineties via the measurements of the luminosity distance of type Ia supernovae \cite{SupernovaSearchTeam:1998fmf, SupernovaCosmologyProject:1998vns}\footnote{See \cite{Rubin:2019ywt} for a recent analysis and careful discussion on the different subtle steps involved leading to the assertion of an accelerated cosmic expansion phase.}, while the later is needed in order to fit the observed rate of structure formation (as well as to account for the anomalies  of the gravitational interaction observed at the galactic scales, a point which will be discussed below).

Therefore, despite its remarkable successes to fit the data and provide a coherent theoretical framework to understand the formation of the light elements, the CMB properties and the formation of the large scale structures, the standard model of cosmology is a giant with feet of clay. Moreover, the growing set of data has underlined the existence of anomalies during the last decades. The most well-known ones are the H$_0$ tension and the related S$_8$ tension, which refer to the mismatch between the values of these cosmological parameters inferred from direct measurements and the values inferred by early cosmology tests. See \cite{DiValentino:2021izs} for a review on the different ways out to alleviate these tensions. While the interpretation of these tensions is still open, it underlines some limitations of our current model of cosmology. Additional cosmological anomalies have also been reported over the last decades. See \cite{Peebles:2022akh} for a recent review. On-going missions such as EUCLID and DESY, and future missions such as LISA will offer new exciting data to shed light on these mysteries. However, from a theoretical point of view, much effort are needed to provide a convincing solution.

This state-of-affairs has motivated the search for alternatives theories of gravity which could i) generate accelerated phases of cosmic expansion, thus accounting for dark energy, ii) alleviate the different cosmological anomalies, and finally iii) quantitatively test deviations from GR at the cosmological scales. In the next section, we shall briefly review the different strategies adopted when constructing these extensions of general relativity.  Yet, at this stage, it is worth pointing that these anomalies on the cosmological scales might also be resolved by trading the isotropic and homogeneous Friedman-Lemaitre-Roberston-Walker (FLRW) background for more complicated exact inhomogeneous solutions of GR  to model the geometry of the universe. See \cite{Celerier:2007jc, Clarkson:2010uz} for reviews. The simplest example illustrating this point is the inhomogeneous Lemaitre-Tolman-Bondi (LTB) cosmology which exhibits a rate of cosmic expansion to be both time-dependent and radially dependent, allowing for an acceleration of the cosmic expansion while agreeing with the CMB and galaxy surveys observations. See \cite{Celerier:1999hp, Clarkson:2009sc, Celerier:2011zh} for details. Thus it appears also crucial, in parallel of the efforts to understand modified theories of gravity, to investigate how to relax our current assumptions at the root of the standard model of cosmology and whether the current cosmological anomalies might be resolved by describing the geometry of the universe with a richer geometrical model allowing for a weaker degree of symmetry\footnote{Let us further expand on this alternative line of research which is much less advertised. It aims at understanding whether non-linear effects in GR could allow one to alleviate and resolve the observed cosmological anomalies. While the large scales geometry of the universe is commonly modeled by a perturbed Friedman geometry, this approximation should break down at some energy scale, prior to inflation. Using instead a fully non-linear radiative and inhomogeneous model to describe the universe such as the Szekeres, Sazfron, Wainwright or Belinski cosmologies \cite{Wainwright:1980jn, Wainwright:1976nv, Szafron:1977zz, Belinsky:1980jh, Carr:1983jzn}, one faces the question whether non-linear radiative and cumulative effects, such as memory effects, could impact the resulting cosmological dynamics. See \cite{Tolish:2016ggo, Belinski:2017luc, Boybeyi:2024aax, Madison:2020xhh, Chakraborty:2024ars, Creminelli:2024qpu, Celerier:2024dvs, Anton:2024ywk} for some preliminary results on this question focusing on different cosmological observables.}. This being said, we now turn to another key observation motivating the exploration for alternatives to general relativity.    \bigskip


\textit{Motivations from the observations of the gravitational field at the galactic scale:} \\
At the galactic scales, the observed dispersion of velocities of stars shows a surprising qualitative transition between large and low accelerations. This transition occurs at a threshold of the order $a_0 =  1,2 \times 10^{-10}$ $m.s^{-2}$. Such transition takes place at the periphery of galaxies, where the gravitational potential is sufficiently weak, such that stars can be treated as non-relativistic objects following the newtonian law.  While this is indeed the case in the regime $a \gg a_0$, the sub-acceleration regime shows a strong deviation with the Keplerian motion. Instead of decaying according to the newtonian prediction in $1/r$, the gravitational  potential settles down to a constant value, giving rise to the observed flat rotation curves for spiral galaxies which translates into an anomalous mass-to-light ratio. This was first reported by Oort \cite{Oort} and confirmed later on with successive accurate measurements by Rubin and Ford in the seventies \cite{Rubin:1970zza, Rubin:1978kmz}. Empirical rules relating the angular velocity $v$ of the stars experiencing the plateau of the potential and the baryonic mass $M_b$ of such spiral galaxy reveals a proportionality relation of the form
\begin{align}
v^4 = 4 a_0 GM_b
\end{align}
Such relation is known as the Baryonic Tully-Fischer law \cite{1977A&A54661T}. See \cite{2012AJ14340M} for a more recent account on this empirical law. This departure from standard newtonian expectations also manifests in other observables at the galactic and meta-galactic scales, where the gravitational potential shows important deviations with the Newtonian expectation. In particular, this can be seen from the motion of galaxies, from the X-rays temperature of the galactic gases and from the weak lensing measurement.

To date, the dominant paradigm to explain this anomaly has been to postulate that galaxies and groups of galaxies are embedded in massive dark matter halos. This new exotic matter component is assumed to interact purely gravitationally with the other components of the universe. Over the last decades, important efforts have been devoted to detect via various ground-based missions the particles which would be the constituents of this hypothetical dark matter, the so-called weakly interacting particles (WIMPS), but so far the direct detection of these WIMPS has remained elusive. See \cite{Schumann:2019eaa, Arcadi:2017kky, Misiaszek:2023sxe} for pedagogical reviews. Another possibility extensively studied recently is that primordial black holes formed in the early universe might constitute a part of dark matter. Depending on the mass range, PBH abundance has been constrained in several ways. Notice that even the absence of PBH would imply strong constraints on the early phase of the universe. See \cite{Villanueva-Domingo:2021spv, Escriva:2022duf} for reviews and \cite{Carr:2024nlv} for an historical account. Nevertheless, it is also fair to point that refining the estimation of their abundance  would require understanding black hole formation in a fully dynamical set-up as well as the outcomes of the Hawking evaporation process, two aspects for which our current theoretical description is limited. 
Moreover, let us emphasize that regardless of what fields constitute dark matter, it also strongly affects the rate of structure formation in the universe and enter as a key ingredient in the $\Lambda$-CDM model. Yet, the DM paradigm also faces limitations when it comes to small scales observables \cite{Bullock:2017xww}. Therefore,  because of the key role it plays at both galactic and cosmological scales, it appears challenging to come up with an alternative to it.

Despite these difficulties, modifications of the Newtonian laws have been proposed to account for the observed discrepancy of the galactic gravitational potential. The MOND proposal which amounts at an ad hoc corrections to the Newtonian law has attracted much attention since its publication \cite{Milgrom:1983ca}. See \cite{Milgrom:2019cle} for a recent review. After more than forty year of efforts, the situation is mixed. Despite concrete verified predictions, the MOND paradigm still faces several difficulties, such as addressing the anomalous behavior observed on different scales and for different types of galaxies. Moreover, to our knowledge, all attempts to embed MOND in a consistent relativistic theory of gravity, either suffer from consistency issues, or have been ruled out (such as TEVES) \cite{Bruneton:2007si}\footnote{See \cite{Blanchet:2017duj} for a recent proposal based on massive bi-gravity}. See \cite{Famaey:2011kh} for a review. On a different front, it has also been proposed that the MOND transition could originate from semi-classical thermal effects\footnote{Consider the two-body problem. Below a given acceleration scale, one would expect that each body starts being sensitive to the thermal bath generated by its partner via the Unruh effect. This transition would occurs below a temperature $T \sim a_0/2\pi$, and one would need to add semi-classical corrections to the classical trajectories of the bodies to take into account this effect.}, or even as hint of the non-relativistic regime of quantum gravity \cite{Smolin:2017kkb}. See also \cite{Ciotti:2022inn} regarding the recent debate about disk models adapted for galaxies in general relativity.

In the last decades, new ideas have emerged to provide test of the dark matter/modified gravity debate. By focusing on binary systems with sufficiently large separation, one can in principle test the newtonian dynamics in the low acceleration regime. Concretely, for a test particle orbiting around the sun, the critical MOND acceleration is reached for separation greater than $7000 $ UA (or equivalently $3,4 \times 10^{-2}$ parsec). These systems are called wide binary systems (WBS) and the key point is that on such scale, the dark matter density is expected to be very small, playing no significant role on the binary dynamics. Thus, by identifying a large number of such WBS, we should be able to decide between the dark matter paradigm (no modification of the gravitational potential for WBS) and the modified gravity one (implying a non-newtonian dynamics for these systems). See \cite{Hernandez:2023qfj, 2024MNRAS.533..729H} for details. WBS have been identified within the catalogues provided by the GAIA mission and the current conclusions are still debated as conflicting claims have been made by different teams \cite{Pittordis:2019kxq, Hernandez:2024bfw}. While only further data will allow the community to conclude, the existence of such tests is fascinating and encouraging.

The different anomalies mentioned above, both at the cosmological and galactic scales, are the key observations signaling the need to go beyond GR and thus motivating the search for alternatives theories of gravity. See \cite{Clifton:2011jh} for a review.

Let us now briefly review the theoretical open questions which, alongside these observational motivations, further fuel the search for such theories.

\subsection{Theoretical motivations}

The theoretical motivations stem from i) the concrete need to have efficient parametrization of the possible deviations to GR to confront the theory to current observations to ii) more fundamental questions related to the description of compact objects and the observational consequences of resolving singularities. Another important motivation behind the exploration of modified gravity is to refine our understanding of the uniqueness properties of GR. In the following, we shall briefly review these different points.   \\ 

\textit{Parametrizing deviations to GR:} With the growing flow of data ranging from the observations of large scale structures at the cosmological scales (DESY, EUCLID), the recent progress in imaging the close environment of super-massive black holes (GRAVITY, EVENT HORIZON TELESCOPE, BLACK HOLE EXPLORER), and the advent of the gravitational wave astronomy (LIGO-Virgo-KAGRA, PTA, LISA, EINSTEIN TELESCOPE), it has become crucial to develop efficient formalisms to confront the theory and its possible extensions to the observations. These data shall allow us to confirm/infirm the predictions of GR and identify a subspace of the viable theories. For this reason, one has to find the most efficient way to capture the phenomenology of modified gravity in a given regime. This task was undertaken first by Nordtvedt, Thorne and Will through the development of the Parametrized Post-Newtonian (PPN) formalism allowing one to translate any modification to GR in a universal language relevant for the weak field regime \cite{Thorne:1971iat}. Since then, other formalisms designed for cosmological tests have been developed, such as the Parametrized Post-Friedman (PPF) formalism. With the development of more complicated theories of modified gravity \cite{Horndeski:1974wa,  Deffayet:2011gz, Gleyzes:2014dya, Langlois:2015cwa, BenAchour:2016fzp} (which will be discussed in this manuscript), this task had to be pushed further to account for the new phenomenology triggered in these extensions. A powerful and unified framework was developed to confront the most general scalar-tensor theories constructed so far \cite{Langlois:2015cwa, BenAchour:2016fzp}, dubbed degenerate higher order scalar-tensor (DHOST) theories, to cosmological observations  in \cite{Gleyzes:2015pma, Langlois:2017mxy}, allowing one to encode the new phenomenology in a set of only six parameters (nine numbers constrained by three consistency conditions). These approaches can be viewed as a type of effective field theory (EFT) approach bringing the theories beyond GR in a ready-to-use language for bench marks tests of modified gravity. They have already revealed useful when confronting DHOST theories on cosmological scales, in particular using the measure of the relative speed between gravitational waves and photons emitted by the GW170817 merger \cite{Langlois:2017dyl}.  Now, important efforts are  devoted to develop the same EFT language for perturbations around compact objects, a task which is more involved due to the lower degrees of symmetry of the involved background \cite{Franciolini:2018uyq, Hui:2021cpm, Mukohyama:2022skk, Mukohyama:2022enj, Khoury:2022zor, Barura:2024uog, Mukohyama:2024pqe}. See also \cite{Langlois:2021aji, Langlois:2022ulw, Roussille:2022vfa} for a new algorithm to study black hole perturbations within DHOST theories. Let us stress that since these formalisms rely on fixing a given background and studying how the new phenomenology manifests in terms of perturbations propagating on this background, they are limited by the very assumptions of the perturbative scheme. New effects induced by the non-linearities of the theory and where no scale separation can be assumed cannot be described and thus constrained using these formalisms. In particular, the description of the possible endpoints of gravitational collapse, and thus the different possible compact objects that can form through the process, have to be studied at the fully non-perturbative level, in particular in finding exact solutions.  \\ 

\textit{Characterizing the endpoints of gravitational collapse:} Within general relativity, the outcome of the gravitational collapse has been beautifully characterized by the no-hair theorem derived first by Israel \cite{Israel:1967wq, Israel:1967za} and then successively refined by Hawking, Carter and Robinson in the seventies \cite{Carter:1971zc, Hawking:1971vc, Robinson:1974nf, Robinson:1975bv}. This well-known theorem stands as a cornerstone of general relativity.  It states that under very general assumptions, the most general stationary and axi-symmetric vacuum four dimensional geometry describing the final state of gravitational collapse is isomorphic to the outer communication region of the Kerr solution. It follows that during the process, the gravitational field describing the collapse radiates all its higher multipoles to settle down to a very simple configuration characterized uniquely by its mass $M$ and angular momentum $J$. These two parameters $(M,J)$ can be understood as asymptotic charges associated with the generators of stationarity and axi-symmetry\footnote{Quite remarkably, the Kerr geometry exhibits powerful symmetries beyond the obvious stationarity and axi-symmetry which further constrain the behavior of test particles orbiting around it and test fields scattering on it.}. It has become standard practice to call any additional number not associated with a symmetry a hair. Within this language, the no-hair theorem states that in general relativity, four-dimensional vacuum stationary and axi-symmetric black holes are hairless. It follows that any fields (scalar, vector and tensor) are either radiated away at infinity or swallowed by the hole, such that asymptotically flat black holes cannot support non-trivial field configurations in GR. See \cite{Chrusciel:1994sn, Mazur:2000pn, Gourgoulhon:2024} for reviews and lectures.

For a long time, the status of the no-hair theorem has remained unchallenged. However, the discovery of asymptotically flat colored rotating black holes from the Einstein-Yang-Mills system have shown that the situation is more subtle \cite{Bartnik:1988am, Bizon:1990sr}. This black hole geometry turns out to be also characterized by an additional number  which cannot be interpreted as an asymptotic charge associated to an (explicit) symmetry. It was further understood that in some case, field with a non-linear coupling can conspire to avoid to be fully radiated away or swallowed by the hole, giving rise to asymptotically flat black holes supporting non-trivial hairs. See \cite{Bizon:1994dh, Chrusciel:2012jk} for reviews. The first example carrying a scalar hair being the (unstable) Bekenstein-Bocharova-Bronnikov-Melnikov black hole \cite{Bocharova:1970skc, Bekenstein:1974sf}. Since then, these findings have triggered important efforts to extend and clarify the status of the no-hair theorem beyond general relativity. The extensions found so far are usually much less restrictive than their GR counterpart but nevertheless they provide key guides to construct exact solutions describing hairy black holes in modified gravity. See \cite{Herdeiro:2015waa, Volkov:2016ehx} for reviews. We shall review these developments in detail and in particular the most recent formulations of the no-hair theorem for DHOST gravity \cite{Capuano:2023yyh} in Chapter 3. Let us further point that another recent line of research has also triggered new questions related to the very definitions of a hair. Indeed, it has been understood that depending on the very notion of boundary conditions one uses, more subtle asymptotic symmetries can be identified which are associated to well-defined asymptotic charges. These soft charges also label the configuration of the gravitational system such as black holes and thus can be considered as soft hairs \cite{Hawking:2016msc, Grumiller:2019fmp}\footnote{Even a simple canonical scalar field can exhibit such soft asymptotic charges, as shown in \cite{Campiglia:2017dpg, Campiglia:2018see}.}. In particular, it has been proposed that these soft charges play a key role in how information can be stored and released during the Hawking evaporation \cite{Haco:2018ske, Flanagan:2021svq}. From that point of view, the very notion of hairy black holes used in the modified gravity community could be too limited. 

When considering hairy black holes, a crucial question is how far does the hairosphere extend? The hairosphere can be defined as the region in which the non-trivial configuration of the field considered as the hair manifests\footnote{In an effective description, it corresponds to the first local minimum of the pressure profile induced by the hair, i.e. computed form the energy-momentum tensor associated to the hair.}. This question finds its answer in the so-called \textit{no-short hair theorem} (NSH) derived first  by Nunez, Quevedo and Sudarsky in \cite{Nunez:1996xv} and further extended in a series of works \cite{Hod:2011aa, Hod:2016ixl, Ghosh:2023kge, Acharya:2024kvv}. Interestingly, it can be shown that under rather general assumptions, the hair of a hairy spherically symmetric black hole has to extend at least up to the photon ring radius. This suggests that if one has access to the near environment of the black hole, one should be able to probe the existence of hair. The NSH theorem holds for a large set of theories and for any dimension, suggesting that it could serve as a less restrictive version of the no-hair theorem for modified gravity. From this perspective, it appears crucial to further investigate the intimate relationship between the existence of a hair and the photon ring structure and clarify how this statement could be contradicted or extended.  \\

\textit{Generalizing the unicity theorems beyond the no-hair theorem:} The no-hair theorem is not the only unicity theorem allowing to characterize the gravitational field of compact objects. GR also enjoys additional fundamental theorems which provide a powerful characterization of  its solution space given some symmetry assumptions. In general, these theorems break down in modified gravity, giving rise to a whole new phenomenology both for compact objects but also for the radiative regime. In the spherically symmetric sector,  the Birkhoff's theorem dramatically constrains the allowed vacuum solutions, stating that any spherically symmetric vacuum solutions of GR (with $\Lambda=0$) are static and asymptotically flat \cite{VojeJohansen:2005nd}. Moreover, in the stationary and axi-symmetric sector, the phase space of GR can be reformulated in terms of the Ernst potentials \cite{Ernst:1967wx} which reveal powerful symmetries\footnote{together with remarkable duality with colliding cylindrical gravitational waves \cite{Chandrasekhar:1986jn}} identified first by Elhers, followed by Matzner, Geroch and later Kinnersley \cite{Ehlers:1957zz, Matzner:1967zz, Geroch:1972yt, Kinnersley:1977pg}. These hidden symmetries allow one to recast the stationary and axi-symmetric phase space of GR as an integrable system, revealing a unique property of this key subsector of the solution space. See \cite{Korotkin:2023lrg} for a recent review. While it is expected that such structures break down in modified gravity, a clear and systematic investigation of their fate beyond GR is still missing to our knowledge. Yet, filling this gap would certainly help understanding the description of compact objects beyond GR.

Besides the beautiful structures governing the vacuum axi-symmetric phase space of GR, one can wonder if one can exhibit additional theorems allowing to characterize gravitational fields with a weaker degree of symmetry. This question was answered in the sixties, during the golden age of general relativity, thanks to the developments of powerful and elegant formalisms. It was recognized that null directions of spacetime are keys tools to decompose, distinguish and  efficiently compare two given gravitational fields. The Newman-Penrose formalism \cite{Newman:1961qr}, based on the decomposition of the kinematical properties of null rays and the Einstein equations onto a null tetrad, together with the Petrov classification based on the algebraic properties of the Weyl tensor \cite{Petrov:1959zfa}, have revealed crucial to shed light on many open questions. In particular they have been crucial tools to study gravitational radiation\footnote{See \cite{Penrose:2024fdd} for a recent historical account on the early research on gravitational radiation}, following the beautiful work of Bondi, Van der Bug, Matzner and Sachs \cite{Bondi:1960jsa, Bondi:1962px, Sachs:1962zza, Sachs:1962wk}, to derive the first examples of the exact non-linear gravitational waves solutions of GR \cite{Robinson:1960zzb, Robinson:1962zz}, to study the structure of null infinity and to characterize asymptotically flat spacetimes \cite{Penrose:1964ge}, identifying on the way new surprising conserved charges even in the fully non-linear radiative regime of GR \cite{Newman:1965ik, Newman:1968uj}, to classify gravitational fields w.r.t their algebraic special character and to derive new exact solutions of GR among which the Kerr solution \cite{Kerr:1963ud}. They have also served to prove a whole family of theorems relating the Petrov type with the existence of hidden symmetries allowing the integrability of wave equations on these geometries\footnote{See \cite{Batista:2020cto} for a recent derivation of the Petrov type allowing Killing-Yano two forms.}. See \cite{Frolov:2017kze} for a review on black hole integrability. This body of results has gradually revealed the rich and fine structure of the GR solution space. Among the different known results, the notion of \textit{algebraic speciality} plays a crucial role. It can be related to the degrees of symmetry of the geometry. Quite remarkably, all relevant exact solutions describing stationary and axi-symmetric black holes in GR, known as the Plebanski-Demianski family \cite{Plebanski:1976gy}, share the same algebraic properties as the Kerr solution, i.e. they are of Petrov type D\footnote{See \cite{Barrientos:2023dlf} for the construction of the enhanced PD family with Petrov type I solutions.}. This is no longer true in modified gravity in general as we shall see in Chapter 3. So far, it has been a challenge to exhibit a deformed Kerr solution which is algebraically special (and even more of Petrov type D). As one is interested in finding exact solutions deviating from GR but mimicking closely the Kerr geometry, this raises the question whether such algebraically special rotating black hole solution exists in modified gravity. The natural extension of this question is what are the criteria to assert that a given solution of a given modified theory of gravity is algebraically special ? Such a criterion exists in vacuum GR and is formalized by the so-called Goldberg-Sachs (GS) theorem \cite{1962AcPPS2213G}. Obtaining a generalization of this GS theorem for a given modified theory of gravity would thus provide a crucial step towards a better understanding of its solution space\footnote{See for example \cite{Batista:2012bk}}.  
\newpage

\textit{Observational consequences of resolving the singularities:} As a final motivation, let us discuss the central problem related to the singularities. It is well-known that even if GR stands today as our best effective field theory of the gravitational field, its domain of applicability is limited and the theory fails to predict the behavior of the gravitational field at sufficiently high energy, near the Planck scale. Indeed, the theory predicts that for sufficiently dense matter, gravitational collapse can lead to the formation of a trapping horizon bounding a trapped region. Provided that the standard energy conditions are satisfied, the collapse inevitably generates pathological singularities which signal the breakdown of GR as a predictive theory at high energies. This statement was formalized through the so called singularity theorems by Penrose in \cite{Penrose:1964wq} and Penrose and Hawking in \cite{Hawking:1970zqf}, showing that this breakdown of the theory is independent of the details of the collapse. It follows that one needs to go beyond general relativity to tackle the difficult questions of the endpoint of gravitational collapse, the final stage of black hole evaporation and the initial conditions of the early universe prior to inflation. 

When constructing regular geometries for compact objects and assuming a pseudo-Riemanian geometry, one quickly realizes that the requirement of regularity at the core of the object can have non-trivial consequences at energy scales well smaller than the Planck scale, in particular by inducing pathological instabilities. For example, it is known that a regular, static, spherically symmetric and asymptotically flat black hole requires, additionally to the outer event horizon, at least one inner horizon. This inner horizon generically leads to an instability \cite{Carballo-Rubio:2018pmi}, known as the mass inflation instability, which quickly induces a divergence of the curvature, spoiling these popular models of regular black holes (except for inner-extremal regular black holes proposed in \cite{Carballo-Rubio:2022kad}). This statement is agnostic of any theory and is based on purely kinematical arguments. A similar result exists for horizonless (circular) compact objects. In that case, one can show that regularity at the core and the absence of horizon implies the existence of a inner \textit{stable} light ring where energy can accumulate, inducing a pathological instability as well \cite{DiFilippo:2024mnc}. This illustrates the challenge in resolving the singularity while providing stable long-lived geometries describing ultra compact objects\footnote{Among the different possibilities, astrophysical black holes might be bouncing compact objects transiting from a black to a white hole in a snapshot for a local observer, but being elapsed in time for an asymptotic far away observer. Attempts to construct consistent models for such bouncing object can be found in \cite{Haggard:2014rza, BenAchour:2020bdt, BenAchour:2020mgu,  BenAchour:2020gon, Cipriani:2024nhx}. See \cite{Carballo-Rubio:2018jzw} for a review.}. Quite remarkably, assuming a pseudo-Riemanian framework, the set of black hole geometries which are geodesically complete is remarkably limited and a classification was provided in \cite{Carballo-Rubio:2019nel, Carballo-Rubio:2019fnb}. Modified theories of gravity allow for such geometries describing regular black holes mimickers (such as wormholes) and provide therefore an interesting platform to explore their phenomenology and their stability \cite{Yoshida:2018kwy, Babichev:2020qpr, Baake:2021jzv}. Then, a key question when studying such geometries is whether the consequences of the resolution of the singularity are confined to the deep interior or whether it can leak outside the photon ring. Answering to this question is crucial for observational purposes\footnote{See \cite{Compere:2019ssx} for a discussion on this point.}. Obtaining exact solutions of theories of gravity beyond GR describing these regular compact objects provides one concrete way to explore this aspect.   

Finally and despite the above discussion, let us stress that classical modified gravity is only a useful tool to study phenomenological consequences of resolving singularities. 
Indeed, provided a given theory allows for an effective energy-momentum tensor to violate the assumptions of the singularity theorem, one can easily exhibit concrete singularity free solutions. Yet these results do not provide any information on how one should describe the gravitational field in this high energy regime at the would-be singularity. Indeed, such description is expected to involve a very different notion of (fuzzy) geometry drastically different from the Riemanian one and where fluctuations of the pre-geometrical degrees of freedom play a key role.  A rigorous treatment of singularities can only be achieved by developing a fully consistent non-perturbative theory of quantum gravity, a task which has remained a tremendous challenge so far.

To be more concrete, let us add few words on this aspect. When investigating the quantum description of the gravitational field, one faces two key challenges: i) identifying the fundamental nature of the microscopic degrees of freedom of the geometry, and ii) understanding the emergence of classical geometries from such microscopic description. From this perspective, classical geometries should be understood as a type of many-body systems built from the collective behavior of these pre-geometric building blocks. That classical gravitational systems can be interpreted as many body systems built from this yet unknown microscopic description of the spacetime fabric is rooted in the fact that the gravitational dynamics admits a thermodynamical interpretation, allowing one to equip black hole \cite{Bardeen:1973gs} but also cosmological horizons \cite{Gibbons:1977mu} (and actually any causal diamond \cite{DeLorenzo:2017tgx, Jacobson:2018ahi}) with an entropy and a temperature.  While there might be different ways to encode the microscopic degrees of freedom depending on the chosen model or theory of quantum gravity, one could expect that their dynamics, and the emergence of classical spacetime geometries in the continuum hydrodynamical approximation, to be governed by universal symmetries\footnote{See \cite{BenAchour:2019ufa, BenAchour:2020njq, Geiller:2020xze, Achour:2021lqq, Achour:2021dtj, Geiller:2021jmg, Geiller:2022baq, BenAchour:2022fif, BenAchour:2023dgj} for efforts to identify such symmetries in mini-superspaces and \cite{BenAchour:2024gir} for their applications in understanding the notion of an emergent cosmology}. From this discussion, modified theories of gravity can at best correspond to an effective description of this mean-field approximation of quantum gravity. But they remain blind to the nature of the fundamental building blocks of geometry and their behavior at the would-be singularities, leaving us with a limited access to the possible phenomenological consequences of resolving singularity. Therefore, arguing that the singularity problem stands as one motivation to search for classical modified gravity as it is often claimed should be taken with caution since the teaching potential of classical theories of modified gravity on this front is limited.

\section{Different paths to modifying general relativity }

Having reviewed the different motivations for exploring alternatives theories of gravity, the natural question is how to start this journey in a systematic way ? Can we find a suitable guide to explore the space of classical theories beyond GR ? 

Quite remarkably, two powerful unicity theorems allow one to start this journey with a rough map of the new territories: the well-known Lovelock theorem \cite{Lovelock:1971yv} and the less popular Hojman-Kuchar-Teitelboim (HKT) theorem \cite{Hojman:1976vp}. As the two theorems do not rely on the same kind of assumptions, we find it useful to state them for completeness. Let us start with the Lovelock theorem.  \\

\textbf{Lovelock theorem:} \textit{Let us assume that the theory of gravity can be described by an action principle. Provided one assumes that
\begin{itemize}
\item The theory is diffeomorphism-invariant
\item The spacetime geometry is described by a lorentzian $d$-dimensional manifold
\item The connection compatible with the metric is torsion and metricity free, i.e. it reduces to the Levi-Civita connection
\item The action functional depends solely on the metric manifold, i.e. $S[g_{\mu\nu}]$
\item The field equations are of second order in derivatives of the metric 
\end{itemize}
then the action functional satisfying these conditions reads
\begin{align}
S[g_{\mu\nu}] = \int \rd^D x \sqrt{|g|} \sum^{n}_{i=0} \beta_i \mathcal{R}^{(i)}  \qquad \text{with} \qquad n = \frac{d-1}{2}
\end{align}
where $\beta_i$ are coupling constants, and $\mathcal{R}^{(i)} $ are curvature invariants of order $i$ in the powers of the Riemann tensor which read
\begin{align}
\mathcal{R}^{(i)} = \frac{k!}{2^k} \delta^{\mu_1 \nu_1 ..... \mu_k \nu_k}_{[\alpha_1\beta_1 .... \alpha_k \beta_k]} \Pi^i_{j=1} R^{\alpha_j \beta_j}{}_{\mu_j \nu_j}
\end{align}
This action functional is unique.}\\

In four dimensions, the action functional reduces to action of general relativity plus a cosmological constant supplemented with the well known Gauss-Bonnet topological term, i.e. 
\begin{align}
S[g_{\mu\nu}] = \int \rd^4 x \sqrt{|g|} \left[ \alpha_1 R + \alpha_0 + \alpha_2 \mathcal{G} \right]
\end{align}
This topological term reads
\begin{align}
 \mathcal{G} = R^2 - 4 R_{\mu\nu} R^{\mu\nu} + R_{\mu\nu\rho\sigma} R^{\mu\nu\rho\sigma} 
\end{align}
and can be safely ignored as it does not affect the field equations. Thus, the assumptions of the Lovelock theorem can be understood as a unicity theorem for four dimensional general relativity. In turn, it also appears as a guiding map to look for alternative theories of gravity. Assuming that we want a theory of gravity different from general relativity with an action principle and where the  spacetime geometry is still described by a lorentzian manifold, one has five different ways to proceed
\begin{itemize}
\item 1) Relax the number of spacetime dimensions to  $d>4$ (the case with $d<4$ is not interesting as in vacuum, gravity exhibit only topological degrees of freedom);
\item 2) Allow for torsion and/or non-metricity;
\item 3) Consider extra fields supplementing the metric;
\item 4) Allow for higher than second order field equations;
\item 5) Break diffeomorphism invariance.
\end{itemize}
Despite its key role in guiding the exploration of alternatives theories beyond GR, Lovelock's theorem relies on the crucial assumption that the theory admits a formulation in terms of an action functional. Yet, one could wonder if instead of starting from a lagrangian formulation, one could derive a unicity theorem for GR starting from an hamiltonian perspective.   

This theorem was derived few years after Lovelock's result by Hojman, Kuchar and Teitelboim (HKT) in the beautiful article \cite{Hojman:1976vp}. The starting point of the HKT theorem is to view the requirement of general covariance as a fundamental property of spacetime dynamics. To fix the idea, consider the $3+1$ foliation of spacetime in terms of spacelike hypersurfaces $\Sigma$ such that the metric can be decomposed as
\begin{align}
\rd s^2 = - N^2 \rd t^2 + \gamma_{ij} (\rd x^i + N^i \rd t) (\rd x^j + N^j \rd t)
\end{align}
where $N$ is the lapse, $N^i \partial_i$ is the shift vector and $\gamma_{ij}$ is the induced metric on the hypersurface $\Sigma$. The geometrodynamics of any spacelike hypersurface $\Sigma$ is severely constrained by the algebraic relations satisfied by the vectors generating the deformations from one slice to another in both the orthognal time-like direction and in the tangential space-like directions. These algebraic relations are fully encoded in the Dirac's hypersurface deformation algebra (DHDA) given by
\begin{align}
\{ H_i[N^i], H_j[M^j] \} & = H_k [N^i \partial_i M^k - M^i \partial_i N^k] \\ 
\{ H_i[N^i], H[N] \} & = - H [ N^i \partial_i N] \\ 
\{ H[N], H[N'] \} & = H_i [ \gamma^{ij} (N \partial_j N' - N' \partial_j N)]  
\end{align}
where $H_a$ is the generator of the tangential deformation while $H$ is the generator of the orthogonal timelike deformation. This algebra is highly non-trivial as it involves structure functions (instead of structure constants) which depend on the dynamical fields defining the geometry of the slice, i.e. the inverse induced metric $\gamma^{ij}$. It is often stated that general covariance, understood as the freedom to freely change coordinates, is an empty concept when it comes to studying alternatives theories of gravity. However, this statement is misleading. At the hamiltonian level, general covariance translates into the statement that the generators of the hypersurface deformation satisfy the DHDA. Then, given a set of assumptions, one can wonder how many different representations of the generators of the DHDA one can exhibit. Proceeding in that way, one gives the prior role to the space-time algebraic structure defining general covariance and classify the different theories, through the form of their first class constraints, which realize these algebraic rules. This is the key idea of the HKT theorem and the meaning behind the title of their seminal article: "geometry regained" (from the DHDA). Concretely, the HKT theorem states the following.  \\

\textbf{Hojman-Kuchar-Teitelboim theorem}: \textit{Consider a theory of gravity defined from its phase space and consider the geometrodynamics of space-like hypersurfaces $\Sigma$. Provided
\begin{itemize}
\item The covariant derivative on $\Sigma$ compatible with $\gamma_{ij}$ is the Levi-Civita connection
\item The only dynamical field involved are the induced metric $\gamma_{ij}$ and its canonical momenta
\item The generators of the deformations involve at most second order derivatives of $\gamma_{ij}$
\item The generators satisfy the Dirac algebra of hypersurface  deformation
\end{itemize} 
then the generators are the canonical scalar and momentum constraints of general relativity \cite{Hojman:1976vp}.} \\

This theorem based on the hamiltonian formulation has not received the same attention as the Lovelock theorem in the modified gravity community. To our knowledge, there are no discussions on its generalization to $d$-dimension, or when including torsion and metricity degrees of freedom. Nevertheless, the HKT theorem suggests another way to go beyond general relativity. Indeed, one can look for deformation of the DHDA by allowing for more complicated structure functions, for example by a more complicated functional dependency on the inverse induced metric $\gamma^{ij}$. This gives rise to the notion of emergent spacetime equipped with a deformed general covariance, deviating from GR only by the structure functions involved in the DHDA. Therefore, it reveals a more subtle conclusion than the Lovelock theorem regarding diffeomorphism invariance: modification of gravity can show up even if one does not break the gauge symmetry but instead deform it. At the moment, this strategy has been limited to symmetry-reduced systems, but it provides yet another interesting path to explore modifications of GR from the universal algebraic structure of spacetime. See \cite{Bojowald:2015zha, BenAchour:2017ivq, BenAchour:2017jof, BenAchour:2018khr, Arruga:2019kyd, Belfaqih:2024vfk} for applications of this idea to investigate singularity resolution and black hole interior geometries. Interestingly, this approach has recently shown interesting results, in particular in reproducing MOND-like corrections to GR \cite{Bojowald:2023xel, Bojowald:2023vvo, Bojowald:2023djr, Bojowald:2024naz}. 

Let us further point one limitation of the HKT theorem. Indeed, one assumption is that the dynamical field is the (induced) metric (and its conjugate momentum). A loophole lies in the fact that one could use different  variables to describe the gravitational field, such as the first order variables (tetrad and connection). In this case, new theories of gravity can be constructed. See for example \cite{Krasnov:2020lku, Krasnov:2020zfi, Freidel:2008ku}.

As we have seen, the Lovelock and the HKT theorems allow one to clearly identify different strategies to develop alternative theories of gravity. By far, the most studied path so far involves adding extra fields and allowing for higher order field equations while retaining most of the properties of general relativity, namely (undeformed) diffeomorphism invariance, a torsion and metricity free connection compatible with the metric and keeping the dimension $d=4$ (even if this last point is not strictly speaking a defining property of GR). This is the case of the theories we shall discuss in this manuscript which involve an additional scalar field coupled to the metric through higher order interaction terms. 

\section{My research in the field}

Having reviewed the key motivations of studying modified theories of gravity and the different paths to construct alternatives to GR, we now describe in more detail the research program carried on in this field during the last years. 
We begin with a hierarchy of general open questions motivating my research program and then discuss the strategy employed to tackle them and split these questions in more concrete projects described in the next chapters. We conclude by a summary of the results obtained which can serve as a guiding map for the rest of the manuscript, the following chapter being based on a selection of our published works.

\subsection{Problematic and open questions}

By far, the most studied alternatives theories of gravity are scalar-tensor theories, i.e. where one allows the gravitational field to be described by both a metric and a scalar companion. They provide the simplest framework to explore modifications to GR. 
The most general scalar-tensor theories constructed so far are dubbed degenerate higher order scalar-tensor (DHOST) theories. As we shall see in Chapter 1, their relevance lies in the fact that even if they allow the scalar companion to couple to the metric via higher order terms in the lagrangian, leading to higher order field equations, they still propagate only three degrees of freedom as expected. The construction of these healthy theories, first identified in \cite{Gleyzes:2014dya} and then systematically constructed and understood in \cite{Langlois:2015cwa} and later extended in \cite{BenAchour:2016fzp}, has led to a whole new landscape of scalar-tensor theories which encompasses all the previous models studied so far, in particular the well-known Horndeski theories introduced in \cite{Horndeski:1974wa} and rediscovered in \cite{Deffayet:2011gz}. 

\textit{Understanding both the key structures ruling the theory space and solution space associated to DHOST gravity is the central objective of the part of my work presented in what follows.} 

As we shall see, understanding the structure of the theory space and in particular the covariance of each degeneracy class under suitable  field redefinition allows one to efficiently study the solution space. This in turn, allows us to study the new phenomenology showing up in these higher-order scalar-tensor theories of gravity with the hope to confront them to current and future observations. This exercise shall allow us to restrict the viable set of  DHOST theories in a given regime. By far, most of the works on the phenomenology of DHOST gravity have focused on its cosmological implications focusing on i) constructing a systematic effective field theory approach to confront the theory to observations\footnote{This has already been achieved by the measured of the relative speed between gravitons and the photons emitted by the merger GW170817 which drastically constrains the allowed DHOST theories, showing the concrete potential of this approach \cite{Langlois:2017dyl}.} \cite{Langlois:2017mxy, LISACosmologyWorkingGroup:2019mwx}, ii) studying possible self-accelerating cosmological solutions in these theories in relation to the dark energy problem \cite{Crisostomi:2017pjs, Crisostomi:2018bsp, Frusciante:2018tvu, Arai:2019zul, Boumaza:2020klg}, iii) investigating energy transfer between the scalar and tensorial sectors and the possible instabilities \cite{Creminelli:2018xsv, Creminelli:2019nok, Creminelli:2019kjy}, iv) studying the effects on the rate of structure formation \cite{Hirano:2019nkz} and more recently v) in confronting the theories to dark-matter \cite{Laudato:2022vmq}. See \cite{Langlois:2018dxi, Lazanu:2024mzj} for reviews on these aspects. 

\textit{Instead, my work focuses on exploring the subsectors of the solution space of DHOST gravity describing i) stationary black holes and more recently ii) (non-linear) radiative spacetimes.}

In order to make progress on this exploration, the most natural strategy is to investigate how the known structures ruling the solution space of GR, i.e. the unicity theorems and the explicit and hidden symmetries, are preserved, deformed or broken in DHOST gravity. Let us start with the subsector of the solution space describing black holes. Among the open questions, the following ones pave the way to explore black holes in DHOST gravity:
\begin{itemize}
\item \textbf{Rigidity theorem:} For asymptotically flat stationary and electro-vacuum spacetime containing a black hole horizon, the rigidity theorem of GR \cite{Hawking:1971vc} asserts that the boundary of the black hole is a Killing horizon. This implies that a stationary rotating black hole must be axi-symmetric. This theorem has been extended to dimensions $d\geqslant 4$ in \cite{Hollands:2006rj} and to effective field theory with higher order curvature terms in \cite{Hollands:2022ajj}. Does a generalization of this theorem exist for DHOST theories ?   
\item \textbf{No-hair theorem:} By far, the most important theorem ruling our understanding of asymptotically flat stationary and axi-symmetric black hole in GR is the no-hair theorem \cite{Carter:1971zc, Robinson:1975bv}. Extensions of this theorem have been proven for the generalized Brans-Dicke theories \cite{Bekenstein:1971hc, Hawking:1972qk, Sotiriou:2011dz}, for K-essence theories \cite{Graham:2014mda}, for shift symmetric Horndeski gravity in \cite{Hui:2012qt} and extended to shift symmetric GLPV models in \cite{Babichev:2017guv}. More recently, a no-scalar monopole theorem has been presented in \cite{Capuano:2023yyh} for shift symmetric DHOST theories. They will be reviewed in Chapter 3. These theorems provide guiding map to construct hairy black holes and therefore are central results in our quest. For higher order scalar-tensor theories, all the known theorems hold under the assumptions of shift symmetry. Can we further relax the various assumptions entering in the known no-scalar hair theorems derived so far for DHOST gravity ? 
\item \textbf{Internal phase space symmetries :} Identifying new exact solutions in GR largely relies on solution-generating maps which stand as internal symmetries of suitable sub-sectors of the solution space.  For instance, the phase space of electro-vacuum solutions of GR admitting a non-null Killing vector exhibits powerful symmetries under the SU$(2,1)$ group known as the Kinnersley transformations \cite{Kinnersley:1977pg}. When one further imposes the existence of a second Killing vector (which commmutes with the first one), the resulting phase space inherits an infinite dimensional symmetry known as the Geroch group \cite{Geroch:1972yt}, making this subsector of GR an integrable system. How does this rich structure is modified in DHOST gravity ? Is it deformed or broken ?  If they exist, what are the conditions on the scalar companion and on the DHOST lagrangians to extend this symmetry of GR ? Notice that these symmetries are not spacetime symmetry but are instead realized as suitable operation on (complex) scalar potentials known as the Ernst potentials. Can we derive Ernst-like formulation for the stationary and axi-symmetric phase space of DHOST gravity ? This systematic approach has already been developed for other modified theories of gravity. See \cite{Astorino:2014mda, Pappas:2014gca, Suvorov:2016kwg} for examples.
\item \textbf{Petrov type D and hidden symmetries (of the solutions):} Obviously, one key goal behind finding new exact black holes solutions of DHOST gravity is to confront them to current and future observations. The properties of a given black hole geometry are usually revealed by first investigating the geodesic motion as well as the propagation of fields (of a given spin) on this geometry. A second aspect involve studying its stability under perturbations and its spectral signature. For the Kerr solution, the motion of test particles and fields is integrable due to the existence of a hidden symmetry found initially in \cite{Carter:1968ks} and characterized later in \cite{Walker:1970un, Carter:1977pq}. It manifests by the existence of rank-$2$ Killing-Yano tensor which is in turn a consequence of the fact that the Kerr geometry is algebraically special of Petrov type D \cite{Batista:2014fpa, Batista:2020cto}. Moreover, as mentioned in the introduction, almost all relevant stationary and axi-symmetric solutions of GR are of Petrov type D. To our knowledge, the few exact solutions describing rotating black holes in DHOST theories fail to be of Petrov type D and thus do not exhibit a Killing-Yano symmetry. We shall illustrate this point in Chapter 4. This raise the question whether DHOST gravity admit stable rotating black hole solutions of Petrov type D (or at least algebraically special) ? 
\item \textbf{Mimicking GR solutions:} On the same foot as the previous question, it appears also important to understand under which conditions DHOST theories admit the GR black hole solutions, i.e. the Kerr geometry. These solutions are dubbed stealth solutions as the scalar companion does not gravitate \cite{Babichev:2013cya}. By construction, these stealth solutions mimic exactly their GR counterpart, except at the level of the perturbations. A great deal of efforts has been devoted to derive the conditions of existence of these solutions and study their perturbations (which are generically pathological). Now, instead of mimicking exactly GR solutions, one would like to identify black hole solutions which deviate from the Kerr geometry only through tiny modifications. In particular, one could wonder whether DHOST gravity can accommodate black holes solutions which while being non-stealth are i) spectrally indistinguishable from Kerr (or Schwarzschild) and/or ii) asymptotically stealth, i.e. which reproduce the Kerr geometry as one approaches the horizon and in the asymptotically flat region ?  Such black holes solutions would be especially interesting as Kerr mimickers. In the following, we shall provide a proposal to construct such solutions.
\item \textbf{Symmetry behind a hair}: Finally, let us discuss the notion of hair. When considering a hairy rotating black hole in a given theory of modified gravity, it is common to understand the hair as a new additional label for the black hole configuration on top of the mass and spin. The later are well understood asymptotic charges related to the two commuting Killing vectors of the geometry. A rather natural question is whether one could understand any additional hair in terms of a hidden symmetry of the theory ? Let us point two examples suggesting that it is indeed the case. First, soft symmetries associated to either asymptotically flat spacetimes or realized at the black hole horizon provide a concrete example of this already in GR. These existence of these soft symmetries translate into an infinite tower of soft charges for the associated spacetime, giving rise to the notion of soft hair. See \cite{Compere:2016gwf, Donnay:2018ckb} for a concrete construction of these soft charges for the Schwarzschild black hole and \cite{Bhattacharjee:2020vfb, Sarkar:2021djs} for investigations on how these soft charges can lead to observational signatures or affect the black hole entropy \cite{Hawking:2016msc}. A second example comes from the well-known exact stationary and axi-symmetric solutions of GR which carry multipole moments such as the Quevedo-Mashhoon solution \cite{Quevedo:1991zz}. The monopole and dipole are well understood in terms of the obvious isometries. However, what are the symmetries associated to the higher multipoles ? In this case, one cannot invoke asymptotic symmetries as no such expansion is used. Interestingly, it appears that the quadrupole moment can be encoded in the Newman-Penrose charges (discussed below). But the case of the higher multipoles for such exact solution is still an open question. Nevertheless, this last example shows that given an exact solution, some of the constants labeling the geometry can be related to obvious symmetries while others turn out to be associated to more subtle charges (not yet understood in term of symmetries). Thus, whether any hair can be understood in terms of an underlying symmetry stands as a key question as it would give more handle to characterize hairy black holes.
\end{itemize}
These open questions provide a (non-exhaustive) list of motivations behind the current research program. One has two different but complementary roads to address them. A first direct approach consists in trying to extend the theorems mentioned above using the DHOST field equations. It goes without saying that it is a fairly complicated task. A second road is to look for exact solutions describing rotating black holes and study their properties. While not systematic, it nevertheless provides interesting lessons towards a more complete picture. For instance, regarding the extension of the rigidity theorem, an exact rotating black hole solution exhibiting a non-killing horizon has already been obtained, suggesting that this theorem might hold only for a subset of theories and scalar profile yet to be determined. The case of solutions mimicking exactly GR black holes has also been extensively studied and the conditions for their existence have been clarified. As we shall see, they have served as key seed solutions for further exploration of the solution space using solution-generating techniques.

Let us now comment on additional key open questions to understand the structure of the solution space of DHOST gravity and in particular when investigating solutions with a weaker degree of symmetry. Indeed, while most of the focus of the literature has been directed towards stationary black holes and stars for obvious reasons, it appears also interesting to investigate fully dynamical geometries and in particular radiative solutions. This would allow one to characterize possibly new signatures related to the production and propagation of gravitational waves in such higher-order scalar tensor theories. Exploring this much wider region of the solution space requires again to understand how unicity theorems derived in GR can be imported or adapted to DHOST gravity. As we shall now discuss, this can also reveal very useful to further understand stationary spacetimes and in particular stationary hairy black holes.
\begin{itemize}
\item \textbf{Characterizing algebraically special geometries:} The solution space of vacuum GR can be divided in two sectors: algebraically special solutions and the ones which are not. This frontier within the GR solution space is ruled by the so called Goldberg-Sachs theorem which states that for vacuum solutions (i.e. $R_{\mu\nu} =0$) admitting a null frame which is geodesic and shearfree (i.e. $\kappa=\sigma=0$), the associated geometry is algebraic special (i.e $\Psi_0 = \Psi_1 =0$). This theorem is of utmost importance when looking for solutions of vacuum GR with or without special symmetries and it played a key role in the derivation of the Kerr solution. See \cite{Batista:2012bk, Ortaggio:2012hc, Batista:2013yka} for generalization and extension of this fundamental theorem. A natural question is what is the counterpart of the Goldberg-Sachs theorem for DHOST theories ? Finding this extension would provide a first key structure of the solution space of DHOST gravity. This theorem being derived in the so called Newman-Penrose formalism, one first needs to provide the complete set of DHOST field equations within the NP formalism, a task which is currently under progress. 
\item \textbf{Newman-Penrose charges:} A fascinating outcome of the Newman-Penrose formalism is the discovery that for asymptotically flat spacetimes satisfying the Peeling theorem, one can exhibit a set of ten exactly conserved charges even if the spacetime is fully non-linear and radiative \cite{Newman:1965ik}. In the linearized theory, one inherits an infinite tower of charges. They have been used to constrain radiative processes, such as the head-on collision of black holes \cite{Dain:2001bh}. Quite remarkably, these charges also play a subtle role for algebraically special stationary spacetimes, such as the Kerr solution. Indeed, it was shown that for such spacetimes, all Newman-Penrose charges vanish \cite{Wu:2006ex}.  Moreover, it has been shown that when these charges do not vanish, they encode physical information related to suitable combination of the monopole, dipole and quadrupole. What are the counterpart of the NP charges for DHOST gravity ? If they exist, do they also signal the failure of being algebraically special for stationary spacetimes ?
\item \textbf{Soft symmetries and gravitational radiation:} The structure of asymptotically flat radiative spacetimes and the gravitational radiation escaping from any bulk process through $\mathcal{I}^{+}$ can be understood in terms of the asymptotic Bondi-Metzner-Sachs (BMS) symmetries derived first in \cite{Bondi:1960jsa, Bondi:1962px, Sachs:1962zza, Sachs:1962wk}. These soft symmetries are associated to flux-balance laws which encode how the different charges (mass, angular momentum, etc ..) measured by an asymptotic observer change as radiation escapes through $\mathcal{I}^{+}$. How is this algebraic structure characterizing gravitational radiation modified in DHOST gravity ? In particular, the possibility for matter to couple to a conformally or disformally related metric can change the falloff of the scalar field profile near $\mathcal{I}^{+}$, while the new higher-order interaction terms beyond GR will modify the expressions of both the charges and fluxes appearing in the flux-balance laws. These modifications were investigated for the Brans-Dicke theory in \cite{Tahura:2020vsa, Hou:2020wbo,Hou:2020tnd,  Hou:2020xme, Seraj:2021qja}. To our knowledge, no higher-order modified gravity theories were investigated along this line so far. A less ambitious target consists in constructing exact radiative solution of modified gravity theories and investigate the new phenomenology showing up in this nonlinear regime \cite{BenAchour:2024tqt, BenAchour:2024zzk}. In particular, a key question is how does the scalar-tensor mixing induced by the higher order terms modify the known phenomenology of GWs ?
\end{itemize}

\subsection{Results and structure of the manuscript}

The above open questions can be addressed in two complementary ways. A first basic approach to progress is based on finding exact solutions of the DHOST field equations. It provides a platform to investigate on a given example the new DHOST phenomenology and guide a more systematic exploration in the future. It goes without saying that obtaining such exact solutions by a brute force resolution of the DHOST field equations is most of the time impossible. Such solutions are usually obtained by exploiting the symmetries and solution-generation techniques.

One central aspect of my work relies on the use of disformal field redefinitions given by
\begin{align}
\left\{g_{\mu\nu}, \varphi \right\} \rightarrow \left\{\tilde{g}_{\mu\nu} = A(\varphi, X) g_{\mu\nu} + B(\varphi, X) \nabla_{\mu} \varphi \nabla_{\nu} \varphi, \varphi \right\}
\end{align}
where $g_{\mu\nu}$ is the seed metric, $\tilde{g}_{\mu\nu}$ the disformed metric, $\varphi$ the scalar companion and $X = \nabla_{\mu} \varphi \nabla^{\mu} \varphi$ its kinetic energy.
The central point is that the DHOST theories are characterized by their degeneracy conditions which split the theory space into three classes. As we shall see, each class is stable under disformal field redefinitions. Therefore, disformal transformations (DTs) provide a swiss-knife to explore the structure of DHOST gravity. In particular, it can be used as a solution-generating technique to construct new exact solutions of the complicated DHOST field equations. While this trick might seem at first as a trivial game, it reveals several subtleties. First, besides circumventing the resolution of the fields equations, since the scalar field remains unaffected by the field redefinition, this trick enables one to obtain a geometry which describes the gravitational field generated by the very same scalar profile within a higher-order scalar-tensor theory. The resulting solution thus offers a simple testbed to analyze the nonlinear phenomenology triggered by the higher-order terms responsible for the new scalar-tensor mixing. In the new exact solution, the effects of these scalar-tensor mixing are repackaged and controlled by the disformal function $B(\varphi, X)$ (often taken to be constant in what follows).

Since they are pure field redefinition which moreover are assumed to be invertible, it might seem that one is simply rewriting the very same geometry in a rather complicated way without introducing new physics. Such misconception is rooted in the fact that when studying the properties of the seed geometry and its disformed counterpart, virtually all observables used in the process require fixing explicitly or implicitly to coupling of matter fields. Concretely, when studying the properties of a seed geometry (resp. of its disformed version), we shall always assume that test matters fields are minimally coupled to the seed metric (resp. to the disformed metric). As we shall see, the fact that a DT does generate new physics provided one chooses appropriately the matter coupling become  even more transparent when working at the level of the tetrad, underlying the role of a DT onto any local rest frame constructed w.r.t a given metric.

Using this trick, we shall present in the following new exact solutions to the DHOST field equations describing spherically stationary symmetric black hole, stationary rotating black hole and a fully non-linear radiative geometry. In each of these cases, we shall discuss in detail how the higher order terms of the DHOST lagrangian affect the geometry and what is the resulting phenomenology. Their properties reveal a rather surprising phenomenology which can further be used as guide to address the questions mentioned above in a more systematic way. They should be considered as preliminary but valuable steps towards that goals. 

For instance, the radiative solution presented in the last chapter and studied in \cite{BenAchour:2024zzk}, the first of this kind in the literature, shows that contrary to GR, tensorial gravitational waves can be generated even in the presence of a purely time-dependent scalar monopole. It suggests that the higher order terms in the DHOST lagrangian should contribute in a highly non-trivial way to the flux-balance law describing the Bondi mass loss in these theories. Another interesting solution constructed so far is the disformed Kerr solution which is of Petrov type I, exhibits a non-circular metric, whose metric functions are non-analytic and which fails to have a Killing horizon despite being stationary and axi-symmetric \cite{BenAchour:2020fgy}. These properties originates from the non-stationarity of its scalar profile, a standard ingredient used so far to construct hairy black hole solutions, especially in spherical symmetry. This suggests that in order to look for a rotating hairy black hole solution which mimics closely Kerr, one should not allow the scalar field to break the isometries, thus reducing the number of assumptions of the no-scalar hair theorem one can break on this journey. At this stage, we stress that one key challenge in using the disformal solution-generating map is to have access to "interesting" seed solutions whose scalar profile satisfies suitable symmetry properties. Such seeds are not always available.

Besides constructing explicit exact solutions for DHOST theories, one can also characterize how disformal field redefinition modify fundamental properties of a given gravitational field. We have seen above the one such property is encoded in the algebraic type of the geometry. Depending on the scalar profile and its symmetries, a DT can modify the Petrov type. In view of the questions raised above, it appears useful to find close formulae to capture this transformation. This question was systematically addressed in \cite{BenAchour:2021pla} where the transformation of the Weyl scalars were computed. Yet, they reveal cumbersome and while they could be used to impose a Petrov type to the target solution, these formula have not been used in this way so far. However, this exercise has led to the derivation of general close formula for the null vectors and their spin coefficients. In turn, these formula have been used to provide necessary and sufficient conditions to generate tensorial gravitational waves at a fully non-linear level through a (constant) DT \cite{BenAchour:2024tqt}. To our knowledge, these two works are the first ones importing the Petrov classification and the Newman-Penrose formalism to study the phenomenology of DHOST gravity.

At the moment, we are following three different lines of research which are direct extensions of the results presented in this manuscript. The first one aims at writing down the DHOST field equations within the Newman-Penrose formalism, opening the road to generalize several key unicity theorems of GR mentioned above. The second objective is to present an extension of the Ernst formalism for stationary and axi-symmetric solutions of DHOST gravity, study how the fate of the GR symmetries and use them to derive new exact solutions for rotating black holes corresponding to a deformed Kerr solution. The third and last objective aims at deriving the asymptotic symmetries and the associated flux-balance laws ruling the gravitational radiation in asymptotically flat spacetimes within DHOST theories. The first and second objectives are under progress in collaboration with Hugo Roussille, Mohammad Ali Gorji, Hongguang Liu.

The rest of the manuscript is structured as follows. Except Chapter \ref{Chapter3}, all the chapters are based on one or two selected publications. We have re-organized some the sections for the flow of the discussion but almost all the material presented below can be found in the original articles. Chapter \ref{Chapter3} is devoted to review some definitions and theorems which are useful for the rest of the results discussed below. No original results of my research are presented in this chapter. I have also included new results which have not been published or which are currently under investigation. This is the case of the last section of Chapter \ref{Chapter4} where we discuss the notion of asymptotically stealth black hole. We have also included some survey of the known solutions for rotating black holes in Horndeski gravity and beyond in the first section of Chapter \ref{Chapter5} and a brief review of the Petrov classification in the first section of Chapter \ref{Chapter6}.
\begin{itemize}
\item \textbf{DHOST theories as disformal gravity:} Chapter \ref{Chapter1} is based on  \cite{BenAchour:2016fzp, BenAchour:2016cay}. We review the most general scalar-tensor theories constructed so far, the so-called degenerate higher order scalar-tensor (DHOST) theories. We briefly comment on their construction, the structure of the theory space and in particular their properties under disformal field redefinitions. This point is central to understand the results that will be discussed in the next chapters. Indeed, this is the existence of this disformal covariance of the degeneracy conditions that will allows us to efficiently explore the solution space of DHOST gravity and derive new exact solutions both for compact objects and radiative spacetimes without solving explicitly their complicated field equations. The interested reader can refer to \cite{Langlois:2018dxi} for a review on DHOST theories.
\item \textbf{No-hair theorems in general relativity and beyond:} Chapter \ref{Chapter3} reviews the key definitions and the central theorems which build our current understanding of black holes. First, we summarize the basic tools to analyze the geometry of a null hypersurface. This allows us to briefly review the notion of Killing horizon and its generalization known as (weakly) isolated horizons. The next section is devoted to a brief historical account on the proof of the no-hair theorems in GR where we state for completeness the modern version of the no-hair theorem in 4d GR for stationary and axi-symmetric black holes following \cite{Gourgoulhon:2024}. After reviewing the main properties of the Kerr black hole, we discuss how a massless minimally coupled scalar field modifies the Schwarzschild and Kerr geometry, presenting the known exact solutions of the Einstein-Scalar system derived in \cite{Janis:1968zz, Mirza:2023mnm}. This allows one to illustrate the no-hair theorem. The last part of this chapter reviews the no scalar hair theorems first in the simplest scalar-tensor theories and then for higher order scalar tensor theories. We discuss the Hui-Nicolis theorem \cite{Hui:2012qt} and the more recent \textit{no scalar monopole theorem} derived in \cite{Capuano:2023yyh}. This chapter serves as an intermediate step before presenting the construction of concrete hairy black holes solutions in DHOST gravity. We refer the interested reader to the beautiful and pedagogical lectures \cite{Gourgoulhon:2005ng, Gourgoulhon:2024} as well as to the related articles for more details.
\item \textbf{On spherically symmetric hairy black holes :} Chapter \ref{Chapter4} focuses on spherically symmetric hairy black holes. To a large extend, these new solutions have been obtained either by allowing the scalar field to be time-dependent or by introducing a coupling to the Gauss-Bonnet term. See \cite{Babichev:2016rlq, Lehebel:2018zga, Babichev:2022awg, Bakopoulos:2022gdv, Babichev:2023psy, Lecoeur:2024kwe} for reviews.
In exploring the spherically symmetric sector of the solution space of DHOST theories, one can distinguish between stealth configurations, i.e. black holes which mimic exactly the GR solutions supporting nevertheless a non-trivial scalar profile, and the non-stealth ones whose geometry are deformed w.r.t the GR solutions. The stealth sector has been extensively studied, both for the Schwarzschild and Kerr cases,  and the conditions on the DHOST lagrangian to exhibit such solutions have been identified by various techniques \cite{Motohashi:2016ftl, Motohashi:2018wdq, BenAchour:2018dap, Takahashi:2020hso}. In the following, we first review one such algorithm and its application to  non shift symmetric DHOST theories following \cite{BenAchour:2018dap}.  Beyond the stealth sector, the exploration of the solution space becomes challenging. A simple and systematic approach to construct new non-stealth solutions of DHOST gravity relies on the structure of the DHOST theory space whose degeneracy classes are covariant under disformal transformations. Thus, it appears natural to take advantage of this property and consider the disformal transformation as a solution generating map to explore the non-stealth sector of the solution space. This point of view was adopted systematically in \cite{BenAchour:2020wiw} and we shall review some of the results obtained in this work. 
Finally, we close this chapter by a new proposal to construct black holes which are not exactly but only asymptotically stealth. This proposal has not been published but it provides a powerful and simple trick to obtain asymptotically flat black hole geometries mimicking GR solutions both in the far region and in the near horizon, while still allowing a deformation of the geometry in region whose size is controlled. Therefore, we includes this proposal for completeness.
\item \textbf{On rotating hairy black holes :} In Chapter \ref{Chapter5}, we review the recent efforts devoted to find exact solutions of the DHOST field equations and describing a stationary rotating black hole. This endavour is a key target of the next years as it would provide a platform to confront these theories to the strong field regime currently observed by the Event Horizon Telescope and the future mission such as the Black Hole Explorer \cite{Berti:2015itd}. We first present a short survey of the exact solutions published so far for Horndeski and beyond Horndeski theories. Then, following the same structure than the previous chapter, we first discuss the effect of a disformal transformation onto the rotating naked singularity of the Einstein-Scalar system found by Bogush and Gal' tsov \cite{Bogush:2020lkp} and presented in \cite{BenAchour:2020fgy}. In the next part, we describe the construction of the stealth Kerr solution presented in \cite{Charmousis:2019vnf} and the new geometry obtained by disforming this rotating stealth solution \cite{BenAchour:2020fgy}. Let us emphasize that the resulting geometry, i.e. the disformed Kerr solution, provides one of the rare exact analytic solution at hand in DHOST theories. The rest of the chapter is devoted to describe its properties (singularities, ergoregions, horizons, geodesic motion) and even more importantly the non-standard ones (non-circularity, non analyticity, absence of Killing horizon, no separability of the geodesic equation). We further present the derivation allowing one to identify which subset of DHOST theories admit this stealth and disformed Kerr geometries as exact solutions.
\item \textbf{Petrov classification and the disformal map}: In Chapter \ref{Chapter6}, we present a first attempt to control how the Petrov type of a given gravitational field changes under a disformal transformation (DT). Provided one obtains close formula for this transformation, one can impose some restriction on the scalar profile depending on the Petrov type one wishes to achieve. However, as we shall see, even with these formula at hand, it is challenging to use them concretely. Nevertheless, this exercise allows one to understand another feature of DT onto a fundamental property of spacetime, namely its algebraic Petrov type. In particular, we shall apply this strategy to further characterize the black hole solutions discussed in the previous chapters \cite{BenAchour:2021pla}.
\item \textbf{Radiative spacetimes and disformal gravitational waves}: Finally, Chapter \ref{Chapter7} is devoted to the study of non-linear radiative solutions of DHOST gravity. We have seen that contrary to conformal transformations, disformal transformations can change the principal null directions of a spacetime geometry. Thus, depending on the frame a gravitational wave (GW) detector minimally couples to, the properties of GWs may change under a disformal transformation. In this chapter, based on \cite{BenAchour:2024tqt, BenAchour:2024zzk}, we provide \textit{necessary} and \textit{sufficient} conditions which determine whether GWs change under disformal transformations or not. Our argument is coordinate-independent and can be applied to any spacetime geometry at the fully non-linear level. As an example, we show that an exact radiative solution of massless Einstein-scalar gravity which admits only shear-free parallel transported frame is mapped to a disformed geometry which does not possess any shear-free parallel transported frame. This radiative geometry and its disformed counterpart provide a concrete example of the possibility to generate tensorial GWs from a disformal transformation at the fully non-linear level. 
This result shows that, at the nonlinear level, the scalar-tensor mixing descending from the higher-order terms in Horndeski dynamics can generate shear out of a pure scalar monopole. We further confirm this analysis by identifying the spin-0 and spin-2 polarizations in the disformed solution using the Penrose limit of our radiative solution. Finally, we compute the geodesic motion and the memory effects experienced by two null test particles with vanishing initial relative velocity after the passage of the pulse. 
This exact radiative solution offers a simple framework to witness nonlinear consequences of the scalar-tensor mixing in higher-order scalar-tensor theories.
\end{itemize}

\def\beq{\begin{equation}}
\def\eeq{\end{equation}}
\newcommand{\bea}{\begin{eqnarray}}
\newcommand{\eea}{\end{eqnarray}}
\def\bi{\begin{itemize}}
\def\ei{\end{itemize}}
\def\ba{\begin{array}}
\def\ea{\end{array}}
\def\bfig{\begin{figure}}
\def\efig{\end{figure}}
\def\d{\delta}
\def\L{{\tilde L}}
\def\tA{\hat A} 
\def\B{{\cal B}}
\def\a{\alpha}
\def\f{f}

\def\aa{\alpha_1}
\def\ab{\alpha_2}
\def\ac{\alpha_3}
\def\ad{\alpha_4}
\def\ae{\alpha_5}
\def\ba{\beta_1}
\def\bb{\beta_2}
\def\tba{\tilde\beta_1}
\def\tbb{\tilde\beta_2}
\def\ka{\kappa_1}
\def\kb{\kappa_2}
\def\kc{\kappa_3}
\def\kd{\kappa_4}
\def\ke{\kappa_5}
\def\tka{\tilde\kappa_1}
\def\tkb{\tilde\kappa_2}
\def\tkc{\tilde\kappa_3}

\def\ga{\gamma_2}
\def\gb{\gamma_1}
\def\gc{\gamma_3}

\def\An{A_*}
\def\dotAn{\dot A_*}
\def\A{{\cal A}}
\def\V{{\cal V}}
\def\M{{\cal M}}
\def\R{R}
\def\tR{\tilde R}
\def\h{h}

\def\C{A} 
\def\D{B} 

\def\n{n} 

\def\tf{\tilde f}
\def\tg{\tilde g}
\def\tnabla{\tilde \nabla}
\def\T{{\cal T}}
\def\tf{\tilde f}
\def\ta{\tilde\alpha}
\def\r{{\gamma}}
\def\tX{\tilde X}
\def\l{\lambda}

\def\fX{f_{,X}}
\def\fphi{f_{,\phi}}

\newcommand{\Gfour}{G_4{}}
\newcommand{\Ffour}{F_4{}}

\chapter{DHOST theories as disformal gravity}
\label{Chapter1}
\minitoc

In this chapter, we review the most general scalar-tensor theories constructed so far, the so-called degenerate higher order scalar-tensor (DHOST) theories. We briefly comment on their construction, the structure of the theory space and in particular their properties under disformal field redefinitions. This point is central to understand the results that will be discussed in the next chapters. Indeed, this is the existence of this disformal covariance of the degeneracy conditions that will allows us to efficiently explore the solution space of DHOST gravity and derive new exact solutions both for compact objects and radiative spacetimes without solving explicitly their complicated field equations. This chapter is based on the two articles published during our first postdoc \cite{BenAchour:2016fzp, BenAchour:2016cay}. The interested reader can refer to \cite{Langlois:2018dxi} for a review on DHOST theories.

\section{Scalar-tensor theories and field redefinition}

To fix the idea, let us briefly review the development of scalar-tensor theories of gravity. The earliest attempt to extend GR with an extra scalar-field is the well-known Brans-Dicke theory, followed by its generalization allowing for a non-minimal coupling and which admit the following lagrangian
\begin{align}
\label{ST-AC}
S_{\text{ST}}[\varphi, \tilde{g}] =  \int \rd^4x \sqrt{|\tilde{g}|} \left( \varphi \tilde{\mathcal{R}} - \frac{\omega(\varphi)}{\varphi} \tilde{g}^{\mu\nu} \partial_{\mu} \varphi \partial_{\nu} \varphi - U(\varphi)\right)  + S_{\text{m}}(\psi, \tilde{g} )
\end{align}
where $\varphi$ is the scalar field, $\psi$ represents matter fields and $\tilde{g}_{\mu\nu}$ is called the Jordan frame metric. The Brans-Dicke theory corresponds to $\omega(\varphi) =\omega_0$. The scalar field plays here the role of a dynamical inverse Newton constant, i.e. $\varphi = G^{-1}_{\text{eff}}$. It is well known that this scalar-tensor action is conformally related to the canonical Einstein-Scalar system through the following conformal field redefinition 
\begin{align}
g_{\mu\nu} =  \varphi \tilde{g}_{\mu\nu} \qquad \rd \phi = \sqrt{\frac{2\omega(\varphi) +3}{16\pi \bar{G}}} \frac{\rd \varphi}{\varphi} \qquad V(\phi) = \frac{U(\varphi)}{\varphi^2}
\end{align}
leading to the standard action
\begin{align}
S_{\text{ST}}[\phi, g, \psi] =  \int \rd^4x \sqrt{|g|} \left( \frac{\mathcal{R}}{16\pi \bar{G}} - \frac{1}{2} g^{\mu\nu} \partial_{\mu} \phi \partial_{\nu} \phi - V(\phi)\right)   + S_{\text{m}}\left(\psi, \phi^{-1}g \right)
\end{align}
It provides one of the simplest example of a field redefinition mapping two theories of gravity. Because of this simple relation, one can investigate easily the properties of the family of scalar-tensor theories described by the action (\ref{ST-AC}). Since their introduction, these theories have attracted a lot of attention and their solution space has been studied in detail. Provided the potential has a minimum, allowing for the scalar field to be in a local equilibrium, a no-hair theorem can be derived, showing that any stationary and axi-symetric black holes solution of these theories reduces to the Kerr geometry \cite{Bekenstein:1971hc, Hawking:1972qk, Bekenstein:1974sf, Sotiriou:2011dz}. The remaining solutions have also been classified, at least in the spherically symmetric sector, giving rise to the four different classes of solutions which describe either naked singularities or wormholes \cite{Faraoni:2016ozb, Faraoni:2018mes}. The conformal field redefinition linking these theories to the canonical Einstein-Scalar system has been a key to achieve these results. On the observational front, these scalar-tensor theories have been gradually challenged by various tests \cite{Bertotti:2003rm, Mariani:2023rca} and the interests in further investigating their structure have fade away. 

With the advent of the dark energy problem and the refinement of the inflationary paradigm, more efforts were devoted to explore alternatives theories of gravity with scalar-tensor mixing terms involving higher order derivatives of the scalar. The development of the covariant Galileon theory \cite{Deffayet:2011gz} led to the rediscovery of the Horndeski theory \cite{Horndeski:1974wa} which has triggered an intense activity. The link between the two was clarified in \cite{Kobayashi:2011nu}. For a long time, the Horndeski gravity stood as a frontier for higher order scalar tensor theories, as it is the most general lagrangian involving second order derivative of the scalar field which nevertheless admits second order field equations. Restricting to its quadratic version, the Horndeski action reads
\begin{align}
S_{\text{Horn}} & = \int\rd^4 x \sqrt{|g|} \left[ G_2(\varphi, X) - G_3(\varphi, X) \Box \varphi \right. \nonumber  \\
& \left. \qquad \qquad \qquad + \;\;G_{4}(\varphi, X) \mathcal{R} + G_{4X} (\varphi, X) \left[ (\Box \varphi)^2 - \varphi_{\mu\nu} \varphi^{\mu\nu}\right]\right]
\end{align}
where the functions $(G_2, G_3, G_4)$ are free functions of the scalar field $\varphi$ and its kinetic term $X = g^{\mu\nu}  \partial_{\mu} \varphi \partial_{\nu} \varphi$.
This theory has been by far the most studied extension of GR both on the cosmological front and for compact objects. Just as the scalar-tensor theories discussed above, a subsector of the Horndeski theory can be related to the Einstein-Scalar system by suitable field redefinition dubbed disformal transformations.

 Much later, the development of new healthy theories dubbed \textit{beyond Horndeski theories} identified in \cite{Gleyzes:2014dya, Gleyzes:2014qga} suggested that more general criteria for building healthy theories beyond GR had yet to be identified. A clear understanding of the healthy character of these new theories was only achieved with a careful hamiltonian analysis, without invoking any gauge choice, which ultimately led to notion of degeneracy conditions as a key tool to construct healthy theories with higher order scalar-tensor mixing terms \cite{Langlois:2015cwa}. The quadratic degenerate higher order scalar-tensor (DHOST) theories was the first example of theories constructed through this systematic algorithm \cite{Langlois:2015cwa}, followed quickly by their cubic generalization \cite{BenAchour:2016fzp}. A key point is that just as their elder sisters, the DHOST theories can also be studied using suitable field redefinition extending the conformal mapping reviewed above and dubbed \textit{disformal transformation} \cite{BenAchour:2016cay}.


 \section{Classification of degenerate theories}
 \label{section_degenerate}

Here, we present the construction of quadratic DHOST theories borrowing from \cite{BenAchour:2016cay}. The case of cubic DHOST follows the very same algorithm and we refer the interested reader  to \cite{BenAchour:2016fzp} for more details.
Thus, let us consider scalar-tensor theories  whose dynamics is governed by an action of the general form
\beq
\label{action}
S=S_{g}+S_\phi,
\eeq
where the first contribution involves the Ricci scalar $\R$ of the metric $g_{\mu\nu}$,
\beq
\label{S_g}
S_g\equiv \int d^4x\,  \sqrt{-g}\,  \f(\phi, X) \, \R \,,
\eeq
and the second contribution  depends quadratically on the second derivatives of the scalar field $\phi$
\beq
\label{S_phi0}
S_\phi \equiv \int  d^4x\,\sqrt{- g}\,   C^{\mu\nu,\rho\sigma}\,  \nabla_\mu\! \nabla_\nu\phi \  \nabla_\rho \!\nabla_\sigma\phi\,,
\eeq
$C^{\mu\nu,\rho\sigma}$ being an arbitrary tensor that depends only on $\phi$ and $\nabla_\mu\phi$. Note that $S_{g}$ reduces to the familiar Einstein-Hilbert action when the function $\f$ is  constant.

We stress that our analysis is also valid if we add to the above action extra contributions that depend at most linearly on $\phi_{\mu\nu}$, i.e. of the form
\beq
\label{S_other}
S_{\rm other}=\int  d^4x\,\sqrt{-g}\,  \left\{P(\phi, X) + Q_1(\phi, X) g^{\mu\nu}\phi_{\mu\nu}
+Q_2(\phi, X) 
\, \phi^\mu \phi_{\mu\nu}\phi^\nu \right
\}\,,
\eeq
where we have used the compact notation $\phi_\mu\equiv \nabla_\mu\phi$ and $\phi_{\mu\nu}\equiv \nabla_\mu\! \nabla_\nu\phi$.  These additional contributions do not  modify the degeneracy conditions derived in \cite{Langlois:2015cwa}, which will be summarized in the next section. For simplicity, we will not include these terms explicitly in our study but one should keep in mind that they can be present. Without loss of generality, we require the tensor $C^{\mu\nu,\rho\sigma}$ in (\ref{S_phi0}) to satisfy the index symmetries 
\beq
C^{\mu\nu,\rho\sigma} = C^{\nu\mu,\rho\sigma}= C^{\mu\nu,\sigma\rho}= C^{\rho\sigma,\mu\nu}\,,
\eeq 
which implies that  the most general form of this tensor  is
\begin{eqnarray}\label{family}
C^{\mu\nu,\rho\sigma} & = &  \frac{1}{2} \aa\, (g^{\mu\rho} g^{\nu\sigma} + g^{\mu\sigma} g^{\nu\rho})+
\ab \,g^{\mu\nu} g^{\rho\sigma} +\frac{1}{2} \ac\, (\phi^\mu\phi^\nu g^{\rho\sigma} +\phi^\rho\phi^\sigma g^{\mu\nu} ) 
\cr
& & \quad +   \frac{1}{4} \ad (\phi^\mu \phi^\rho g^{\nu\sigma} + \phi^\nu \phi^\rho g^{\mu\sigma} + \phi^\mu \phi^\sigma g^{\nu\rho} + \phi^\nu \phi^\sigma g^{\mu\rho} ) +  \ae\, \phi^\mu \phi^\nu \phi^\rho \phi^\sigma \label{four} \,,
\end{eqnarray}
where the $\a_I$ are five arbitrary functions of $\phi$ and $X$. Defining the five  elementary Lagrangians quadratic in second derivatives
\begin{eqnarray}
L_1^\phi&\equiv& \phi^{\mu\nu} \phi_{\mu\nu}\,, \quad 
L_2^\phi\equiv (\phi_\mu^{\ \mu})^2\,, \quad 
L_3^\phi\equiv  \phi_\mu^{\ \mu} \, \phi^\rho\phi_{\rho\sigma}\phi^\sigma\,,
\nonumber
\\
L_4^\phi &\equiv& \phi^\mu \phi_{\mu\nu}\phi^{\nu\rho}\phi_\rho
\,, \quad 
L_5^\phi\equiv (\phi^\rho\phi_{\rho\sigma}\phi^\sigma)^2
\,, 
\end{eqnarray}
 the action $S_\phi$  in (\ref{S_phi0}) now reads
\beq
\label{S_phi}
S_\phi = \int d^4x \sqrt{- g}\, \left(\aa L_1^\phi +\ab L_2^\phi+\ac L_3^\phi+\ad L_4^\phi
+\ae L_5^\phi
\right)\equiv  \int d^4x \sqrt{- g}\, \a_I L_I^\phi\,,
\eeq
where the summation over the index $I$ ($I=1,\dots, 5$) is implicit in the last expression. It is not difficult to see  that the general action (\ref{action}) also includes terms of the form
\beq
\label{S_Ricci}
S_{\rm Ricci}\equiv \int d^4x \sqrt{-g} \, h(\phi, X) \, \R_{\mu\nu}\phi^\mu \phi^\nu\,,
\eeq
where $h$ is an arbitrary function.
Indeed, using the definition of the Ricci tensor and the properties of the Riemann tensor, one can write
\begin{eqnarray}
\phi^\mu \R_{\mu\nu} \phi^\nu&=& -\phi^\mu g^{\rho\sigma}\R_{\rho\mu\nu\sigma} \phi^\nu=
-\phi^\mu g^{\rho\sigma}\left(\nabla_\rho\nabla_\mu-\nabla_\mu\nabla_\rho\right)\phi_\sigma
\cr
&=&
-\phi^\mu\nabla_\mu \nabla_\nu \phi^\nu+\phi^\mu\nabla_\nu \nabla_\mu \phi^\nu\,.
\end{eqnarray}
Substituting this into the action (\ref{S_Ricci}), one gets, after integration by parts,
\beq
S_{\rm Ricci}\equiv \int d^4x \sqrt{-g}\left\{ \, -h \left(L_1^\phi-L_2^\phi\right) +2 h_X\left(L_3^\phi-L_4^\phi\right)+h_\phi\left(X\phi^\mu_\mu -\phi^\mu \phi_{\mu\nu}\phi^\nu\right)
\right\}\,,
\eeq
where the contribution proportional to $h_\phi$ is of the form (\ref{S_other}).

Now let us  summarize  the main results obtained in  \cite{Langlois:2015cwa}, as well as some additional elements derived in \cite{Langlois:2015skt}, and present all  the quadratic DHOST theories, i.e. all the theories of the form (\ref{action}) which are degenerate.
 
 \subsection{Degeneracy conditions}
 In order to study the degeneracy of (\ref{action}), it is useful to introduce the auxiliary field $A_\mu\equiv \nabla_\mu\phi$. For an arbitrary foliation of spacetime by spacelike hypersurfaces $\Sigma(t)$, endowed with spatial metric $\h_{ij}$, the metric in ADM form reads 
 \beq
 ds^2=-N^2 dt^2 +\h_{ij} (dx^i+N^i dt)(dx^j+N^jdt)\,,
 \eeq
 where $N$ is the lapse and $N^i$ the shift vector. 
 The $(3+1)$ decomposition of the action (\ref{action}) leads to a kinetic term of the form~\cite{Langlois:2015cwa}
 \beq
 \label{S_kin}
S_{\rm kin}   =  \int  dt \, d^3x \, N \sqrt{\h} \left[\frac{1}{N^2}{\cal A} \, A_*^2 +
\frac{2}{N} {\cal B}^{ij} A_* K_{ij} + {\cal K}^{ijkl} K_{ij} K_{kl} \right]\,,
\eeq
where we have introduced  the quantity
\beq
A_*\equiv\frac1N (A_0-N^i A_i)\,, 
\eeq
and the extrinsic curvature tensor
\beq
K_{ij}\equiv \frac{1}{2N}\left(\dot\h_{ij}-D_i N_j -D_jN_i\right)\,.
\eeq
The coefficients that appear in (\ref{S_kin}) depend on the six arbitrary functions $\f$ and $\a_I$ of (\ref{action}). They are explicitly given by~\cite{Langlois:2015cwa, Langlois:2015skt}
\bea
\label{A}
\A & = & \aa+\ab-(\ac+\ad)\An^2+ \ae \An^4\,, \qquad
{\cal B}^{ij}  = \ba \h^{ij} + \bb \tA^i \tA^j \,,
\\
{\cal K}^{ij,kl} & = &  \ka \h^{i(k}\h^{l)j} +\kb\, \h^{ij} \h^{kl}+  \frac12 \kc\left(\tA^i\tA^j \h^{kl}+\tA^k \tA^l \h^{ij}\right)
    \cr
  &&
  +\frac 12 \kd  \left(\tA^i\tA^{(k}\h^{l) j}+\tA^j\tA^{(k} \h^{l)i}\right) 
  +\ke \tA^i \tA^j\tA^k\tA^l\,,
\eea
 with 
 \begin{eqnarray}
&& \ba=\frac{\An}{2}(2\ab - \ac \An^2 + 4\fX) \,, \quad \bb=\frac{\An}{2} (2\ae \An^2 - \ac - 2\ad)  \,,
\\
&&\ka=\aa \An^2 + f\,, \, \kb=  \ab \An^2-f\,, \,    \kc=- \ac\An^2+4\fX\,, \,  \kd=- 2\aa\,,\,  \ke=\ae\An^2-\ad\,.
\label{k}
 \end{eqnarray}
 The three-dimensional vector $\tA^i$ is defined by $\tA_i\equiv A_i$ and $\tA^i\equiv h^{ij}\tA_j$.
 
  By choosing an appropriate basis of  the six-dimensional vector space of symmetric $3\times 3$ matrices, where the $K_{ij}$ take their values,  the kinetic matrix associated with (\ref{S_kin}) can  be written as a $7\times 7$ block diagonal symmetric matrix of  the form~\cite{Langlois:2015skt}
 \beq
 \left(
\begin{array}{cc}
{\cal M} & \bf{0}\\
\bf{0} & {\cal D}
\end{array}
\right)\,,
\eeq
with the $3\times 3$ matrix 
\bea\label{M}
{\cal M}\equiv \left(
\begin{array}{ccc}
{\cal A} & \frac{1}{2}(\beta_1 + \tA^2 \beta_2) & \frac{1}{\sqrt{2}} \beta_1 \\
\frac{1}{2}(\beta_1 + \tA^2 \beta_2) & \ka + \kb + \tA^2(\kc + \kd) + (\tA^2)^2 \ke  &\sqrt{2}(\kb + \frac{1}{2} \tA^2 \kc) \\
\frac{1}{\sqrt{2}} \beta_1 & \sqrt{2}(\kb + \frac{1}{2} \tA^2 \kc) & \ka + 2 \kb
\end{array}
\right),
\eea
and the diagonal matrix 
\beq
{\cal D}={\rm Diag}\left[\ka, \ka, \ka+\frac12\tA^2\kd ,\ka+\frac12\tA^2\kd \right]\,.
\eeq
The coefficients in the first line (or first row)  of ${\cal M}$ describe the kinetic terms associated with  the scalar field related variable $\An$, including its mixing with the metric sector. As for the metric sector alone, it  is described by the right lower $2\times 2$ submatrix of ${\cal M}$, which we will call ${\cal M}_K$,  together with ${\cal D}$. As our goal is to eliminate the extra degree of freedom due to the higher derivatives of the scalar field, we are looking for a degeneracy of the kinetic matrix that arises from the scalar sector. As a consequence, we will be interested in theories such that ${\cal M}$ is degenerate, while ${\cal M}_K$ and ${\cal D}$ remain nondegenerate in order to preserve the usual tensor structure of gravity.

Requiring the determinant of the matrix ${\cal M}$ to vanish\footnote{Note that we have not used the same matrix in \cite{Langlois:2015cwa} but another, non symmetric, matrix constructed  by solving for null eigenvectors of the kinetic matrix. The two methods are obviously equivalent.}
 yields an expression of the form
\beq
\label{determinant}
D_0(X)+D_1(X) \An^2+D_2(X) \An^4=0\,,
\eeq
where we have substituted the expressions (\ref{A})-({\ref{k}) into (\ref{M}) and replaced all $\tA^2$ by $X+\An^2$. 
The functions $D_0$, $D_1$ and $D_2$ depend on the six arbitrary functions $\tf$ and $\a_I$ of the initial Lagrangian:
\beq
\label{D0}
D_0(X)\equiv -4 (\aa+\ab) \left[X \f (2\aa+X\ad+4\f_X)-2\f^2-8X^2\f_X^2\right]\,,
\eeq
\begin{eqnarray}
D_1(X)&\equiv& 4\left[X^2\aa (\aa+3\ab)-2\f^2-4X\f \ab\right]\ad +4 X^2\f(\aa+\ab)\ae 
\cr
&&
+8X\aa^3-4(\f+4X\f_X-6X\ab)\aa^2 -16(\f+5X \f_X)\aa \ab+4X(3\f-4X \f_X) \aa\ac 
\cr
&&
-X^2\f \ac^2 +32 \f_X(f+2X f_X) \ab-16\f \f_X \aa-8\f (\f-X\f_X)\ac+48\f \f_X^2 \,,
\end{eqnarray}
\begin{eqnarray}
D_2(X)&\equiv& 4\left[ 2\f^2+4X\f \ab-X^2\aa(\aa+3\ab)\right]\ae  + 4\aa^3+4(2\ab-X\ac-4\f_X)\aa^2+3X^2 \aa\ac^2
\cr
&&
-4X\f \ac^2+8 (\f+X\f_X)\aa\ac -32 \f_X \aa\ab+16\f_X^2\aa
+32\f_X^2\ab-16\f\f_X\ac\,.
\label{D2}
\end{eqnarray}
Since the determinant must vanish for any value of $\An$, we deduce that degenerate theories are characterized by the three conditions
\beq
D_0(X)=0, \qquad D_1(X)=0, \qquad D_2(X)=0\,.
\eeq
By solving these three conditions, one can determine and classify all  DHOST theories, as discussed in \cite{Langlois:2015cwa}.

\subsection{Degenerate theories}
\label{Section_classification}
The  condition $D_0(X)=0$ is the simplest of all three and allows to distinguish several  classes of theories. Indeed,  $D_0$ can vanish either if 
 $\aa+\ab=0$, which defines our first class of solutions, or  if the  term between brackets in (\ref{D0}) vanishes, which defines our second class, as well as our third class corresponding to the special case where $f=0$.

\subsubsection{Class I  ($\aa+\ab=0$)}
\label{subsectionA}  
This class is characterized by the property
\beq
\aa=-\ab\,.
\eeq
One can then  use the conditions  $D_1(X)=0$  and  $D_2(X)=0$ to express, respectively, $\ad$  and $\ae$ in terms of $\ab$ and $\ac$, provided $f+X\ab\neq 0$. This defines the subclass Ia, characterized by
\begin{eqnarray}
\label{a4_A}
\ad&=&\frac{1}{8(f+X\ab)^2}\left[16 X \ab^3+4 (3\f+16 X\f_X)\ab^2
+(16X^2 \f_X-12X\f) \ac\ab-X^2\f \ac^2
\right.
\cr
&&\qquad\qquad \qquad 
\left.
+16 \f_X(3\f+4X\f_X)\ab+8\f (X\f_X-\f)\ac+48\f \f_X^2\right]
\end{eqnarray}
and
\beq
\label{a5_A}
\ae=\frac{\left(4\f_X+2\ab+X\ac\right)\left(-2\ab^2+3X\ab\ac-4\f_X \ab+4\f \ac\right)}{8(f+X\ab)^2}\,.
\eeq
Degenerate theories in class Ia thus depend on three arbitrary functions $\ab$, $\ac$ and $f$. 

In the special case $f+X\ab= 0$,  we find another subclass of solutions characterized by
\beq
\aa=-\ab=\frac{\f}{X}\,, \qquad \ac=\frac{2}{X^2}\left(f-2X f_X\right),\qquad ({\rm Class\  Ib}),
\eeq
where $\f$, $\ad$ and $\ae$ are arbitrary functions. In the following, we will not explore this class much further because  the metric  sector is degenerate. Indeed,   the last two eigenvalues of ${\cal D}$, which are equal  to $f-\aa X$,  vanish in this case.

\subsubsection{Class II}
The condition $D_0(X)=0$  can also be  satisfied if 
\beq
\label{cond33}
X \f (2\aa+X\ad+4\f_X)-2\f^2-8X^2\f_X^2=0 \,.
\eeq
We can then proceed as previously by solving $D_1(X)=0$ and $D_2(X)=0$ to express $\ad$ and  $\ae$ in terms of the three other functions. Substituting the obtained expression for $\ad$ into the condition (\ref{cond33}),  one finally gets
\beq
 (X\aa -f)\left[(4f^2 + X f (8\ab  + 2 \aa  + X\ac-4\f_X)-4X^2 f_X(\aa+3\ab)\right]=0\,.
\eeq
Assuming that $f- X\aa \neq 0$, this   leads to the expressions
\begin{eqnarray}
   \ac &=&\frac{1}{X^2 f}\left[-4 f (f - X f_X )
   -2 X (f - 2 X f_X )\aa+4 X (-2 f + 3 X f_X )
   \right] \,,
   \\
   \ad & = & \frac{2}{X^2 f}\left[f^2 - 2 f X f_X  + 4 X^2 f_X^2- X f  \aa\right]\,,
   \\
   \ae & = & \frac{2}{f^2 X^3}\left[4 f (f^2 - 3 f X f_X  + 2 X^2 f_X^2)+(3 X f^2  - 8 X^2 f f_X  + 6 X^3  f_X^2)\aa
   \right. 
   \nonumber
   \\
&& \left. \qquad \qquad   +2 X (2 f - 3X  f_X )^2\ab\right]\,,
   \label{classII}
\end{eqnarray}
while  $\tf$, $\aa$ and $\ab$ are arbitrary. This describes our class IIa, characterized by three arbitrary functions.

The case $f=X\aa$ defines another class, similar to class Ib, which we will call class IIb, described by
\begin{eqnarray}
 \aa&=&\frac{f}{X}\,,\qquad \ad=4 f_X\left(2\frac{f_X}{f}-\frac{1}{X}\right)\,,
 \\
    \ae&=&\frac{1}{4 X^{3} f ( f + X \ab ) }\left[
    8 X ( 4 X f_{X} f - f^{2} - 4 X^{2} f^{2}_{X} ) \ab + X f ( 8 X^{2} f_{X} + X^{3} \ac - 4 f) \ac 
     \right. 
   \nonumber
   \\
   && \left. \qquad \qquad \qquad \qquad 
    + 4 ( X f_{X} f^{2} - 2 X^{3} f^{3}_{X} + 2 X^{2} f^{2}_{X} f - f^{3} )
    \right]\,,
   \label{classIIa}
\end{eqnarray}
where $\ab$ and $\ac$ are arbitrary functions. Like class Ib,  the metric sector is degenerate for these theories and we will not consider them further  in the following.

\subsubsection{Class III ($f=0$)}
Finally, we devote a special class to the case $f=0$, which also leads automatically to $D_0=0$. Using $D_1=0$ and $D_2=0$ to determine $\ad$ and $\ae$, one gets 
\beq
\ad = -\frac{2}{X}\aa\,, \qquad  \ae=\frac{4 \aa^2+8 \aa
   \ab-4 \aa \ac X+3 \ac^2 X^2}{4 X^2 (\aa+3
   \ab)}  \qquad {\rm (class\  IIIa)}\,,
 \eeq
 while $\aa$, $\ab$ and $\ac$ are arbitrary, provided $\aa+3
   \ab\neq 0$. This  defines our class IIIa. 
   Note that the intersection of IIIa with the class Ia is described by
  \beq
  \ab=-\aa\,, \quad \ad = -\frac{2}{X}\aa\,, \qquad \ae=\frac{(2\aa-X\ac)(2\aa+3 X\ac)}{8 X^2 \aa}\,, \qquad ( {\rm   IIIa} \cap {\rm  Ia} )\,,
  \eeq
  which depends on two arbitrary functions, $\aa$ and $\ac$, and  includes the Lagrangian $L_4^{\rm bh}$ (for which $\aa/X=\ac/2=F_4$). 
   
The case $\aa+3\ab= 0$ yields another subclass,  
\beq
f=0\,,\qquad \aa=\frac32 X\ac,\qquad \ab=-\frac{X}{2}\ac  \qquad {\rm (class\  IIIb)}\,,
\eeq
which in general leads to a degenerate metric sector. 
Another special case  corresponds to the class 
\beq
f=0\,,\qquad \aa=0\,,   \qquad {\rm (class\  IIIc)}
\eeq
which depends on four arbitrary functions. Since $f-\aa X=0$, this class  is also degenerate in the metric sector.

We can now turn to the key properties of the degeneracy class, namely their stability under a disformal field redefinition.

\section{Disformal transformations}
\label{section_disformal}
We now study the effect of conformal-disformal transformations, or generalized disformal transformations, introduced in \cite{Bekenstein:1992pj}, in which the ``disformed''  metric $\tg_{\mu\nu}$ is expressed in terms of $g_{\mu\nu}$ and $\phi$ as
\beq
\label{disformal}
\tg_{\mu\nu}=\C(X, \phi) g_{\mu\nu}+\D(X, \phi) \, \phi_\mu\, \phi_\nu\,.
\eeq
Via this transformation,  any action $\tilde S$ given as a functional  of  $\tg_{\mu\nu}$ and $\phi$ induces a new   action $S$  for  $g_{\mu\nu}$ and $\phi$, when one substitutes the above expression for  $\tg_{\mu\nu}$ in $\tilde S$: 
 \beq
S[\phi, g_{\mu\nu}]\equiv\tilde S\left[\phi, \tg_{\mu\nu}=\C \,g_{\mu\nu}+\D \, \phi_\mu\phi_\nu\right]\,.
\eeq
We will  say that the actions $S$ and $\tilde{S}$ are related by the disformal transformation (\ref{disformal}). Starting from an action $\tilde S$ of the form (\ref{action}), 
\beq
\tilde S=\tilde S_g+\tilde S_\phi=\int d^4x\sqrt{-\tg}\left[\, \tf\, \tilde R+\ta_I \tilde L_I^\phi\right]\,,
\eeq
we show below that the action $S$, related to $\tilde S$ via a disformal transformation, is also of the form (\ref{action}), up to terms of the form (\ref{S_other}), and we compute explicitly the relations between the functions that appear in the two Lagrangians.
Interestingly, if the disformal transformation is invertible, in the sense that the metric $g_{\mu\nu}$ can be expressed in terms of $\tg_{\mu\nu}$, then the number of degrees of freedom associated with $S$ and $\tilde S$ should be the same. See \cite{Domenech:2015tca,Takahashi:2017zgr} for a detailed discussion on this point. One thus expects that the disformal transformations of all the degenerate theories described in the previous section are also degenerate. 
We will also discuss the special case where the transformation is non invertible in the last subsection.

\subsection{General formulas}
In order to write explicitly the above action in terms of $g_{\mu\nu}$ and $\phi$, we will need the expression of the inverse metric 
\beq
\tg^{\mu\nu}=\C^{-1}\left(g^{\mu\nu}-\frac{\D}{\C+\D X}\nabla^\mu\phi\nabla^\nu\phi\right)\,.
\eeq
Contracting this relation with $\phi_\mu\, \phi_\nu$ gives  $\tilde X$ as a function of $X$:
\beq
\label{X_tilde}
\tilde X=\frac{X}{\C+\D X}\,.
\eeq
It is also useful to introduce the ratio
\beq
{\cal J}_g\equiv \frac{\sqrt{-\tg}}{\sqrt{-g}}=\C^{3/2}\, \sqrt{\C+\D X} \,.
\eeq
The difference between the two covariant derivatives $\tilde\nabla$ and $\nabla$, associated respectively to the two metrics $\tg_{\mu\nu}$ and $g_{\mu\nu}$, is fully characterized by the  difference of their respective Christoffel symbols, 
\beq
C_{\mu\nu}^\lambda\equiv\tilde\Gamma_{\mu\nu}^\lambda-\Gamma_{\mu\nu}^\lambda\,, 
\eeq
which defines a tensor. In particular, the relation between the respective second order covariant derivatives of $\phi$ reads
\beq
\tnabla_\mu\tnabla_\nu\phi=\nabla_\mu\nabla_\nu\phi-C_{\mu\nu}^\lambda\phi_\lambda\,.
\eeq
The explicit expression for $C_{\mu\nu}^\lambda$  is given by
\begin{eqnarray}
\label{C}
C_{\mu\nu}^\lambda&=&\frac{\C_X}{\C}\left[2\d^\lambda_{(\mu}\phi_{\nu)\sigma}\phi^\sigma-\phi^{\lambda\sigma}\phi_\sigma g_{\mu\nu}+\frac{\D}{\C+\D X}\left(-2\phi^\lambda\phi_{(\mu}\phi_{\nu)\sigma}\phi^\sigma+\phi^\lambda \phi^\rho\phi_{\rho\sigma}\phi^\sigma  g_{\mu\nu}\right)\right]
\nonumber
\\
&&+\D_X\left[-\frac{1}{\C}\phi_\mu\phi_\nu\phi^{\lambda\sigma}\phi_\sigma
+\frac{1}{\C+\D X}\left(2\phi^\lambda\phi_{(\mu}\phi_{\nu)\sigma}\phi^\sigma+\frac{\D}{\C}\phi^\rho\phi_{\rho\sigma}\phi^\sigma \phi^\lambda\phi_\mu\phi_\nu\right)\right]
+\frac{\D}{\C+\D X}\phi^\lambda\phi_{\mu\nu}
\nonumber
\\
&&
+\frac{A_\phi}{2 A }\left[ \d^\lambda_\mu \phi_\nu +\d^\lambda_\nu\phi_\mu- \frac{1}{A+B X}(A \phi^\lambda g_{\mu\nu}+ 2B \phi^\lambda \phi_\mu\phi_\nu)\right]+\frac{B_\phi}{2(A+B X)}\phi^\lambda \phi_\mu \phi_\nu\,.
\end{eqnarray}
The last line does not depend on second derivatives of $\phi$. As we will see, this implies that the terms in $A_\phi$ and $B_\phi$ appear only in the transformed action as terms of the form (\ref{S_other}), which we will not compute explicitly.
Notice also that when $(A,B)$ are constants, the above formula reduces to
\beq
C_{\mu\nu}^\lambda = \frac{\D}{\C+\D X}\phi^\lambda\phi_{\mu\nu}
\eeq
This simplification will be useful when studying black holes solutions.

Let us first concentrate, in $\tilde S$,  on the term depending on the Ricci scalar of $\tg_{\mu\nu}$. Following  the derivation presented in \cite{Zumalacarregui:2013pma},
the Ricci scalar $\tilde R$   can be written in terms of the tensor $C_{\mu\nu}^\lambda$ and of the metric $g_{\mu\nu}$, according to the expression
\beq
\tilde R\equiv \tg^{\mu\nu}\tilde R_{\mu\nu}
=\C^{-1}\left(g^{\mu\nu}-\frac{\D}{\C+\D X}\phi^\mu\phi^\nu\right)
\left(R_{\mu\nu}+C_{\mu\rho}^\sigma C_{\nu\sigma}^\rho-C_{\mu\nu}^\rho C_{\rho\sigma}^\sigma\right)+\tnabla_\rho \xi^\rho\,,
\eeq
with 
\beq
\xi^\rho\equiv \tg^{\mu\nu}C_{\mu\nu}^\rho -\tg^{\rho\mu} \, C_{\mu\nu}^\nu\,.
\eeq
All the terms quadratic in $C_{\mu\nu}^\lambda$ can be rewritten in terms of the elementary Lagrangians 
$L^\phi_I$. One finds
\beq
\label{CC}
\C^{-1}\left(g^{\mu\nu}-\frac{\D}{\C+\D X}\phi^\mu\phi^\nu\right)\left(C_{\mu\rho}^\sigma C_{\nu\sigma}^\rho-C_{\mu\nu}^\rho C_{\rho\sigma}^\sigma\right)=\sum_I \r_I L_I^\phi +(\dots)\,,
\eeq
with 
\bea
\r_1 &=& \r_2=0\,, \qquad \r_3=-\frac{B \left(B X A_X+A \left(2
   A_X+X B_X+B\right)\right)}{A^2
   (A+B X)^2}\,,
   \\
\r_4&=& \frac{(6 A^2 + 8 A B X + 2 B^2 X^2) A_X^2+4 A X (A + B X) A_X B_X+A^2 B (B + X B_X)
}{A^3 (A+B
   X)^2}  \,,
 \\
\r_5&=& -\frac{2 A_X \left(B A_X+2 A
   B_X\right)}{A^3 (A+B X)}\,.
\eea
The dots in (\ref{CC}) indicate terms that are  at most linear in $\phi_{\mu\nu}$, i.e. of the form (\ref{S_other}), which we will not write down explicitly. 

The total derivative  $\tnabla_\rho \xi^\rho$ can be ignored if the function $\tilde f$ multiplying the scalar curvature is a constant. Otherwise, one also needs to reexpress this term as a function of $g_{\mu\nu}$ and $\phi$. This can be done after an integration by parts so that one gets  
\beq
\int d^4 x\sqrt{-\tg} \, \tf \tnabla_\mu \xi^\mu=- \int d^4 x\sqrt{-\tg}\, \xi^\mu \nabla_\mu \tf
=- 2\int d^4 x\sqrt{-\tg}\,  \tf_{\tilde X} \, \tX_X \, \xi^\mu\phi_{\mu\nu}\phi^\nu +(\dots)\,,
\eeq
with 
\beq
\tX_X\equiv \frac{\partial\tilde X}{\partial X} =\frac{A}{(A+B X)^2}\,.
\eeq
Since $\xi^\mu$ contains second derivatives of $\phi$, the scalar quantity $\xi^\mu\phi_{\mu\nu}\phi^\nu$ can be decomposed as a combination of the elementary terms $L_I^\phi$. One finds 
\beq
\xi^\mu\phi_{\mu\nu}\phi^\nu=\l_I L_I^\phi +(\dots)\,,
\eeq
with 
\begin{eqnarray}
\l_1 &=& \l_2=0\,,\qquad 
\l_3= \frac{B}{A^2+A B X}\,, 
\\
\l_4&=&-\frac{4 B X A_X+A \left(6 A_X+2 X
   B_X+B\right)}{A^2 (A+B X)}\,, \qquad 
\l_5 = \frac{2 \left(2 B A_X+A B_X\right)}{A^2
   (A+B X)}\,,
\end{eqnarray}   
and the dots stand as usual for the terms at most linear in $\phi_{\mu\nu}$.   
   
Putting everything together, one finds that the scalar curvature term yields
\beq
\int d^4x\sqrt{-\tg}\, \tf\, \tilde R=\int d^4x\sqrt{-g} \, {\cal J}_g\left\{\frac{\tf}{\C}\left[\, R
-\frac{\D}{\C+\D X} R_{\mu\nu}\, \phi^\mu\phi^\nu \right]+\left(\r_I-2 \tX_{X}\tf_{\tilde X} \l_I\right) L_I^\phi\right\}+(\dots)\,,
\eeq
where the term in $R_{\mu\nu}\, \phi^\mu\phi^\nu$ is of the form (\ref{S_Ricci})
with the function
\beq
\label{function_h}
h=-{\cal J}_g\, \frac{\D}{\C(\C+\D X)} \, \tf \,,
\eeq
and the dots correspond to terms of the form (\ref{S_other}).

Let us now consider the terms quadratic in second derivatives  of the scalar field. Each of the  five terms in $\tilde S_\phi$ can be decomposed, after  substitution of (\ref{disformal}), into the five terms that appear in the final action $S_\phi$. 
Let us illustrate this with the first term $\tilde L_1^\phi\equiv\tilde\phi_{\mu\nu}\, \tilde\phi^{\mu\nu}$, which can be decomposed as follows:
\begin{eqnarray}
\tilde L_1^\phi&=&\tg^{\mu\rho} \, \tg^{\nu\sigma} \,\tnabla_\mu\!\tnabla_\nu\phi\,  \tnabla_\rho\!\tnabla_\sigma\phi   
\\
&=& \C^{-2}\left(g^{\mu\rho}-\frac{\D}{\C+\D X}\phi^\mu\phi^\rho\right)\left(g^{\nu\sigma}-\frac{\D}{\C+\D X}\phi^\nu\phi^\sigma\right)\left(\phi_{\mu\nu}-C_{\mu\nu}^\lambda\phi_\lambda\right)\left(\phi_{\rho\sigma}-C_{\rho\sigma}^\tau\phi_\tau\right)
\nonumber
\\
&=&
\T_{11}\, L_1^\phi+\T_{13}\, L_3^\phi+\T_{14}\, L_4^\phi+\T_{15}\, L_5^\phi +(\dots)\,,
\end{eqnarray}
where the coefficients are determined explicitly by substituting the expression (\ref{C}) for $C_{\mu\nu}^\lambda$. Note that the term $L_2^\phi$ does not appear in the decomposition.

Proceeding similarly with all the other terms, one finally gets five similar decompositions, which can be summarized by the  expression
\beq
\tilde L^{\phi}_I=\T_{IJ} L^\phi_J +(\dots)\,,
\eeq
where the summation with respect to the index $J$ is implicit.  The nonvanishing coefficients $\T_{IJ}$ are given by
\begin{eqnarray}
\label{T11}
\T_{11}&=& \frac{1}{(A+B X)^2}\,, \quad \T_{13}=\frac{2 A_X}{A (A+B X)^2}\,,\quad  
\T_{14}=\frac{2 \left(X \left(A_X+X
   B_X\right){}^2-A \left(2 \left(A_X+X
   B_X\right)+B\right)\right)}{A (A+B
   X)^3}, 
\nonumber
\\
\T_{15}&=&\frac{1}{A^2
   (A+B X)^4}\left[2 A^3 B_X+A^2 \left(-2 X A_X
   B_X+2 A_X^2+B^2-X^2 B_X^2+4 B X
   B_X\right)+3 B^2 X^2 A_X^2
   \right.
   \cr
   &&\left. \qquad \qquad \qquad \qquad
   -2 A B X
   \left(2 X A_X B_X+A_X \left(B-2
   A_X\right)+X^2 B_X^2\right)\right]
\nonumber
\\
\T_{22}&=&\frac{1}{(A+B X)^2}\,, \quad  
 \T_{23}=-\frac{2 \left(A \left(-2 A_X+X B_X+B\right)-3 B X A_X\right)}{A (A+B X)^3}\,, 
 \nonumber
 \\
   \T_{25}&=& \frac{\left(A \left(-2 A_X+X
   B_X+B\right)-3 B X
   A_X\right){}^2}{A^2 (A+B X)^4}\,,
 \nonumber
 \\
     \T_{33}&=& \frac{A-X \left(A_X+X B_X\right)}{(A+B
   X)^4} \,,\quad
\T_{35}= \-\frac{\left(A \left(-2 A_X+X
   B_X+B\right)-3 B X A_X\right)
   \left(A-X \left(A_X+X
   B_X\right)\right)}{A (A+B X)^5}\,,
 \nonumber
   \\
    \T_{44}&=& \frac{\left(A-X \left(A_X+X
   B_X\right)\right){}^2}{A (A+B X)^4}\,,\quad  
   \T_{45}=-\frac{B \left(A-X \left(A_X+X B_X\right)\right){}^2}{A (A+B X)^5}\,, 
   \nonumber
   \\
   \T_{55}&=&\frac{\left(A-X \left(A_X+X B_X\right)\right){}^2}{(A+B X)^6}\,.
\end{eqnarray}
It can be noticed that these coefficients form a triangular matrix. 

Collecting all the results obtained above, one can now write the functions that appear in the action $S$ in terms of the functions $\tf$ and $\ta_I$ of $\tilde S$. 
We find
\begin{eqnarray}
\label{f_gen}
f&=& {\cal J}_g\, \C^{-1} \tf\,,
\\
\aa&=&-h +{\cal J}_g \,\T_{11} \,\ta_1\,,
\label{a1_gen}
\\
\ab&=&h +{\cal J}_g \,\T_{22} \,\ta_2\,,
\\
\ac&=& 2 h_X+{\cal J}_g \left[ \tf \r_3 
-2\tX_X \tf_{\tX}\l_3+\T_{13} \,\ta_1+\T_{23} \,\ta_2+\T_{33} \,\ta_3
\right]\,,
\\
\ad&=& - 2 h_X+{\cal J}_g \left[ \tf \r_4 
-2\tX_X \tf_{\tX}\l_4+\T_{14} \,\ta_1+\T_{44} \,\ta_4
\right]\,,
\\
\ae&=&{\cal J}_g \left[ \tf \r_5
-2\tX_X \tf_{\tX}\l_5+\T_{15} \,\ta_1+\T_{25} \,\ta_2+\T_{35} \,\ta_5+\T_{45} \,\ta_5+\T_{55} \,\ta_5
\right]\,.
\label{a5_gen}
\end{eqnarray}
By substituting all the formulas given in the previous subsections, one obtains the explicit expressions of $\f$ and $\a_I$ in terms of $\tf$, $\ta_I$, $\C$ and $\D$. One can verify that the $\f$ and $\a_I$ satisfy the degeneracy conditions (\ref{D0})-(\ref{D2}). In fact, it turns out that this is a very efficient way to check the expressions for $\f$ and $\a_I$. A first conclusion is thus that all quadratic DHOST theories transform into quadratic DHOST theories.

\subsection{Stability of the degeneracy classes}
\label{section_classI}
  
First of all, let us note that if $\tf=0$ then necessarily $f=0$. Therefore, the transformed version  of  theories in class III remains in class III. 
As a consequence of $\T_{11}=\T_{22}$, we also   find the relation
\beq
\aa+\ab={\cal J}_g \,\T_{11} \, (\ta_1+\ta_2)\,,
\eeq
which shows that the property $\aa+\ab=0$ (or $\aa+\ab\neq 0$) is unchanged by disformal transformations. 
This implies  that class I, characterized by $\aa+\ab=0$, is  stable under disformal transformations. Therefore all the three main classes are stable. 
We study more precisely  the impact of disformal transformations  in the next two sections.
 
If we now consider class Ia theories such that $\tf\neq 0$, which depend on three arbitrary functions, it is natural to expect that generic theories  can be ``generated'' from the subset of (quadratic) Horndeski theories, characterized by a single arbitrary function $\tf$, via general disformal transformations, which depend on two arbitrary functions.  We can check that this is indeed the case\footnote{The same calculation has been performed independently in the recent paper \cite{Crisostomi:2016czh}.}, by starting from the quartic Horndeski Lagrangian expressed in terms of the metric $\tg_{\mu\nu}$ and of the scalar field $\phi$,
\beq
\label{S_tilde}
\tilde S[\phi, \tg_{\mu\nu}]=\int d^4 x\sqrt{-\tg} \left\{\tf (\tilde X,\phi) \tR - 2 \tf_{,\tilde X}(\tilde X, \phi)\left[ (\tnabla^\mu \tnabla_\mu\phi)^2 - \tnabla_\mu\tnabla_\nu\phi\, \tnabla^\mu \tnabla^\nu \phi\right]
\right\}\,.
\eeq
Substituting (\ref{disformal}), we obtain an action $S$  for $g_{\mu\nu}$ and $\phi$, which is characterized by the functions
\beq
\label{f_gen_I}
f=A^{1/2}\,  \sqrt{A+B X}\, \tf\,,
\eeq
and
\begin{eqnarray}
\aa&=&-\ab = -\frac{2 A^{3/2} }{(A+B X)^{3/2}} \, [ ( B + XB^{2} )\tf + 2 \tf_{\tX} \, ]
\label{a1_gen_I}
\\
\ac&=& -\frac{2 \left(B A_X+A
   B_X\right)}{A^{1/2} (A + BX)^{1/2}}\tf 
   +\frac{4 ( X B_X- A_X) A^{1/2}}{(A+B X)^{3/2}}
\\
\ad&=& \frac{2 \left(A^2 B_X+A A_X \left(2
   X B_X+B\right)+A_X^2 (3 A+B
   X)\right)}{A^{3/2} ( A + BX )^{1/2}}\tf
   \cr
&&   -\frac{4 
   \left(-A_X \left(A-2 X^2
   B_X\right)+A X B_X+2 X
   A_X^2\right)}{A^{1/2}(A+B X)^{3/2}}\tf_{\tX}
\\
\ae &=&-\frac{2 A_X 
   \left(B A_X+2 A B_X\right)}{A^{3/2}
   (A+B X)^{1/2}}\tf
   +\frac{4 A_X 
   \left(2 X B_X-A_X\right)}{A^{1/2} (A+B
   X)^{3/2}}\tf_{\tX}
    \label{a5_gen_I}
\end{eqnarray}
If one starts from a generic theory in Class I, defined by the functions $\f$ and $\alpha_I$, it is possible to determine two functions $\C$ and $\D$ such that this theory is  disformally related to  Horndeski, as we now show.  According to (\ref{f_gen_I}), the Horndeski function $\tf$ is related to $f$, $\C$ and $\D$ by 
\beq
\label{G2f}
\tf=A^{-1/2} (A + BX)^{-1/2}\,f\,.
\eeq
Substituting this expression  for  $\tf$ into  (\ref{a1_gen_I}) yields
\beq
\aa = -\ab= \frac{2 A f_X- f(2 A_X+X B_X)}{A-XA_X-X^2 B_X},
\eeq
which one can solve to find $B_X$ in terms of $\ab$, $f$ and $A$:
\beq
\label{B2a2}
B_X=\frac{(2f_X+\ab) A- (2 f+X \ab) A_X}{X(f+X\ab)}\,.
\eeq
Substituting (\ref{G2f}) and (\ref{B2a2}) in $\ac$ gives 
\beq
\label{A2a3}
\frac{A_X}{A}=\frac{4f_X+2\ab+X\ac}{4(f+X\ab)}\,.
\eeq
Finally, by substituting successively (\ref{G2f}), (\ref{B2a2}) and (\ref{A2a3}), one can rewrite $\ad$ and $\ae$ 
in terms of $f$, $\ab$ and $\ac$ and check that one recovers exactly the expressions (\ref{a4_A}) and (\ref{a5_A}). This proves that generic  theories in class Ia  are ``generated''  from the Horndeski quadratic Lagrangians  via disformal transformations (\ref{disformal}).

In analogy with the choice between the  ``Jordan frame'' and ``Einstein frame'' for traditional scalar tensor theories, the above construction shows that  theories belonging to class Ia with $f\neq0$ can be defined either in the ``Jordan frame'',
where the metric is minimally coupled to  matter, 
\beq
S_{\rm total}=\int d^4 x\sqrt{-g}\left[\f R +\a_I L_I^\phi\right]+S_m[g_{\mu\nu},\Psi_m]\,,
\eeq
or in the ``Horndeski frame'', where the gravitational part of the action is described by Horndeski,
\beq
\tilde S_{\rm total}=\int d^4 x\sqrt{-\tg} \left\{\tf  \tR - 2 \tf_{,\tilde X}\left[ (\tilde\square\phi)^2 - \tilde\phi_{\mu\nu}\tilde\phi^{\mu\nu}\right]
\right\}+(\dots) +S_m[g_{\mu\nu},\Psi_m]\,.
\eeq
In the ``Horndeski frame'', the matter action is nonminimally coupled, but can be expressed explicitly in terms of the ``Horndeski metric'' by inverting the transformation (\ref{disformal}).

Note that the Einstein-Hilbert Lagrangian, with $\tf$ constant and $\ta_I=0$,  is a particular case of Horndeski. It generates, via disformal transformations, the family characterized by the expressions (\ref{f_gen_I})-(\ref{a5_gen_I}) with  $\tf_{\tX}=0$. If the disformal transformation is invertible, one thus gets a family of scalar-tensor theories which are in fact general relativity in disguise and, as such, 
are doubly degenerate and contain only two tensor modes. Of course, one can always add another term of the form (\ref{S_other}) in the action,  which does not modify the quadratic part of the action (\ref{action}),  in order to break the second degeneracy. One then obtains a degenerate scalar-tensor theory with one scalar mode and two tensor modes. This is precisely how a theory ``beyond Horndeski'' was constructed in \cite{Zumalacarregui:2013pma}.

We have already pointed out the stability, under disformal transformations, of the sign (including zero)  of $f$ and of  $\a_1+\a_2$, which guarantes the stability of the classes I, II and III separately. We now consider the criteria that distinguish the subclasses within these classes. 
One can first notice the relation 
 \beq
 f-\aa X= {\cal J}_g \left[\frac{1}{A+B X}\tf-\frac{X}{(A+BX)^2} \,\ta_1\right]=
 \frac{ {\cal J}_g }{A+B X}\left(\tf-\tX \,\ta_1\right) \,,
 \eeq
 where we have substituted the expression (\ref{function_h}) for $h$ and the coefficient $\T_{11}$ in (\ref{a1_gen}).
If we start from a theory in Class Ib or in Class IIb, characterized by $\ta_1=\tf/\tX$, the above relation implies that the  disformally transformed theory verifies $\aa=\f/X$ and thus belongs to the same subclass, either Ib or IIb, as the  original theory. Therefore, the classes Ia, Ib, IIa and IIb are separately stable.  We find the same properties for the subclasses in class III. Indeed, when $f=0$, we have $\aa={\cal J}_g \T_{11}\ta_1$ and $\ab={\cal J}_g \T_{11}\ta_2$. Therefore the signs of $\aa$ and $\aa+3\a_2$ which distinguish the subclasses IIIa, IIIb and IIIc are conserved in a disformal transformation.

In summary, all the classes and subclasses that we have distinguished are separately stable under disformal transformations. In particular, the intersections of two classes or subclasses, when  non empty, are also 
stable. This applies for instance to the intersection of Ia and IIIa, which contains $L_4^{\rm bh}$.

\subsection{Non-invertible disformal transformation and mimetic gravity}

Let us now briefly comment on the consequences of considering non-invertible DT which leads to the so called mimetic gravity theories introduced in \cite{Chamseddine:2013kea}. These theories were extensively studied, starting from \cite{Lim:2010yk, Chamseddine:2014vna, Hammer:2015pcx} and further investigated in \cite{Firouzjahi:2017txv, Gorji:2017cai, Takahashi:2017pje, BenAchour:2017ivq, Bodendorfer:2018ptp, Chamseddine:2018gqh, Casalino:2018wnc, Ganz:2018mqi, Langlois:2018jdg, Gorji:2020ten, Jirousek:2022kli}.

So far, we have  assumed that the disformal transformation \eqref{disformal} is invertible, in the sense  
that one can also  express the metric $g_{\mu\nu}$ in terms of $\tilde{g}_{\mu\nu}$. Now consider a non-invertible
disformal transformation of the  Einstein-Hilbert action.  By differentiating the expression
\bea
\tilde{g}_{\mu\nu} = A(\phi,X) g_{\mu\nu} + B(\phi,X) \phi_\mu \phi_\nu\,,
\eea
one obtains
\bea
\label{d_tg}
\delta \tilde{g}_{\mu\nu} = F_{\mu \nu} \, \delta \phi + H_{\mu\nu}^\alpha \, \nabla_\alpha \delta \phi + 
J_{\mu\nu}^{\alpha\beta} \, \delta g_{\alpha\beta} 
\eea
with
\bea
F_{\mu \nu} & = & A_\phi g_{\mu \nu} + B_\phi \phi_\mu \phi_\nu \; ,\\
H_{\mu\nu}^\alpha & = & 2(A_X  g_{\mu\nu} + B_X  \phi_\mu \phi_\nu) \phi^\alpha + 
B (\phi_\nu \delta_\mu^\alpha + \phi_\mu g_\nu^\alpha) \,,
\\
J_{\mu\nu}^{\alpha\beta} &=&
A \delta_{(\mu}^\alpha \delta_{\nu)}^\beta - \phi^\alpha \phi^\beta (A_X g_{\mu\nu} +
B_X \phi_\mu \phi_\nu)\,.
\eea
As discussed in \cite{Zumalacarregui:2013pma}, the  disformal transformation 
is non invertible, i.e. $g_{\mu\nu}$ cannot be determined from $\tilde{g}_{\mu\nu} $, if
 the determinant of  the Jacobian matrix  $J_{\mu\nu}^{\alpha\beta} \equiv \frac{\partial \tilde{g}_{\mu\nu}}{\partial g_{\alpha\beta}} $
vanishes.  This happens when $J_{\mu\nu}^{\alpha\beta} $
admits a  null vector $v_{\alpha\beta}$  such that 
\beq
J_{\mu\nu}^{\alpha\beta} v_{\alpha\beta} = 0 \,. 
\eeq
It is straightforward to check that the combination
\bea\label{nullvectors}
v_{\alpha\beta}=A_X g_{\alpha\beta} + B_X \phi_\alpha\phi_\beta
\eea
is a null vector of the Jacobian matrix, provided the functions $A$ and $B$ verify 
\beq
\label{cond}
B_X= \frac{A-XA_X}{X^2}\,.
\eeq
After integration, this yields 
\beq
\label{B_mimetic}
B= - \frac{A}{X} + \mu(\phi)\,,
\eeq
corresponding to the disformal transformation
\beq\label{disformal mimetic}
\tilde{g}_{\mu\nu} = A(\phi, X)\left(g_{\mu\nu} - \frac1X \phi_\mu \phi_\nu\right)+ \mu(\phi)\,  \phi_\mu \phi_\nu\,.
\eeq
Note that if we insert (\ref{B_mimetic}) into (\ref{X_tilde}), one gets $\tX=1/\mu(\phi)$, which shows that $\tX$ does not depend on $X$.

Now, if we  start from an action of the form
\bea\label{generalactionfordisformal}
\tilde{S}[\phi,\tilde{g}_{\mu\nu}] = \int d^4x \; \sqrt{-\tg} \left( \tf(\phi) \tilde{R} + 
\alpha_I(\phi) \tilde{L}_I^\phi \right)\,,
\eea
and substitute (\ref{disformal mimetic}), we obtain a new action $S$, given  as a functional of $g_{\mu\nu}$ and $\phi$. 
This leads to a  subclass of our DHOST theories with  particular properties. This procedure has been used  in  \cite{Chamseddine:2013kea} for the Einstein-Hilbert action, i.e. $\tf=1$ and $\ta_I=0$, with the disformatl transformation characterized  by $A=X$ and $B=0$, to introduce the model of mimetic dark matter.  It has been extended in \cite{Deruelle:2014zza}  to a general non-invertible transformation, with (\ref{B_mimetic}). 
In contrast with the generic case where the disformal transformation is invertible,  the number of degrees of freedom is not necessarily the same for $\tilde{S}$ and $S$. In particular, if $\tilde{S}$ is the Einstein-Hilbert action, with only two degrees of freedom, one ends up with three degrees of freedom for $S$, as discussed in  
\cite{Chamseddine:2013kea} and \cite{Deruelle:2014zza}.

Interestingly, the mimetic action $S$  is invariant under the local symmetry
\bea\label{mimetic symmetry}
\delta \phi = 0 \qquad \text{and} \qquad 
\delta g_{\mu\nu} =\varepsilon\,  v_{\mu\nu}= \varepsilon (A_X g_{\mu\nu} + B_X \phi_\mu\phi_\nu)\,,
\eea
where $\varepsilon$  is an infinitesimal space-time function. This symmetry follows immediately from (\ref{d_tg}), together with the property that $v_{\mu\nu}$ is a null eigenvector of the Jacobian matrix.

In the Hamiltonian framework, such a symmetry implies the existence of  an extra first class 
constraint in addition to the usual Hamiltonian and momentum constraints associated with diffeomorphism invariance. This is in contrast with  the standard quadratic DHOST theories, for which the extra constraints are 
generically second class. One of these second class constraints is necessary to eliminate the Ostrogradski ghost and we  thus  expect that, even though mimetic theories contain three degrees of freedom, the Ostrogradski ghost is still present.

This is indeed the case for the simplest model of mimetic gravity, obtained from Einstein-Hilbert  with $A=X$ and $B=0$. 
In that case,  the symmetry \eqref{mimetic symmetry} reduces to an invariance under conformal transformations of $g_{\mu\nu}$.
In the Hamiltonian description, this symmetry is necessarily associated to a first class constraint. 
Following the analysis of \cite{Langlois:2015skt}, and introducing the conjugate momenta $\pi^{ij}$ and $p_*$ of  $\h_{ij}$ and $\An$, respectively, one can show that the primary constraint reduces to
\bea
\Psi \equiv \gamma_{ij} \pi^{ij} - \frac{1}{2} p_*\,,
\eea
which is indeed the generator of infinitesimal conformal transformations. As a consequence, the primary constraint is first class and it Poisson commutes with the Hamiltonian and momentum constraints. Hence, there is no  secondary  constraint that eliminates the Ostrogradski ghost. This has already been noticed   in  \cite{Chaichian:2014qba} and  we expect this to remain true for any mimetic-like theory.

To conclude let us briefly mention several recent works focusing on generalized disformal transformations. Entering in the details of each of these works goes beyond the scope of this chapter, and we refer the interested reader to the cited articles.
\begin{itemize}
\item In \cite{ Jirousek:2022rym}, it was shown that the non-invertibility condition (\ref{cond}) admits more general solutions than (\ref{B_mimetic}), revealing an extended set of transformation to construct mimetic gravity theories with Weyl invariance. In \cite{Jirousek:2022jhh}, it was shown explicitly that even an invertible but singular DT can change the number of degrees of freedom. This was confirmed by an hamiltonian analysis and implemented both on mechanical examples and on gravitational theories such as K-essence.
\item In \cite{ Domenech:2023ryc}, the authors have upgraded disformal transformation to a symmetry,  building invariant tensors under DT and providing a disformal invariant action. Relations to mimetic gravity and U-DHOST have been discussed.
\item More recently, several authors have considered higher order DTs. One key challenge in doing so is to ensure that the transformation is invertible, which is usually not the case. Obviously, finding more complex versions of the standard DT would be valuable as it could potentially lead to new interesting theories and their exact solutions. See \cite{Takahashi:2021ttd, Takahashi:2022mew, Takahashi:2023vva, Takahashi:2023jro, Alinea:2024gjn, Alinea:2024jrf, Babichev:2024eoh} for details.
\end{itemize}


\chapter{No-hair theorems in general relativity and beyond}
\label{Chapter3}

\minitoc

In this chapter, we review the key definitions and the central theorems which build our current understanding of black holes. First, we summarize the basic tools to analyze the geometry of a null hypersurface. This allows us to briefly review the notion of Killing horizon and its generalization known as (weakly) isolated horizons. The next section is devoted to a brief historical account on the proof of the no-hair theorems in GR where we state for completeness the modern version of the no-hair theorem in 4d GR for stationary and axi-symmetric black holes following \cite{Gourgoulhon:2024}. After reviewing the main properties of the Kerr black hole, we discuss how a massless minimally coupled scalar field modifies the Schwarzschild and Kerr geometry, presenting the known exact solutions of the Einstein-Scalar system derived in \cite{Janis:1968zz, Mirza:2023mnm}. This allows one to illustrate the no-hair theorem. The last part of this chapter reviews the no scalar hair theorems first in the simplest scalar-tensor theories and then for higher order scalar tensor theories. We discuss the Hui-Nicolis theorem \cite{Hui:2012qt} and the more recent \textit{no scalar monopole theorem} derived in \cite{Capuano:2023yyh}. This chapter serves as an intermediate step before presenting the construction of concrete hairy black holes solutions in DHOST gravity. We refer the interested reader to the beautiful and pedagogical lectures \cite{Gourgoulhon:2005ng, Gourgoulhon:2024} as well as to the related articles for more details.


\section{On the notion of horizons}


In order to review the main definitions and properties of a black hole horizon at equilibrium, we first need to review few key properties of a null hypersurface. We shall follow \cite{Gourgoulhon:2005ng, Gourgoulhon:2024}. Consider a spacetime $(\mathcal{M}, g)$ and a null tetrad $(\ell, k, m, \bar{m})$ satisfying the orthonormality relations
\begin{align}
\ell^{\mu} k_{\mu} =-1 \qquad m^{\mu} \bar{m}_{\mu} = 1
\end{align}
while all other scalar products vanish, i.e. $(\ell, k, m, \bar{m})$ are null vectors. The metric can be decomposed as 
\begin{align}
g_{\mu\nu} = - \ell_{\mu} k_{\nu} -  k_{\mu} \ell_{\nu} + q_{\mu\nu} \qquad q_{\mu\nu} =  m_{\mu} \bar{m}_{\nu} + \bar{m}_{\mu} m_{\nu} 
\end{align}
Consider thus a null hypersurface $\mathcal{N}$ whose normal vector is $\ell$. Since the null nature of $\ell$ is invariant under a rescaling $\ell \rightarrow \alpha \ell$, the null hypersurface has to be understood as an equivalence class under such rescaling. We denote this equivalence class $[\ell]$. We assume that this null hypersurface splits into $\mathcal{N} = \mathbb{R}\times S^2$ where $S^2$ is the topological $2$-sphere.

Now, in order to discuss the intrinsic geometry of $\mathcal{N}$, we introduce the projector onto the cross-sections of $\mathcal{N}$ given by
\begin{align}
q^{\mu}{}_{\nu} = \delta^{\mu}{}_{\nu} + \ell^{\mu} k_{\nu} + k^{\mu} \ell_{\nu} 
\end{align}
such that $q^{\mu}{}_{\nu} \ell^{\nu} = 0 = q^{\mu}{}_{\nu} k^{\nu} = 0$. The projection of the covariant derivative of $\ell$, dubbed the deformation tensor, can be decompose as follows
\begin{align}
\Theta^{(\ell)}_{\alpha\beta} = q^{\mu}{}_{\alpha} q^{\nu}{}_{\beta} \nabla_{\mu} \ell_{\nu} = \frac{1}{2} \theta_{\ell} \; q_{\alpha\beta} + \sigma_{\alpha\beta} + \omega_{\alpha\beta}
\end{align}
where $\theta_{\ell}$ is the expansion of $\ell$ defined by
\begin{align}
\theta_{\ell} &  = q^{\mu\nu} \nabla_{\mu} \ell_{\nu} = \frac{1}{2} q^{\mu\nu} \mathcal{L}_{\ell} q_{\mu\nu} 
\end{align}
while the traceless symmetric tensor $\sigma_{\mu\nu}$ encodes its shear and the anti-symmetric tensor $\omega_{\mu\nu}$ corresponds to its rotation. They are respectively given by
\begin{align}
\sigma_{\mu\nu} & = q^{\mu}{}_{\alpha} q^{\nu}{}_{\beta} \nabla_{(\mu} \ell_{\nu)} - \frac{1}{2} \theta\; q_{\mu\nu} \\
\omega_{\mu\nu} & = q^{\mu}{}_{\alpha} q^{\nu}{}_{\beta} \nabla_{[\mu} \ell_{\nu]}
\end{align}
Notice that these three objects are defined intrinsically, i.e they refer only to the geometry of $\mathcal{N}$. 
Now, if one defines $\mathcal{N}$ as the null hypersurface corresponding to constant level of a scalar field $\Phi$, the null normal $\ell$ can be decomposed as
\begin{align}
\label{level}
\ell_{\alpha} =  - \mu \nabla_{\alpha} \Phi
\end{align}
It follows that $\ell$ is a geodesic vector, satisfying 
\begin{align}
\ell^{\mu} \nabla_{\mu} \ell_{\alpha} =  \kappa \ell_{\alpha} \qquad \text{with} \qquad \kappa = - \ell^{\beta} \nabla_{\beta} \mu
\end{align}
The function $\kappa$ is called to inaffinity coefficient. In terms of the inaffinity, the expansion of $\ell$ can be recast into 
\begin{align}
\theta_{\ell} = \nabla_{\mu} \ell^{\mu} - \kappa
\end{align}
Now, one can show that since $\ell$ is hypersurface-orthogonal to the hypersurface of level $\mu$, then its flow is hypersurface orthogonal which implies that the rotation tensor vanishes, i.e. $\omega_{\alpha\beta} =0$. Then, the evolution of the expansion $\theta_{\ell}$ along the geodesic flow generated by $\ell$ is encoded in the null Raychaudri equation given by
\begin{align}
\ell^{\alpha} \nabla_{\alpha} \theta_{\ell} = \kappa \theta_{\ell} - \frac{1}{2} \theta_{\ell}^2 - \sigma^{\alpha\beta} \sigma_{\alpha\beta}  - R_{\alpha\beta} \ell^{\alpha} \ell^{\beta}
\end{align}
This is the key equation from which several fundamental results related to black hole horizon can be extracted.
Having review the main quantities of interest to describe the geometrical properties of a null hypersurface, we are now in position to review the different notions of horizons. 

For a long time, the properties of a black hole horizon have been drawn from the few exact globally stationary black holes solutions known in GR, i.e. the Schwarzschild and Kerr solutions. In such stationary spacetime geometries which exhibit (at least) two commuting killing vectors, one generating the stationarity and the second the azimuthal symmetry, the notion of Killing horizon (KH) describes very well the black hole boundary.\\

\textbf{Killing horizon:} \textit{Consider a spacetime $(\mathcal{M}, g)$ admitting a killing vector $\xi^{\mu} \partial_{\mu}$, i.e. such that $\mathcal{L}_{\xi} g_{\mu\nu} = 2 \nabla_{(\mu}\xi_{\nu)} =0$. A Killing horizon $\mathcal{N}$ is a connected null hypersurface of $(\mathcal{M}, g)$ such that, on $\mathcal{N}$, the killing vector $\xi$ is normal to $\mathcal{N}$  and thus $\xi^{\mu}\xi_{\mu}|_{\mathcal{N}} =0$.} \\
 
Notice that the property $\xi^{\mu}\xi_{\mu}|_{\mathcal{N}} =0$ is not shared by all the Killing vectors which can be either null of spacelike on $\mathcal{N}$. Nevertheless, among the different Killing vectors, one can construct a linear combination of them such that the resulting Killing vector $\xi$ belongs to the equivalence class of null normals $[\ell]$ to $\mathcal{N}$, i.e. $\xi \in [\ell]$. In that case, we say that the horizon is Killing w.r.t $\xi$.

A direct consequence of this definition is that on a KH, the deformation tensor associated to the null normal $\ell$ which coincides with $\xi$ vanishes. Indeed, if one assumes that the geodesic flow generated by $\ell$ is hypersurface orthogonal, i.e. one has $\Theta^{(\ell)}_{[\alpha\beta]} = \omega_{\alpha\beta} =0$. Then, the above definition implies that
\begin{align}
\Theta^{(\ell)}_{\alpha\beta} = q^{\mu}{}_{\alpha} q^{\nu}{}_{\beta} \nabla_{\mu} \ell_{\nu} = q^{\mu}{}_{\alpha} q^{\nu}{}_{\beta} \nabla_{(\mu} \xi_{\nu)} =0
\end{align}
It follows from this that the expansion of the Killing vector vanishes on $\mathcal{N}$, i.e. 
\begin{align}
\theta_{\xi}  |_{\mathcal{N}}  = \theta_{\ell} = q^{\alpha\beta} \Theta^{(\ell)}_{\alpha\beta} = 0
\end{align}
This stands as a key property of the black hole horizon. Moreover, one can show that provided the dominant energy condition (DEC) is satisfied on the KH, the inaffinity coefficient $\kappa$ of the null normal $\ell$ coinciding with the killing vector $\xi$ defining $\mathcal{N}$ is constant over $\mathcal{N}$. This stands as the zeroth law of black hole mechanics. The second law states that  first and second laws encode the qualitative and quantitative changes of the horizon area in terms of the change of the asymptotic charges (mass and angular momentum defined measured by an asymptotic observer) \cite{Bardeen:1973gs}. The importance of the notion of Killing horizon cannot be over emphasized. Indeed, a major result in black hole theory are the rigidity theorems derived by Hawking \cite{Hawking:1971vc} and refined in \cite{Hawking:1973uf, Chrusciel:1996bj}. The weak version of this theorem , derived first by Carter in \cite{Carter:1969zz}, states that if a stationary and axi-symmetric spacetime $(M, g)$ contains a black hole horizon, then this horizon is Killing. A consequence of this theorem is that the black hole horizon rotates rigidly, hence the name of the theorem. In the strong version of the theorem, the assumption on axi-symmetry is relaxed but the metric functions are assumed to be analytic. This central result is not only crucial in proving the no-hair theorem of 4d GR, but it also plays a crucial role in the derivation of the no scalar hair theorems obtained first by Bekenstein. See \cite{Hollands:2006rj} for an extension of the rigidity theorem for higher dimensional axi-symmetric black holes.

In the following, it will be useful to introduce the following distinction between the different possible KHs. First, since $\kappa$ is constant on $\mathcal{N}$, we distinguish between 
\begin{itemize}
\item  the non-degenerate KH with $\kappa\neq 0$ 
\item  the degenerate KH with $\kappa =0$. 
\end{itemize}
Now, consider the stationary killing vector $\xi$ tangent to $\mathcal{N}$ (which does not always belong to $[\ell]$). Since it can be either spacelike or null on $\mathcal{N}$, one distinguishes between 
\begin{itemize}
\item  non-rotating KH where $\xi$ is null, i.e. $\xi^{\mu} \xi_{\mu} |_{\mathcal{N}} =0$
\item  rotating KH where  $\xi$ is spacelike, i.e. $\xi^{\mu} \xi_{\mu} |_{\mathcal{N}} > 0$
\end{itemize}
These different cases will be useful when stating the no-hair theorem in the next section.  So far, we have reviewed the most restrictive notion of horizon which relies on the existence of at least one killing symmetry in the whole spacetime geometry. This is obviously the case for the Schwarzschild and Kerr black hole solutions of GR. 

Nevertheless, while the notion of KH is well suited for globally stationary black holes, it fails to describe a trapping horizon if the bulk geometry does not possess any symmetry. Yet, such configuration is known in GR. For example, the Robinson-Trautman family of solutions are radiative solutions describing a collapse which settles down to the Schwarzschild black hole at late time. In this case, the fully dynamical spacetime geometry has no killing vector fields anywhere and the notion of a Killing horizon is useless. For this reason, important efforts have been devoted to characterize black hole horizons without referring to the bulk geometry. These efforts have led to the notion of weakly isolated horizon which provides a much weaker notion of trapping horizon than the Killing one. The idea is to extract from the definition of a KH the minimal ingredient such that only the intrinsic geometry of the null hypersurface is constrained, leaving a complete freedom on the bulk geometry. This fully quasi-local approach was initiated twenty five years ago in \cite{Ashtekar:1998sp, Ashtekar:1999yj} and has led to major results in black hole mechanics, numerical general relativity and black hole state counting. The interested reader can refer to \cite{Ashtekar:2004cn} for a pedagogical review.  Let us first introduce the notion of a non-expanding horizon before presenting the one of a weakly isolated horizon. \\
 
\textbf{Non expanding horizon}: A null hypersurface $\mathcal{N}$ of a spacetime $(\mathcal{M}, g)$ is a non-expanding horizon if 
\begin{itemize}
\item $\mathcal{N}$ is null and  topologically $S^2 \times \mathbb{R}$
\item Any null normal $\ell$ of $\mathcal{N}$ has a vanishing expansion, i.e $\theta_{\ell} =0$
\item All equations of motion hold at $\mathcal{N}$ and the stress energy tensor is such that $- q^{\mu}{}_{\nu} T^{\nu}{}_{\alpha} \ell^{\alpha}$ is future-causal for any future directed null normal $\ell$
\end{itemize} 

With this definition, the energy condition together with the positivity of the squared shear implies that $\sigma_{\mu\nu}=0$ and $R_{\mu\nu} \ell^{\mu} \ell^{\nu} =0$ independently. The vanishing of the shear and the expansion further imposes that $q^{\mu}{}_{\alpha} q^{\nu}{}_{\beta} \nabla_{(\mu} \ell_{\nu)} = 0$. With this property, one can define a unique metric compatible connection $\mathcal{D}$ on $\mathcal{N}$ such that
\begin{align}
\mathcal{D}_{(\alpha} \ell_{\beta)} =  \frac{1}{2} \mathcal{L}_{\ell} q_{\alpha\beta} = 0
\end{align}
Therefore, the above definition implies that the intrinsic metric on $\mathcal{N}$ is invariant under the geodesic flow of $\ell$, i.e. that $\ell$ is a symmetry generator on $\mathcal{N}$. Therefore, the hypersurface $\mathcal{N}$ can be considered as time-independent along the flow of $\ell$. Notice that this does not impose a restriction on the bulk geometry since this condition is purely intrinsic to $\mathcal{N}$ such that the whole spacetime need not be globally stationary. Yet, while this definition allows one to extract the minimal properties from the more familiar KH, one needs to further impose some condition in order to derive the first and second laws of black hole thermodynamics. This is achieved by introducing the notion of weakly isolated horizon. \\

\textbf{Weakly isolated horizon}:  The structure $(\mathcal{N}, [\ell], \mathcal{D})$ is a weakly isolated horizon provided $\mathcal{N}$ is a NEH and each null normal $\ell$ satisfies
\begin{align}
\left( \mathcal{L}_{\ell} \mathcal{D}_{\alpha} - \mathcal{D}_{\alpha} \mathcal{L}_{\ell} \right) \ell^{\beta} =0
\end{align} 
This last step imposes that the connection $\mathcal{D}_{\alpha} \ell^{\beta}$ be time-independent. Finally, the notion of isolated horizon is achieved by extending the above property to any tangent vector to $\mathcal{N}$ and not only to the represents of the equivalence class of null vectors belonging to $[\ell]$.  \\

Having review the definition of a black hole horizon, from the most restrictive one, i.e. the KH, to the more general and fully quasi-local definition encoded in the isolated horizon, we now turn to the key theorem of general relativity, the no-hair theorem. The key target is to understand how generic are the black holes solutions known in GR and whether they represent the unique end points of a generic gravitational collapse.

\section{Uniqueness of black holes in $4d$ general relativity}

Gravitational collapse is probably the physical mechanism where GR manifests the most striking deviations compared to newtonian gravity through the formation of horizons and singularities, the most simple dynamical solution being the spherically symmetric dust collapse derived in \cite{Oppenheimer:1939ue}. In the mid sixties, important efforts were devoted to further characterize the different possible endpoints of this highly non-linear process. These efforts predated and motivated the beautiful work initiated by Werner Israel which led to the very first no-hair theorem \cite{Israel:1967wq}. There it was shown that under a list of technical assumptions, the Schwarzschild black hole was the unique spherically symmetric and asymptotically flat black hole solution of GR in $d=4$. Over the next decades, the different assumptions regarding the properties of the horizon, the analyticity of metric functions, and the dimension of spacetime were slowly relaxed step by step by various authors, giving rise to a very general and strong unicity theorem for spherically symmetric asymptotically flat black holes, the unique solution in $d$-dimension being known as the Schwarzschild-Tangherlini black hole. In parallel, the role of the notion of Killing horizon reviewed above, first introduced by Carter, was gradually recognized as a key concept to characterize stationary black hole geometries and derive their uniqueness properties.

The case of the spherically symmetric electro-vacuum black hole solutions was tackled around the same time first by Israel \cite{Israel:1967za} and then generalized by different authors in the next decades \cite{Robinson:1974nf}. In $d=4$, the theorem states that, under generic assumption, the unique black hole solutions of the Einstein-Maxwell system consist in either the Reissner-Nordstrom black hole or the Majumdar-Papapetrou solution \cite{Majumdar:1947eu}. In $d\geqslant 5$, the allowed solutions reduces to the Reissner-Nordstrom-Tangherlini black hole, a $d$-dimensional version of the Reissner-Nordstrom geometry.

Finally, the case of the stationary and axi-symmetric black hole turns out to be more involved. As already mentioned earlier, the strong rigidity theorem derived by Hawking implies that a black hole horizon in any stationary geometry whose metric functions are analytic has to be Killing and rotates rigidly \cite{Hawking:1971vc}. Removing the assumption of analyticity was achieved in certain cases but it is still an open question whether this assumption can be removed completely. As we shall see later, one can exhibit explicit solutions in DHOST gravity for which the properties of circularity and analyticity are lost, leading to interesting consequences. Around the same time, Carter and few years later Robinson succeeded to show that the only stationary and axi-symmetric black hole solutions in $d=4$ GR reduces to the $2$-parameter family described by the Kerr black hole \cite{Carter:1971zc, Robinson:1975bv}. Just as for the spherically symmetric case, the assumptions used in the initial proof were slowly relaxed and the generalization of this theorem to the electro-vacuum case, demonstrating the uniqueness of the Kerr-Newman solution, were achieved only ten years later as additional techniques were needed to tackle the extensions to the Einstein-Maxwell system.
We refer the interested reader to \cite{Chrusciel:1994sn, Mazur:2000pn, Gourgoulhon:2024} for pedagogical reviews and lectures on the different steps involved in establishing the no-hair theorem. 

Having briefly reviewed the historical development of the efforts towards to proof of the no-hair theorem of four dimensional general relativity, we now state it for completeness. Following \cite{Gourgoulhon:2024}, the no-hair theorem can be expressed as follows. \\

\textbf{The no hair theorem:} \textit{If a spacetime $(\mathcal{M}, g_{\mu\nu})$ is 
\begin{itemize}
\item four dimensional
\item  asymptotically flat
\item stationary
\item analytic
\item satisfies the electro-vacuum Einstein equations
\item contains a connected event horizon that is either i) rotating or ii) non-rotating and non-degenerate
\item does not contains any closed causal curve in the domain of outer communications 
\end{itemize}
then $(\mathcal{M}, g_{\mu\nu})$ is isomorphic to the domain of outer communications of a Kerr-Newman black hole spacetime labelled by four parameters: the mass $M$, the angular momentum $J$, the electric charge $Q$ and the magnetic monopole charge $P$. The Kerr black hole  corresponds to $Q=P=0$. }\\

Since the Kerr geometry is the key object to which one has to confront any new hairy rotating black holes, let us briefly review some of its property. Within the Boyer-Lindquist coordinates $(t,r, \theta, \phi)$, the Kerr metric \cite{Kerr:1963ud} is given by
\begin{align}
\rd s^2 & = - \left( 1 - \frac{2m r}{\rho^2}\right) \rd t^2 - \frac{4 a m r \sin^2{\theta}}{\rho^2} \rd t \rd \phi + \rho^2 \left( \frac{\rd r^2}{\Delta} + \rd \theta^2 \right) \\
& \;\;\;+ \left( r^2 + a^2 + \frac{2a^2 m r \sin^2{\theta}}{\rho^2}\right) \rd \phi^2
\end{align}
where the functions $\rho$ and $\Delta$ takes the form
\begin{align}
\rho^2 = r^2 + a^2 \qquad \Delta = r^2 - 2M r + a^2 = (r- r_{+}) (r- r_{-})
\end{align}
with
\begin{align}
\label{horr}
r_{\pm} = m \pm \sqrt{m^2-a^2}
\end{align}
See \cite{Kerr:2007dk} for an historical account on the discovery of this central solution. Let us mention the main properties of the Kerr solution. The Kreshman scalar is given by
\begin{align}
K = R_{\alpha\beta\mu\nu} R^{\alpha\beta\mu\nu} = \frac{48 m^2}{\rho^{12}} \left( r^6 - 15 r^4 a^2 \cos^2{\theta} + 15 r^2 a^4 \cos^4{\theta} - a^6 \cos^6{\theta}\right)
\end{align}
which reveals a (annular) singularity at $\rho \rightarrow 0$. 

Next the geometry admits two commuting Killing vectors $\xi^{\mu} \partial_{\mu} = \partial_t$ and $\chi^{\mu} \partial_{\mu} = \partial_{\phi}$. Since $\xi^{\mu} \xi_{\mu} < 0$ as $r\rightarrow +\infty$, the Killing vector $\xi$ generating stationarity is asymptotically timelike and the Kerr geometry is stationary. However, $\xi^{\mu} \partial_{\mu}$ is not normal to the hypersurfaces $t=\text{constant}$, i.e it is not hypersurface-orthogonal. The norm of the stationary Killing vector $\xi^{\mu} \partial_{\mu}$ is given by
\begin{align}
\xi^{\mu} \xi_{\mu} =  \frac{2mr - r^2 - a^2 \cos^2{\theta}}{r^2 + a^2 \cos^2{\theta}}
\end{align}
It shows that the KV becomes null on the two hypersurfaces located at
\begin{align}
r^{\pm}_E = m \pm \sqrt{m^2 - a^2 \cos^2{\theta}}
\end{align}
These two hypersurfaces are called the ergospheres. The KV is spacelike for $r^-_E < r < r^+_E$ and timelike outside. The two ergospheres are timelike hypersurfaces and a particle can thus enter/exit these frontiers.

Finally, let us focus on the hypersurfaces of constant radius. Their normal is $\rd r$ and its (rescaled) dual vector reads
\begin{align}
s^{\alpha} \partial_{\alpha} = 2mr \partial_t + \Delta \partial_r + a \partial_{\phi}
\end{align}
In order to investigate the causal nature of these hypersurfaces, let us compute the norm of the normal vector which reads
\begin{align}
s^{\mu} s_{\mu} = \rho^2 \Delta
\end{align}
Therefore, the $r$-constant hypersurface becomes null precisely when $\Delta =0$, i.e. at the locus $r_{\pm}$ given by (\ref{horr}). Let us denote these null hypersurfaces $\mathcal{N}_{\pm}$. Focusing on the outer one located at $r_{+}$, the normal vector reduces to 
\begin{align}
\label{KV}
s^{\alpha} \partial_{\alpha} |_{\mathcal{N}_{+}}= 2mr_{+} \partial_t + a \partial_{\phi} = 2mr_{+} \left( \xi^{\mu}\partial_{\mu} + \frac{a}{2m r_{+}} \chi^{\mu}\partial_{\mu} \right) = 2mr_{+} \zeta_{+}^{\mu} \partial_{\mu}
\end{align}
which is proportional to a linear combination of the stationary and azimuthal Killing vectors. Therefore, $\mathcal{N}_{+}$ is null and the Killing vector (\ref{KV}) is normal to it. Thus, it fullfils the definition of a Killing horizon w.r.t the Killing vector $\zeta_{+}^{\mu}\partial_{\mu}$. To complete the description of $\mathcal{N}_{+}$, let us compute the inaffinity $\kappa_{+}$ associated to $\zeta_{+}^{\mu}\partial_{\mu}$ and the norm of the stationary killing vector $\xi^{\mu}\partial_{\mu}$ on $\mathcal{N}_{+}$. They read
\begin{align}
\kappa_{+} = \frac{r^2_{+} - a^2}{4m r^2_{+}} \neq 0 \qquad \xi^{\mu} \xi_{\mu} |_{\mathcal{N}_{+}} = \frac{a^2 \sin^2{\theta}}{r^2_{+} + a^2 \cos^2{\theta}} >0
\end{align}
Therefore, according to the definitions introduced above, the outer horizon $\mathcal{N}_{+}$ of Kerr is a rotating and non-degenerate Killing horizon. Let us finally point that the Kerr geometry can be characterized by a tensor known as the Simon-Mars tensor \cite{Mars:1999yn} which vanishes exactly only for the Kerr solution, allowing one to capture any (stationary) deviation to Kerr in a  covariant manner.

Having review the no-hair theorem stating the uniqueness of the Kerr solution in 4d vacuum GR, together with a brief description of the Kerr black hole geometry, the very first question one can ask is how does a canonical self-gravitating scalar field modify the Kerr geometry and its non-rotating Schwarzschild limit ?


\section{Scalar vacuum solutions in general relativity}

In the following, we present several exact solutions describing a static or stationary self-interacting scalar field in GR, demonstrating the absence of black holes horizons in this family of scalar vacuums. This will help to contrast with the hairy black holes solutions we shall discuss in the next chapters. The system corresponds to the action
\begin{align}
S[g, \varphi] = \frac{1}{16\pi} \int \rd^4x \sqrt{|g|} \left( \mathcal{R} - 8\pi g^{\mu\nu} \nabla_{\mu}\varphi \nabla_{\nu} \varphi\right)
\end{align}
The field equations related to the above action are
\begin{align}
R_{\mu\nu} & = 8\pi \nabla_{\mu}\varphi \nabla_{\nu} \varphi \qquad \Box \varphi =0
\end{align}
Notice that the scalar field is not self-interacting,  i.e. there is no potential $V(\varphi)$.
The geometries presented below are exact solutions of this system.

\subsection{Spherical symmetric configuration}

\label{naked}

The form  of the gravitational field generated by a static spherically symmetric self-gravitating scalar field in GR was first derived by Janis, Newman and Winicour in the late sixties \cite{Janis:1968zz}. It is given by the metric
\begin{align}
\label{JNW}
 d s^2 = - f(r)^{\gamma} dt^2 + f(r)^{-\gamma} dr^2 + r^2 f(r)^{1-\gamma} d\Omega^2\,, \qquad f(r) = 1- \frac{r_s}{r}\,.
\end{align}
where $\rd \Omega^2 = \rd \theta^2 + \sin^2{\theta} \rd \phi$ and the associated scalar profile and its kinetic energy read
\begin{align}
\label{scalarprofile}
\phi(r) = \frac{S}{r_s\sqrt{4\pi}} \log\left( 1-\frac{r_s}{r}\right)\,, \qquad 
X(r) = g^{\mu\nu} \phi_{\mu} \phi_{\nu} = \frac{S^2}{4\pi} \frac{1}{r^4} \left( 1-\frac{r_s}{r}\right)^{\gamma-2}\,.
\end{align}
The parameters of the solution are related to the mass $M$ of the geometry and the scalar charge $S$. They are both related to the two constants $(\gamma, r_s)$ by
\begin{align}
\gamma = \frac{2M}{r_s} \,, \quad r_s = 2 \sqrt{M^2 + S^2}\,,\qquad \rightarrow \qquad 0 \leqslant \gamma \leqslant 1
\end{align}
Notice that when the scalar charge vanishes, i.e $S=0$, $r_s= 2M$ and $\gamma=1$, then one recovers the standard Schwarzschild vacuum geometry. 

Let us now briefly discuss how the JNW geometry differs from its Schwarzschild limit.
The curvature invariants are given by
\begin{align}
  R (r) & =  \frac{1-\gamma^2}{2} \frac{r_s^2 f(r)^{\gamma-2}}{r^4}\,,  \\
  R_{\mu\nu\rho\sigma} R^{\mu\nu\rho\sigma} (r)& = \frac{1}{12} \frac{{r_s}^2 f(r)^{2 \gamma -4 } \Delta_1(r)}{ r^8}\,, \\
  C_{\mu\nu\rho\sigma} C^{\mu\nu\rho\sigma} (r)& = \frac{1}{3} \frac{r_s^2 f(r)^{2\gamma-4}\Delta_2 (r)}{ r^8}\,,
  \end{align}
where the functions $\Delta_{1,2}(r)$ are given by
\begin{align}
\Delta_1(r) & = 4 \left[(\gamma +1) (2 \gamma +1) r_s-6 \gamma  r \right]^2 + 5 r_s^2 (1-\gamma^2)^2\,,\\
\Delta_2(r) & = \left[(1+\gamma)(1+2\gamma)r_s - 6\gamma r\right]^2\,.
\end{align}
From the power of $f(r)$ in each of the three curvature invariants, it is clear that the geometry has a curvature singularity at $r= r_s$ where each invariant blows up unless $\gamma = 1$.
The line element describes therefore the geometry for $r\in ]r_s, + \infty[$. Notice that when $\gamma=1$, the above expressions reduce as expected to the standard Schwarzschild case, the three curvature invariants are finite at $r=r_s$ and the singularity is located only at $r=0$. See \cite{Virbhadra:1995iy, Virbhadra:1998dy, Virbhadra:1998kd, Harada:2001nj} for details. 

Let us now show the absence of horizon for this geometry. There are several ways to identify the presence of horizons in a given geometry, the most direct one being to compute the expansions of (twist free) nulls rays. In spherical symmetry, this is encoded in the norm of the so-called Kodama vector. The Kodama vector is an inbuilt property of any spherically symmetric spacetime geometry. The key point for our purposes is that the vanishing of its norm signals the existence of a trapping or anti-trapping horizon. See \cite{Kodama:1979vn, Abreu:2010ru, Faraoni:2016xgy} for details on the Kodama vector. Computing the norm of the Kodama vector, we find that
\begin{align}
K_{\alpha} K^{\alpha} =  - \frac{1}{r^2 f}  \left[ r- \left( 1+ \gamma\right) \frac{r_s}{2}\right]^2\,.
\end{align}
It is clear that for $0\leqslant\gamma < 1$, the Kodama vector remains time-like in the whole region $r \in [ r_s, + \infty [$ and thus there is no horizon. However, for $\gamma=1$, one finds an horizon at $r=r_s$, which corresponds to the standard Schwarzschild event horizon as expected. Let us point out additionally that the scalar field satisfies the weak energy condition. Using the time-like vector
\begin{align}
u^{\alpha} \partial_{\alpha} =\frac{1}{ \sqrt{ f^{\gamma}}} \partial_t \qquad \text{with} \qquad  g_{\alpha\beta} u^{\alpha} u^{\beta} =-1
\end{align}
the energy density associated to the scalar field and measured by this observer satisfies
\begin{align}
T_{\mu\nu} u^{\mu} u^{\nu} = \frac{1}{2} X(r) \geqslant 0 \qquad \forall \;\; r \geqslant r_s
\end{align}
It is worth pointing out that while this solution provides a simple example of a naked singularity in GR, it is nevertheless not a counterexample to the cosmic censorship conjecture which apply to dynamical collapse. Finally, let us mention that the properties of the JNW solution have been investigated in details, in particular its gravitational lensing, the motion of particle around it, as well as its stability under scalar perturbations \cite{Virbhadra:2002ju, Chowdhury:2011aa, Sadhu:2012ur, Shaikh:2019hbm}. Other static solutions of the Einstein-Scalar system are known. See for example \cite{Cadoni:2015gfa, Cadoni:2015qxa, Cadoni:2018pav, Yu:2020bxd}.

However, relaxing the assumption of an asymptotic time-like Killing vector, one inherits few exact solutions which describes the formation of dynamical apparent horizons and thus trapped (or anti-trapped) regions embedded in an inhomogeneous cosmology \cite{Husain:1994uj, Fonarev:1994xq}. See \cite{Faraoni:2018xwo} for a review on these types of solutions. Moreover, the Einstein-Scalar system have been a crucial toy model in investigating the process of gravitational collapse. It led in particular to the Choptuik's conjecture \cite{Choptuik:1992jv} which now plays a key role in our understanding of the conditions under which a small mass black hole (such as primordial black hole) can form. See \cite{Gundlach:2007gc} for a review on this last point.
Now let us describe the rotating case. 

\subsection{Axi-symetric configurations}

Contrary to the spherically symmetric case, rotating scalar vacuum solutions have revealed challenging to derive in GR. The main strategy employed in the literature has been to implement the Janis-Newman trick which successfully generates the Kerr solution out of the Schwarzschild one in vacuum GR \cite{Newman:1965tw}. See \cite{Erbin:2016lzq} for a review on this algorithm. However, this trick is not harmless when matter is present, and several misleading results have been published over the last decades \cite{Agnese:1985xj}. See \cite{Pirogov:2013wia, Hansen:2013owa} for a critical analysis of this approach. A careful analysis was undertaken using different and more powerful methods by Bogush and Gal'tsov who provided a new exact rotating scalar vacuum solution \cite{Bogush:2020lkp}. More recently, Mirza, Kangazian and Sadeghi have succeeded to derive a four parameters family of scalar vacuum solutions\footnote{See also \cite{Azizallahi:2023rrv}} \cite{Mirza:2023mnm}. This family encompasses the BG solution and for the first time the rotating version of the JNW geometry, reducing to the Kerr family when the scalar field vanishes. In the following, we present this family. Using the coordinates $(t, \phi, x, y)$, the metric of this family of scalar vacuums takes the form
\begin{align}
\rd s^2 = - f (\rd t - \omega \rd \phi)^2 + \frac{\sigma^2}{f} \left[ e^{2\eta} (x^2 - y^2) \left( \frac{\rd x^2}{x^2-1} + \frac{\rd y^2}{1-y^2}\right) + (x^2-1) (1-y^2) \rd \phi^2\right]
\end{align}
where $\sigma$ is a positive constant and the functions $(f, \omega, \eta )$ depend solely on $(x,y)$. These coordinates are related to the standard spherical coordinates $(r, \theta)$ via
\begin{align}
x = \frac{r-m}{\sigma} \qquad y = \cos{\theta}
\end{align}
The functions $(f, \omega, \eta )$ can be decompose as
\begin{align}
f = \frac{A}{B} \qquad \omega = - 2 \left( a + \frac{\sigma C}{A}\right) \qquad e^{2\eta} = \frac{A}{4 (x^2 -1)^{1+q}} \left( 1+ \frac{m}{\sigma}\right)^2 \left( \frac{x^2-1}{x^2-y^2}\right)^{1-\nu}
\end{align}
where $(m, a, q, \nu)$ are some real parameters which will be discussed in a moment and the functions $(A,B,C)$ are given by
\begin{align}
A & = c_{+} c_{-} + b_{+} b_{-} \\
B & = c^2_{+} + b^2_{+} \\
C & = (x+1)^q \left[ x (1-y^2) (\lambda + \xi) c_{+} + y (x^2 -1) (1-\lambda \xi) b_{+}\right]
\end{align}
where $(c_{\pm}, b_{\pm})$ are given by
\begin{align}
c_{\pm}  = (x\pm 1)^q \left[ x (1-\lambda \xi) \pm (1+\lambda \xi)\right] \qquad b_{\pm}  = (x\pm 1)^q \left[ y (\lambda + \xi) \mp (\lambda - \xi)\right] 
\end{align}
while $(\lambda, \xi, \alpha, \sigma)$ read
\begin{align}
\lambda  = \alpha (x^2 -1)^{-q} (x+y)^{2q} \qquad \xi = \alpha (x^2-1)^{-q} (x-y)^{2q} \qquad \alpha  = - \frac{a}{m+\sigma} \qquad \sigma = \sqrt{m^2 -a^2}
\end{align}
Finally, the scalar field profile is given by
\begin{align}
\varphi(r) & = \sqrt{\frac{1-\gamma^2 - \nu}{16\pi}} \log\left[ \frac{r-m - \sqrt{m^2 -a^2}}{r-m + \sqrt{m^2 -a^2}}\right]
\end{align}
and the parameter $q$ is related to $\gamma$ by $q = \gamma -1$.
The different parameters of the solutions have to satisfy 
\begin{align}
m \geqslant  a \qquad S =1-\gamma^2 - \nu \geqslant 0
\end{align}
This family of exact scalar vacuum solutions has several interesting limits which we now describe. For now on, we always consider $a\neq 0$.
\begin{itemize}
\item For $\nu= 0$ and $\gamma=1$, the scalar charge $S$ vanishes and one recovers as expected the Kerr black hole solution of vacuum GR.
\item For $\gamma =1$ but leaving $\nu$ a free parameter, one recovers the Bogush-Gal'tsov scalar vacuum solution.
\item For $\nu =0$ but $\gamma \neq 1$ a free parameter, one recovers the rotating version of the Janis-Newman-Winicour solution described in the previous section.
\end{itemize}
This family of exact solutions is to date the most general one describing the gravitational field of a minimally coupled self-interacting scalar field. By inspection, one can easily show that for $S\neq 0$, all the solutions describe naked singularities, in agreement with the no-hair theorem. Recently, these rotating scalar vacuum solutions have been understood in a systematic way and a self-complete presentation have been presented in \cite{Barrientos:2025abs}. This concludes the review of the scalar vacuum solutions within GR when there is no self-interaction\footnote{ Let us also mention the case of the conformally coupled scalar field. The spherically symmetric BBMB solution found initially in \cite{Bocharova:1970skc, Bekenstein:1974sf} was extended to a rotating one in \cite{Astorino:2014mda}.}. Let us now discuss the no-hair theorems derived in modified gravity.

\section{No scalar hair theorems}

In view of the power of the no-hair theorem in 4d GR, it is natural to wonder whether similar uniqueness theorems can be derived in modified gravity theories. This is indeed possible and we shall briefly review in this section the two main no-hair theorems. The key point is that they serve a guiding map to build new hairy black holes solutions and study their phenomenology. Notice that contrary to GR, the way one formulate these theorems is slightly different. Since GR is the current paradigm for the theory of the gravitational field, the question is whether given some generic assumptions, the end-points of gravitational collapse in modified gravity can be described by other black hole configurations than the Kerr black hole ? As we shall see, this possibility is highly constrained.

\subsection{Brans-Dicke and its extensions}

We start by reviewing the no-hair theorem derived for the simplest scalar-tensor theories generalizing the Brans-Dicke theory\footnote{A modern classification of the four classes of solutions in Brans-Dicke theory can be found in \cite{Faraoni:2016ozb}}. Since this family of scalar-tensor theories is related to the canonical Einstein-Scalar system by a suitable conformal field redefinition, one can use this trick to explore the solution space of these theories. Therefore, one can first ask the question whether the canonical Einstein-Scalar system with a self-interacting potential can support other black hole configuration than the Kerr black hole with a trivial constant scalar field. Then, using this result, one can translates this conclusion thanks to the simple conformal field redefinition relating the two set of theories. The case of the canonical Einstein-Scalar system was first tackled by Bekenstein in 1972 in a series of works\footnote{See \cite{Penney:1968zz, Chase:1970omy} for earlier results in this direction.} \cite{Bekenstein:1971hc, Bekenstein:1974sf} and further generalized by Hawking to Brans-Dicke theory the same year \cite{Hawking:1972qk}. The generalization to scalar-tensor theories conformally related to the canonical Einstein-Scalar was undertaken by Sotiriou and Faraoni in \cite{Sotiriou:2011dz}. In the following, we briefly review this last step. Consider the canonical Einstein-Scalar system with a self-interacting potential. It is described by the action
\begin{align}
S_{\text{ES}}[\phi, g] =  \int \rd^4x \sqrt{|g|} \left( \frac{\mathcal{R}}{16\pi \bar{G}} - \frac{1}{2} g^{\mu\nu} \partial_{\mu} \phi \partial_{\nu} \phi - V(\phi)\right) 
\end{align}
The only equation we need to the proof is the scalar equation of motion which reads
\begin{align}
\label{eoom}
\Box \phi = V'(\phi)
\end{align}
We are interested in stationary axi-symmetric black hole configurations which could describe the end-points of gravitational collapse in this system. Such geometry possess two Killing vectors:  $\xi^{\mu}\partial_{\mu}$ which is asymptotically time-like and which generates stationarity, and $\chi^{\mu} \partial_{\mu}$ which is space-like and generates the axi-symmetry.  This configuration is assumed to be asymptotically flat, which requires that at spatial infinity, the metric approaches Minkowski while the scalar field settles down to a constant value, i.e. $\phi =\phi_0$. Plugging this in (\ref{eoom}) implies that at spatial infinity, one has $V_{\phi}(\phi_0) =0$, i.e. the scalar field is in a local extremum of its potential. Let us multiply the e.o.m (\ref{eoom}) by $V_{\phi}(\phi)$ and integrate over a given region of spacetime of volume $\mathcal{V}$, one obtains
\begin{align}
\int_{\mathcal{V}} \sqrt{|g|} V_{\phi}  \Box \phi = \int_{\mathcal{V}} \sqrt{|g|} V^2_{\phi}  
\end{align} 
An integration by part allows one to recast this equality as
\begin{align}
\label{eqq}
\int_{\mathcal{V}} \sqrt{|g|} \left[ V_{\phi \phi} X + V^2_{\phi} \right]  = \oint_{\partial \mathcal{V}} \sqrt{|h|} V_{\phi}  \mathcal{L}_s \phi 
\end{align}
where $s^{\mu} \partial_{\mu}$ is the normal to the boundary of the region $\partial \mathcal{V}$. Let us choose the region of integration to extend from the black hole horizon to the asymptotically flat region such that its boundary $\partial \mathcal{V}$ is the union of two partial Cauchy slices $\Sigma_1$ and $\Sigma_2$ with the black hole horizon $\mathcal{N}$ and a time-like hypersurface $\mathcal{T}$. Concretely, one can split the r.h.s of (\ref{eqq}) as
 \begin{align}
 \oint_{\partial \mathcal{V}} =  \int_{\Sigma_1} + \int_{\Sigma_2}  + \int_{\mathcal{T}} + \int_{\mathcal{N}}
 \end{align}
 Then, the first two integrals compensate each other. The third integral on $\mathcal{T}$ vanishes since we have assumed that in the asymptotically flat region, the scalar field is a constant such that $ \mathcal{L}_s \phi |_{\mathcal{T}}=0$. Finally, the rigidity theorem implies that the black hole horizon $\mathcal{N}$ is a Killing horizon. Thus the normal vector $s^{\mu} \partial_{\mu}$ is null on $\mathcal{N}$ and that it is a symmetry generator, such that $ \mathcal{L}_s \phi |_{\mathcal{N}}=0$. Therefore, the r.h.s of (\ref{eq}) vanishes and one obtains the constraint
 \begin{align}
\label{eq}
\int_{\mathcal{V}} \sqrt{|g|} \left[ V_{\phi \phi} X + V^2_{\phi} \right]  =0 
\end{align}
Let us analyze the integrand. One has bviously $V^2_{\phi} \geqslant 0$. Then, the symmetry of the scalar field, i.e. $\xi^{\mu} \nabla_{\mu} \varphi =0$, imposes that the gradient of the scalar field is either a spacelike or a null vector, showing that $X \geqslant 0$. Now, if one demands that within the region $\mathcal{V}$, there are no unstable extremum of the potential, it imposes that 
\begin{align}
V_{\phi \phi} \geqslant 0
\end{align}
Then, (\ref{eq}) is satisfied only for a trivial constant scalar profile, i.e. $\phi=\phi_0$ everywhere, and  $V_{\phi} (\phi_0)=0$. Under this assumption, the only stationary black hole with a stable scalar field supported by this system is the standard Kerr solution with a constant scalar field. How does this no-hair theorem translates for scalar-tensor theories ?

Consider the generalized Brans-Dicke theories of gravity described by the action
\begin{align}
\label{ST-BD}
S_{\text{ST}}[\varphi, \tilde{g}] =  \int \rd^4x \sqrt{|\tilde{g}|} \left( \varphi \tilde{\mathcal{R}} - \frac{\omega(\varphi)}{\varphi}\tilde{X} - U(\varphi)\right) 
\end{align}
where $\varphi$ is the scalar field, $\tilde{g}_{\mu\nu}$ is called the Jordan frame metric and $\tilde{X} = \tilde{g}^{\mu\nu} \partial_{\mu} \varphi \partial_{\nu} \varphi$. The equation of the scalar field is given by
\begin{align}
\label{eooom}
(2\omega +3) \tilde{\Box}\varphi = - \omega_{\varphi} \tilde{X} + \varphi U_{\varphi} - 2 U
\end{align}
It is related to the canonical Einstein-Scalar system by the following field redefinition
\begin{align}
g_{\mu\nu} =  \varphi \tilde{g}_{\mu\nu} \qquad \rd \phi = \sqrt{\frac{2\omega(\varphi) +3}{16\pi \bar{G}}} \frac{\rd \varphi}{\varphi} \qquad V(\phi) = \frac{U(\varphi)}{\varphi^2}
\end{align}
Let us assume that $\omega(\varphi) \neq -3/2$ and that the relation $\phi(\varphi)$ is well behaved.
Now, one can easily check that 
\begin{align}
V_{\phi} = \sqrt{\frac{2\omega (\varphi) + 3}{16\pi }} \varphi^{-1} \left[ \varphi U_{\varphi} - 2 U\right]
\end{align}
Therefore, demanding a black hole configuration which is asymptotically flat in these scalar-tensor theories requires that at spatial infinity, the metric approaches Minkowski and that the scalar field settles down to a constant, i.e. $\varphi = \varphi_0$. From (\ref{eooom}), it follows that in the asymptotically flat region, one has 
\begin{align}
 \varphi_0 U_{\varphi}(\varphi_0) - 2 U(\varphi_0) =0
\end{align}
which is equivalent to $V_{\phi} (\phi_0) =0$ in the Einstein frame where $\phi(\varphi_0)= \phi_0$. But we just proof that within the Einstein frame, any asymptotically flat stable configuration of that kind is nothing else than the Kerr solution endowed with a trivial constant scalar field everywhere. Therefore, since the metrics in the two frames are related through $g_{\mu\nu} =  \varphi_0 \tilde{g}_{\mu\nu}$, we conclude that the only asymptotically flat stationary black hole with a Killing horizon admitted by these generalized scalar-tensor theories is the Kerr black hole with a constant scalar field. .

The same strategy can be adopted to generalize the above no-hair theorem to scalar-tensor theories with non-linear kinetic terms dubbed K-essence theories. A first attempt was presented by Bekenstein in \cite{Bekenstein:1995un} although with some restrictions on the applicability of the resulting theorem. Much later, Graham and Jha revisited the proof, generalizing its results to asymptotically flat stationary black holes \cite{Graham:2014mda}. We do not present this proof since in the next section, we shall review a much more general theorem which apply to all shift symmetric DHOST theories.

\subsection{DHOST gravity}

The first no-hair theorem derived for a higher-order scalar tensor theory was provided by Hui and Nicolis in the context of Horndeski gravity \cite{Hui:2012qt}. Let us briefly review this theorem before discussing a more recent generalization. Consider the Horndeski action
\begin{align}
S_{\text{Horn}} = \int \rd^4 x \sqrt{|g|} \left( L_2 + L_2 + L_3 + L_4+ L_5\right)
\end{align}
where
\begin{align}
L_2 & = G_2(\varphi, X) \\
L_3 & = - G_3 (\varphi, X) \Box \varphi \\
L_4 &= G_4(\varphi, X) \mathcal{R} + G_{4X} \left[ (\Box \varphi)^2 -\varphi_{\mu\nu} \varphi^{\mu\nu}\right] \\
L_5 & = G_5(\varphi, X) G_{\mu\nu} \phi^{\mu\nu} - \frac{1}{6} G_{5X} \left[ (\Box \varphi)^3 - 3 \Box \varphi \phi_{\mu\nu} \phi^{\mu\nu} + 2 \phi_{\mu\nu} \phi^{\nu\rho} \phi_{\rho}{}^{\mu}\right]
\end{align} 
The theorem can be stated as follows. \\

\textbf{No hair theorem for Horndeski gravity: } \\ \textit{Consider a spacetime $(M, g)$ ruled by a Horndeski theory. Provided
\begin{itemize}
\item i) The geometry is static, spherically symmetric and asymptotically flat and contains a black hole horizon
\item ii) The scalar field shares the same symmetry as the geometry
\item iii) The Horndeski action is shift-symmetric
\item iv) The norm of the Noether current associated to the shift symmetry $J_{\mu}$ is finite on the horizon 
\item v) The free functions $G_i(X)$ are analytic at the point $X=0$
\item vi)  is a canonical term $X \subseteq G_2$
\end{itemize}
then the scalar profile $\varphi$ is constant everywhere and the only black hole solution is locally isometric to the Schwarzschild geometry.} \\

Early efforts to identify hairy configurations in Horndeski theories led to a whole new class of black hole solutions supporting non-trivial scalar profiles \cite{Rinaldi:2012vy, Babichev:2012re, Anabalon:2013oea, Minamitsuji:2013ura, Kobayashi:2014eva, Charmousis:2014zaa}. Yet , the above theorem naturally provides a guiding map to construct and explore hairy black hole in Horndeski theories in a systematic way. A rather natural way to bypass this no-hair theorem is to relax condition ii), allowing for example the scalar field to be time dependent. This strategy was explored in \cite{Babichev:2013cya} leading a whole new set of black holes solutions with non-trivial and regular scalar configurations. In particular, focusing on a profile linear in time, it was possible to construct a Schwarzschild black hole dressed with a non-gravitating scalar field. This configuration dubbed \textit{stealth} has played an important role in that it allows one to mimic exactly the unique spherically symmetric solution of GR while still allowing for a non-trivial scalar profile. A systematic investigation of how to explicitly break each of the assumptions of the Hui-Nicolis theorem was undertaken in \cite{Babichev:2017guv}, allowing the authors to identify different static and asymptotically flat black holes solutions, among which new stealth configurations. In particular, new black holes solutions were found for theories which do not satisfy conditions v) or condition vi). The interested reader can refer to the following reviews and thesis \cite{Babichev:2016rlq, Lehebel:2018zga, Babichev:2022awg, Bakopoulos:2022gdv, Babichev:2023psy, Lecoeur:2024kwe}. Now let us briefly discuss some of the extensions of the above theorem.

Indeed, despite its interesting role as a guiding map to construct new hairy black holes, the Hui-Nicolis theorem is rather restrictive, applying only to shift symmetric (beyond) Horndeski theories and to static, spherically symmetric and asymptotically flat black holes.  Yet, the key target is to explore and identify exact rotating hairy black hole configurations in modified gravity in order to confront them to the current and future observations. Moreover, as reviewed in the first chapter, the Horndeski theories are not the only healthy higher-order scalar-tensor theories one can construct as the landscape of DHOST theories is now well understood. This begs for a generalization of the above theorem including stationary and axi-symmetric black holes and applying to the DHOST landscape. A first extension\footnote{See \cite{Barausse:2017gip,Barausse:2015wia} for earlier works in this direction. } partially fulfilling these objectives was recently discussed by Capuano, Santoni and Barausse in \cite{Capuano:2023yyh}. In the following, we briefly review the main arguments leading to this recent no-hair theorem for DHOST gravity.

Consider the shift symmetric DHOST action. The shift symmetry $\varphi \rightarrow \varphi + c$ ensures that the scalar field equation can be recast into the local conservation of a vectorial current, i.e.  $\nabla_{\mu} J^{\mu} =0$. Integrating over a region of spacetime of volume $\mathcal{V}$, one obtains
\begin{align}
\label{bound}
\int_{\mathcal{V}} \sqrt{|g|} \nabla_{\mu} J^{\mu} = \oint_{\partial \mathcal{V}} \sqrt{h} \; \chi^{\mu} J_{\mu} = 0
\end{align}
where $\chi^{\mu} \partial_{\mu}$ is the normal to the boundary of the region of integration. Now, consider an circular, asymptotically flat, stationary and axi-symmetric spacetime for which the most general line-element is given by the Weyl-Papapetrou form
\begin{align}
\rd s^2 = - f \left( \rd t - \Omega \rd \phi\right)^2 + f^{-1} \left(  g (\rho^2 + \rd z^2) + \rho^2 \rd \phi^2\right)
\end{align}
The functions $(f, \Omega, g)$ depend only on the coordinates $(\rho, z)$. The geometry has therefore two commuting Killing vectors $\xi^{\mu} \partial_{\mu} = \partial_t$ and $\zeta^{\mu} \partial_{\mu} = \partial_{\phi}$. Let us assume that the spacetime contains black hole horizon $\mathcal{N}$. Choosing the region $\mathcal{V}$ of integration extending from the black hole horizon $\mathcal{N}$ to a timelike hypersurface $ \mathcal{T}$ in the asymptotically flat region, the boundary integral (\ref{bound}) can again be split as
 \begin{align}
 \oint_{\partial \mathcal{V}} =  \int_{\Sigma_1} + \int_{\Sigma_2}  + \int_{\mathcal{T}} + \int_{\mathcal{N}}
 \end{align}
 where $\Sigma_{1,2}$ are two partial Cauchy slices intersecting the black hole horizon $\mathcal{N}$ and $ \mathcal{T}$. Now, one has to analyze the scalar product $\chi^{\mu} J_{\mu}$ on each of the four hypersurface.
 The key result obtained in \cite{Capuano:2023yyh} is that provided the scalar field admits the same symmetries as the metric, i.e. $\mathcal{L}_{\xi} \varphi = \mathcal{L}_{\chi} \varphi =0$, then one can show that for any shift symmetric DHOST theories, one has
 \begin{align}
 J_t = J_{\phi} =0
 \end{align}
 This property dramatically simplifies the different contributions. 
 \begin{itemize}
 \item On the two partial Cauchy slices, the unit normal vector $\xi^{\mu} \partial_{\mu}$ is timelike and given by $\chi^{\mu} = \delta^{\mu}{}_t$, such that the scalar product $\chi^{\mu} J_{\mu} = J_t =0$. Therefore, both contributions vanish.
 \item To analyze the contribution coming from the black hole horizon, let us slightly deviate from the argument presented in \cite{Capuano:2023yyh}. Assuming that the black hole horizon $\mathcal{N}$ is Killing\footnote{The rigidity theorem implies that this is guaranteed in GR, but also in effective theory \cite{Hollands:2022ajj}. However, there is no prof of the theorem for higher order scalar-tensor theories, therefore, we introduce this assumption for safety.}, it follows that the normal vector to $\mathcal{N}$ coincides with a linear combination of the Killing vectors $\xi^{\mu} \partial_{\mu} = \partial_t$ and $\zeta^{\mu} \partial_{\mu} = \partial_{\phi}$. This automatically implies that $\chi^{\mu} J_{\mu} =0$ on $\mathcal{N}$. 
\end{itemize}
At this stage, one is left with the last contribution coming from the asymptotically flat region. Using the assumption of asymptotic flatness, the unit normal vector to the timelike hypersurface $\mathcal{T}$ is given by $\chi^{\mu} \partial_{\mu} = \partial_r$. One can further assume that in this region, the conserved current takes the form $J_{\mu} = -  \partial_{\mu} \varphi$. Expanding the scalar field in multipoles, i..e writing $\Psi = \sum_{\ell} \Psi_{\ell}(r) P_{\ell}(\cos{\theta})$ where $P_{\ell}$ are the Legendre polynomials and $\Psi_{\ell}(r) = a_{\ell} /r^{\ell+1}$, one obtains
\begin{align}
\oint_{ \mathcal{T}} \sqrt{h} \; \chi^{\mu} J_{\mu} = r^2  \int^{t_1}_{t_0} \rd t \oint \rd \Omega \; \chi^{r} J_{r} = 4\pi (t_1-t_0) a_0 =0 
\end{align}
where we have neglected all the higher multipoles. Therefore, under the above assumptions, no monopole scalar charge can be supported by the black hole geometry. Let us summarize this statement. \\

\textbf{No-monopole-scalar-charge theorem:} \\\textit{Consider a spacetime $(M, g)$ ruled by a DHOST theory. Provided  
\begin{itemize}
\item The geometry is stationary, axi-symmetric and circular
\item The scalar field shares the same symmetry as the geometry
\item The DHOST action is shift-symmetric
\item The Noether current associated to the shift symmetry $J_{\mu}$ reduces to $J_{\mu} = -\partial_{\mu} \varphi$ in the asymptotically flat region
\item The black hole horizon contained in the geometry is Killing / or the norm of the conserved current $J_{\mu} J^{\mu}$ is finite on the horizon
\end{itemize}
then the monopole scalar charge vanishes. }\\

Le us briefly expand on why the last assumptions appears in two different versions. The authors of the theorem have used the second version to prove the above theorem. The interesting point is that if one can show that a version of the rigidity theorem holds in shift symmetric DHOST gravity (which is an open question to my knowledge even if extension of the rigidity theorem to higher order effective field theory exists \cite{Hollands:2022ajj}), namely that if a spacetime is stationary and axi-symmetric and if it contains a black hole horizon, then this horizon has to be Killing, one can remove the step invoking the finiteness of $J_{\mu} J^{\mu}$ in the proof presented in \cite{Capuano:2023yyh}. Moreover, if such extended version of the rigidity theorem exists, the first version of the last assumption becomes a consequence of the very first assumption, lowering the numbers of assumptions required for the proof. 

Let us further comment on this last theorem. First, just as the Hui-Nicolis theorem, the Capuano-Santoni-Barausse theorem is again restricted to shift symmetry. Second, it requires the stationary and axi-symmetric spacetime to be circular. While the generalized Papapetrou theorem \cite{ Papapetrou:1966zz} ensures that any stationary and axi-symmetric vacuum solution of GR is indeed circular, this is no longer true beyond GR. The assumption of circularity is thus an important restriction. See \cite{Gourgoulhon:2010ju} for a detailed discussion of the notion of circularity. Finally, let us point that while the theorem indeed applies to rotating black holes and to DHOST, it is also weaker than the Hui-Nicolis theorem in that the above result does not prevent the scalar field to exhibit subleading multipole charges. Therefore, the scalar profile does not have to be a constant everywhere.

Before going to the next section, let us summarize the three main points learned from the review of the no-hair theorems derived so far. The three key paths to derive new hairy black hole solutions bypassing the above theorems are
\begin{itemize}
\item Allowing the scalar profile to break the symmetry of the black hole geometry 
\item Giving up the assumption of shift-symmetry
\item For stationarity and axi-symmetric back holes, giving up the assumption of circularity
\end{itemize}
In the next chapters, we shall present exact black holes solutions which do break all these assumptions, illustrating the different (or combined) ways to bypass these no-hair theorems.

\chapter{On spherically symmetric hairy black holes}
\label{Chapter4}

\textit{"Si tu mets un drift, spectralement, ca passe!"\\ 
Entendu à la machine à caf\'e}
\bigskip

\minitoc

In the previous chapter, we have reviewed the key no-hair theorems constraining the black hole solutions in Horndeski and DHOST theories. They are also guiding maps to construct new hairy black holes solutions within these theories. To a large extend, these new solutions have been obtained either by allowing the scalar field to be time-dependent or by introducing a coupling to the Gauss-Bonnet term. See \cite{Babichev:2016rlq, Lehebel:2018zga, Babichev:2022awg, Bakopoulos:2022gdv, Babichev:2023psy, Lecoeur:2024kwe} for reviews.
In exploring the spherically symmetric sector of the solution space of DHOST theories, one can distinguish between stealth configurations, i.e. black holes which mimic exactly the GR solutions supporting nevertheless a non-trivial scalar profile, and the non-stealth ones whose geometry are deformed w.r.t the GR solutions. 

The stealth sector has been extensively studied, both for the Schwarzschild and Kerr cases,  and the conditions on the DHOST lagrangian to exhibit such solutions have been identified by various techniques \cite{Motohashi:2016ftl, Motohashi:2018wdq, BenAchour:2018dap, Takahashi:2020hso}. In the following, we first review one such algorithm and its application to  non shift symmetric DHOST theories following \cite{BenAchour:2018dap}. 

Beyond the stealth sector, the exploration of the solution space becomes challenging. A simple and systematic approach to construct new non-stealth solutions of DHOST gravity relies on the structure of the DHOST theory space whose degeneracy classes are covariant under disformal transformations. Thus, it appears natural to take advantage of this property and consider the disformal transformation as a solution generating map to explore the non-stealth sector of the solution space. This point of view was adopted systematically in \cite{BenAchour:2020wiw} and we shall review some of the results obtained in this work. 

We close this chapter by a new proposal to construct black holes which are not exactly but only \textit{asymptotically} stealth. This proposal has not been published but it provides a powerful and simple trick to obtain asymptotically flat black hole geometries mimicking GR solutions both in the far region and in the near horizon, while still allowing a deformation of the geometry in region whose size is controlled. 

Finally, let us stress that several important subjects won't be addressed in this chapter, namely the question of stability against perturbations, the screening mechanism and the new DHOST phenomenology found for stars. The reader is referred to the references provided at the end of the chapter.

\section{Stealth black holes beyond shift symmetry}

\label{A}

Since the field equations of DHOST theories are quite complicated, one needs to find suitable algorithms to explore the solution space. A powerful one was presented in \cite{Babichev:2016kdt}. It allows one, given some known geometry, to determine the DHOST lagrangian which admits this geometry as solution. In the following we apply this algorithm to stealth black holes and the DHOST theories beyond shift symmetry. This section is based on \cite{BenAchour:2018dap}. 

\subsection{From the solution to the lagangian}

Consider the family of DHOST theories given by the action
\begin{align}
\label{ACTION}
 S_{\text{vDHOST}} [g, \phi] & =  \int d^4x \sqrt{|g|} \;\sum_I \mathcal{L}^I \left( g, \phi \right)
\end{align}
where the different lagrangians read
\begin{align}
 \mathcal{L}_2 & = P(\phi, X) \\
\mathcal{L}_3 & = Q \left( \phi, X\right) \Box \phi \\
\mathcal{L}_4 & = F \left( \phi, X\right) \mathcal{R} \\
\mathcal{L}_5 & = A_3 \left( \phi, X\right)  \phi^{\mu} \phi^{\nu} \phi_{\mu\nu} \Box \phi+ A_4 \left( \phi, X\right) \phi^{\mu} \phi_{\lambda} \phi_{\mu\nu} \phi^{\nu\lambda} + A_5( \phi, X) \left( \phi_{\mu\nu} \phi^{\mu} \phi^{\nu}\right)^2 
\end{align}
where the six potentials $\left(P, Q, F, A_I \right)$ with $I \in \{ 3,4,5 \}$ are free functions of $\phi$ and its kinetic term $X$. We have adopted the notation of \cite{Langlois:2017dyl}. This family of theories corresponds to the quadratic DHOST theories amputated from the lagrangians $\mathcal{L}_{1,1} =\left(\Box \phi\right)^2$ and $\mathcal{L}_{1,2} = \phi_{\mu\nu} \phi^{\mu\nu}$, namely $A_1 = A_2 =0$ in the standard notation \cite{Langlois:2015cwa}.

This class of DHOST theories can be made consistent with the recent observational constraint from GW170817 which imposes that the speed of gravitons equal the speed of light (up to deviations of order $10^{-15}$), at least on cosmological scales \cite{Langlois:2017dyl}. In order to satisfy this constraint, the last two functions $A_4( \phi, X)$ and $A_5\left( \phi,  X\right)$ are related to $F( \phi, X)$ and $A_3\left( \phi,  X\right)$ through
\begin{align}
\label{A4}
& A_4 = \frac{1}{8F} \left( 48 F^2_{X} - 8(F - X F_{X}) A_3 - X^2 A^2_3\right)  \\
\label{A5}
& A_5 = \frac{1}{2F} \left( 4F_{X} + X A_3\right) A_3
\end{align}
The potentials $A_3( \phi, X)$ and $F\left(\phi,   X\right)$ remain free functions, and the viable DHOST theories contain thus only four free potentials $(P, Q, F, A_3)$.


In order to solve the field equations, we adopt the elegant strategy presented in \cite{Babichev:2016kdt}. Starting from the model (\ref{ACTION}), we derive the field equations that we write in a compact way
\begin{align}
\delta \mathcal{L} = \mathcal{E}^{(g)}_{\alpha\beta} \delta g^{\alpha\beta} + \mathcal{E}^{(\phi)} \delta \phi
\end{align}
The field equations being rather complicated, we do not write them explicitly here. Instead, the equation of motion with respect to the metric $g_{\alpha\beta}$ can be written in the simple form
\begin{align}
\label{EOM1}
 F  G_{\alpha\beta} & = \mathcal{T}_{\alpha\beta} \\
 \label{EOM2}
\left( Q_{\phi} -  Q_{X} -  P_{X}\right) \Box \phi & = \zeta  \;\;\; \; 
\end{align}
where $G_{\alpha\beta}$ is the standard Einstein tensor and where $\mathcal{T}_{\alpha\beta}$ and $\zeta$ account for all the other terms obtained from the variation of the action respectively w.r.t $g_{\alpha\beta}$ and $\phi$, containing therefore all the higher order terms. 

Looking for stealth black hole solutions implies that the scalar field does not gravitate. This can be translated in (\ref{EOM1}) by $\mathcal{T}_{\alpha\beta} =0$. Then, one can solve the l.h.s of the equation of motion using a GR black hole solution such that $G_{\alpha\beta}=0$. A common strategy is to assume for example a constant kinetic term for the scalar field profile, such that $X= X_{\ast}$. Then, under some specific conditions on the potentials of the Lagrangian, the effective energy momentum tensor $\mathcal{T}_{\alpha\beta}$ can be written as
\begin{align}
\label{emt}
\mathcal{T}_{\alpha\beta} = f(X) T_{\alpha\beta}
\end{align}
such that $f(X_{\ast})=0$, and example of which being $f(X) = \log{\left(X_{\ast}/X\right)}$. Below, we should derive the condition on the DHOST lagrangian (\ref{ACTION}) to admit such stealth black hole solutions. We consider therefore a static spherical symmetric metric which reads
\bea
ds^2 = - e^{\nu(r)} dt^2 + e^{\lambda(r)} dr^2 + r^2 d\Omega^2
\eea
and we choose a linear time-dependent profile for the scalar field
\begin{align}
\label{lin}
\phi(t,r) =  \dot{\phi}_c t + \psi(r)
\end{align}
where $\dot{\phi}_c$ is assumed to be a constant. Following \cite{Babichev:2016kdt}, we introduce the notation $\dot{\phi}_c = M q$ which implies that the kinetic term reads
\begin{align}
X = - \frac{g^{\alpha\beta} \partial_{\beta} \phi \partial_{\alpha} \phi}{M^2} =  q^2 e^{-\nu}  - e^{-\lambda} \frac{\left(\psi'\right)^2}{M^2}
\end{align}
In order to further (drastically) simplify the field equations, \textit{we also assume that the kinetic term is constant everywhere, such that $X = X_{\ast} = q^2$}. Notice that the kinetic energy $-M^2X$ is negative since the gradient of the scalar field is a time-like vector. With this assumption, all the unknown potentials, commonly denoted $f(\phi, X)$, can now be written as function
\begin{align}
f \left( \phi, X\right) = f \left( qt + \psi(r), X_{\ast} \right)
\end{align}
In the following, we restrict further to potentials $f(\phi, X)$ satisfying
\begin{align}
\label{condition}
f_{\phi} (\phi, X_{\ast}) =0 \Rightarrow \;\; \partial^n_{\phi} f(\phi, X_{\ast}) =0 \;\;\;\; \forall n \in \mathbb{N}
\end{align}
This will allow us to simplify our conditions in the beyond shift symmetric case. Notice that while this condition is quite general, it is still possible to find counter-example in principle, and thus, we are potentially restricting the set of allowed potentials.

The third simplification, inherited from the assumption of a constant kinetic term, lies in that the radial dependent part of the scalar field is given by
\begin{align}
\label{psi}
\psi' = M q \sqrt{ e^{\lambda} \left( 1 + e^{-\nu} \right) }
\end{align}
Hence, $\psi'$ is directly known in term of the metric components, as well as its higher order derivatives: $\psi'', \psi'''$ etc. This can be plugged back in the field equations to further simplify the expression.

In the end, the field equations becomes lengthy expressions depending on the radial coordinate $r$. Now these field equations have to be satisfied at any point of space-time, and thus at any couple $(t,r)$. The elegant strategy followed in \cite{Babichev:2016kdt} is to expand the resulting field equations around a given $r_{\ast}$ and check the resulting conditions between the unknown potential $A_I(qt + \psi(r), q^2)$ and their derivatives. Denoting $\epsilon = r - r_{\ast}$, the expansion  of the equations of motion can be written as
\bea
& \mathcal{E}_{tt} (r,t) = \sum^{m}_{n } \mathcal{E}^{(n)}_{tt}(t,r) \big{|}_{r_{\ast}} \epsilon^{n} + \mathcal{O}(\epsilon^m) = 0 \;\;\;\;\;\\
&  \mathcal{E}_{rr} (r,t) = \sum^{m}_{n } \mathcal{E}^{(n)}_{rr}(t,r)\big{|}_{r_{\ast}} \epsilon^{n} + \mathcal{O}(\epsilon^m) = 0 \;\;\;\;\; \\
\label{ephi}
& \mathcal{E}_{\phi} (r,t) = \sum^m_{n} \mathcal{E}^{(n)}_{\phi} (t,r) \big{|}_{r_{\ast}} \epsilon^{n}  + \mathcal{O}( \epsilon^m) = 0 \;\;\;\;\;
\eea
The conditions we obtain out of this procedure are of the form
\begin{align}
& \mathcal{E}^{(n)}_{tt}(t,r) \big{|}_{r_{\ast}} = \mathcal{E}^{(n)}_{rr}(t,r)\big{|}_{r_{\ast}}  = \mathcal{E}^{(n)}_{\phi} (r,t) \big{|}_{r_{\ast}} = 0 
\end{align}
but these conditions are not all independent. Moroever, they are only valid when evaluated at $X= X_{\ast} = q^2$. Once conditions on the potentials $\left( P, Q, F, A_I\right)$ (and their derivatives) are obtained at a given order, we inject them back in the full field equations and expand once more around the same $r_{\ast}$ to obtain new conditions. The algorithm closes when we obtain enough conditions between the potentials such that the full field equations are completely satisfied. 

Notice that this perturbative algorithm is rather general, and especially useful when working with such complicated Lagrangian. Owing to the large freedom in the potentials $A_I(\phi, X)$, the search for black hole solutions in these theories is somehow reversed, since one can start with any black hole metric and scalar profile, and using this algorithm, look for a specific Lagrangian which admits this ansatz as solution of its field equations.

Obviously, one can in principle proceed to the expansion around any value of $r$. But in practice, some specific values will allow to close the algorithm in a quicker way.
In the following, we shall expand the field equations around $r_{\ast} =0$.  The set of conditions we obtained being quite involved to reduce, we emphasize that, once the full conditions on the potentials have been found, we have checked the consistency of our solution by injecting it directly in the full (spherically symmetric reduced) field equations and check that there are identically vanishing. Having review the method of resolution of the field equations borrowed from \cite{Babichev:2016kdt}, we present now our result. We consider the Schwarzschild-(A)dS metric given by
\begin{align}
e^{\nu} = e^{-\lambda} = 1 - \frac{2m}{r} - \Lambda r^2
\end{align}
The radial dependent part of the scalar field is straitforwardly obtained by integrating (\ref{psi}) and reads
\begin{align}
\psi '(r) = M q \sqrt{ \frac{ \left( 2m + \Lambda r^3 \right) r }{\left( 2m - r + \Lambda r^3 \right)^2}}
\end{align}
and one observes that $\psi'(r) \rightarrow 0$ when $r \rightarrow 0$. We can now inject this in the equations of motion and proceed to the expansion around $r_{\ast}=0$.

\subsection{Application to shift symmetry breaking theories}

Applying the algorithm reviewed above, we obtain a complete set of conditions on the potentials which fully solve the field's equations. These conditions, valid only when evaluated at the value $X_{\ast} =  q^2$, read
\begin{align}
\label{c} 
 Q_{\phi} \big{|}_{X_{\ast}} = F_{\phi}  \big{|}_{X_{\ast}} & =  0 \\
  \label{co}
 Q_{X}\big{|}_{X_{\ast}} & = - \frac{X}{2} A_{3\phi}\big{|}_{X_{\ast}} \\
 \label{cc} 
 P \big{|}_{X_{\ast}} & = - 6 \Lambda F \big{|}_{X_{\ast}} \\
 \label{ccc} 
A_3 \big{|}_{X_{\ast}} & =  \frac{2}{9 X \Lambda} \left( P_{X} + 12 \Lambda F_{X}\right) \big{|}_{X_{\ast}}
\end{align}
At this stage, there are no condition on the potentials $A_4$ and $A_5$, and the class solution of our conditions depend still on sic free potentials $(P, Q, F, A_I)$.
With this conditions (\ref{c}) to (\ref{ccc}), we have obtained a subset of DHOST theories, larger than the sector staisfying $c_{\text{grav}} = c_{\text{light}}$, which admits a stealth Schwarzschild-(A)dS solution dressed with a linear time dependent scalar field (\ref{lin}), assuming a constant kinetic term $X= X_{\ast} =  q^2$.

As a last step, let us provide a concrete examples of potentials solutions of our conditions. An example of solution of our conditions (\ref{c})-(\ref{ccc}) is given by
\begin{align}
	& F (\phi, X )= f_1(\phi, X)  \log{\left(\frac{q^2}{X}\right)} + f_2(X)\\
	& P(\phi, X ) = f_3(\phi, X)  \log{\left(\frac{q^2}{X}\right)} -6 \Lambda  f_2(X)\\
	& A_3(\phi, X ) = \frac{2}{9 X \Lambda} \left[ 6 \Lambda f_{2X}(X) -   \frac{f_{3}(\phi, X)  + 12 \Lambda f_1(\phi, X)  }{X}  \right]  \\
	& Q(\phi, X ) = - \frac{1}{9 \Lambda} \left[ f_{3 \phi}(\phi, X)  + 12 \Lambda f_{1 \phi}(\phi, X)  \right] \log{\left(\frac{q^2}{X}\right)}  + \int dX \log{\left(\frac{q^2}{X}\right)} f_4(X)
\end{align}
where $(f_1,f_3)$ are free potentials depending on both $\phi$ and $X$ while $(f_2,f_4)$ are free potentials depending only on $X$. Imposing the observational constraints and the absence of graviton decay. 

Focusing on the GLPV subclass satisfying the observational constraint $c_{\text{grav}} = c_{\text{light}}$, a set of potentials solutions of our conditions is given for example by
\begin{align}
	& F (\phi, X )= f_1(\phi, X)  \log{\left(\frac{q^2}{X}\right)} + f_2(X)\\
	& P(\phi, X )  =  \left[- 30 \Lambda f_1(\phi, X) + 24 \Lambda   X f_{2X}(X) \right] \log{\left(\frac{q^2}{X}\right)}   -6 \Lambda  f_2(X)    \\
	& A_3(\phi, X )  = - \frac{4}{X} \left[ f_{1X}(\phi, X)  \log{\left(\frac{q^2}{X}\right)} - \frac{ f_{1}(\phi, X)}{X}  + f_{2X}(X) \right]\\
	& Q(\phi, X ) = 2 \left[ f_{1 \phi}(\phi, X)  \right] \log{\left(\frac{q^2}{X}\right)}  +  \int dX \log{\left(\frac{q^2}{X}\right)} f_{4}(X)
\end{align}
which depends again on four free potentials. Additional examples can be found in \cite{BenAchour:2018dap} which satisfy the observational constants and the absence of graviton decay. These results illustrate the power of this algorithm and the ability to implement it beyond shift symmetric DHOST theories.

Let us now discuss the disformal solution generating map and review some selected results obtained with this approach.

\newpage

\section{On disformal transformations}

As explained above, each of the three different subclasses of DHOST theories corresponding to a given degeneracy condition is stable under a general disformal transformation.  Therefore, one natural way to generate new solutions for each class is to perform a disformal mapping on the known exact solution for a given sub-sector of DHOST theories. The goal of this section is to discuss the outcome of this procedure. This section is borrowed from \cite{BenAchour:2020wiw, BenAchour:2024tqt}.

\subsection{Disformal transformation as a solution-generating technique}

In the following, we shall be interested in invertible disformal mapping between exact solutions of the form
\begin{align}
\left( \tilde{g}_{\mu\nu} , \phi \right) \; \qquad \rightarrow \qquad \left( g_{\mu\nu} , \phi  \right)\,.
\end{align}
such that
\begin{align}
\tilde{g}_{\mu\nu} = A(\phi, X) g_{\mu\nu} + B(\phi, X) \phi_{\mu} \phi_{\nu}
\end{align}
The kinetic term of the scalar field transform as
\begin{align}
\label{Xtilde}
\tilde{X} = \frac{X}{A + B X}\,. 
\end{align}
The invertibility of this mapping requires that the Jacobian determinant is non-zero, 
\begin{align}
\left|\frac{\partial \tilde{g}_{\mu\nu}}{\partial g_{\mu\nu}} \right| \neq 0\,. 
\end{align}
This is equivalent to the requirement that all eigenvalues of the Jacobian be non-vanishing, resulting in~\cite{Zumalacarregui:2013pma}
\begin{equation}
 A - X A_X - X^2 B_X \ne 0\,, \quad A \ne 0\,. \label{eqn:invertibility}
\end{equation}
The first condition can be rewritten as $\partial\tilde{X}/\partial X \ne 0$, where $\tilde{X}$ is considered as a function of $\phi$ and $X$ through the relation (\ref{Xtilde}).

If $\partial_{\mu}\phi$ is timelike before and after the transformation then there is a simple interpretation of the invertibility condition (\ref{eqn:invertibility})\footnote{One can easily find a similar interpretation if $\partial_{\mu}\phi$ is spacelike before and after the transformation.}. In this case one can take the unitary gauge where $\phi=t$, and in this gauge the lapse functions for the two metrics $\tilde{g}_{\mu\nu}$ and $g_{\mu\nu}$ are $\tilde{N}=1/\sqrt{-\tilde{X}}$ and $N=1/\sqrt{-X}$, respectively. Therefore the first condition in (\ref{eqn:invertibility}) is nothing but the invertibility of the mapping between the two lapse functions. On the other hand, the mapping between the spatial metrics in the unitary gauge is a conformal transformation and the second condition in (\ref{eqn:invertibility}) tells that the conformal factor for this conformal transformation should be non-vanishing. 

Now, at the level of the solution space, a given exact solution to a DHOST theory $(g_{\mu\nu}, \phi)$ is mapped to another exact solution $(\tilde{g}_{\mu\nu}, \phi)$ of the DHOST theory belonging to the same degeneracy class. Let us emphasize that at this level, the disformal transformation is a pure field redefinition and does not contain any new physics. \textit{The key point enters when choosing to which metric the matters gets minimally coupled.} Indeed, for test fields minimally coupled to the disformal metric $\tilde{g}_{\mu\nu}$, the behavior will be different from test fields minimally coupled to the seed metric $g_{\mu\nu}$. Notice that fixing the coupling of matter fields can also be done implicitly, for example when one constructs a parallel transported frame w.r.t a given geometry \cite{BenAchour:2024tqt}. Therefore, although it is a pure field redefinition, a disformal transformation actually changes the causal structure, the algebraic properties such as the Petrov type \cite{BenAchour:2021pla}, and all the observables which implicitly assume fixing the coupling of matter are generically modified. We shall come back on this point in Chapter 4.

Thus, this simple method allows one to construct exact solutions of degenerate higher-order scalar tensor theories without having to solve for their complicated field equations \cite{BenAchour:2020wiw}. See \cite{Faraoni:2021gdl, Bakopoulos:2022csr} for the construction of new black hole solutions beyond the stealth sector in DHOST gravity, among which exact rotating black holes solutions \cite{BenAchour:2020fgy, Anson:2020trg,Baake:2021kyg}. Another interesting feature of the disformal transformation is that the scalar profile $\phi$ remains unchanged. Thus, it allows to contemplate, for a given scalar profile, the modification on the metric sector induced by the higher-order terms in the action $\tilde{S}[\tilde{g}_{\mu\nu}, \phi]$. The parameter $B_0$ can be interpreted as a dimensionless coupling constant encoding the strength of the higher-order modifications. 

Before applying this trick to construct new solutions, let us discuss why a simple field redefinition as the disformal transformation can actually induce new physics depending on how one fixes the matter coupling. This is a subtle point which has led to confusion and it worths clarifying this point before going any further.
	
	\subsection{Disformal transformation on a local rest frame and matter coupling}
	
Following \cite{BenAchour:2024tqt} and anticipating on the last chapter, we now study how the disformal transformation applies at the level of the tetrad.  This allows one to discuss the effect of a disformal transformation on a local rest frame (represented by a parallel transported null frame) and contemplate why new physics shows up by implicitly fixing w.r.t which metric the equivalence principle is realized. Consider a spacetime geometry with metric $g_{\mu\nu}$ and let us introduce a set of tetrads $\theta^{\mu}{}_a$ at any point of the spacetime which allows us to project the metric $g_{\mu\nu}$ into the local rest frame of the observer as
\begin{align}\label{g-Tetrads0}
&{ g}_{\mu\nu} = \eta_{ab}\,{\theta }^a{}_{\mu} { \theta}^b{}_{\nu} \,,
&&
\eta_{ab} = { g}_{\mu\nu}{ \theta}^{\mu}{}_a \, { \theta}^{\nu}{}_b \,,
\end{align}
where $a,b=0,1,2,3$ are the Lorentz indices and $\eta_{ab}$ is the Minkowski metric. ${\theta }^a{}_{\mu}$ is the inverse of $\theta^{\mu}{}_a$ such that
\begin{align}\label{Tetrads-complete}
&{ \theta}^a{}_{\mu} {\theta}^{\mu}{}_b = \delta^a{}_b \,, &&
{ \theta}^{\mu}{}_a { \theta}^a{}_{\nu} = \delta^\mu{}_\nu \,.
\end{align}
In order to work with the Newman-Penrose formalism \cite{Newman:1961qr}, we set a null tetrad basis which is given in terms of four complex null vectors
\begin{align}\label{Tetrads}
&{ \theta}^a{}_{\mu} = \big( - { n}_\mu, - { \ell}_\mu, {\bar m}_\mu, { m}_\mu \big) \,, 
&&
{ \theta}^{\mu}{}_a = \big( { \ell}^\mu, { n}^\mu, {m}^\mu, {\bar m}^\mu \big) \,,
\end{align}
where ${ m}^\mu$ and ${ {\bar m}}^\mu$ are complex conjugate to each other, and the Minkowski metric takes the following null form
\begin{eqnarray}\label{eta}
\eta_{ab} = \eta^{ab} \doteq \begin{pmatrix}
0 & -1 & 0 & 0 \\
-1 & 0 & 0 & 0 \\
0 & 0 & 0 & 1 \\
0 & 0 & 1 & 0 
\end{pmatrix} \,.
\end{eqnarray}
 In this null basis, the metric $g_{\mu\nu}$ can be expressed as
\begin{equation}\label{g}
{ g}_{\mu\nu} = - { \ell}_\mu { n}_\nu - { n}_\mu { \ell}_\nu
+ { m}_\mu {\bar { m}}_\nu + {\bar { m}}_\mu { m}_\nu \,,
\end{equation}
where the null vectors satisfy the normalization and orthogonality conditions
\begin{align}\label{null}
&{ g}_{\mu\nu}{ \ell}^\mu{ n}^\nu = -1 \,,
&&{ g}_{\mu\nu}{ m}^\mu{\bar{ m}}^\nu = 1 \,,
\end{align}
and all other contractions between null vectors vanish.

We consider $\ell$ to be geodesic, i.e.
\begin{align}\label{geodesic-ll}
D\ell^{\mu} = 0 \,,
\end{align}
where $D \equiv \ell^{\alpha} \nabla_{\alpha}$ is the covariant derivative along $\ell$. The properties of the null congruence are characterized by the twelve complex Newman-Penrose spin coefficients
$\kappa, \epsilon, \pi, \alpha, \beta, \rho, \sigma, \lambda, \nu, \tau, \mu, \gamma$ which are defined in Chapter \ref{Chapter6}. They describe how the bundle of light rays expand, accelerate, shear and twist along each of the null directions ${ \theta}^{\mu}{}_a = \big( { \ell}^\mu, { n}^\mu, {m}^\mu, {\bar m}^\mu \big)$. The vector $\ell$ being geodesic \eqref{geodesic-ll}, a parallel transported frame (PTF) can be constructed by further demanding that
\begin{align}\label{PTF}
&D n^\mu = 0 \,, 
&&D m^\mu = 0 \,, 
&&D \bar{m}^\mu = 0 \,,
\end{align}
which in terms of the spin coefficients takes the form
\begin{align}\label{PT0-conditions}
&\kappa = -m^{\alpha}D\ell_{\alpha} = 0 \,, 
&& \epsilon = \frac{1}{2}(\bar{m}^{\alpha} Dm_{\alpha} - n^{\alpha}D\ell_{\alpha}) = 0 \,, 
&&\pi = \bar{m}^{\alpha}Dn_{\alpha} = 0 \,.
\end{align}
The tetrads are defined up to Lorentz transformations and the  six degrees of freedom in the Lorentz transformations make it possible to always achieve the above six conditions.

With the PTF \eqref{PT0-conditions} at hand, one effectively realizes a set of null Fermi coordinates \cite{Guedens:2012sz}. The remaining spin coefficients associated to this PTF provide the physical quantities characterizing the bundle of light rays that a freely falling observer (following a light-like trajectory) will measure.

Now, let us investigate how null tetrads change under a DT
\begin{align}\label{Tetrads-dis0}
{ \theta}^a{}_{\mu} = \big( -n_\mu , - { \ell}_\mu, {\bar m}^\mu, { m}^\mu \big)
&\xrightarrow{\mbox{DT}} 
{\tilde \theta}^a{}_{\mu}
= \big(- {\tilde n}_\mu,- {\tilde \ell}_\mu, \tilde{{\bar m}}^\mu, \tilde{ m}^\mu \big) \,.
\end{align}
For the sake of simplicity, we restrict our analysis to a pure constant DT with 
\begin{align}\label{DT0}
&C(\phi, X) =1 \,,
&&B(\phi, X) = B_0 \,,
\end{align}
where $B_0$ is a constant. The equivalence principle guaranties that there should exist a set of disformed null tetrads ${\tilde \theta}^{\mu}{}_a$ such that
\begin{align}\label{g-T-Tetrads0}
&{\tilde g}_{\mu\nu} = \eta_{ab}\,{\tilde \theta }^a{}_{\mu} {\tilde \theta}^b{}_{\nu} \,,
&&
\eta_{ab} = {\tilde g}_{\mu\nu}{\tilde \theta}^{\mu}{}_a \, {\tilde \theta}^{\nu}{}_b \,.
\end{align}
This tetrad ${\tilde \theta}^{\mu}{}_a$ provides at each point of the geometry the projection onto the local rest frame where the equivalence principle is realized. 
It can be easily shown that under the DT~\eqref{DT0}, the relation between the disformed tetrads and original tetrads is given by  \cite{BenAchour:2021pla}
\begin{align}\label{Tetrads-disform0}
{ \theta}^a{}_\mu&\xrightarrow{\mbox{DT}} {\tilde \theta}^a{}_\mu
= J^a{}_b \, { \theta}^b{}_\mu \,,
\end{align}
where
\begin{align}\label{J-def0}
&J^a{}_b = \delta^a{}_b + {\cal B}\, \phi^a \phi_b \,;
&&
{\cal B}(X) \equiv \frac{1}{X}\big[\sqrt{1+B_0 X}-1\big] \,,
\end{align}
in which 
\begin{align}
&\phi_a \equiv \phi_{\alpha} \theta^\alpha{}_a \,,
&&\phi^a \equiv g^{\alpha\beta}\phi_{\alpha} \theta^a{}_\beta \,,
\end{align}
such that $\phi_\mu = \phi_a \theta^a{}_\mu$ and $\phi^\mu = \phi^a \theta^\mu{}_a$. Substituting \eqref{J-def0} in \eqref{Tetrads-disform0} we find the disformal transformation can be written at the level of the tetrad as follows
\begin{align}\label{Tetrads-disform-explicitt}
{\tilde \theta}^a{}_\mu
= { \theta}^a{}_\mu + {\cal B}\, \phi^a \phi_{\mu} \,.
\end{align}
The vector $\phi_\mu$ can be expressed in terms of the null basis as
\begin{align}
\label{}
\phi_\mu = - \phi_n \ell_\mu - \phi_\ell n_\mu 
+ \phi_{\bar m} m_\mu + \phi_m {\bar m}_\mu \,.
\end{align}
Substituting this in \eqref{Tetrads-disform-explicitt}, we find the explicit expression of the disformed null vectors in terms of the seed null vectors.

Equipped with the disformed null vectors, it is straightforward to compute the disformed spin coefficients. In particular, we find
\begin{align}\label{PTF-diss}
\begin{split}
\kappa &\xrightarrow{\mbox{DT}} \tilde{\kappa} = \kappa + \kappa_{\rm DT} \,,
\\
\epsilon &\xrightarrow{\mbox{DT}} \tilde{\epsilon} = \epsilon + \epsilon_{\rm DT} \,,  
\\
\pi &\xrightarrow{\mbox{DT}} \tilde{\pi} = \pi + \pi_{\rm DT} \,,
\end{split} 
\end{align}
such that $\kappa_{\rm DT}=\epsilon_{\rm DT}=\pi_{\rm DT}=0$ for $B_0=0$. The explicit form of these spin coefficients to first order in $B_0$ are given in Chapter \ref{Chapter7}. The results \eqref{PTF-diss} clearly show that, in general, the PTF \eqref{PT0-conditions} does not remain a PTF after performing the DT since 
\begin{align}\label{PT-conditions-DT}
&\tilde{\kappa} \neq 0 \,, 
&&\tilde{\epsilon} \neq 0 \,, 
&&\tilde{\pi} \neq 0 \,,
\end{align} 
as  in general $\kappa_{\rm DT}\neq0$, $\epsilon_{\rm DT}\neq0$, $\pi_{\rm DT}\neq0$.
In particular, it shows that the DT induces a deviation w.r.t the geodesic motion (see Fig.~\ref{Fig}). Let us further expand on this point.

As we shall see, it reveals instructive to contemplate the effects of the disformal transformation on the local rest frames $\tilde{\eta}_{ab}$ (resp. $\eta_{ab}$) associated to $\tilde{g}_{\mu\nu}$ (resp. $g_{\mu\nu}$). This reveals the key property of the disformal mapping and the role of implicitly fixing the matter coupling when choosing w.r.t which local rest frame the equivalence is realized, i.e. the matter is minimally coupled.

To see this, recall that the local Minkowski metric is invariant under the Lorentz transformations (LTs) 
\begin{align}
\eta_{ab} \Lambda^a{}_{c} \Lambda^b{}_{d} = \eta_{cd} \,,
\end{align}
where $\Lambda^a{}_b$ is any Lorentz transformation which are characterized by six parameters. Substituting the above relation in \eqref{g-Tetrads0} and \eqref{g-T-Tetrads0}, we see that both seed tetrads $\theta^a{}_\mu$ and disformed tetrads ${\tilde \theta}^a{}_\mu$ are defined up to Lorentz transformations (LTs). 

Let us rewrite the first equation in \eqref{g-T-Tetrads0} as
\begin{align}\label{g-T-Tetrads-LT}
&{\tilde g}_{\mu\nu} 
= \eta_{ab}\,{\tilde \theta }^a{}_{\mu} {\tilde \theta}^b{}_{\nu} 
= \eta_{ab} J^a{}_{c} J^b{}_{d} \, { \theta }^c{}_{\mu} { \theta}^d{}_{\nu}  \,,
\end{align}
Comparing the above result with \eqref{g-Tetrads0}, we see that the disformed metric does not take the local Minkowski form when we express it in terms of the original tetrads since, in general, one has 
\begin{align}
\eta_{ab} J^a{}_{c} J^b{}_{d} \neq \eta_{cd} \,.
\end{align}
First, this shows that the effects of DT would be completely redundant in the particular case when $J^a{}_{b}\subset\Lambda^a{}_{b}$. For example, from \eqref{J-def0} we find $J_{ab} = \eta_{ac} J^c{}_b = \eta_{ab} + {\cal B}\, \phi_a \phi_b$ and, for infinitesimal LTs $\Lambda_{ab} = \eta_{ab} + \omega_{ab}$ with $\omega_{ab}=-\omega_{ba}$ while $\phi_a\phi_b = \phi_b \phi_a$. 

Secondly, the appearance of $\eta_{ab} J^a{}_{c} J^b{}_{d} \neq \eta_{cd}$ in \eqref{g-T-Tetrads-LT} shows that if the equivalence principle is considered w.r.t. $g_{\mu\nu}$, a free fall observer in $g_{\mu\nu}$ is not a free fall observer in ${\tilde g}_{\mu\nu}$. This is because the two tetrads/observers ${ \theta}^a{}_{\mu}$ and ${\tilde \theta}^a{}_{\mu}$ are not related to each other through LTs $\Lambda^a{}_b$ but through the $J$-map $J^a{}_b$ defined in \eqref{Tetrads-disform0}. The matter minimally couples to the metric w.r.t which the equivalence principle is realized, i.e. which can take locally the Minkowski form. \textit{Thus, the choice of tetrad w.r.t which the equivalence principle is realized implicitly fixes the coupling of matter fields.} The matter minimally couples to $g_{\mu\nu}$ (resp. to ${\tilde g}_{\mu\nu}$) if one works with the tetrad ${\theta}^a{}_{\mu}$ (resp. ${\tilde \theta}^a{}_{\mu}$). \textit{Hence, depending with which tetrads one works with, one deals with a different theory as by choosing tetrads, one implicitly fixes the matter coupling.} 

In particular, that is why the Petrov type changes under a DT \cite{BenAchour:2021pla}. From this discussion, the $J$-map ${\tilde \theta}^a{}_\mu = J^a{}_b \, { \theta}^b{}_\mu$, that is defined in Eq. \eqref{Tetrads-disform0}, can be understood as the operation encoding the departure from a local Minkowski frame in which the equivalence principle is realized, i.e. the effective distortion of the local freely falling frame induced by the DT. Obviously, one has always the choice to decide to assign the free fall and thus the realization of the equivalence principle to ${\tilde \theta}^a{}_\mu$ or to ${ \theta}^b{}_\mu$.

\section{From naked singularity to black hole: A no go}

As a first step, let us consider the minimal static spherically symmetric solution of the canonical Einstein-Scalar system obtained by Janis, Newman and Winicour \cite{Janis:1968zz}. Performing the disformal transformation on (\ref{JNW}), one obtains a new family of exact solutions of a subsector of DHOST theories. The metric of the geometries belonging to this new family reads
\begin{align}
\label{JNWdhost}
ds^2 = g_{\mu\nu}dx^{\mu}dx^{\nu} = \frac{1}{A(r)} \left\{ - F^{\gamma} dt^2 + G F^{-\gamma} dr^2 +  r^2 F^{1 - \gamma}d\Omega^2\right\}\,,
\end{align}
where the functions $F(r)$ and $G(r)$ are given by
\begin{align}
F(r) & = 1-\frac{r_s}{r}\,,\\
\label{g}
G(r) & = 1 - \frac{q^2}{4\pi} \frac{F^{\gamma -2}}{r^4} B\left(\phi(r), X(r) \right)\,.
\end{align}
The disformal transformation leaving the scalar field unaffected, the scalar profile is given again by its GR expression (\ref{scalarprofile}).
Using the above metric and the scalar profile, one obtains the form of the kinetic term in the DHOST frame which reads
\begin{align}
\label{XX}
X(r) = g^{\mu\nu} \phi_{\mu} \phi_{\nu} = \frac{q^2}{4\pi} \frac{A(r)}{G(r)} \frac{1}{r^4} F^{\gamma-2}\,.
\end{align}
For the JNW initial solution, one can then check the consistency relation (\ref{Xtilde}) between (\ref{X}) and (\ref{XX}), which is trivially satisfied. This concludes the disformal transformation of the JNW solution of the Einstein-Scalar system. Before discussing the possibility to have black hole solutions, let us point that considering perturbations on top of these new solutions might lead to pathological behavior, as some subsectors of the perturbations are known to be invariant under a general disformal mapping \cite{Motohashi:2015pra, Tsujikawa:2015upa}. Hence, the instability of the seed solution due to the presence of a naked singularity would persist in the DHOST frame. 

Now, we would like to investigate whether the DHOST higher order terms allow, through the free potentials $A(r)$ and $G(r)$, to modify the causal structure of the naked singularity at $r=r_s$. Under a set of assumptions that simplifies the analysis, we shall investigate if it is possible to change the asymptotically flat naked singularity in the Einstein frame to a black hole horizon in the DHOST frame. In order to track the existence of a black hole horizon at $r=r_s$ in the DHOST frame, we need to require that
\begin{itemize}
\item (a) the Kodama vector is timelike, null and spacelike for $r>r_s$, $r=r_s$ and $r<r_s$, respectively,
\item (b) the curvature invariants remain regular at $r=r_s$,
\item (c) the metric is Lorentzian for both $r>r_s$ and $r<r_s$ and its determinant is finite at $r=r_s$,
\end{itemize}
and for simplicity we assume that $A$ and $G$ behave as $A \sim c_1 F^{\alpha}$ and $G \sim c_2 F^{-\beta}$ near $r=r_s$, where $c_1$, $c_2$, $\alpha$ and $\beta$ are constants\footnote{ Notice that condition (a-b) imply that $r=r_s$ is a null hypersurface since codama vector is the normal vector to surface $const = R[r] := \sqrt{g_{\theta \theta}}$.}. Rescaling the coordinates $t$ and $r$ to absorb $c_1$ and $c_2$ and neglecting subleading contributions, we have 
\begin{equation}\label{agforms}
 A = F^{\alpha}\,, \quad G = F^{-\beta}\,. 
\end{equation}

Let us begin the proof of the no-go result. 
In order to impose the condition (a), we compute the norm of the Kodama vector as
\begin{align}
\label{kodd}
\mathcal{K}_{\alpha} \mathcal{K}^{\alpha}  = - r^2 \frac{F}{G} \left[ \frac{1}{r} + \frac{1-\gamma}{2} \frac{F'}{F} - \frac{A'}{2A}\right]^2\,.
\end{align}
We then compute the determinant of the metric (\ref{JNWdhost}) which will be useful to demand the condition (c). It reads
\begin{align}
\label{deeet}
- \text{det}(g) = \frac{G F^{2(1-\gamma)}}{A^4} r^4 \sin^2{\theta}\,.
\end{align}
Finally, we can compute the three curvature invariants in order to impose the condition (b). 

In the rest of this section we assume { $A$ and $G$ behave as (\ref{agforms})}. 
We consider the following three cases separately: 
\begin{itemize}
\item (i) $\beta \leq -1$; 
\item (ii) $-1 < \beta \leq 1$; 
\item (iii) $\beta > 1$.
\end{itemize}

For (i), it is obvious that $\mathcal{K}_{\alpha} \mathcal{K}^{\alpha}$ does not vanish at $r_s$ and thus the condition (a) cannot be fulfilled. In the following we thus consider the cases (ii) and (iii) only.

For (ii), the condition (a) requires that 
\begin{align}
\alpha = 1 - \gamma\;, \qquad \text{and} \qquad \beta = 0\,.
\end{align}
In this case, computing the curvature invariants, one finds that the condition (b) demands that $\gamma = 1/2$ or $\gamma = 1$. The latter ($\alpha = 0$, $\beta = 0$, $\gamma = 1$) corresponds to the Schwarzschild geometry without a scalar hair and is not of our interest. On the other hand, the former ($\alpha = 1/2$, $\beta = 0$, $\gamma = 1/2$) does not satisfy the condition (c) 
since $- \text{det}(g) \propto F^{-1}$ is negative in the region $r<r_s$. Therefore, the case (ii) does not work.

We now consider the case (iii). The condition (a) gives $\mathcal{K}_{\alpha} \mathcal{K}^{\alpha} \sim -(\alpha + \gamma -1)^2F^{\beta-1}/4$ near $r=r_s$ for $\alpha + \gamma \ne 1$ and $\mathcal{K}_{\alpha} \mathcal{K}^{\alpha} \sim -F^{\beta+1}$ near $r=r_s$ for $\alpha + \gamma = 1$. The condition (b) is equivalent to $\alpha + \gamma \geq 1$. Since the norm of the Kodama vector has to change sign at the horizon, it implies that $\beta = 2n$ where $n \in \mathbb{N}$. The condition (c) is then restated as $2\alpha + \gamma = m$ ($m=0,\pm 1, \pm 2, \cdots$) and $1-m-n \geqslant0$ so that $- \text{det}(g) \propto F^{2(1-m-n)}$ is positive for both $r>r_s$ and $r<r_s$ and is finite at $r=r_s$. Also, excluding the trivial Schwarzschild case ($\gamma=1$), it follows that $0 \leq \gamma < 1$. In summary the conditions (a)-(c) are satisfied if and only if
\begin{equation}
 \beta = 2n\,, \quad (n=1,2,\cdots\,) \,, 
\end{equation}
and 
\begin{equation}
 2\alpha + \gamma = m\,, \quad 1- m -n \geq 0\,,  \quad \alpha + \gamma \geq 1\,, \quad 0 \leq \gamma < 1\,,
\end{equation}
Expressing $\gamma$ in term of $\alpha$ and $m$, the above conditions can be recast as
\begin{eqnarray}
\alpha \geqslant 0\,, \quad m- \alpha \geqslant 1\;, \quad \frac{m-1}{2} <\alpha \leq \frac{m}{2}
\end{eqnarray}
for which there is no admissible solutions satisfying also $\beta =2 n >1$.


Let us now show that this conclusion extends to any horizon located $r_{\ast} \neq r_s$. Considering the norm of the Kodama vector (\ref{kodd}), it is easy to see that the spacetime admits an horizon at $r_{\ast} \neq r_s$ provided the function $G$ changes sign, since inside the trapped region, the Kodama vector has to be spacelike. However, if $G$ changes sign at $r_{\ast}$, then the determinant (\ref{deeet}) will change sign too, preventing the spacetime from being Lorentzian both outside and inside such horizon. In the end, this prevents one to dress the naked singularity with a well defined black hole horizon at some radius $r_{\ast} > r_s$, completing therefore the no go result.

To conclude, we have shown that starting from the Janis-Newman-Winicour naked singularity as a seed solution, one can indeed obtain a new family of exact solutions of the quadratic DHOST theories. However, there is no admissible choices for the free potentials $A(r)$ and $B(r)$ such that this new family of exact solutions contains asymptotically flat black hole geometries. As the tractable aymptotically flat spherically symmetric and static solution of the canonical Einstein-Scalar system represent naked singularity, it seems hopeless to use these kind of geometries as seed solution to generate new exact asymptotically flat black hole solutions for DHOST theories. Nevertheless, we emphasize that naked singularity geometries can still provide interesting geometries, in modeling high energy phenomena, such as gamma ray burst emission. See \cite{Chakrabarti:1994ig, Singh:1998vw, Patil:2011ya, Patil:2011aw, Patil:2011aa}. Let us now discuss the disformal transformation of exact black hole solution of DHOST theories.

\section{Constructing new hairy black holes}

As a first step, we intend to present the outcome of the disformal transformation on a general metric ansatz. In the following, we make the adopt the following restriction
\begin{align}
A:=A(X) \qquad B:=B(X)
\end{align} 
Consider therefore the seed metric given by
\begin{align}
\label{seedm}
d\tilde{s}^2 & = \tilde{g}_{\mu\nu} dx^{\mu} dx^{\nu}  = - F(r) dr^2 + \frac{dr^2}{F(r)} + r^2 d\Omega^2\,,
\end{align}
which encompasses the two exact solutions we shall use as seed geometries. Moreover, we work with the general scalar profile
\begin{align}
\label{seedscal}
\phi(t,r) = qt + \psi(r)\,.
\end{align}
Performing a disformal transformation on the general metric (\ref{seedm}) associated to the profile (\ref{seedscal}), we obtain the new exact solution
\begin{align}
ds^2 = g_{\mu\nu}dx^{\mu}dx^{\nu} =  A^{-1} \left[ - F G_1 dt^2 - 2 N_r dt dr + \frac{G_2}{F} dr^2 + r^2 d\Omega^2\right]\,,
\end{align}
where the new functions $G_1$, $G_2$ as well as the shift are related to the free potential $B(r)$ through
\begin{align} 
\label{eqn:G1G2Nr-B}
G_1(r) & = 1 + \frac{q^2 B(r) }{F}\,, \\
\label{eqn:G1G2Nr-BB}
 G_2 (r) & = 1 - \frac{Z B(r) }{F}\,, \\
 \label{eqn:G1G2Nr-BBB}
 N_r & = \pm  \frac{qB(r)}{F} \sqrt{Z }\,,
\end{align}
where $Z(r) = \tilde{X}(r) F(r) + q^2$ in general. The plus and minus signs for $N_r$ correspond to a scalar profile that regularly penetrates the black hole horizon and the white hole horizon, respectively. We can now apply this general construction to the two seed solutions reviewed in the previous section. 

\subsection{Starting from a stealth solution}

\label{secstealth}

As a first example, we consider the Schwarzschild seed solution with a constant kinetic term described by
\begin{align}
ds^2 = - F(r) dt^2 + F(r)^{-1} dt^2 + r^2 d\Omega^2\,, \quad F(r) = \left( 1 - \frac{r_s}{r} \right)\,.
\end{align}
Under suitable conditions derived in \cite{BenAchour:2018dap, Motohashi:2019sen}, quadratic DHOST theories admit such geometry as an exact solution associated to the scalar profile
\begin{align}
\label{tdscal}
\phi(t,r) = q t + \psi(r)
\end{align}
while its kinetic term remains constant, such that
\begin{align}
\tilde{X}_{\circ} = \tilde{g}^{\mu\nu} \phi_{\mu} \phi_{\nu}  = - \frac{q^2}{F} + F \left( \psi'\right)^2\,.
\end{align}
As mentioned above, the resolution of the strong coupling issue requires that $\partial_{\mu}\phi$ be timelike. We thus suppose that
\begin{equation}
  -q^2 \leq \tilde{X}_{\circ} < 0\,,
\end{equation}
where the first inequality follows from the reality of $\psi'$. Let us introduce for simplicity the notation
\begin{align}
\label{def}
Z = \tilde{X}_{\circ} F+ q^2 \;, \qquad Z_{\circ} = \tilde{X}_{\circ}+q^2\,.
\end{align}
This allows us to fix the form of the spatial gradient of the scalar field to 
\begin{align}
\psi'(r) = \pm \sqrt{\frac{Z}{F^2}}\,. 
\end{align}
When $r\rightarrow +\infty$, it asymptotes to $\psi' \sim \pm \sqrt{Z_{\circ}}$.
Integrating this equation, one obtains the expression for the function $\psi(r)$ which reads
\begin{align}
 \psi(r) & = \psi_{\circ} \pm \left\{ r\sqrt{Z}   + \frac{\left( 2q^2 + \tilde{X}_{\circ}\right)r_s}{\sqrt{Z_{\circ}}} \log{\left[ \sqrt{r Z_{\circ}} \left( \sqrt{Z_{\circ}} + \sqrt{Z} \right)\right]}   + q r_s \ln \left|\frac{q-\sqrt{Z}}{q+\sqrt{Z}}\right| \right\}\,.
\end{align}
This profile blows up going from $\phi \rightarrow + \infty$ for $r\rightarrow + \infty$ to $\phi \rightarrow - \infty$ at $r=r_s$. The assumption of the constant kinetic term fully fixes the form of the linear time-dependent scalar profile, up to the choice of the sign of $\psi'$ and the three integration constants $\tilde{X}_{\circ}$, $q$ and $\psi_{\circ}$.

As a first remark, notice that since we have assumed that $A:=A(X)$ and $B:= B(X)$ (otherwise they would become time-dependent for $q\neq0$), the right hand side of (\ref{Xtilde}) is a function of $X$ only. Hence, whenever (\ref{Xtilde}) is invertible with respect to $X$, $X$ can be expressed as a function of $\tilde{X}$, which is set to be the constant $\tilde{X_{\circ}}$ for the stealth seed solution considered here. Therefore, restricting to invertible disformal transformations implies that $X$, $A$ and $B$ are constant
\begin{align} 
X = X_{\circ} \;, \qquad A = A_{\circ}\;, \qquad B = B_{\circ}\,.
\end{align}
From this, it is straightforward to see that 
\begin{align} 
X_{\circ} 
 = \frac{A_{\circ}\tilde{X}_{\circ}}{1-B_{\circ} \tilde{X}_{\circ}}\,.
\end{align}
 Moreover, the form of the $G_2(r)$ modification is fixed to
\begin{align} 
G_2 (r)= 1 - \frac{Z(r) B_{\circ} }{F(r)}
\end{align}
Therefore, under the above assumptions, the metric we obtain from the stealth Schwarzschild black hole as a consequence of the disformal transformation reads
\begin{align}
\label{newwsol}
ds^2&  =  \frac{1}{A_{\circ}}\left\{ - F \left( 1 + \frac{q^2 B_{\circ} }{F}   \right) dt^2   \mp \frac{2 q  \sqrt{Z} B_{\circ} }{F} dt dr + F^{-1} \left( 1 - \frac{ZB_{\circ} }{F}\right) dr^2 + r^2 d\Omega^2 \right\}\,.
\end{align}
Notice that this geometry exhibits a deficit solid angle whenever $Z_{\circ}B_{\circ} \ne 0$. Indeed, the area radius is $r/\sqrt{A_{\circ}}$ and the $rr$-component of the the metric in the region $r\rightarrow + \infty$ becomes $g_{rr} \sim  ( 1 - Z_{\circ}B_{\circ})/A_{\circ}$. The norm of the Kodama vector is given by
\begin{align} 
\mathcal{K}_{\alpha} \mathcal{K}^{\alpha} = - \frac{r+q^2B_{\circ}r-r_s}{r(1-\tilde{X}_{\circ}B_{\circ})}\,,
\end{align}
which implies that the horizon is located at
\begin{align} 
r_{\ast} = \frac{r_s}{1+ q^2 B_{\circ}}\,. 
\end{align}
Depending on the sign of $B_{\circ}$, the new horizon radius can be either greater of smaller than $r_s$. 
Computing moreover the Misner-Sharp mass, we obtain
\begin{align} 
M_{\text{MS}} = \frac{r_s - Z_{\circ}B_{\circ}r}{2\sqrt{A_{\circ}}(1-\tilde{X}_{\circ}B_{\circ})}\,.
\end{align}
This shows that the above solution reduces to the standard Schwarzschild only when $B_{\circ}Z_{\circ}=0$, i.e. when either $B_{\circ}=0$ or $\tilde{X}_{\circ} = - q^2$. 
Finally, computing the determinant of the metric, it reads
\begin{align} 
 \text{det}(g) = -r^4 \sin ^2(\theta) \left(1-B_{\circ} \tilde{X}_{\circ} \right)A_{\circ}^{-4}\,,
\end{align}
which keeps the same sign in the whole spacetime. Let us summarize the outcome of the disformal transformation of the standard stealth Schwarzschild solution. Assuming that $-q^2\leq \tilde{X}_{\circ} < 0$ and $B_{\circ}\ne 0$, one can classify the solutions after the disformal transformation into the following two cases. 
\begin{itemize}
 \item Case 1: $-q^2 < \tilde{X}_{\circ} < 0$. This provides a solution which describe an asymptotically locally flat non-stealth black hole with a deficit solid angle. The deficit solid angle is traced back to the non-zero value of $\psi'$ at $r=\infty$, while the non-stealth character of the solution can be recognized from the non-trivial $r$-dependence of its Misner-Sharp energy. 
 \item Case 2: $-q^2 = \tilde{X}_{\circ} < 0$. In this case, the solution after the transformation is stealth and corresponds to a Schwarzschild black hole with a shifted mass and horizon radius. 
\end{itemize}

Finally, let us point that such disformal transformation of the Schwarzschild stealth solution provides a straightforward way to build minimal hairy deformation of vacuum GR solution which are exact solution of a DHOST theory\footnote{Such disformal transformation of the stealth Schwarzschild black hole has already been considered in the literature, in particular to study the change in the speed of gravitational waves \cite{Babichev:2017lmw}, but it seems that the systematic solution-generating point of view has remained largely ignored.}. Up to now, such minimal deformation of the vacuum solutions of GR have been introduced for phenomenological investigation and remain largely ad hoc. Using the new exact solution (\ref{newwsol}), and performing a linear expansion in term of the parameters $A_{\circ}$ and $B_{\circ}$, our solution generating method allows instead to endow these phenomenologically interesting geometries within the effective approach of DHOST theories. Consider therefore infinitesimal version of the disformal transformation used above where
\begin{align} 
A_{\circ}=1+\epsilon_1\;, \qquad B_{\circ}=\epsilon_2\;,  
 \qquad
 \text{with} \qquad |\epsilon_1|, \; |\epsilon_2| \ll 1\,.
\end{align}
Keeping only the leading terms in $\epsilon_1$ and $\epsilon_2$, one can generate using the disformal transformation a new exact solutions of DHOST theories of the form
\begin{align}
 ds^2 & = - \left[ (1-\epsilon_1)\left(1-\frac{r_s}{r}\right) + q^2\epsilon_2 + \mathcal{O}(\epsilon_{1,2}^2) \right] dt^2  \mp 2 q \epsilon_2 \sqrt{\tilde{X}_{\circ} + q^2 - \frac{\tilde{X}_{\circ} r_s}{r}} \left( 1-\frac{r_s}{r}\right)^{-1} dt dr \nonumber  \\
 &  + \left( \left( 1  - \epsilon_1 - \tilde{X}_{\circ} \epsilon_2 \right) \left( 1-\frac{r_s}{r}\right)^{-1}  - q^2 \epsilon_2  \left( 1-\frac{r_s}{r}\right)^{-2}  + \mathcal{O}(\epsilon_{1,2}^2)\right) dr^2  + \left( 1-\epsilon_1 + \mathcal{O}(\epsilon_1^2)\right) r^2 d\Omega^2
\end{align}
We can now analyse the properties of this geometry. First of all, the horizon structure of this geometry is given by
\begin{equation}
\label{kodeps}
 \mathcal{K}_{\alpha} \mathcal{K}^{\alpha} (r_{\ast})  =
 \frac{(1+q^2\epsilon_2)r_{\ast}-r_s}{r(1-\tilde{X}_{\circ}\epsilon_2)}  + \mathcal{O}(\epsilon^2) = 0\,,
\end{equation}
such that the position of the horizon is slightly shifted and reads
\begin{align} 
r_{\ast} \simeq \left( 1- q^2 \epsilon_2 \right) r_s + \mathcal{O}(\epsilon^2) \,.
\end{align}
From (\ref{kodeps}), it is straightforward to see that the Misner-Sharp mass,
\begin{equation}
M_{\text{MS}} = \frac{r_s-Z_{\circ}\epsilon_2r}{2\sqrt{1+\epsilon_1}(1-\tilde{X}_{\circ}\epsilon_2)} + \mathcal{O}(\epsilon^2)\,,
\end{equation}
is not constant if $Z_{\circ}\epsilon_2 \ne 0$, where $Z_{\circ}$ is defined in (\ref{def}). Hence, the Misner-Sharp energy does not coincide with the horizon radius divided by 2, as it is the case for any stealth Schwarzschild solution, and therefore, the new geometry is not stealth in general.

We can now compute the curvature invariants for this new solutions. At leading order, they coincide with the Schwarzschild's ones, such that
\begin{align}
\mathcal{R} & = -\frac{2 \epsilon_2 \left(q^2+\tilde{X}_0 \right)}{r^2} + \mathcal{O}(\epsilon^2)\,, \\
\mathcal{R}^{\mu\nu\rho\sigma} \mathcal{R}_{\mu\nu\rho\sigma} = \mathcal{C}^{\mu\nu\rho\sigma} \mathcal{C}_{\mu\nu\rho\sigma} & = \frac{12 r_s^2}{r^6}  -\frac{8 r_s \left(r \epsilon_2 \left(q^2+\tilde{X}_0 \right)-3 r_s (\tilde{X}_0 \epsilon_2+ \epsilon_1)\right)}{r^6} + \mathcal{O}(\epsilon^2)\,.
\end{align}
Finally, computing the kinetic term, one obtains that it remains constant at all order. The leading term reads
\begin{align} 
X_{\circ} \simeq  \left( 1 + \epsilon_1 + \epsilon_2 \tilde{X}_{\circ} \right) \tilde{X}_{\circ} + \mathcal{O}(\epsilon^2)\,.
\end{align}
This concludes the construction of the minimally modified Schwarzschild black hole in the context of DHOST theories. We can now turn to the disformal transformation of non-stealth seed solution of DHOST theories.

\subsection{Starting from a non-stealth solution}

Now let us consider a non-stealth seed solution. We choose the exact solution obtained in \cite{Babichev:2017guv} and whose metric is given by
\begin{align}
ds^2 = - F(r) dr^2 + \frac{dr^2}{F(r)} + r^2 d\Omega^2\,,
\end{align}
with
\begin{align} 
F(r) = 1 - \frac{\mu}{r} + \frac{\eta}{4\zeta r} \int dr r^2 \tilde{X}\,.
\end{align}
The scalar profile is time-dependent and has the same form as in (\ref{tdscal}) although with the different radial contribution $\psi(r)$. The associated kinetic term is given by the equation
\begin{align} 
\label{fonc44}
\left( \sqrt{\frac{1}{2}\tilde{X}}\right)^2 \left( 1- \frac{\eta}{\beta} r^2 \sqrt{\frac{1}{2}\tilde{X}}\right) = - \frac{q^2}{2}\,,
\end{align}
The discriminant being negative\footnote{ The discriminant reads
\begin{align} 
\Delta = - \frac{q^2}{2} \left( 4 + \frac{27 \eta^2}{\beta^2} r^2\right) <0
\end{align}
which implies that $\sqrt{\tilde{X}/2}$ has two conjugated complex solutions and only one real solution.} for any $r$, there is only one single real root for $\sqrt{\tilde{X}/2}$ which is given by
\begin{align}
\label{XX}
\sqrt{\frac{1}{2}\tilde{X}} & = \frac{\beta}{3\eta r^2} \left( 1 +  K ^{1/3} + K^{-1/3}\right)\,,\\
 K & = \frac{1}{2} \left[ 2 + \frac{27 q^2 \eta^2 r^4}{2\beta^2} - \sqrt{- 4 + \left( \frac{27 q^2 \eta^2 r^4}{2\beta^2} + 2\right)^2} \right]\,.
\end{align}
The function $K(r)$ is always positive, never vanishes and remain finite everywhere\footnote{In (3.17) of \cite{Babichev:2017guv}, in order for its right hand side to be real, it should be understood that $(-1)^{1/3}=-1$ instead of $(-1)^{1/3}=(1+i\sqrt{3})/2$. Also, $K$ is denoted as $-A$ in \cite{Babichev:2017guv}.} and the kinetic term $\tilde{X} = \tilde{g}^{\mu\nu} \partial_{\mu} \phi \partial_{\nu} \phi>0$ for all $r$ for such configuration. This shall have important consequence in the following. Using this solution, one can show that the geometry admits a single horizon located at a radius that we shall denote $r_{\circ}$ in what follows. Moreover, its asymptotic behavior is given by
\begin{align}
F(r) \sim \frac{3\eta}{10\zeta} \left( \frac{q^2 \beta}{2\eta}\right)^{2/3}  r^{2/3}\;, \qquad \text{when} \qquad r \rightarrow + \infty\,.
\end{align}
The radial contribution to the scalar profile can be obtained by integrating 
\begin{align}
F^2 \left(\phi'\right)^2 = q^2 + \frac{2\beta^2}{9\eta^2 r^4} F \left( 1+ K^{1/3} + K^{-1/3}\right)\,.
\end{align}
The crucial point is that this exact black hole solution, although quite complicated, provides one of the few known examples which describe a non-stealth black hole geometry, with a non constant kinetic term and a time-dependent scalar profile. From the point of view of the three properties stated earlier in introduction, this solution satisfies all the three. As we shall see in the next section, it is therefore not a surprise that it serves as a seed solution allowing for the most general modifications when constructing new black hole solutions using a disformal mapping.  Before presenting this construction, let us present its static limit. 

In the static case, where we have $q=0$, the solution becomes
\begin{align}
\label{fonc1}
F(r) = 1 - \frac{\mu}{r} - \frac{\beta^2}{2\zeta \eta r^2}\,.
\end{align}
Notice that because $\zeta >0$ and $\eta >0$, the metric has the same form as a Reissner-Nordstrom metric with an imaginary electric-like charge. 
The associated scalar profile is given by
\begin{align}
\label{fonc2}
\phi(r) & = \pm 2 \sqrt{\frac{\zeta}{\eta}} \left\{ \text{Arctan} \left[ \frac{\beta^2 + \zeta \eta \mu r}{\beta \sqrt{2\zeta \eta r \left( r-\mu\right) -\beta^2}}\right]  - \text{Arctan} \left( \frac{\mu}{\beta} \sqrt{\frac{\zeta \eta}{2}}\right) \right\}
\end{align}
when $ \beta >0$ and $\eta > 0$, while it is given by 
\begin{align}
\label{fonc3}
\phi(r) & = \pm 2 \sqrt{\frac{\zeta}{\eta}} \left\{ \text{Arcth} \left[ \frac{\beta^2 + \zeta \eta \mu r}{\beta \sqrt{2\zeta \eta r \left( r-\mu\right) -\beta^2}}\right]  - \text{Arcth} \left( \frac{\mu}{\beta} \sqrt{\frac{\zeta \eta}{2}}\right) \right\}
\end{align}
when $ \beta <0$ and $\eta < 0$. 
Finally, the kinetic term of the scalar field is not constant and decays as $r^{-4}$, such that
\begin{align}
\label{fonc4}
\tilde{X} = \frac{2\beta^2}{\eta^2 r^4}\,.
\end{align}
This exact solution provides therefore an interesting example of a non-stealth configuration with a rather simple profile for the kinetic term. Indeed, when $q\neq 0$, one can easily write the radius $r$ in term of $\tilde{X}$ using (\ref{fonc44}), while when $q=0$, the relation (\ref{fonc4}) is straightforward to invert. This ensures that one can realize any $r$-dependence of the potentials $A$ and $B$ through their dependence on $\tilde{X}$. Finally, when working with this seed solution, we shall introduce the generalization of (\ref{def}) which reads
\begin{align}
Z = \tilde{X} F + q^2\,.
\end{align}
This concludes the presentation of the non-stealth seed solution. 

Now let us study its disformed version. Consistency imposes that the function $G_1$ and the shift $N_r$ are related to the free function $G_2$ through
\begin{align}
\label{g1}
G_1(r) & = 1 + \frac{q^2 \left( 1-G_2\right) }{Z }\,,  \\
N_r (r) & =  \pm \frac{q \left( 1-G_2\right)}{\sqrt{Z  }}\,.
\end{align}
Therefore, the freedom is encoded in the functions $A(r)$ and $G_2(r)$. Using the above relations, we obtain finally the new exact solution of DHOST theories given by
\begin{align}
\label{newsol}
ds^2&  = A^{-1} \left[  - F \left( 1 + \frac{q^2  \left( 1-G_2 \right)}{Z } \right) dt^2  \mp \frac{2 q   \left( 1-G_2\right)}{\sqrt{Z }} dt dr + \frac{G_2}{F} dr^2 + r^2 d\Omega^2 \right]\,.
\end{align}
This new class of geometries provide, by construction, a new family of exact solutions for quadratic DHOST theories. Let us now compute the profile of the kinetic term in this new geometry. It reads 
\begin{align}
X & = \frac{A \tilde{X}}{1- \tilde{X} B} = \frac{ A \tilde{X} Z }{\tilde{X} FG_2+q^2}\,.
\end{align}
This concludes the presentation of this new family of exact solutions of quadratic DHOST theories. We now can turn to the analysis of its properties. The condition for asymptotic flatness is easily obtained as
\begin{align}
\lim_{r\to\infty} A (r) = A_{\infty}\,, \quad \lim_{r\to\infty} G_2(r) = 1\,,
\end{align}
where $A_{\infty}$ is a positive constant. Let us now compute the norm of the Kodama vector and the determinant of the metric. They reads
  \begin{align}
\label{kod4}
\mathcal{K}^{\alpha} \mathcal{K}_{\alpha} & = - \frac{F \left[r  A' +2 A \right]^2 \left(q^2 (1-G_2) + Z\right)}{4 \left(\tilde{X} F G_2+q^2\right) A^2}\,,\\
\label{det4}
 \det( g) & = -\frac{r^4 \sin ^2(\theta) \left(\tilde{X} F G_2+q^2\right)}{A^4 Z}  = - \frac{r^4 \sin^2{\theta}}{A^3} \frac{\tilde{X}}{X}\,.
  \end{align}
Using (\ref{kod4}) and (\ref{det4}), the necessary conditions for having a well defined modified geometry, with possibly new horizons, while preserving at the same time the sign of the determinant, are given by
\begin{align}
\label{choice0}
0<|A(r)| < \infty\,, \qquad \text{and} \qquad \forall r \geq r_s,\; |G_2(r)| < \infty\,, 
  \end{align}
and $q^2 >0$. 
As we are going to see now, preserving the determinant does not prevent from adding additional horizons to the seed geometry when $q\neq 0$. However, this is no longer true in the static case $q=0$.

Moreover, in order to have a well defined hairy black hole solution, it is required that the kinetic term of the scalar field $X$ remains regular on any horizons. It is straightforward to see that the behavior of $X$ is related to the behavior of the determinant $ \det( g) $ through (\ref{det4}). Since $\tilde{X}$ is always positive, then provided $ \det( g)$ does not vanish on any horizons, the kinetic term $X$ remains also regular on these horizons.

Consider therefore the case with $q\neq0$, and let us introduce a choice of the potentials $A(r)$ and $G(r)$ which satisfies the above conditions and generate a new black hole horizon. One interesting example is given by the following potentials
\begin{align}
\label{choice}
A(r)=1\,, \qquad \text{and} \qquad G_2(r) = \left(1 - \frac{r_{\ast}}{r}\right)^{2}\,,
\end{align}
which introduce a new arbitrary scale $r_{\ast}$. 
Depending on the value of $r_{\ast}$, one can introduce a new horizon on top on the one pre-existing at $r_{\circ}$ in the seed solution. This is made possible because the norm of the Kodama vector $\mathcal{K}^{\alpha} \mathcal{K}_{\alpha} =0$ can now have several roots. However, this modification introduces new singularities, which nevertheless remain hidden inside the inner horizon. The metric being now quite involved, we investigate the new structure of the solution only numerically. To this end, we shall plot the norm of the Kodama vector, the determinant and the $00$-component of the new metric as well as its Ricci scalar for the new solution for two set of the parameters $\left( \eta, \zeta, \beta, \mu, r_{\ast} \right)$. In what follows, we work with $\eta / (2\zeta) = 10^{-4}$ and $\mu = 1$ such that $r_{\circ}\sim 1$. Moreover, we set  $r_{\ast}/\mu = 4$. The properties of the seed solution are depicted in yellow, while the properties of the modified solution are depicted in blue.

As a first example, we consider the case $\eta/ \beta =2$. 
The plots of the three key quantities are given in Figure-\ref{fig1}. First, one can observe that the norm of the Kodama vector vanishes only one time in the seed solution, while it has two zeros in the modified geometry, signaling thus two horizons. The plot of the $00$-component of the metric shows that it vanishes precisely at the locus of each horizons, showing that they are indeed light-like horizons. Moreover it is clear from the sign of the norm of the Kodama vector that it becomes space-like in two regions, which signals the existence of a trapped region, as expected. The Ricci scalar remains regular on each horizon, and diverges in the interior region, where there are two new singularities, represented by the orange vertical lines. Nevertheless, they affect only the interior region bounded by the inner horizon, such that the geometry is well defined only up to the first singularity. In this range of $r$, the determinant is negative and keeps the same sign in the trapped region, ensuring that one can safely cross each horizons. Moreover, as the determinant does not vanish on any horizon, it implies from (\ref{det4}) that the kinetic term of the scalar field remains also regular on these horizons, providing a well defined hairy configurations.
\begin{figure}[h!]
  \centering
  \includegraphics[scale=0.5]{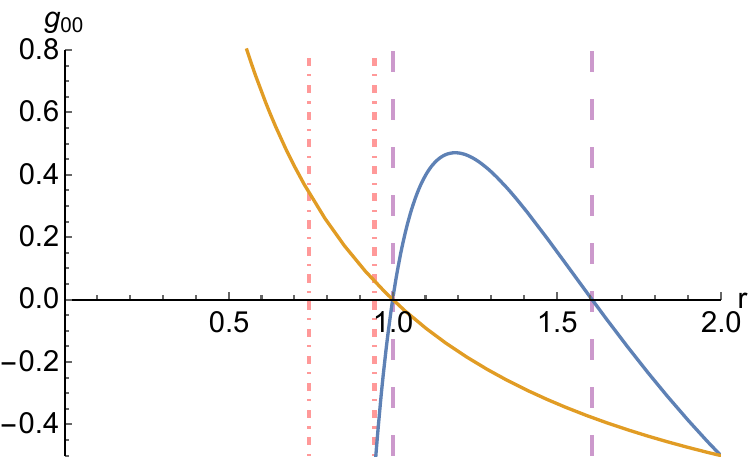}
 \; \includegraphics[scale=0.5]{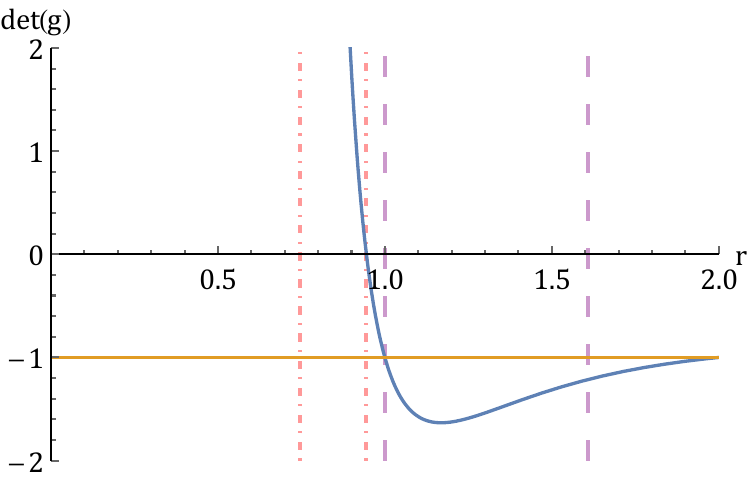}
 \includegraphics[scale=0.5]{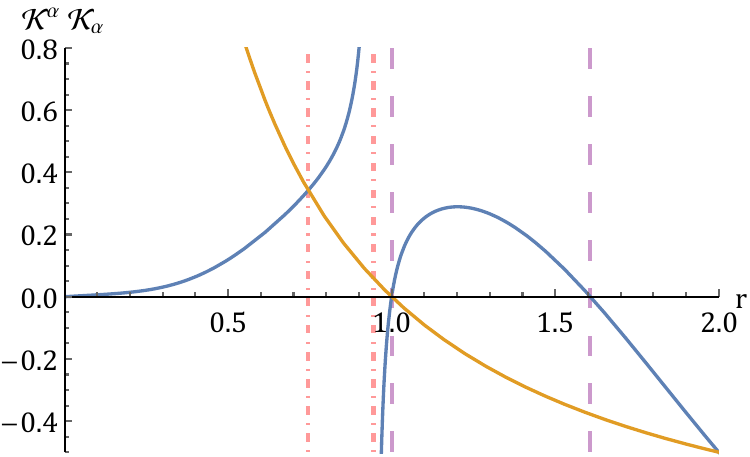}\,\,
 \includegraphics[scale=0.5]{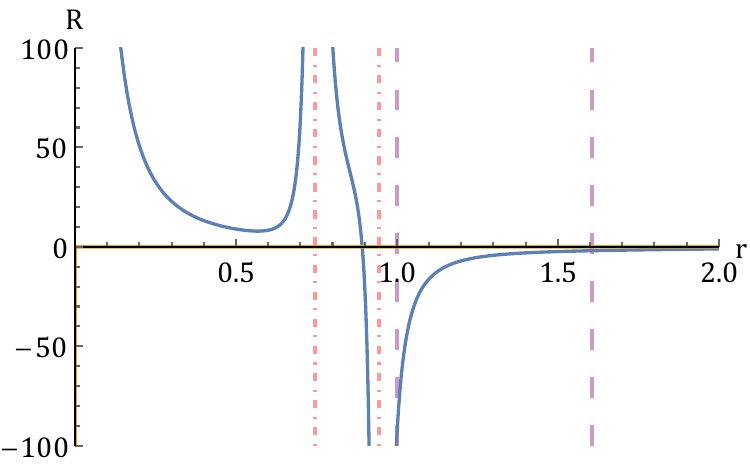}
  \caption{The determinant of the new metric $\text{det}(g)$, the $00$-components of the metric $g_{\alpha\beta}$,  the norm of the Kodama vector $\mathcal{K}^{\alpha} \mathcal{K}_{\alpha}$ as well as the Ricci scalar $R$ for the specific case $ \mu = 1$, $\eta/ \beta = 2$,  $\eta / (2\zeta) = 10^{-4}$ and $r_{\ast} /\mu= 4$. The yellow lines indicate the quantities related to the seed solution, while the blue lines indicate the behaviour of the modified geometry. The violet large dashed lines indicate the location of the two horizons, while the orange dashed lines indicate singularities.}
    \label{fig1}
\end{figure}
Notice that the position of the singularity can be shifted by tuning the value of the ration $\eta/\beta$. This is depicted on the Figure-\ref{fig2}, which corresponds to a ratio $\eta/\beta = 200$ while keeping the same values for the other parameters. For this second example, the singularity is pushed in the deep interior of the black hole. 
\begin{figure}[h!]
  \centering
  \includegraphics[scale=0.5]{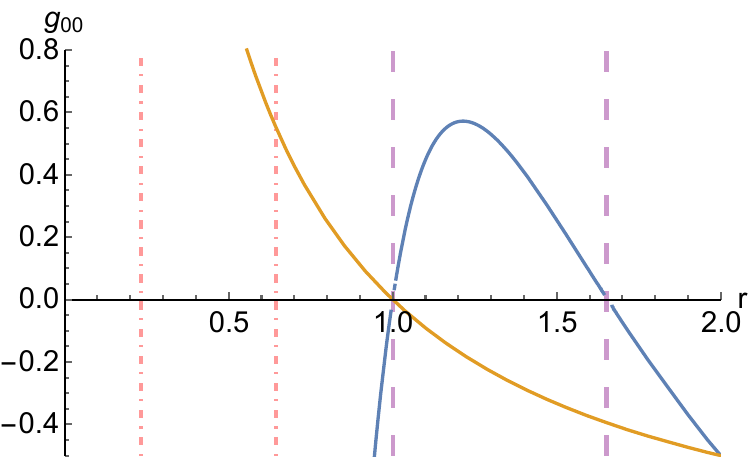}
 \; \includegraphics[scale=0.5]{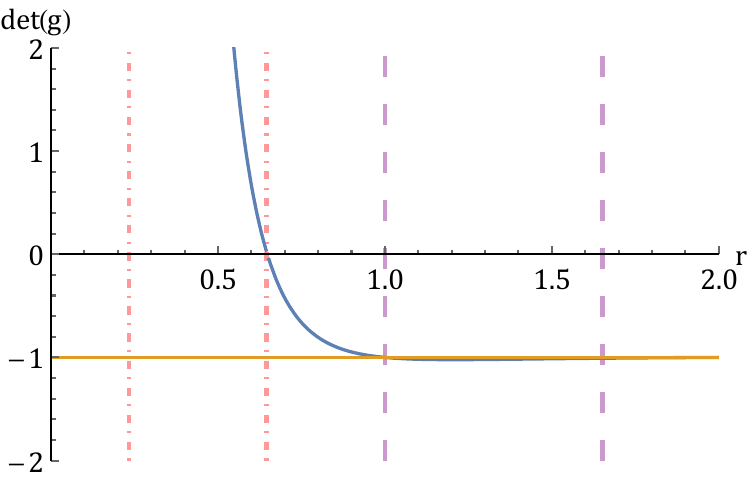}
 \includegraphics[scale=0.5]{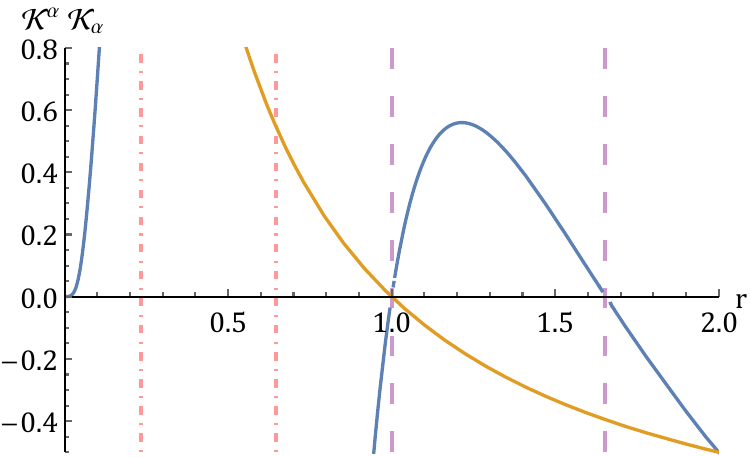}\,\,
 \includegraphics[scale=0.5]{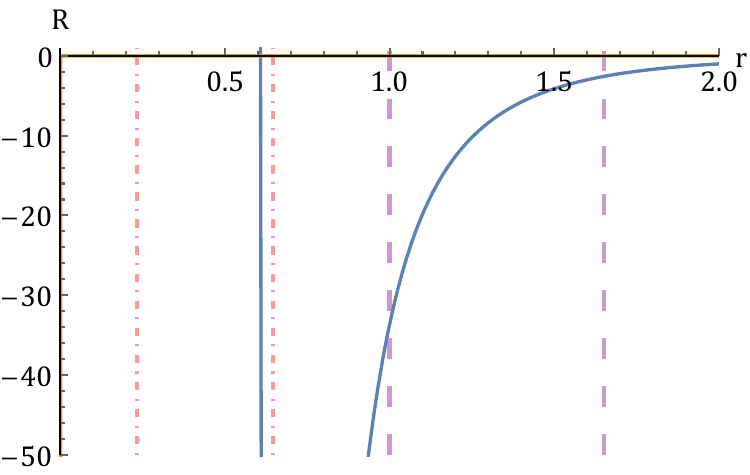}
  \caption{The determinant of the new metric $\text{det}(g)$, the $00$-components of the metric $g_{\alpha\beta}$,  the norm of the Kodama vector $\mathcal{K}^{\alpha} \mathcal{K}_{\alpha}$ as well as the Ricci scalar $R$ for the specific case $ \mu = 1$, $\eta/ \beta = 200$,  $\eta / (2\zeta) = 10^{-4}$ and $r_{\ast} /\mu= 4$. The yellow lines indicate the quantities related to the seed solution, while the blue lines indicate the behaviour of the modified geometry. The violet large dashed lines indicate the location of the two horizons, while the orange dashed lines indicate singularities.}
    \label{fig2}
\end{figure}

As a final remark, notice that one can use (\ref{fonc44}) to write the potential $G_2(r)$ as a function of the kinetic term $\tilde{X}$ of the seed solution. We obtain therefore
\begin{align}
G_2(r(\tilde{X})) = \left[ 1 -  r_{\ast} \sqrt{\frac{ \eta}{\sqrt{2}\beta} \frac{\tilde{X}^{3/2}}{1 + q^2\tilde{X}}}\right]^2\,.
\end{align}
Let us now investigate the case with $q=0$. In this static limit, the above modifications cannot be introduced anymore without spoiling the behavior of the determinant. One encounters the same situation as when starting from the JNW seed solution, as both are characterized by a non-constant kinetic term but a static scalar profile.
To see this, let us set $A=1$ and investigate possible corrections through the function $G_2(r)$. Now the norm of the Kodama vector and the expression of the determinant of the metric are given by the simple expressions
\begin{align}
\mathcal{K}_{\alpha} \mathcal{K}^{\alpha} & = - \frac{ F(r)}{G_2(r)}\,,\\
\text{det}(g)&  = - r^4 \sin^2{\theta} \; G_2(r)\,.
\end{align}
Therefore, the function $G_2(r)$ is constrained to satisfy
\begin{align}
0 < G_2(r) < + \infty\,.
\end{align} 
Moreover, the asymptotic flatness requires that $G(r) \sim 1$ when $r\rightarrow + \infty$.  While these conditions leave a large freedom in the choice of $G_2(r)$ to modify the exterior geometry of the black hole, they cannot generate a new horizon, since it would generate some inconsistencies in the behavior of the determinant. This illustrates the importance, additionally to the existence of a non-constant kinetic term, of a time-dependent scalar profile for the scalar field to introduce interesting modification of the seed solution through a disformal transformation. 

Notice that for this sub-case, the kinetic term of the scalar field takes a fairly simple form, i.e  $\tilde{X} \propto r^{-4}$ which allows to re-write straightforwardly the disformal potential in term of $\tilde{X}$. 
Once the function $G_2(r)$ is chosen, one can use the profile of the kinetic term of the seed solution to write
\begin{align}
G_2(r) = G_2\left( \left(\frac{2\beta^2}{\eta^2\tilde{X}}\right)^{1/4}\right)\,,
\end{align} 
which ensures that the transformation is invertible. This concludes the investigation of the disformal transformation of the last example of seed solutions discussed first in \cite{BenAchour:2020wiw}.

\section{Asymptotically stealth Schwarzschild black hole}

We have seen that even if the disformal transformation indeed allows one to construct non-stealth black hole slutions out of the stealth configuration, the outcome is not satisfactory. Because of the asymptotic behavior of the radial gradient of the scalar field profile, the geometry inherits a deficit angle. Stealth black hole configurations are also known to be generically unstable under perturbations, these later exhibiting a strong coupling problem. This raises the question whether one can still construct black hole solution carrying primary hair, but relaxing the stealth property such that the black hole geometry is only \textit{asymptotically} stealth. It would allow one to exhibit a black hole with a primary hair which reduces to the Schwarzschild (or Kerr) geometry in the near horizon and in the asymptotically flat regions. Morever, since these two regions are the ones used to fix the boundary conditions for the wave analysis, in particular for the stability analysis under perturbations, the hope is that such asymptotically stealth black hole could also mimic perturbations of GR black holes at least in the relevant regime.

In the following, we propose one simple mechanism to construct such asymptotically stealth black hole using non-shift symmetric disformal transformation. Consider a stealth black hole solution, say the Schwarzschild stealth solution where the scalar profile diverges both at the horizon and at spatial infinity. Then consider the following disformal transformation given by
\begin{align}
\tilde{g}_{\mu\nu} = g_{\mu\nu} + B_0 e^{-  \frac{(\varphi - \varphi_0)^2}{\sigma^2}} \varphi_{\mu} \varphi_{\nu}
\end{align}
Since the scalar field is diverging at the Schwarzschild horizon $r_s$ and at $r\rightarrow +\infty$, one obtains a geometry which deviates from the GR solution $g_{\mu\nu}$ only on a finite region of spacetime. The extend of this region is controlled by the two parameters of the Gaussian $(\sigma, \varphi_0)$. Yet, in the near horizon and at asymptotic infinity, the effects of the scalar field completely disappear, giving rise to the notion of asymptotically stealth.

Obviously, a first question is whether the use of a disformal function $B(\varphi)$ generates pathological divergences in some crucial quantities. The inverse metric, the jacobian and the kinetic energy of the scalar field transform as
\begin{align}
\tilde{g}^{\mu\nu} & = g^{\mu\nu} - \frac{B(\varphi)}{1+ B(\varphi) X} \phi^{\mu} \phi^{\nu} \qquad \tilde{X} = \frac{X}{1+ B(\varphi) X} \qquad \frac{\sqrt{|\tilde{g}|}}{\sqrt{|g|}} = \sqrt{1+ B(\varphi) X} 
\end{align}
One can see that even when $B(\varphi)$ vanishes, which is the critical behavior leading to potential divergence, none of these three quantities blow up.  
Consider now the correction to the Chritoffel symbol induces by the DT which reads 
\begin{align}
C_{\mu\nu}^\lambda = \frac{B(\varphi)}{1+B(\varphi) X}\phi^\lambda\phi_{\mu\nu} +\frac{B_{\varphi}(\varphi)}{1+B(\varphi) X}\phi^\lambda\phi_{\mu} \phi_{\nu} 
\end{align}
A potential divergence might shows up because of the second term, where
\begin{align}
B_{\varphi}(\varphi) = - 2 B_0\kappa (\varphi - \varphi_0) e^{- \kappa (\varphi - \varphi_0)^2}
\end{align}
Yet this contribution remains always bounded even when the scalar field profile tends to $\pm \infty$ since the exponential always dominates. It follows that any quantities constructed from the metric, its inverse and the Christoffel won't blow up when the scalar profile diverges. This is in particular the case of any curvature invariant.  Therefore, this broad analysis suggests that one can use this non-shift symmetric disformal transformation to construct a non-pathological geometry describing an asymptotically stealth black hole solution in DHOST gravity. Let us now discuss how to proceed in more detail.

We start by implementing our idea on the simplest spherically symmetric case, namely the stealth Schwarzschild black hole.
We first need to derive the seed solution which will be constructed using the method introduced in \cite{Charmousis:2019vnf}. Then, we shall construct the asymptotically stealth Schwarzschild black hole. Consider the Schwarzschild geometry
\begin{align}
\rd s^2 = - f(r) \rd t^2 + \frac{\rd r^2}{f(r)} + r^2 \rd \Omega^2 \qquad f(r) = 1 - \frac{r_s}{r}
\end{align}
When considering the geodesic motion, one can formulate it in terms of iso-action levels. This foliation in terms of the geodesic action function $S_{\text{geo}}$ gives a straightforward way to dress the black hole geometry with a scalar field by simply identifying the geodesic action with the scalar field, i.e. such that 
\begin{align}
S_{\text{geo}} = \varphi
\end{align}
Moreover, the integral form of the geodesic action function can be derived by means of the Hamilton-Jacobi method. Given the hamiltonian of the geodesic motion of a test particle of mass $m$, the action function satisfies 
\begin{align}
-\frac{\partial S_{\text{geo}}}{\partial \tau} =  H(x^{\mu}, \frac{\partial S_{\text{geo}}}{\partial x^{\mu}}, \tau) = g^{\mu\nu} \frac{\partial S_{\text{geo}}}{\partial x^{\mu}} \frac{\partial S_{\text{geo}}}{\partial x^{\nu}} = - m^2
\end{align}
where the momenta are given by
\begin{align}
p_{\mu} =  \frac{\partial S_{\text{geo}}}{\partial x^{\mu}}
\end{align}
Since spherical symmetry implies that given a fixed $\theta_0$, the trajectory remains in the same plane, let us assume that without loss of generality, we have $\theta= \pi/2$ and $\dot{\theta} =0$. For the Schwarzschild metric, one obtains 
\begin{align}
- f^{-1} \left( \frac{\partial S_{\text{geo}}}{\partial t}\right)^2 + f \left( \frac{\partial S_{\text{geo}}}{\partial r}\right)^2 + \frac{1}{r^2} \left( \frac{\partial S_{\text{geo}}}{\partial \phi}\right)^2 = -m^2
\end{align} 
Now, since $(t, \phi)$ are cyclic variables, the solution for $S$ takes the form
\begin{align}
S_{\text{geo}}(t, \phi, r) = - p_t t + p_{\phi} \phi + S_r(r)
 \end{align}
 where $p_t = E$ and $p_{\phi} = L$ are the constant of motions associated to the time-like and spacelike killing symmetries. It gives
 \begin{align}
 \left( \frac{\partial S_{\text{geo}}}{\partial r}\right)^2  = \frac{1}{f} \left[   \frac{E^2}{f} -m^2 - \frac{L^2}{r^2}\right] = \frac{1}{f^2} \left[  E^2 - f \left( m^2 + \frac{L^2}{r^2}\right)\right]
\end{align} 
Finally, the form of $S_r(r)$ is given by
 \begin{align}
 S_r(r) = \int \frac{\rd r}{f} \sqrt{E^2 - f \left( m^2 + \frac{L^2}{r^2}\right)}
 \end{align}
 This provide with the profile of the scalar field which reads
 \begin{align}
 \varphi(t, r, \phi) = - E t + L \phi + \int \frac{\rd r}{f} \sqrt{E^2 - f \left( m^2 + \frac{L^2}{r^2}\right)}
 \end{align}
With this solution, the scalar field diverges at $t\rightarrow \pm \infty$. More importantly, it also diverges at the horizon $r=r_s$ and at large $r$.
 Therefore, the stealth Schwarzschild black hole constructed with this method provides the required seed metric. Now, we can construct the asymptotically stealth Schwarzschild black hole. The metric is given by
\begin{align}
\rd s^2 & = - \left( f + B_0 E^2 e^{- \kappa (\varphi - \varphi_0)^2} \right) \rd t^2 -  \frac{B_0 E}{f} \sqrt{E^2 - f \left( m^2 + \frac{L^2}{r^2}\right)} e^{- \kappa (\varphi - \varphi_0)^2} \rd t \rd r  \\
& \;\;\;+ \frac{1}{f} \left( 1 + \frac{B_0}{f}  e^{- \kappa (\varphi - \varphi_0)^2} \left[  E^2 - f \left( m^2 + \frac{L^2}{r^2}\right)\right] \right) \rd r^2  \\
& \;\;\; + r^2 \left[ \rd \theta^2 + \left( \sin^2{\theta} + B_0 L^2 e^{- \kappa (\varphi - \varphi_0)^2} \right) \rd \phi^2 \right]
\end{align}
while the scalar field is given by
 \begin{align}
 \varphi(t, r, \phi) = - E t + L \phi + \int \frac{\rd r}{f} \sqrt{E^2 - f \left( m^2 + \frac{L^2}{r^2}\right)}
 \end{align}
 They are four parameters to this solution: $(B_0, m, E, L)$ along with the Schwarzschild mass. By construction, this black hole is solution of a given class of DHOST theories and it reproduces the Schwarzschild geometry both near the horizon and in the asymptotically flat region.
 This construction is currently under investigation to see wether it can provide an interesting algorithm to construct phenomenologically interesting black holes solution in DHOST gravity.\\ \\
 \bigskip
 
 To conclude, let us mention several key points which go beyond the scope of this chapter. An important challenge when identifying new hairy black holes is to test their stability against perturbations. Addressing this question is challenging and important efforts are currently devoted to this task. See \cite{Takahashi:2019oxz, Takahashi:2021bml, Tomikawa:2021pca, Hui:2021cpm, Mukohyama:2022enj, Barura:2024uog, Mukohyama:2024pqe} for an EFT approach to this problem. See also \cite{Langlois:2021aji, Langlois:2022ulw, Roussille:2022vfa} for a new algorithm to study black hole perturbations within DHOST theories. See also \cite{Charmousis:2019fre}. Several crucial points have not been addressed in this chapter. The reader interested in the screening mechanism and in the new phenomenology arising in models of stars can refer to \cite{Crisostomi:2017lbg, Boumaza:2022abj} and references therein.

\def\be{\begin{equation}}
\def\ee{\end{equation}}
\def\beq{\begin{equation}}
\def\eeq{\end{equation}}
\def\bi{\begin{itemize}}
\def\ei{\end{itemize}}
\def\ba{\begin{array}}
\def\ea{\end{array}}
\def\bfig{\begin{figure}}
\def\efig{\end{figure}}

\newcommand\alambda{{\alpha_\lambda}}
\newcommand\cA{\mathcal{A}}
\newcommand\cB{\mathcal{B}}
\newcommand\cD{\mathcal{D}}
\newcommand{\CI}{{\cal C}_{\rm I}}
\newcommand{\CII}{{\cal C}_{\rm II}}
\newcommand{\CU}{{\cal C}_{\rm U}}
\newcommand\cL{{\mathcal{L}}}
\newcommand\cS{{\mathcal{S}}}
\newcommand\cT{{\mathcal{T}}}
\newcommand\cJ{{\mathcal{J}}}
\newcommand\cK{{\mathcal{K}}}
\newcommand\cR{{\mathcal{R}}}
\newcommand\cO{{\mathcal{O}}}
\newcommand\cC{{\mathcal{C}}}
\def\tA{\hat A} 
\def\B{{\cal B}}
\def\a{\alpha}
\def\f{f}

\newtheorem{theorem}{Theorem}
\newtheorem{corollary}{Corollary}[theorem]
\newtheorem{lemma}[theorem]{Lemma}

\chapter{On rotating hairy black holes}
\label{Chapter5}

\textit{"The old problem of constructing rigorously the field of a finite rotating body is as yet unsolved, even as to its exterior part"}\\ \\
\bigskip
\textit{J. Elhers and W. Kundt, 1962, Review - Exact solutions of the gravitational field equations}

\minitoc

In this chapter, we review the recent efforts devoted to find exact solutions of the DHOST field equations and describing a stationary rotating black hole. This endavour is a key target of the next years as it would provide a platform to confront these theories to the strong field regime currently observed by the Event Horizon Telescope and the future mission such as the Black Hole Explorer \cite{Berti:2015itd}. We first present a short survey of the exact solutions published so far for Horndeski and beyond Horndeski theories. Then, following the same structure than the previous chapter, we first discuss the effect of a disformal transformation onto the rotating naked singularity of the Einstein-Scalar system found by Bogush and Gal' tsov \cite{Bogush:2020lkp} and presented in \cite{BenAchour:2020fgy}. In the next part, we describe the construction of the stealth Kerr solution presented in \cite{Charmousis:2019vnf} and the new geometry obtained by disforming this rotating stealth solution \cite{BenAchour:2020fgy}. Let us emphasize that the resulting geometry, i.e. the disformed Kerr solution, provides one of the rare exact analytic solution at hand in DHOST theories. The rest of the chapter is devoted to describe its properties (singularities, ergoregions, horizons, geodesic motion) and even more importantly the non-standard ones (non-circularity, non analyticity, absence of Killing horizon, no separability of the geodesic equation). We further present the derivation allowing one to identify which subset of DHOST theories admit this stealth and disformed Kerr geometries as exact solutions.

\section{A survey of rotating solutions in Horndeski and beyond}

We begin with a brief catalogue of the attempts to construct hairy rotating black hole solutions in Horndeski and beyond Horndeski theories. There are only few such solutions available in the literature.
\begin{itemize}
\item \cite{Maselli:2015yva} : The authors investigate slowly rotating black holes in Horndeski gravity by following the approximation scheme of Hartle, i.e. including the effect of rotating at first order in the black hole angular velocity. The scalar profile is considered time and radially dependent and the coupling to the Gauss-Bonnet term is taken into account. See also \cite{Delgado:2020rev} for a spinning solution in the Einstein-Gauss-Bonnet model.
\item \cite{Babichev:2017guv} : By investigating how to explicitly break each assumptions of the Hui-Nicolis no-hair theorem, the authors identified a set of lagrangian allowing for static asymptotically flat black holes. Among them, a stealth Kerr black hole is presented which, to our knowledge, is the first example found in Horndeski and beyond Horndeski gravity.
\item \cite{Charmousis:2019vnf} : A new stealth Kerr solution was presented in this work. The key idea is to identify the scalar field with the level of the iso-action hypersurface for the geodesic motion. The scalar field then follows the Hamilton-Jacobi equation on the GR background and does not gravitate. The key properties of this new solution is that one can select a profile which is regular on both the past and future Kerr horizons. This solution and its construction will be review in the next section.
\item \cite{VanAelst:2019kku}: This solution was obtained numerically in the context of the cubic galileon theory. It provides an interesting example of a stationary axisymmetric black hole which i) is non-circular, ii) exhibits a Killing horizon but iii) for which the zeroth law of black hole thermodynamics is no longer valid, i.e. the surface gravity is not constant on the horizon \cite{Grandclement:2023xwq}. Equatorial geodesic motion around this hairy black hole have been studied in \cite{VanAelst:2021uem}. Moreover, the asymptotic fall-off in $1/r$ turns out to be too strong, leading to a vanishing ADM mass.
\item \cite{BenAchour:2020fgy, Anson:2020trg}: This new solution was constructed independently by two groups by performing a constant disformal transformation onto the stealth Kerr solution found in \cite{Charmousis:2019vnf}. It provides a first exact and analytic non-stealth rotating black hole solution in DHOST gravity. Its properties have been explored by several groups, among which \cite{Long:2020wqj, Chagoya:2020bqz, Babichev:2024hjf, Babichev:2024hjf}. Despite being given in analytic form, the identification of its horizons has not been completed yet. Moreover one can show that the horizon, if it exists, is not Killing. This chapter is devoted to a review of this solution, following \cite{BenAchour:2020fgy}.
\item \cite{Baake:2021kyg}: The disformal trick is further used to construct a rotating stealth black hole in 5d, i.e. a stealth Myers-Perry black holes solutions. 
\item \cite{Walia:2021emv}: The authors have constructed a new rotating black hole in Horndeski gravity by implementing the Janis-Newman algorithm on the spherically symmetric hairy black hole solution presented in \cite{Bergliaffa:2021diw}. This new rotating black hole has been studied in \cite{Afrin:2021wlj}.
\item \cite{Babichev:2023mgk}: A non-stealth rotating black hole solution of DHOST which is asymptotically FLRW is presented and studied. The solution is obtained by a conformal transformation onto the stealth Kerr-de Sitter solution found in \cite{Charmousis:2019vnf}.
\end{itemize}
As one can see from this short list, exact solutions describing rotating black hole beyond the stealth sector rely on either numerical efforts or on the use of solution-generating maps based on field redefinitions. While new structures of the solution space of DHOST theories might be found in the near future, it underlines the crucial role played by the disformal solution-generating technique to construct analytical solutions whose phenomenology can be studied. In the following we shall review the outcome of disforming the stealth Kerr black hole with a time-dependent scalar profile. Yet, we stress that at the moment, we are lacking a more systematic approach to study rotating compact objects in DHOST gravity. What one would need to go further is to understand how does the phase space of stationary and axi-symmetric geometries ruled by DHOST theories deform or break the known symmetries of its GR counterpart. This would require investigating how to reformulate this phase space in terms of Ernst-like equations for both the metric and the scalar field. This is an on-going project.

\section{A rotating naked singularity solution in DHOST}
\label{AppB}

In the previous chapter, we have reviewed a no-go result, demonstrating that under generic assumptions, the spherically symmetric naked singularity solution of the massless Einstein-Scalar system cannot be turned into a black hole by means of a (simple) disformal transformation. In this first section, we show that the same result holds when implementing a disformal transformation on the rotating naked singularity solution found by Bogush and Gal' tsov in \cite{Bogush:2020lkp}. 

Nevertheless, this GR solution provides an interesting seed as it allows to work with a generalized Kerr geometry associated to a scalar source whose kinetic term is no longer constant. Moreover, the static limit of this solution reduces to the well known quadrupolar metric, also known as the $q$-metric, which describes a non-rotating but axi-symmetric quadrupolar deformation of the Schwarzschild black hole. Here, the quadrupole charge and the scalar charge play the same role at the level of the metric. Therefore, it allows us to generate a DHOST generalization of this quadrupolar-like deformation of the Schwarzschild geometry.

\subsection{The Bogush-Gal' tsov naked singularity}

Let us first describe the seed solution of the massless Einstein-Scalar system whose action reads
\be
\cS = \int \rd^4x \sqrt{|g|} \left( \frac{R}{16 \pi G} -  \frac{1}{2} g^{\mu\nu} \phi_{\mu} \phi_{\nu}  \right) \, .
\ee
The field equations are given by
\begin{align}
G_{\alpha\beta}  = 8\pi G \left(\phi_\alpha \phi_\beta - \frac{1}{2} \phi_{\mu} \phi^{\mu} \, g_{\alpha\beta}  \right)  \;, \qquad \Box \phi  = 0 \, .
\end{align}
As usual, the equation of the scalar field is a consequence of the Bianchi identity and the Einstein equation.
Recently, an exact stationary solution of these  equations was constructed in \cite{Bogush:2020lkp}. This solution was obtained by exploiting the well known hidden symmetries of vacuum axisymmetric solutions of GR \cite{Ehlers:1957zz, Matzner:1967zz, Geroch:1972yt} as well as solution-generating method based on the Einstein-Maxwell system \cite{Galtsov:1995mb, Clement:1997tx}. The resulting geometry provides a generalization of the Kerr geometry with a scalar source. Explicitly, the metric  takes the form
\begin{align}
ds^2 = - f \left( dt - \omega d\psi \right)^2 +  \frac{h_{ij}}{f} dx^i dx^j  \, ,
\end{align}
where the functions $f, \omega$ and $\Delta $ entering in the metric coefficients are given by
\begin{align}
f \equiv 1 - \frac{2Mr}{\rho}\;, \qquad 
 \omega  \equiv - \frac{2a M r \sin^2{\theta}}{\Delta - a^2 \sin^2{\theta}}\;, \qquad  \Delta  \equiv \left(r- M\right)^2 - b^2 \, ,
\end{align}
with $b \equiv \sqrt{M^2 - a^2}$ and $\rho = r^2 + a^2 \cos^2{\theta}$. The remaining spatial part of the line element is explicitly given by,
\be
h_{ij} dx^i dx^j \equiv H \left( dr^2 + \Delta \, d\theta^2 \right) + \Delta \sin^2{\theta} \, d\psi^2 \, 
\ee
where the function $H$ is given by
\be
H \equiv 
\frac{\rho}{\Delta}f \zeta \;, \qquad \text{with} \qquad \zeta = \left( 1 + \frac{b^2}{\Delta} \sin^2{\theta}\right)^{- \Sigma^2 /b^2}.
 \ee
Finally, the associated scalar field is given by
\be
\label{scalprof}
\phi(r) = \phi_{0} + \frac{\Sigma}{2b} \log{\left[ \frac{r - M + b}{r-M-b}\right]} \, .
\ee
This  solution depends on three parameters: the mass $M$, the (rescaled) angular momentum $a$ and the scalar charge $\Sigma$. When the scalar charge vanishes, i.e $\Sigma =0$, the metric reduces to the Kerr geometry as expected. The effect of the massless scalar field is encoded in the function $\zeta(r,\theta)$. The singularities of the solution can be tracked by computing the Ricci scalar which reads
\be
R = \frac{2\Sigma^2 }{\Delta \zeta \left( r^2 + a^2 \cos^2{\theta}\right)}  \, .
\ee
At constant angle $\theta$ when $\Delta \rightarrow 0$, we see that $R \sim \Sigma^2 r^{-2} \Delta^{-1-\Sigma^2/b^2}$.
Hence the hypersurface defined by $\Delta =0$ is singular when $\Sigma \neq 0$, which signals that the Kerr outer horizon has been turned into a curvature singularity because of the presence of the scalar field. If $M \geqslant a$, the location of this singular hypersurface is at
\be
r_{+} = M + M \sqrt{1 -  J^2} \leqslant 2M \, ,
\ee
where $J \equiv a /M$.  Consequently, this solution is only defined for $r \in \; ] r_{+}, + \infty [$ and describes the gravitational field of rotating compact object endowed with a scalar charge.

Taking the non-rotating limit $a\rightarrow 0$, the geometry reduces to a deformed Schwarzschild metric which corresponds to a sub-class of the Zipoy-Voorhees (ZV) metric with scalar source which was first derived in \cite{Fisher:1948yn, Janis:1969ivo}. See \cite{Astorino:2014mda, Turimov:2018guy, Chauvineau:2018zjy} for more recent generalizations. The vacuum ZV metric represents the simplest static and axi-symmetric vacuum solution of GR \cite{Voorhees:1970ywo}. See \cite{Toktarbay:2014yru, Quevedo:2012ttw} for details. We shall now investigate the disformal transformation of this Einstein-Scalar exact solution. 

\subsection{Disformed naked singularity and $q$-like solution}

Performing the same constant disformal transformation as the one used in the core of the paper, we obtain a new exact solution in  DHOST theories whose metric still takes the form 
\begin{align}
ds^2 = g_{\mu\nu} dx^{\mu} dx^{\nu} =   - f \left( dt - \omega d\psi \right)^2 + \frac{h_{ij}}{f} dx^i dx^j \, ,
\end{align}
with the same scalar profile (\ref{scalprof}).
The only modification, compared to the previous undeformed solution, shows up in the $g_{rr}$ components.  
Indeed, the spatial metric $h_{ij}$ reads now
\be
h_{ij} dx^i dx^j = H \left( G dr^2 + \Delta \, d\theta^2 \right) + \Delta \sin^2{\theta}\, d\psi^2
\ee
with
\begin{align}
H  =  \frac{\rho f }{\Delta}\left( 1 + \frac{b^2}{\Delta} \sin^2{\theta}\right)^{- \Sigma^2 /b^2} \;, \qquad 
 G = 1 -  \frac{B_{0} f}{H} \left( \phi' \right)^2 \, .
\end{align}
where a prime denotes derivative w.r.t the radial coordinate $r$.
As expected, the new solution has now four parameters: the mass $M$, the rescaled angular momentum $a$, the scalar charge $\Sigma$ and the disformal parameter $B_0$. We shall again only consider sufficiently small values of this parameter which ensures that the deformation function $G$ does not generate any singularity.
We can now investigate the properties of this new exact DHOST solution.

\medskip

First, we study the static limit of the disformed geometry. When the angular momentum vanishes, i.e.\ $a\rightarrow 0$, the DHOST solution reduces to the following metric
\begin{align}
ds^2 & = - \left( 1 - \frac{2M}{r}\right) dt^2 + r (r-2M) \sin^2{\theta} \, d\psi^2 \nonumber \\
& + \left( 1 - \frac{2M}{r}\right)^{-1 + \Sigma^2/M^2} \left( 1 - \frac{2M}{r} + \frac{M^2}{r^2} \sin^2{\theta}\right)^{-\Sigma^2/M^2} \left( G \, dr^2 + r(r- 2M) \, d\theta^2 \right)
\end{align}
while the scalar profile (\ref{scalprof}) reduces to
\be
\label{scal}
\phi(r) = \phi_{0} - \frac{\Sigma}{2M} \log{\left[ 1-\frac{M }{r}\right]} \, .
\ee
The explicit form of the function $G$ which contains the disformal parameter is given by
\be
\label{g}
G(r) = 1 - B_{0} \left( 1 - \frac{2M}{r}\right)^{1 + \Sigma^2/M^2} \left( 1 - \frac{2M}{r} + \frac{M^2}{r^2} \sin^2{\theta}\right)^{-\Sigma^2/M^2} \left( \phi' \right)^2 \, .
\ee
Interestingly, this axisymmetric but static solution provides a DHOST generalization of the well known vacuum $q$-metric of GR \cite{Quevedo:2012ttw} and its recent extension with a scalar source presented in \cite{Turimov:2018guy}.
When the scalar charge vanishes, i.e.\ $\Sigma = 0$, the solution reduces to the Schwarzschild metric. As such, this new solution provides an new static axi-symmetric deformation of the Schwarzschild geometry with a quadrupole momentum in  DHOST theories. Let us also mention that, if one computes the gradient of the scalar field which enters in (\ref{scal}), we show that $G(r) \sim 1$ when $r\rightarrow +\infty$, and there is therefore no conical defect.

\medskip

We  now  investigate the causal structure. As the disformal parameter appears only in the $g_{rr}$ and $g_{\theta\theta}$ components of the metric, one can easily compute principal null directions given by
\begin{align}
\ell_{+}^{\alpha}\partial_{\alpha} & = \frac{\Delta}{\rho} \left( 
 \frac{r^2 + a^2}{\Delta} dt + \frac{dr}{G^{1/2}\zeta^{1/2}} +
 \frac{a}{\Delta} d\psi  \right) \, ,\\
\ell_{-}^{\alpha}\partial_{\alpha} & = 
  \frac{r^2 + a^2}{\Delta} dt - \frac{dr}{G^{1/2}\zeta^{1/2}} + 
 \frac{a}{\Delta} d\psi  \, ,
\end{align}
such that $g_{\alpha\beta}\ell_{\pm}^{\alpha} \ell^{\alpha}_{\pm} = 0 $ and $g_{\alpha\beta} \ell_{+}^{\alpha} \ell^{\beta}_{-} =-2$. 
The associated expansions  are given by
\begin{align}
\Theta_{+} & = \frac{ r \left( r-2M\right) +a^2 }{4  \sqrt{G \zeta}} \ \left( \frac{4 r}{\rho} +\frac{\zeta'}{\zeta} \right) \, ,\\
\Theta_{-} & = - \frac{1}{4\sqrt{G \zeta}} \left(  \frac{4 r}{\rho} +\frac{\zeta'}{\zeta} \right) \, .
\end{align}
Therefore, the product of the expansions reads
\be
\Theta_{+} \Theta_{-} =- \frac{ r \left( r-2M\right) +a^2 }{4 |G|| \zeta|} \left( \frac{4 r}{\rho} +\frac{\zeta'}{\zeta} \right)^2 \, .
\ee
The effect of the disformal transformation appears through the function $G$ which depends explicitly on $B_{0}$. We observe that this function appears either as a square root in the individual expansion or as an absolute value in the product $\Theta_{+} \Theta_{-} $. Therefore the disformal transformation cannot change the global sign of this quantity and the causal structure remains the same as the GR one in the DHOST frame. The new solution is horizonless and thus describes a rotating naked singularity. 

\section{Stealth Kerr solution}

Let us now review the construction of the stealth Kerr solution obtained in \cite{Charmousis:2019vnf}. When considering the geodesic motion, one can formulate it in terms of iso-action levels. This foliation in terms of the geodesic action function $S_{\text{geo}}$ gives a straightforward way to dress the black hole geometry with a scalar field by simply identifying the geodesic action with the scalar field, i.e. such that 
\be
S_{\text{geo}} = \varphi
\ee
Moreover, the integral form of the geodesic action function can be derived by means of the Hamilton-Jacobi method. Given the hamiltonian of the geodesic motion of a test particle of mass $m$, the action function satisfies 
\begin{align}
-\frac{\partial S}{\partial \tau} =  H(x^{\mu}, \frac{\partial S}{\partial x^{\mu}}, \tau) = g^{\mu\nu} \frac{\partial S}{\partial x^{\mu}} \frac{\partial S}{\partial x^{\nu}} = - m^2
\end{align}
where the momenta are given by
\be
p_{\mu} =  \frac{\partial S}{\partial x^{\mu}}
\ee
Following this strategy, it was shown in \cite{Charmousis:2019vnf} that one can dress a Kerr-(A)dS black hole with a scalar profile which is well behaved on both past and future horizons.
In Boyer-Lindquist coordinates $(t,r,\theta,\psi)$, the Kerr-(A)dS metric reads
\begin{align}
ds^2 = - \frac{\Delta_r}{\Xi^2 \rho} \left( dt - a \sin^2{\theta} \, d\psi \right)^2 + \rho \left( \frac{dr^2}{\Delta_r} + \frac{d\theta^2}{\Delta_{\theta}}\right) + \frac{\Delta_{\theta} \sin^2{\theta}}{\Xi^2 \rho} \left( a dt - \left(r^2 + a^2 \right) d\psi \right)^2
\label{Kerrmetric}
\end{align}
where $\Xi \equiv 1 + {a^2}/{\ell^2}$ is a constant, and the different functions entering in the metric are defined by
\begin{align}
\Delta_r & = \left( 1 - \frac{r^2}{\ell^2}\right) \left( r^2 + a^2\right) - 2 Mr \;, \qquad  \Delta_{\theta}  = 1 + \frac{a^2}{\ell^2} \cos^2{\theta} \;,\quad \rho  = r^2 + a^2 \cos^2{\theta} \, ,
\end{align}
while $M$ is the mass of the black hole and $a$ the angular momentum parameter satisfying the condition $a \leq M$.

Applying the strategy explained above where the scalar profile coincides with the
Hamilton-Jacobi potential associated to the Kerr geodesic equation, one obtains that
\bea
\label{profilekerr}
\phi(t,r,\theta) = -Et + S_r(r) + S_{\theta} (\theta) \, , \qquad
S_r \equiv  \pm \int dr \, \frac{\sqrt{\cR}}{\Delta_r} \;, \quad S_{\theta} \equiv \pm \int d\theta \, \frac{\sqrt{\Theta}}{\Delta_{\theta}} 
\eea
where $E$ is a constant while the two functions $S_r$ and $S_\theta$ are defined, up to a sign ambiguity, as integrals involving the
radial and angular functions
\be
\label{func}
\cR(r) \equiv m^2 \left( r^2 + a^2\right) \left[ \eta^2 \left( r^2 + a^2\right) - \Delta_r\right]\;, \qquad \Theta(\theta) \equiv a^2 m^2 \sin^2{\theta} \left( \Delta_{\theta} - \eta^2 \right) \, ,
\ee
with
\begin{align}
X_0=-m^2 \qquad \eta \equiv \Xi \frac{E}{m}
\end{align} 
In fact, there are four different branches for the scalar field $\phi$ because of the freedom to choose the signs
of $S_r$ and $S_\theta$. It has been shown in \cite{Charmousis:2019vnf}  that one can make use of these branches to construct a scalar field solution which is regular and finite in an untrapped region as well as a trapped region (either a black hole region or a white hole region but not both), and in particular on the the black hole horizon (as well as on the cosmological horizons when $\ell^2 > 0$).  

In the particular case where there is no cosmological constant $\ell \rightarrow \pm \infty$, we have $\eta = 1$, then $\Theta$ vanishes, and
finally the scalar field does not depend on the variable $\theta$ anymore. As we are going to see in the following section, this is the case we 
will focus on when we consider disformal Kerr solutions to avoid several issues. Furthermore, the radial function $\cR(r)$ simplifies as well and becomes
\be
\label{funcR0}
\cR(r) \equiv 2 M m^2  r \left( r^2 + a^2\right)\; ,
\ee
with the condition $E=m$ which identifies the kinetic energy $X_0=-m^2$ to $E^2$.

It was derived for theories within the class Ia with no cubic galileon term where gravitational waves propagate at the speed of light ($c_{\rm GW}=c$), i.e.\
\bea
\label{cequal1}
A_1=A_2=0 \, , \qquad Q=0 \, .
\eea
Here, the condition $c_{\rm GW}=c$ implies $A_1=0$~\cite{Langlois:2017dyl}, and we have used the degeneracy condition $A_2=-A_1$. 
In this case, the stealth conditions are given by
\bea
P(X_0) + 2 \Lambda F(X_0) = 0 \, , \quad
P_X(X_0) + 4 \Lambda F_X(X_0)  = 0 \, , \quad
A_3(X_0) = 0 \, .
\eea
The metric is the usual  Kerr solution of GR, or the de Sitter (dS)/Anti-de Sitter (AdS) Kerr solution

\section{Beyond circularity: the disformed Kerr solution}

We now describe the construction of the disformed Kerr black hole presented in \cite{BenAchour:2020fgy, Anson:2020trg}.
Consider the previous stealth Kerr-(A)dS seed solution with a constant kinetic term. We turn now to generate a new non-stealth solution via the disformal solution generating map.  The disformed metric takes the form
\bea
\label{disKerrformal}
g_{\mu\nu} \; = \; \tilde{g}_{\mu\nu} - B_0 \, \phi_\mu \phi_\nu \, ,
\eea
where $\tilde{g}_{\mu\nu} $ is the Kerr metric \eqref{Kerrmetric}. Without loss of generality, we have fixed $A_0=1$, otherwise the metric would simply get a global physically irrelevant constant conformal factor. If the scalar field $\phi$ depends on the angular variable $\theta$, then 
the  disformed Kerr metric \eqref{disKerrformal} acquires new components, among which
\bea
g_{t \theta} =  \pm B_0 E \frac{\sqrt{\Theta}}{\Delta_\theta} \, , 
\eea
where the expression of $\Theta(\theta)$ and $\Delta_\theta(\theta)$ has been recalled  in \eqref{func}.
Such a  term depends on the radial variable $r$ and then they do not vanish at infinity. As a consequence, one cannot expect
that the disformed metric is asymptotically flat, dS or AdS. To avoid this pathological behavior, we require that the scalar field does not
depend on $\theta$ which implies necessarily the vanishing of the cosmological constant $\ell \rightarrow \pm \infty$, then $\eta=1$ and $E=m$ (as
a consequence of $\Theta=0$). Hence, from now on, we consider only this case which, as recalled before \eqref{funcR0},  corresponds to scalar field of the form
\bea
\label{profilekerr}
\phi(t,r) = -m t + S_r(r)  \, , \qquad
S_r =  \pm \int dr \, \frac{\sqrt{\cR}}{\Delta} \, , \qquad
\Delta = r^2+a^2-2Mr \; , 
\eea
where, for simplicity, we have omitted the subscript $r$ in $\Delta$ (as there is no more possible ambiguity).
Thus, the disformal transformation (with $A_{0} =1$) leads to the new solution 
\bea
\label{newsol}
ds^2 &=&- \frac{\Delta}{\rho} \left( dt - a \sin^2{\theta} \, d\psi \right)^2 + \frac{\rho}{\Delta} {dr^2} + \rho \, {d\theta^2} + \frac{\sin^2{\theta}}{\rho} \left(a \, dt - \left(r^2 + a^2 \right) d\psi \right)^2 \nonumber \\
&& + \alpha \left( dt  \pm  {\sqrt{2Mr(r^2+a^2)}}/{\Delta} \, dr\right)^2 \, ,
\eea
with $\alpha \equiv -B_0 m^2$ while the scalar field profile remains unchanged (\ref{profilekerr}). 
The inverse disformed Kerr metric is given by
\bea
\label{inversedisf}
g^{\mu\nu} \, = \, \tilde{g}^{\mu\nu} + \frac{\alpha}{m^2(1-\alpha)} \phi^\mu \phi^\nu \, ,
\eea
where $\tilde{g}^{\mu\nu}$ is the inverse Kerr metric while the only non-vanishing components of $\phi^\mu = \tilde{g}^{\mu\nu} \phi_\nu$ 
are,
\bea
\phi^t =  \frac{m}{\Delta}\left( r^2+a^2 + \frac{2M r a^2 \sin^2\theta}{\rho}\right) \, , \quad
\phi^r =   m \frac{\sqrt{2M r (r^2+a^2)}}{\rho}\,, \quad
\phi^\varphi =m\frac{2aMr }{\Delta \rho} \, .
\eea
Therefore, the disformal transformation of the stealth Kerr solution provides a new \textit{non-stealth} exact solution which is parametrized, in addition to the  mass and angular momentum parameters $(M, a)$ of the Kerr family, by one new deformation parameter $ \alpha$ which encodes precisely the deviations from GR. The apparent $\pm$ ambiguity in \eqref{newsol} can be absorbed thanks to simple redefinitions of 
$t$ and $a$ which are replaced by $\pm t$ and $\pm a$. Hence, we can safely fix the sign to $+$ from now on without loss of generality. 

\subsection{Singularities and asymptotic behavior}

In this section, we quickly discuss geometrical properties of the disformed Kerr space-time. First of all,  we say a few words on its singularities. Even though the metric is 
singular in Boyer-Lindquist coordinates when $\Delta=0$ (at the values $r_\pm = M \pm \sqrt{M^2-a^2}$ of the radial coordinate), this is 
not a physical singularity but only a coordinate singularity exactly as in the usual Kerr black hole. Indeed, this can be seen immediately from the expressions of the curvature invariants,
\begin{align}
R & =  \frac{ \alpha }{1-\alpha} \frac{6 a^2 M r}{\rho^3} \left(\cos^2{\theta} - \frac{1}{3} \right)\, , \\
\label{CI2}
R_{\mu\nu} R^{\mu\nu} & = - \frac{\alpha^2}{\left( 1-\alpha \right)^2} \frac{18 a^4 M^2 }{\left( r^2 + a^2\right)\rho^6} \; P_1(r, \theta, a) \, ,\\
\label{CI3}
R_{\mu\nu\rho\sigma} R^{\mu\nu\rho\sigma} & = \frac{48 M^2}{ \left( 1-\alpha \right)^2 \left( r^2 + a^2\right) \rho^6} \; P_2 \left( r, \theta,a\right) \, ,
\end{align}
where  the functions $P_1(r, \theta, a)$ and $P_1(r, \theta, a)$ admit polynomial expressions given in \cite{BenAchour:2020fgy}.
As for the Kerr metric, the disformed Kerr geometry is singular at $\rho=0$ only. Nonetheless, one important difference between the disformed and the usual Kerr metric is that the disformed geometry ($\alpha \neq 0$) is, interestingly, no longer Ricci flat.

At infinity where $r\rightarrow + \infty$, the disformed Kerr metric becomes equivalent to,
\bea
\label{asymptomet}
ds^2 &\simeq& -\left( 1-\frac{2M_1}{r}\right)dt^2 + \left( 1-\frac{2M_2}{r}\right)^{-1} dr^2 + r^2 (d\theta^2 + \sin^2\theta \, d\varphi^2) \nonumber \\
&+& 2\alpha \sqrt{\frac{2M_1}{r}} dr\,dt + {\cal O}\left( \frac{1}{r^2}\right) \, , 
\eea
where we introduced the notations,
\bea
M_1 \equiv \frac{M}{1-\alpha} \, , \quad M_2\equiv(1+\alpha)M \, ,
\eea
and we rescaled the time coordinate $t$ by $\sqrt{1-\alpha}$. Note that the coefficient $\alpha$ modifies the black hole mass in the matrix elements $g_{tt}$ and $g_{rr}$ in a different way in Schwarzschild coordinates as $M_1 \neq M_2$. These masses agree at the first order in the parameter $\alpha$. Moreover, while the cross term $g_{tr}dt dr$ induced by the disformal transformations decays in the asymptotic regime, one can show that it cannot be removed by a coordinate change without introducing new off-diagonal terms, such that the new solution is not circular. This property appears a the key novelty of this new exact solution. Hence, the metric is asymptotically flat but, contrary to the Kerr metric, the disformed one is not equivalent to the Schwarzschild metric at infinity
essentially because the difference between the masses $M_1$ and $M_2$. Nevertheless, the deviations introduced by the presence of the cross term proportional to $dr dt$ in the metric become manifest only at next-to-leading order in the asymptotic expansion \cite{Charmousis:2019vnf}.

\subsection{Ergoregions and horizons}

Then, we see immediately that the disformed geometry admits the two same Killing vectors $\xi_t \equiv \partial_t$ and $\xi_\varphi \equiv \partial_\varphi$ as in the Kerr black hole because none of the
coefficients of the metric depend on $t$ and $\varphi$. As a consequence, we can look at the positions of the ergospheres, i.e.\ the hypersurfaces where the Killing vector field $\xi_t$ is null, i.e.\
\bea
\xi_t \cdot \xi_t = g_{tt} = 0  \qquad \Longleftrightarrow \qquad r^2 -2 {M_1} r + a^2 \cos^2\theta = 0 \, ,
\eea
where $M_1=M/(1-\alpha)$ as above \eqref{asymptomet}.
As a consequence, the disformed Kerr metric admits, as the usual Kerr metric, an outer and an inner ergospheres denoted 
respectively by ${\cal E}^+$ and ${\cal E}^-$ whose positions are given by the same formulae as the Kerr ones,
\bea
r=r_{{\cal E}^\pm}(\theta)= M_1 \pm \sqrt{M_1^2 - a^2 \cos^2\theta} \, ,
\eea
with the difference that the mass of the black hole has now been rescaled.  The ergoregions are defined similarly and one expects the possibility for a Penrose process (with an energy extraction mechanism) to exist in this geometry as well.

Now, let us consider the null directions. Indeed, computing the null directions is particularly interesting  to understand the causal structure of a metric and to see whether a metric $g_{\mu\nu}$ describes a black hole (or more generally possesses  horizons). These vectors enable us, in particular, to compute  light rays (the principal null geodesics) in the space-time and also to characterize the properties of horizons. The normalized (future directed) principal null vectors are denoted by
$\ell_\pm^\mu$ and satisfy the normalization conditions,
\bea
g_{\mu\nu}  \, \ell_\pm ^\mu \, \ell_\pm^\nu \; = \; 0 \; , \qquad g_{\mu\nu}  \, \ell_+^\mu \,  \ell_-^\nu \; = \; - 1 \, .
\eea
These conditions do not define completely (and then uniquely) the null vectors which can be rescaled according to $\ell_\pm \rightarrow \mathcal{N}^{\pm 1} \ell_\pm$ where $\mathcal{N}$ is an arbitrary (non-vanishing) function. 

In the case of the Kerr metric $\tilde{g}_{\mu\nu}$, the null vectors are well-known and are given by,
\bea
\tilde{\ell}_+^\mu \partial_\mu \equiv \frac{r^2+a^2}{\Delta} \partial_t + \partial_r + \frac{a}{\Delta} \partial_\varphi \,, \qquad
\tilde{\ell}_-^\mu \partial_\mu \equiv  \frac{r^2+a^2}{2 \rho} \partial_t - \frac{\Delta}{2\rho}\partial_r + \frac{a}{2 \rho} \partial_\varphi \, .
\eea
Interestingly, we see that, at the horizons where $\Delta=0$, the null vector field $\tilde{\ell}_+$ is proportional to the Killing vector
\bea
\label{lolo}
\frac{\Delta}{2(r^2+a^2)} \tilde{\ell}_+ =\partial_t + \Omega_H \, \partial_\varphi \, , 
\eea
where $\Omega_H=a/(2M r_\pm)$ is a constant whose value depends whether we are considering the outer ($r=r_+$) or the inner ($r=r_-$)
horizon\footnote{We are grateful to Eric Gourgoulhon for pointing us a mistake in Eq~(\ref{lolo}) in the first version of this work}. 

One can easily see that the principal directions $\ell_\pm$ of the disformed metric $g_{\mu\nu}$ are also ``disformed'' in the sense
that they are now given by,
\bea
\ell_\pm^\mu = \tilde{\ell}_\pm^\mu + \beta  (\phi_\alpha    \tilde{\ell}_\pm^\alpha) \, \phi^\mu \, , \qquad \beta \equiv \frac{(1- B_0  X)^{-1/2} -1}{X} =  \frac{1-(1- \alpha)^{-1/2} }{m^2} \, ,
\eea
where $\phi^\mu = \tilde{g}^{\mu\nu} \phi_\nu$ and $X=-m^2$ here. Notice that these formulae generalize immediately to any disformal transformation (when $X$ and $B_0$ are not necessarily constant) of  an arbitrary metric $g_{\mu\nu}$.  
Interestingly the effect of the disformal transformation on the null directions is a shift of the usual Kerr null vectors in the direction of the gradient of the scalar field $\phi^\mu$. Everything happens as if the scalar field is somehow drifting the light rays. As we shall see, this close formula for the disformed null direction can be derived using the $J$-map introduced in \cite{BenAchour:2021pla}. This is discussed in the next chapter when addressing the question of the effect of a disformal transformation on the Petrov type.

The explicit expressions of the null directions of the 
disformed Kerr metric can be easily written from the relations
\bea
\phi_\alpha \tilde \ell_+^\alpha & = & -\frac{m \sqrt{r^2+a^2}}{\sqrt{r^2+a^2} + \sqrt{2Mr}} \, , \qquad
\phi_\alpha \tilde \ell_-^\alpha =- m\frac{r^2+a^2 + \sqrt{2Mr(r^2+a^2)}}{2 \rho} \, ,
\eea
which enable us to obtain, after a direct calculation, 
\bea
\ell_+ & = & \left[ \frac{r^2+a^2}{\Delta} + \beta m (\phi_\alpha \tilde \ell_+^\alpha) \left(1+ \frac{\cR}{m^2 \rho \Delta} \right)\right] \partial_t \nonumber \\
&+& \left[ 1 + \beta (\phi_\alpha \tilde \ell_+^\alpha) \frac{\sqrt{\cR}}{\rho}\right] \partial_r  - \frac{a}{\Delta}\left[1 + \beta m (\phi_\alpha \tilde \ell_+^\alpha) \frac{2  M r}{\rho}  \right] \partial_\varphi \, ,   \\
\ell_- & = & \left[ \frac{r^2+a^2}{2\rho} + \beta m (\phi_\alpha \tilde \ell_-^\alpha) \left(1+ \frac{\cR}{m^2 \rho \Delta} \right)\right] \partial_t
\nonumber \\
&+ & \left[ -\frac{\Delta}{2 \rho} + \beta (\phi_\alpha \tilde \ell_-^\alpha) \frac{\sqrt{\cR}}{\rho}\right] \partial_r - \frac{a}{2 \rho}\left[1 + \beta m (\phi_\alpha \tilde \ell_-^\alpha) \frac{2  M r}{\Delta}  \right] \partial_\varphi   .
\eea
If we proceed as in the Kerr case, we would look at the regions where $\ell_+$ becomes proportional to Killing vectors. For 
$\ell_+$, we obtain the condition,
\bea
\Delta \left[ 1 + \beta (\phi_\alpha \tilde \ell_+^\alpha) \frac{\sqrt{\cR}}{\rho}\right] \; = \; 0 \, ,
\eea
which fixes $r$. Interestingly, $r_\pm$ are solutions but there are extra non-trivial solutions which are given by $r=F(\theta)$, i.e. $r$
is a function of $\theta$ and whose limit $\alpha \rightarrow 0$ is not defined. 
For $\ell_-$, we obtain the condition,
\bea
  -{\Delta} + 2 \beta (\phi_\alpha \tilde \ell_-^\alpha) \sqrt{\cR} \; = \; 0 \, ,
\eea
which also fixes $r$ at some non trivial function of $\theta$. In both cases, both null vectors reduce to a vector field proportional to,
\bea
\partial_t + \Omega_\pm \, \partial_\varphi \, ,
\eea
where $\Omega_\pm$ is no more a constant and depends on $\theta$. 
Therefore, none of the principal null directions reduce to  Killing vectors in some hypersurfaces. \textit{Hence if they exist, the horizons cannot be Killing}. This is an important difference with the Kerr geometry. 
Furthermore, one can easily check that the hypersurfaces of constant $r$ are not null because the norm of their normal vector $\partial_r$ is given by \eqref{inversedisf}
\bea
g^{rr}= \frac{\Delta}{\rho} + \frac{\alpha}{1-\alpha}\frac{2Mr(r^2+a^2)}{\rho^2}  \, ,
\eea
and therefore depends on $\theta$ through $\rho=r^2+a^2 \cos^2\theta$. As a consequence, the horizons of the disformed Kerr metric cannot be obtained in the way we get the
Kerr horizons. 

The problem of finding event horizons seems complicated. A first attempt was presented in \cite{Anson:2020trg} where candidates  has been proposed and analyzed. The basic idea is rather simple and consists in looking at null hypersurfaces defined by an equation of the form $F(r,\theta)=0$  with a $\theta$-dependency contrary to the Kerr case. We assume that we can locally solve $r$ as a function of $\theta$ and restrict ourselves to separable functions of the form $F(r,\theta)=r+F(\theta)$.
 The condition that such an hypersurface is null implies that its normal vector  
$(0,1, \partial_\theta F,0)$ is also null (by definition), which leads to a non-linear differential equation for $F(\theta)$,
\bea
g^{rr}  + g^{\theta \theta} \left(\frac{dF}{d\theta}\right)^2 \, = \, 0 \,  \Longleftrightarrow \, \left[{\Delta} + \frac{\alpha}{1-\alpha}\frac{2Mr(r^2+a^2)}{\rho} \right] +  \left(\frac{dF}{d\theta}\right)^2 \, = \, 0 \, ,
\eea
where we used \eqref{inversedisf} for the coefficients of the inverse disformed metric and $r=-F(\theta)$ everywhere in this equation.  It is the 
same equation as Eq.(23) in  \cite{Anson:2020trg}. The geometry of this null hypersurface is subtle as shown by the detailed analysis  in \cite{Anson:2020trg}. Computing its expansions may provide an alternative check and we hope to study this issue in details in a future work. Nevertheless, it is worth emphasizing that the characterization of quasi-local horizon through the expansions of the null directions is slicing dependent, and the choice of the null directions and thus of the $2$-surface foliating our geometry is therefore ambiguous as different choices might allow to identify different quasi-local (not necessarily null) horizons. See \cite{Faraoni:2016xgy} for detailed discussions on this point.

\subsection{Geodesic motion}

We finish with a quick discussion on the geodesic equations in the disformed Kerr background. Following the same method as in the case
of the Kerr metric, the geodesic equations can be obtained from the Hamilton-Jacobi equation for the ``action'' $S$,
\bea
H(x^\mu, {\partial_\mu S}) + \frac{\partial S}{\partial \lambda} = 0 \, , \qquad H(x^\mu,p_\mu) \equiv \frac{1}{2} g^{\mu\nu} p_\mu p_\nu \, ,
\eea
where $\lambda$ is the affine parameter along the geodesic. Due to the invariance of the disformed metric (whose components do not depend neither on $t$ nor on $\varphi$), the action $S$ takes the form
\bea
S(t,r,\theta,\varphi) \; = \; \frac{1}{2} \mu^2 \lambda + p_t t + p_\varphi \varphi + \Phi(r,\theta) \, ,
\eea
where $\mu$, $p_t$ and $p_\varphi$ are the standard constants of motion. A straightforward calculation shows that $\Phi(r,\theta)$
satisfies the differential equation,
\begin{align}
0&=\left[ \mu^2 r^2 + \Phi_r^2 \Delta - \frac{(r^2+a^2)^2}{\Delta} - \frac{4Mra}{\Delta} p_t p_\varphi - \frac{a^2}{\Delta} p_\varphi^2 \right] \nonumber \\
& + \left[ \mu^2 a^2 \cos^2\theta + \Phi_\theta^2 + a^2 p_t^2 \sin^2\theta + \frac{p_\varphi^2}{\sin^2\theta}\right] \label{HJeq} \\
&-\frac{\alpha}{(1-\alpha m^2)\rho} \left[ \frac{m}{\Delta} (r^2+a^2)^2 p_t + \sqrt{\cR} \Phi_r - ma^2 p_t \sin^2\theta \right] ^2 \, ,
\nonumber
\end{align}
where $\Phi_r\equiv \partial \Phi/\partial r$ and $\Phi_\theta\equiv \partial \Phi/\partial \theta$.
In the case where $\alpha=0$, the equation is clearly separable as the first line depends only on $r$ while the second one depends on $\theta$ only. This makes the geodesic equation integrable. Furthermore, the separability of the Hamilton-Jacobi equation is intimately linked to the existence of the famous  Carter constant and of a hidden symmetry of the Kerr metric (associated to a Killing tensor)~\cite{Carter:1968ks, Carter:1977pq}. 

When $\alpha \neq 0$, the Hamilton-Jacobi equation is no more separable in the Boyer-Lindquist coordinates and it is very likely that the geodesic equation is no more integrable and one cannot find a ``disformed'' Carter constant associated to the
disformed Kerr metric. Interestingly, there is an obvious solution of the equation \eqref{HJeq} given by 
\bea
\Phi= z \int dr \, \frac{\sqrt{\cR}}{\Delta} \, ,
\eea 
where $z$ is a
constant when the integration constants $\mu,p_t$ and $p_\varphi$ coincide with those of the scalar field according to,
\bea
p_t=-zm \, , \quad
p_\theta=0 \, , \quad
p_\varphi=0 \, , \quad
\mu^2 = \frac{z^2m^2}{1-\alpha m^2} \, .
\eea
In that case, the geodesic follows exactly the gradient of the scalar field. 

\section{Theories supporting the stealth and disformed Kerr solutions}

Having analyzed the the properties of the disformed Kerr black hole, we now establish which subset of DHOST theories admit such geometry as an exact stationary and axi-symmetric solution of their field equations.
The most general theory of quadratic DHOST theory \cite{Langlois:2015cwa} is described by the action 
\begin{equation}
\label{DHOST}
S=\int d^4x\sqrt{-g}\left(P(X,\phi)+Q(X,\phi)\, \Box \phi+F(X,\phi)\,R+\sum_{i=1}^{5}A_{i}(X,\phi)\, L_{i}\right)
\end{equation}
where the functions $A_{i},\,F,\,Q$ and $P$ depend on the scalar
field $\phi$ and its kinetic term $X\equiv \phi_\mu \phi^\mu$ with $\phi_\mu \equiv \nabla_{\mu}\phi$,
and $R$ is  the Ricci scalar.
The five elementary Lagrangians $L_{i}$ quadratic in second derivatives of $\phi$  are defined by 
\begin{eqnarray}
&&L_1 \equiv \phi_{\mu\nu} \phi^{\mu\nu} \, , \quad
L_2 \equiv (\Box \phi)^2 \, , \quad
L_3 \equiv \phi^\mu \phi_{\mu\nu} \phi^\nu\Box \phi \, , \quad \nonumber \\
&&L_4 \equiv  \phi^\mu  \phi_{\mu\nu} \phi^{\nu\rho} \phi_\rho \, , \quad
L_5 \equiv (\phi^\mu \phi_{\mu\nu} \phi^\nu)^2 \, ,
\end{eqnarray}
where we are using the standard notations $\phi_\mu \equiv \nabla_\mu \phi$ and $\phi_{\mu\nu} \equiv \nabla_\nu \nabla_\mu \phi$ for the first and second (covariant) derivatives of $\phi$. Theories in class Ia, the only ones which are viable interesting candidates theories, are labelled by the three free functions $F,A_1$ and $A_3$ (in addition to $P$ and $Q$) and 
the three remaining functions are given by the relations
\cite{Langlois:2015cwa}
\begin{eqnarray}
A_{2} & = & -A_{1} \, ,\label{deg1}
\\
A_{4} & = & \frac{1}{8\left(F+XA_{2}\right)^2}\Bigl(A_{2}A_{3}\left(16X^{2}F_{X}-12XF\right)+4A_{2}^{2}\left(16XF_{X}+3F\right) \nonumber\\
 &  & +16A_{2}\left(4XF_{X}+3F\right)F_{X} +16XA_{2}^{3}+8A_{3}F\left(XF_{X}-F\right)-X^{2}A_{3}^{2}F+48FF_{X}^{2}\Bigr) \, \label{funA4},\\
A_{5} & = & \frac{1}{8\left(F+XA_{2}\right)^2}\Bigl(2A_{2}+XA_{3}-4F_{X}\Bigr)\Bigl(3XA_{2}A_{3}-4A_{2}F_{X}-2A_{2}^{2}+4A_{2}^{3}F\Bigr) \, ,
\label{deg3}
\end{eqnarray}
The above relations (\ref{deg1}-\ref{deg3}) are a direct consequence of the three degenerate conditions that
guarantee only one scalar degree of freedom is present \cite{Langlois:2015cwa,Langlois:2015skt}. In conclusion, this means that all the DHOST theories we study here are characterized by five free functions of $X$ and $\phi$, which are $P$, $Q$, $F$, $A_1$ and $A_3$. Notice that we have implicitly supposed the condition 
\begin{align}
F+XA_{2}\neq 0
\end{align}
Finally, coupling to  external fields (perfect fluids, scalar fields, vector fields, etc.) can be done  by adding to the DHOST action an action $S_m$ where the external degrees of freedom are minimally coupled to the metric $g_{\mu\nu}$ (which is assumed to be the physical one). 

\subsection{Conditions for stealth Kerr}

Here we consider the following assumptions. First, we impose shift-symmetry which means that the DHOST action \eqref{DHOST} is unchanged by the transformation $\phi \rightarrow \phi+c$ where $c$ is a constant, and thus all the functions entering in the definition of \eqref{DHOST} depend on $X$ only.  Second, one
assumes the solution is such that  $X=X_0$ is a constant which drastically simplifies the modified Einstein equations. 
And finally, one looks for  
stealth solutions where the metric $g_{\mu\nu}$ is also a solution of the vacuum Einstein equations with a cosmological constant $\Lambda$,
\bea
G_{\mu\nu}+\Lambda g_{\mu\nu} = 0\, ,
\eea
where $G_{\mu\nu} \equiv R_{\mu\nu} - R g_{\mu\nu}/2$ is the Einstein tensor. One can go further
and requires that a given DHOST theory admits all GR solutions, and not only some of them, as the metric part of stealth solutions. This is the case if the following conditions hold \cite{Takahashi:2020hso},
\bea
\label{condStealth}
P + 2 \Lambda F = 0 \, , \quad
P_X + \Lambda (4 F_X - X_0 A_{1X}) = 0 \, , \quad
Q_X=0 \, , \quad
A_1 = 0 \, \quad
A_3 + 2 A_{1X} = 0 \, ,
\eea
where all these functions are evaluated on the solution $X=X_0$.
These conditions have been recently generalized to non-shift symmetric theories and to the case where matter is coupled to gravity minimally
\cite{Takahashi:2020hso}. 

Notice that these conditions are very strong and drastically restrict the set of DHOST theories. 
It was also argued that some stealth solutions lead to a problem of strong coupling \cite{Minamitsuji:2018vuw,deRham:2019gha} at the level of linear perturbations. 
Further, using the effective field theory framework it was shown in \cite{Motohashi:2019ymr} that perturbations about stealth solutions are strongly coupled, for de Sitter background in the decoupling limit, and for the Minkowski background even away from the decoupling limit, so long as we require evolution equation of perturbations to be second order.
Thus, in general the strong coupling is inevitable for asymptotically de Sitter or flat stealth solutions.  
Moreover, even if spacetime is different from de Sitter or Minkowski on superhorizon scales, the strong coupling is inevitable on subhorizon scales where the spacetime is nearly flat and hence the analysis of \cite{Motohashi:2019ymr} applies.
However, we can introduce a controlled detuning of the degeneracy condition, dubbed the scordatura mechanism in \cite{Motohashi:2019ymr}, to render the perturbations weakly coupled all the way up to a sufficiently high scale, as in the ghost condensate~\cite{Arkani-Hamed:2003pdi}. 
The Ostrogradsky ghosts associated with the scordatura is adjusted to show up only above the cutoff scale of the effective field theory.
It is also important to note that the scordatura does not change the properties of the stealth solutions of degenerate theories at astrophysical scales (similarly to the stealth solution \cite{Mukohyama:2005rw} in the ghost condensate). 
Thus, below we focus on stealth solutions in degenerate theories.

The stealth Kerr solution in DHOST theories was derived for theories within the class Ia with no cubic galileon term where gravitational waves propagate at the speed of light ($c_{\rm GW}=c$), i.e.\
\bea
\label{cequal1}
A_1=A_2=0 \, , \qquad Q=0 \, .
\eea
Here, the condition $c_{\rm GW}=c$ implies $A_1=0$~\cite{Langlois:2017dyl}, and we have used the degeneracy condition $A_2=-A_1$ in \eqref{deg1}. 
Therefore, the stealth conditions \eqref{condStealth} simplify and become,
\bea
P(X_0) + 2 \Lambda F(X_0) = 0 \, , \quad
P_X(X_0) + 4 \Lambda F_X(X_0)  = 0 \, , \quad
A_3(X_0) = 0 \, .
\eea
We now look for the theories which admit the disformed Kerr geometry as solution.

\subsection{Conditions for disformed Kerr}

Disformal transformations on the metric induce transformations on DHOST actions. Given an action $\tilde S[\tilde{g}_{\mu\nu},\phi]$,
one defines a new action $S[g_{\mu\nu},\phi]$ by the identification,
\bea
S[g_{\mu\nu},\phi] = \tilde S[A(X) g_{\mu\nu} + B(X) \phi_\mu \phi_\nu,\phi] \, .
\eea
Interestingly, DHOST theories are stable under disformal transformations \cite{BenAchour:2016cay} and the transformation
rules between the functions  (of $\tilde{X} \equiv \tilde{g}^{\mu\nu} \phi_\mu \phi_\nu$) $\tilde{P}$, $\tilde{Q}$,
$\tilde{F}$ and $\tilde{A}_I$ entering in the definition of the action $\tilde{S}[\tilde{g}_{\mu\nu},\phi]$ on one side, 
and  the functions (of $X = g^{\mu\nu} \phi_\mu \phi_\nu$) $P$, $Q$, $F$ and $A_I$ defining $S[g_{\mu\nu},\phi]$  on the other side
are given in \cite{BenAchour:2016cay}. 

As it turns out, these rules, which are rather complicated, simplify drastically when one considers 
constant disformal transformations where $A=A_0$ and $B=B_0$ do not depend on $X$ anymore. As pointed above, this is the case when considering invertible and shift symmetric disformal mapping of seed solution with constant kinetic term. After a straightforward
calculation, one shows that the k-essence, the cubic galileon and the Ricci terms transform as follows,
\bea
P = \tilde{P} \, , \qquad
Q =  A_0  \int dX \, N {\tilde Q}_X \, , \qquad F  =\frac{A_0}{N}\tilde{F}  \, ,
\eea
while the functions $A_I$ entering in the quadratic part of the Lagrangian transform as,
\bea
&&A_1 = N (B_{0} \tilde{F} + N^2 \tilde{A}_1) \, , \quad
A_2  = N(- B_{0} \tilde{F} + N^2 \tilde{A}_2) \, ,\nonumber \\
&& A_3  = \frac{N}{A_0} \left[  - 4 A_0 B_0 \tilde{F}_X - 2 B_0 N^4\tilde{A}_2 +N^6 \tilde{A}_3\right] \, , \nonumber  \\
&&A_4  = \frac{N}{A_0} \left[- N^2 B_0^2 \tilde{F} + 4 A_0 B_0 \tilde{F}_X - 2 N^4 B_0 \tilde{A}_1+ {N^6} \tilde{A}_4 \right] \, , \nonumber  \\
&&A_5  = \frac{N^7}{A_0^2}\left[ B^2_{0} ( \tilde{A}_1 + \tilde{A}_2) + N^2 B_{0}( \tilde{A}_3 - \tilde{A}_4) +N^4 \tilde{A_5} \right] \, ,
\eea
where we introduced the factor
\bea
N \equiv {A_0}^{1/2}{(A_0 + X B_0)}^{-1/2} \, .
\eea
We recall that ``tilde'' functions $\tilde{P}$, $\tilde{Q}$, $\tilde{F}$, $\tilde{A}_I$, in the right-hand side of the previous equations are viewed as functions of $X$ via the relation,
\bea
\label{X}
\tilde{X} = \frac{X}{ A_0 + X B_0} \, .
\eea
Now, we assume that the theory $\tilde{S}[\tilde{g}_{\mu\nu},\phi]$ satisfies the conditions \eqref{condStealth} to have a stealth 
solution where $\tilde{X}_0$ is constant and  $\tilde{g}_{\mu\nu}$ is the Kerr metric recalled above \eqref{Kerrmetric}. Then, we 
want to translate these conditions in terms of the functions entering into the action $S[g_{\mu\nu},\phi]$. 
We first remark that under constant disformal transformation $X_0$ is also constant when $\tilde{X}_0$ is constant.
After a direct calculation, from \eqref{condStealth} we obtain
\bea
&&P + \frac{2 \Lambda N}{A_0} F = 0 \, , \qquad
{\partial_X} \left[ P + \frac{\Lambda}{A_0} \left( 4 + \frac{B_0 X_0}{N}\right) F - \frac{\Lambda X_0}{N^3} A_1\right] = 0  \, , \label{cond1}\\
&& Q_X = 0 \, , \qquad A_1 - \frac{N^2 B_0}{A_0} F = 0 \, , \label{cond2}
\eea
together with the remaining more complicated condition 
\bea
\frac{A_0}{2N} A_3 +  \left( 2 B_0 N_X + \frac{B_0 N^2}{A_0}\right)F + 2 B_0 N F_X + B_0 N A_2 + \frac{N^8}{A_0}
\partial_X \left( \frac{A_1}{N^3} - \frac{B_0}{A_0} \frac{F}{N}\right)=0 \, , \label{cond3}
\eea
which comes from the last equation of \eqref{condStealth}. Let us recall that these equations holds only when they  are
evaluated on $X_0$. As a consequence, any DHOST theories which satisfy all these conditions admit disformal stealth solutions, which
are in general non-stealth. Obviously, these conditions reduce to \eqref{condStealth} for a trivial disformal transformation where
$A_0=1$ and $B_0=0$.

In the special case where the theory $\tilde{S}$ satisfies, in addition, the conditions \eqref{cequal1} which insure that gravitational waves
propagate at the speed of light, the disformed theory $S$ satisfies in turn, 
\bea
A_1= - A_2 = N^2 \frac{B_0}{A_0} F \, , \qquad Q=0 \, .
\eea
Hence, \eqref{cond2} are automatically satisfied and the first equation in \eqref{cond1} is unchanged. 
The last  conditions in  \eqref{cond1}  and \eqref{cond3} are also simplified according to
\bea
\partial_X \left( P + \frac{4 \Lambda}{A_0} F\right)=0 \, , \qquad
\frac{A_0}{2 N B_0} A_3 + \left(2N_X + \frac{N^2}{A_0}(1-NB_0) \right) F + 2N F_X =0 \, .
\eea 
In that case, the theory $S$ admits the disformed Kerr black hole that we have described above.

\section{Summary of results}

We now summarize the main properties of this new exact stationary and axi-symmetric solution of (a subset of) shift symmetric DHOST theories identified above:
\begin{itemize}
\item \textit{Non-circular geometry}: the disformal transformation breaks the circularity of the Kerr metric. It manifests in the off diagonal term in $g_{tr} \neq 0$. Notice that circularity is guaranteed for stationary axi-symmetric vacuum GR solution. Moreover, it is one of the assumptions of the CSB theorem \cite{Capuano:2023yyh}. Thus this solution evades the CSB theorem by at least two assumptions: the time-dependency of its scalar profile and the non-circularity of its metric.
\item \textit{Failure of the multipoles decomposition}: It is standard practice to identify a spacetime with its tower of multipoles defined at (conformal) spatial infinity. One standard way to proceed is to build asymptotically cartesian and mass-centered (ACMC) coordinates. Such asymptotic coordinates system only exists under suitable conditions \cite{Mayerson:2022ekj}. In particular, the metric functions need to be analytic which is not the case for the  disformed Kerr metric due to the off diagonal term $g_{tr} \propto \sqrt{r}$. Thus, at least using Thorne approach, one fails to assign a tower of multipole to the disformed Kerr geometry. Multipole moments for compact objects in scalar-tensor theories have been studied in \cite{Pappas:2014gca}.
\item \textit{Non-separability of the geodesic motion}: the disformal transformation breaks the separability of the geodesic equation in Boyer-Lindquist coordinates which implies a loss of the well-known integrability of the geodesic motion (and wave dynamics) on Kerr \cite{Carter:1968ks, Carter:1977pq}. This integrability being related to the existence of a rank-$2$ Killing-Yano tensor on the Kerr geometry, which in turn is known to be related to either Petrov type D or Petrov type N spacetimes, it suggests that the disformal transformation modifies the algebraic type of the seed Kerr solution if not completely breaking its algebraic specialness. As we shall see in the next chapter, this is precisely what happens as the disformed Kerr solution turns out be of Petrov type I \cite{BenAchour:2021pla}. A natural question raised by this observation is whether there are exact stationary and axi-symmetric \textit{non-stealth} black hole solutions of DHOST gravity which are algebraically special ? To our knowledge, a concrete example of such solution is still missing in the literature. Beyond finding one given example, a more systematic way would be to understand whether one can provide a generalization of the Goldberg-Sachs theorem of GR to DHOST gravity. Current works are currently in progress in order to address these key questions.
\item \textit{Location of the horizons still missing}: Finally, as discussed above, it has revealed challenging to identify the analytic position of the horizons in the disformed Kerr geometry. To date, only one attempt has been provided in \cite{Anson:2020trg} but without providing a close formula for the position of the horizons. Moreover, these horizons are not Killing and do not rotate rigidly contrary to the Kerr horizons.  
\end{itemize}
Having sum up the main pathological properties of the disformed Kerr solution, we now suggest one simple idea to construct a black hole solution by a disformal transformation without spoiling its key symmetries. 

\section{Preserving the symmetries}

A rather natural question is whether one can perform a DT on a stealth Schwarzschild or Kerr solutions without spoiling the crucial symmetries of the GR solution ? Let us mention two types of symmetries.
\begin{itemize}
\item \textbf{Killing tensor symmetry and separability of geodesic and wave equations:} Indeed, beyond the obvious isometries of the Schwarzschild and the Kerr background, these geometries also exhibit hidden symmetries. For the Kerr spacetime, the existence of a Killing tensor (which descends from the existence of a Killing-Yano two form) stands as a crucial ingredient to separate both the geodesic equation but also the various waves equations. Concretely, for the geodesic motion, this rank-$2$ Killing tensor $K_{\mu\nu} \rd x^{\mu} \rd x^{\nu}$ generates a conserved quantity quadratic in the momenta, which is related to the Carter constant and which is in involution with the three other constants of motion coming from the two Killing and the hamiltonian. In the non-rotating limit, this constant reduces to the square of the angular momentum. For the scalar wave,  the Killing tensor induces an operator, i.e. $\nabla_{\mu} K^{\mu\nu} \nabla_{\nu}$, which commutes with the wave operator for a scalar field, leading again to the separability of the wave equation. Therefore it plays a key role in our ability to study analytically the properties of the Kerr geometry. 
\item \textbf{Conformal near-horizon symmetry:} Additionally, it is well known that in the near-horizon region of the Schwarzschild and Kerr black holes, a subtle conformal symmetry emerges. It can be understood be looking at the structure of the scalar wave operator in this region. 
\end{itemize}
These two symmetries are crucial in our investigations of the black holes properties. While the first one is well-known, the second one is more subtle. Let us briefly review how it appears.
Consider the Schwarzschild metric
\begin{align}
ds^2 = - \left( 1 - \frac{2M}{r}\right) dt^2 + \left( 1 - \frac{2M}{r}\right)^{-1} dr^2 + r^2 d\Omega^2
\end{align}
with a coordinate singularity at the locus of the event horizon, $r = r_s= 2M$. Consider the scalar wave equation
\begin{align}
\Box \phi  = \frac{1}{\sqrt{|g|}} \partial_{\alpha}\left( \sqrt{|g|} g^{\alpha\beta} \partial_{\beta} \phi\right)
\end{align}
It is well known that the wave operator is separable in this coordinate, such that one can write the scalar field in the form
\begin{align}
\Phi\left( t, r , \theta, \phi\right) = e^{-i\omega t} R(t) Y^{\ell}_m\left( \theta, \phi\right)
\end{align}
One obtains then
\begin{align}
& \left[ \partial_r \Delta \partial_r  + \frac{\omega^2 r^4 }{\Delta}  + \frac{1}{\sin{\theta}} \partial_{\theta} \left( \sin{\theta} \partial_{\theta}  \right) + \frac{1}{\sin^2{\theta}} 
 \partial^2_{\phi} \right] \Phi = 0
\end{align}
where we have introduced
\begin{align}
\Delta = r^2 - 2M r = r \left( r- r_{+}\right)
\end{align}
Using the $su{(2)}$ Casimir on the sphere
\begin{align}
\triangle_{\cS^2} Y^{\ell}_m = \left[  \frac{1}{\sin{\theta}} \partial_{\theta} \left( \sin{\theta} \partial_{\theta}  \right) + \frac{1}{\sin^2{\theta}} 
 \partial^2_{\phi} \right] Y^{\ell}_m = - \ell\left( \ell +1\right) Y^{\ell}_m
\end{align}
leads to the simple wave equation valid for the whole exterior geometry
\begin{align}
\left[ \partial_r \Delta \partial_r  + \frac{\omega^2 r^4 }{\Delta} \right] \Phi = \ell \left( \ell +1 \right) \Phi
\end{align}
We would like now to investigate the near horizon regime of this wave operator. To this end, we need to develop the second term of the l.h.s. Denoting $\epsilon = r-r_{+}$, we obtain in the regime $\epsilon \rightarrow 0$ that
\begin{align}
\frac{\omega^2 r^4}{\Delta} & = \frac{\omega^2}{\epsilon} \left( \epsilon +r_{+}\right)^3 \simeq  \frac{\omega^2 r^3_{+}}{\epsilon} \left[ 1 + \frac{3\epsilon}{r_{+}} + \cO\left( \frac{\epsilon^2}{r^2}\right)\right]
\end{align}
The leading term behaves as $\epsilon^{-1}$ and blows up at the horizon, while the second term is of order $\cO(1)$, all the other terms being negligible. We can therefore keep only the leading term and write
\begin{align}
\frac{\omega^2 r^4}{\epsilon} \simeq \frac{\omega^2 r_{+}^3}{\left( r-r_{+}\right)} = \frac{\omega^2 r_{+}^4}{\Delta}
\end{align}
such that the wave equation becomes in the near horizon limit
\begin{align}
\left[ \partial_r \Delta \partial_r  + \frac{\omega^2 r_{+}^4 }{\Delta} \right] \Phi = \ell \left( \ell +1 \right) \Phi
\end{align}
The statement is that this wave operator can be written as an $\text{SL}(2,\mathbb{R})$ Casimir operator, exhibiting therefore a conformal structure. Concretely one can write 
\begin{align}
 \Box \Phi & =  \left[ \Delta \partial^2_r + 2 \left( r - 2M\right) \partial_r - \frac{16 M^4}{\Delta} \partial^2_t \right] \Phi\\
 & =  \left[ - \zeta^2_{\circ} + \frac{1}{2} \left( \zeta_{+} \zeta_{-} + \zeta_{-} \zeta_{+} \right) \right] \Phi
\end{align}
where vectors fields are given by
\begin{align}
\zeta_{\pm} & = \pm i e^{\pm 2\pi T_{H} t} \left[ \Delta^{1/2} \partial_r + 4 M \left( r- M\right) \Delta^{-1/2} \partial_t\right]\\
\zeta_{\circ} & = - 4 i M \partial_t
\end{align}
where $T_H$ is the Hawking temperature
\be
T_{H} = \frac{1}{8\pi M}
\ee
These vectors satisfy a sl$(2,\mathbb{R})$ algebra given by
\begin{align}
\{ \zeta_{\circ}, \zeta_{\pm} \} = \mp i \zeta_{\pm1} \qquad \{ \zeta_{+}, \zeta_{-}\} = 2i \zeta_{\circ}
\end{align}
which are the generators of the conformal symmetry in the near-horizon region.

From this discussion, it appears that the properties of these test fields directly teach us the key structure of underlying spacetime and are very useful to underline hidden symmetries. For this reason, it appears useful to investigate how such test fields transform under a disformal mapping. Consider for example the canonical free scalar field. The key question is whether there is a subset of disformal transformation which while changing the metric, would leave invariant the scalar wave equation. It turns out that this class of disformal transformation has been identified. 

Indeed, in \cite{Falciano:2011rf}, it was shown that given a disformal transformation characterized by the two functions $A(\phi, X)$ and $B(\phi, X)$, one can identify a subset of them which leaves the canonical scalar wave equation invariant. Since this will be useful when discussing black hole solutions, we briefly review this result here. Consider two metrics related by a disformal transformation
\begin{align}
\tilde{g}_{\mu\nu} =  A g_{\mu\nu} + B \phi_{\mu} \phi_{\nu} \qquad \tilde{g}^{\mu\nu} = A^{-1} \left[ g^{\mu\nu} - \frac{B}{A + B X} \phi^{\mu} \phi^{\nu}\right]
\end{align}
The square root of the determinant of the two metrics is related through
\begin{align}
\sqrt{|\tilde{g}|} = \sqrt{A^3(A+BX)} \sqrt{|g|}
\end{align}
Then, the scalar wave transforms as 
\begin{align}
\tilde{\Box} \phi & = \frac{1}{\sqrt{|\tilde{g}|}} \partial_{\mu} \left( \sqrt{|\tilde{g}|} \tilde{g}^{\mu\nu} \phi_{\nu} \right) \\
& = \frac{1}{\sqrt{A^3(A+BX)} \sqrt{|g|}} \partial_{\mu} \left( \sqrt{\frac{A^3}{A+BX}} \sqrt{| g|} g^{\mu\nu} \phi_{\nu} \right)
\end{align}
Therefore, provided the disformal function is given by
\begin{align}
B(\phi, X) = \frac{\alpha A^3(\phi,X) - A(\phi, X)}{X}
\end{align}
the scalar wave transforms as
\begin{align}
\tilde{\Box} \phi = \frac{1}{\alpha A^3} \Box \phi =0
\end{align}
and the solutions of the scalar wave on the two disformally related metrics are the same. Notice that it does not depend on the form of the conformal factor $A(\phi, X)$ which is arbitrary. 

To conclude, if one uses such DT, it could be in principle possible to build new hairy black holes solutions enjoying the same spectral properties (at least for the scalar wave) of the seed solution, despite allowing for non-trivial deformation. The second outcome is that if the scalar wave remains separable, it could provide a way to identify the hidden Killing tensor in the new disformed geometry.

\newcommand{\mat} [2] {\left ( \begin{array}{#1}#2\end{array} \right ) }

\chapter{Petrov classification and disformal transformation}
\label{Chapter6}

\textit{The light you shall follow! \\
Yoda}
\bigskip

\minitoc

In this chapter, we present a first attempt to control how the Petrov type of a given gravitational field changes under a disformal transformation (DT). Provided one obtains close formula for this transformation, one can impose some restriction on the scalar profile depending on the Petrov type one wishes to achieve. However, as we shall see, even with these formula at hand, it is challenging to use them concretely. Nevertheless, this exercise allows one to understand another feature of DT onto a fundamental property of spacetime, namely its algebraic Petrov type. In particular, we shall apply this strategy to further characterize the black hole solutions discussed in the previous chapters \cite{BenAchour:2021pla}.

\section{Review of the Petrov classification}

The Petrov classification introduced in \cite{Petrov:1959zfa} is based on the algebraic properties of the Weyl tensor. There are several ways to present this classification \cite{Chandrasekhar:1985kt, Stephani:2003tm}. In the following, we shall follow the approach adopted in \cite{Chandrasekhar:1985kt} and adopt its choice of signature for the metric which is $(+,-,-,-)$. 
Consider a spacetime $(\mathcal{M}, g)$ and a null tetrad $\theta^{\mu}_A$ where $\mu \in \{ 1,2,3,4\}$ is a spacetime index and $A \in \{ 1,2,3,4\}$ is a Lorentz index such that
\begin{align}
g^{\mu\nu} = e^{\mu}{}_A e^{\nu}{}_B \eta^{AB} \qquad \text{with} \qquad \eta^{AB} = \mat{cccc}{0& 1 & 0&0 \\ 1& 0 &0 &0 \\ 0& 0 & 0 &-1 \\ 0 & 0 & -1 & 0}
\end{align}
One can introduce the following four null vectors $(\ell, n, m, \bar{m})$ defined at each spacetime points with the standard orthogonality relations
\begin{align}
\ell^{\mu} n_{\mu} = 1 \qquad m^{\mu} \bar{m}_{\mu} = - 1
\end{align}
while all other scalar products vanish. In terms of the tetrad components, one has
\begin{align}
e^{\mu}_1 = \ell^{\mu} \qquad e^{\mu}_2 = n^{\mu} \qquad e^{\mu}_3 = m^{\mu} \qquad e^{\mu}_4 = \bar{m}^{\mu}
\end{align}
This tetrad is not invariant under Lorentz transformations and there are six degrees of freedom with which one can rotate the tetrad. They correspond to the six parameters of the Lorentz group. It is convenient to split them in terms of their effect on the different vectors of the null basis.
One can distinguish
\begin{itemize}
\item rotations of class I leaving $\ell$ unchanged and labelled by the complex parameter $a$
\item rotations of class II leaving $n$ unchanged and labelled by the complex parameter $b$
\item rotations of class III leaving the couple $(\ell, n)$ unchanged but rotate $m$ and $\bar{m}$ by an angle $\theta$ in the $(m, \bar{m})$ plane. They are labelled by the real conformal rescaling parameter $A$ and the real angle $\theta$.
\end{itemize}
Now consider the Weyl tensor 
\begin{align}
C_{\alpha\beta\mu\nu} = R_{\alpha\beta\mu\nu} + \frac{1}{2} \left( g_{\alpha\mu} R_{\beta\nu} - g_{\beta\mu} R_{\alpha\nu} - g_{\alpha\nu} R_{\beta\mu} + g_{\beta\nu} R_{\alpha\mu}\right) + \frac{1}{6} \left( g_{\alpha\mu} g_{\beta\nu} - g_{\alpha\nu} g_{\beta\mu}\right) R
\end{align}
It shares the same symmetries as the Riemann tensor but it is moreover traceless, i.e. $C^{\alpha}{}_{\beta\mu\alpha} =0$. Using these symmetries and decomposing it on the null basis, one finds that all the information contained in the Weyl tensor can be encoded into five complex scalar given by
\begin{align}
\Psi_0 & = - C_{\alpha\beta\mu\nu} \ell^{\alpha} m^{\beta} \ell^{\mu} m^{\nu} \\
\Psi_1 & =  - C_{\alpha\beta\mu\nu} \ell^{\alpha} n^{\beta} \ell^{\mu} m^{\nu} \\
\Psi_2 & =  - C_{\alpha\beta\mu\nu} \ell^{\alpha} m^{\beta} \bar{m}^{\mu} n^{\nu} \\
\Psi_3 & =  - C_{\alpha\beta\mu\nu} \ell^{\alpha} n^{\beta} \bar{m}^{\mu} n^{\nu} \\
\Psi_4 & =  - C_{\alpha\beta\mu\nu} n^{\alpha} \bar{m}^{\beta} n^{\mu} \bar{m}^{\nu} 
\end{align}
Obviously, these complex scalars depend on the choice frame orientation. Under a class I transformation, the $\Psi$'s transform as
\begin{align}
& \Psi_0 \rightarrow \Psi_0 \\
& \Psi_1 \rightarrow \Psi_1 + \bar{a} \Psi_0 \\
& \Psi_2 \rightarrow \Psi_2 + 2 \bar{a} \Psi_1 + \bar{a}^2 \Psi_0 \\
& \Psi_3 \rightarrow \Psi_3 + 3 \bar{a} \Psi_2 + 3\bar{a}^2 \Psi_1 + \bar{a}^3 \Psi_0  \\
& \Psi_4 \rightarrow \Psi_4 + 4 \bar{a} \Psi_3 + 6\bar{a}^2 \Psi_2 + 4\bar{a}^3 \Psi_1 +  \bar{a}^4 \Psi_0 
\end{align}
Under a transformation of class II, the $\Psi$'s transform as
\begin{align}
\label{main}
& \Psi_0 \rightarrow \Psi_0 + 4 b \Psi_1 + 6 b^2 \Psi_2 + 4 b^3 \Psi_3 + b^4 \Psi_4 \\
& \Psi_1 \rightarrow \Psi_1 + 3b \Psi_2 + 3 b^2 \Psi_3 + b^3 \Psi_4  \\
& \Psi_2 \rightarrow \Psi_2 + 2 b \Psi_3 + b^2 \Psi_4 \\
& \Psi_3 \rightarrow \Psi_3 + b \Psi_4   \\
& \Psi_4 \rightarrow \Psi_4 
\end{align}
while under a transformation of class III, one has
\begin{align}
& \Psi_0 \rightarrow A^{-2} e^{2i\theta} \bar{\Psi}_0  \\
& \Psi_1 \rightarrow  A^{-1} e^{i\theta}\Psi_1   \\
& \Psi_2 \rightarrow \Psi_2  \\
& \Psi_3 \rightarrow A e^{-i\theta}\Psi_3    \\
& \Psi_4 \rightarrow A^2 e^{-2i\theta}\Psi_4 
\end{align}
\textit{The central question at the core of the Petrov classification is how many and which $\Psi$'s can be set to zero by a suitable choice of orientation ?} Answering to this question can be achieved by a suitable sequence of Lorentz transformations of each class. Once this frame has been identified, the associated null directions are called the \textit{principal null directions}. Let us summarize how this work. Details can be found in \cite{Chandrasekhar:1985kt}.

Consider a transformation of class II. From (\ref{main}), one can made $\Psi_0$ to vanish by imposing 
\begin{align}
\Psi_0 + 4 b \Psi_1 + 6 b^2 \Psi_2 + 4 b^3 \Psi_3 + b^4 \Psi_4 =0
\end{align}
It provides a polynomial equation for the complex Lorentz parameter $b$. Since one can always assume that $\Psi_4 \neq 0$ (because one can always generate a non-vanishing $\Psi_4$ from a class I transformation), it follows that this polynomial equation has always four roots.
The Petrov type of a given gravitational field depends on the multiplicity of these roots.

\begin{itemize}
\item Petrov type I: The four roots are distinct. Then one can set $\Psi_0 =0$ by a class II and $\Psi_4 =0$ by a class I. Since this is always possible, such spacetime are said not algebraically special. The only non-vanishing Weyl scalars $(\Psi_1, \Psi_2, \Psi_3)$ cannot  be made to vanish by a class III transformation. See \cite{Bini:2021aze} for some subtleties on the Petrov type I.
\item Petrov type II:  There is a double root. It implies that by the same class II transformation, one can set both $\Psi_0=\Psi_1 =0$. By a class I, one can also set $\Psi_4 =0$, leaving only $(\Psi_2, \Psi_3)$ non-vanishing.
\item Petrov type III: There are three roots coinciding. Then, by the same class II transformation, one can set $\Psi_0= \Psi_1 = \Psi_2 =0$, while setting $\Psi_4 =0$ by a class I, leaving only $\Psi_3$ non-vanishing
\item Petrov type D: There are two distinct double roots. In this case, by the same class II transformation, one can set $\Psi_0 =\Psi_1 =0$, then by the same class I transformation, one can set $\Psi_3 =\Psi_4 =0$, leaving only $\Psi_2$ non-vanishing. This Petrov type is of outer most importance as it contains all the relevant black holes solutions known so far in 4d GR.
\item Petrov type N: All the four roots coincide. By the same class II transformation, one can set $\Psi_0 =\Psi_1=\Psi_2=\Psi_3 =0$, leaving only $\Psi_4$ non-vanishing.
\end{itemize}
Notice that the vanishing of the Weyl scalars only holds in a given frame, the frame built from the PND. In general this frame is not the best suited one to analyze the properties of a given solution, such that one can often work within a frame where $\Psi_0\neq 0$ even if this quantity can always be set to zero by a Lorentz transformation. In particular, it should be emphasize that when considering a parallel transported frame which is well adapted to capture the properties measured by a freely falling observer, the Weyl scalars which are turned on are in general very different from the ones which survive in the PND frame. We shall met this situation in the next chapter when discussing the generation of disformal gravitational wave.

Another way to proceed is to consider the complex combination defined by
\begin{align}
\tilde{C}_{\alpha\beta\mu\nu} = C_{\alpha\beta\mu\nu} - i \epsilon_{\alpha\beta}{}^{\rho\sigma} C_{\rho\sigma}{}_{\mu\nu}
\end{align}
and the two Lorentz invariants defined as
\begin{align}
I & = \frac{1}{32} \tilde{C}_{\alpha\beta\mu\nu} \tilde{C}^{\alpha\beta\mu\nu} = \Psi_0 \Psi_4 - 4 \Psi_1 \Psi_3 + 3 \Psi^2_3 \\
J & = \frac{1}{384}  \tilde{C}_{\alpha\beta\mu\nu} \tilde{C}^{\mu\nu}{}_{\rho\sigma} \tilde{C}^{\rho\sigma\alpha\beta} = \Psi_0 \Psi_1 \Psi_4 - \Psi^2_1 \Psi_4 - \Psi_0 \Psi^2_3 + 2 \Psi_1 \Psi_2 \Psi_3 - \Psi^3_2 
\end{align}
The first step of the algorithm consists in computing the speciality index defined by
\begin{align}
S = I^3 - 27 J^2
\end{align}
If $S\neq 0$, the geometry is of type I, i.e. it is not algebraically special. If $S=0$, the geometry is algebraically special and one has to evaluate news scalars to identify to which Petrov type the geometry belongs. Consider the following scalars
\begin{align}
K = \Psi_1 \Psi^2_4 - 3 \Psi_4 \Psi_3 \Psi_2 + 2 \Psi^3_3 \qquad L = \Psi_2 \Psi_4 - \Psi^2_3 \qquad N = 12 L^2 - \Psi^2_4 I
\end{align}
The key point is that these scalars are invariant under a class II transformation which one uses in the first step to set $\Psi_0 =\Psi_1 =0$ for an algebraically special spacetime, i.e. for geometry with $S=0$. However, they are not invariant under the remaining class I and III transformations. Now the different algebraic special type can be identified as follows
\begin{itemize}
\item Petrov type II: $I\neq 0$, $J\neq 0$, $K\neq 0$ and/or $N\neq 0$
\item Petrov type D: $I\neq 0$, $J\neq 0$ while $K=N=0$
\item Petrov type III: $I= J= 0$ while $L\neq 0$ and/or $K\neq 0$
\item Petrov type N: $I= J=L = K = 0$
\end{itemize}
These two equivalent algorithm allows one to compare different gravitational fields through their algebraic properties. The key point is that the algebraic type plays a key role in understanding the underlying symmetries of a given gravitational field. For instance, it is known that spacetime geometries admitting a rank-$2$ Killing-Yano tensor are restricted to Petrov type D and N. In turn, the existence of this rank-$2$ Killing-Yano tensor gives rise to the existence of a Killing tensor which can be used to separate the geodesic and wave equations on the geometry. This is the key properties of the Kerr geometry, but it extends to any Petrov type D geometries. Therefore, in view of the powerful symmetries associated to Petrov type D spacetime, it appears natural to wonder whether one can construct black hole solution of Petrov type D in DHOST gravity. Obviously, one is interested in finding a rotating black hole solution of this type which deviates from Kerr. Whether this solution exists is an open question. Yet, in order to address it, preliminary works are required. 

First one should identify the Petrov type of the known solutions derived so far. Second, it would be useful to understand how the Petrov type change under a disformal transformation, depending on the properties of the scalar profile used in the transformation. Then, one would need to derive explicit and if possible tractable formula which relate the PNDs and the relevant combination of the Weyl scalars under a generic disformal transformation. This might provide a first step towards building a systematic guide to construct new solutions with a specific Petrov type. These preliminary steps were implemented in \cite{BenAchour:2021pla}. This is the work we shall review in this chapter. 

\section{Disformal transformation on the tetrad}

Since the Petrov classification is based on the decomposition of the Weyl tensor onto a null frame, one first has to reformulate the effects of a disformal transformation at level of the tetrad. As we shall see, this will also reveal useful to derive the disformal transformation of the optical scalars characterizing a bundle of light rays. At the level of the tetrad, the disformal transformation can be encoded through the so-called $J$-map.

\subsection{The $J$-map}

We now consider a modified theory of gravity on ${\cal M}$ with metric ${\tilde g}$ which includes scalar and vector degrees of freedom on top of the two tensor degrees of freedom associated to the metric ${\tilde g}$. In particular we consider two scalar fields $A$ and $B$ which lives in ${\cal M}$ and also a one form with components $V_\mu$ which lives on the cotangent space $T^\ast_p(\cal M)$. To be concrete, we assume that ${\tilde g}$ is the metric in Jordan frame to which the matter minimally couples. Then, we consider a disformal transformation between the Jordan frame and a seed frame (the seed frame being usually the seed frame) as follows
\begin{equation}\label{DTT}
{\tilde g}_{\mu\nu} = A\, {g}_{\mu\nu} + B\, V_\mu V_\nu \,,
\end{equation}
where $g_{\mu\nu}$ is the metric in the seed frame. In the particular case of $V_\mu = \partial_\mu \phi$ for a scalar field $\phi$ and $A$ and $B$ be functions of $\phi$ and $g^{\mu\nu}\partial_\mu\phi \partial_\nu\phi$, we find the well-known disformal transformation which is widely used in the context of the scalar-tensor theories. Here, however, we keep the setup as general as possible. The inverse metric defined by ${\tilde g}^{\mu\nu} {\tilde g}_{\mu\rho}= \delta^\nu{}_\rho$ is given by
\begin{equation}\label{inv-DT-metric}
{\tilde g}^{\mu\nu} = \frac{1}{A} \Big[ {g}^{\mu\nu} - \frac{B}{A+B Y} 
(g^{\mu\alpha} V_\alpha) (g^{\nu\beta}V_\beta) \Big] \,, \hspace{1cm}
Y = g^{\mu\nu}V_\mu V_\nu \,,
\end{equation}
where we have used the fact that ${g}^{\mu\nu} {g}_{\mu\rho}= \delta^\nu{}_\rho$.

We can move between Jordan and seed frames by means of the transformation \eqref{DT} and the inverse disformal transformation $g_{\mu\nu} = A^{-1} ( {\tilde g}_{\mu\nu} - B V_\mu V_\nu)$. After performing disformal transformation Eq.~\eqref{DTT}, the disformal term $B$ changes the null directions while the conformal term preserves the null directions as it is a well-known fact. Therefore, the null directions ${\tilde l}^\mu$ and ${\tilde k}^\mu$ in the Jordan frame, that we presented in the previous subsection, are no longer null in the seed frame. In this subsection, our aim is to find the new null directions in the seed frame. Similar to what we did for the metric in Jordan frame ${\tilde g}$ in the previous subsection, we can now expand the metric in seed frame in terms of the new null tetrad basis as
\begin{equation}\label{g-T}
{ g}_{\mu\nu} = \eta_{ab}\,{ \theta}^a{}_{\mu} { \theta}^b{}_{\nu} \,, \hspace{1cm}
\eta_{ab} = { g}_{\mu\nu}{ \theta}^{\mu}{}_a \, { \theta}^{\nu}{}_b \,,
\end{equation}
where
\begin{equation}\label{T}
{ \theta}^a{}_{\mu} = \big( - { k}_\mu, - { l}_\mu, {\bar m}_\mu, { m}_\mu \big) \,, \hspace{1cm}
{ \theta}^{\mu}{}_a = \big( { l}^\mu, { k}^\mu, {m}^\mu, {\bar m}^\mu \big) \,,
\end{equation}
denote the new tetrad basis in the seed frame.

Substituting from Eqs. \eqref{T} and \eqref{eta} in the first Eq. in \eqref{g-T}, we can express the metric components in the seed frame in terms of the new tetrad components as
\begin{equation}\label{g}
{ g}_{\mu\nu} = - { l}_\mu { k}_\nu - { k}_\mu { l}_\nu
+ { m}_\mu {\bar { m}}_\nu + {\bar { m}}_\mu { m}_\nu \,,
\end{equation}
and from the second Eq. in \eqref{DTT} we also find the following conditions
\begin{eqnarray}\label{null}
&&{ g}_{\mu\nu}{ l}^\mu{ l}^\nu = 0 \,, \hspace{3cm}
{ g}_{\mu\nu}{ k}^\mu{ k}^\nu = 0 \,, \hspace{2.7cm}
{ g}_{\mu\nu}{ l}^\mu{ k}^\nu = -1 \,, \\
&&{ g}_{\mu\nu}{ l}^\mu{ m}^\nu = 0 \,, \hspace{1cm}
{ g}_{\mu\nu}{ l}^\mu{\bar{ m}}^\nu = 0 \,, \hspace{1cm}
{ g}_{\mu\nu}{ k}^\mu{ m}^\nu = 0 \,, \hspace{1cm}
{ g}_{\mu\nu}{ k}^\mu{\bar{ m}}^\nu = 0 \,, 
\nonumber \\ \label{m}
&&{ g}_{\mu\nu}{ m}^\mu{ m}^\nu = 0 \,, \hspace{2.5cm} 
{ g}_{\mu\nu}{\bar{ m}}^\mu{\bar{ m}}^\nu = 0 \,, \hspace{2.5cm} 
{ g}_{\mu\nu}{ m}^\mu{\bar{ m}}^\nu = 1 \,.
\end{eqnarray}
The disformal transformation Eq. \eqref{DTT} is a field redefinition between metrics ${\tilde g}$ and $g$ and obviously it is not a coordinate transformation. This field redefinition relation can be also thought as a redefinition of the corresponding tetrad basis associated to the metric ${\tilde g}$ and $g$ as follows
\begin{equation}\label{T-map}
{\tilde \theta}^a{}_{\mu} \equiv J^a{}_b(A,B,V)\, {\theta}^b{}_{\mu} \,,
\end{equation}
where $J^a{}_b(A,B,V)$ is the transformation matrix which depends on the scalar fields $A$, $B$, and the vector field $V$. We can also express the components of the vector field $V$ in either Jordan frame tetrad basis \eqref{T-map} or seed frame tetrad basis \eqref{T}. But, it is important to note that the components of the vector field in different frames \eqref{T-map} and \eqref{T} are not related to each other through the transformation matrix defined in Eq. \eqref{T-map}. The reason is that transformation \eqref{T-map} is the tetrad counterpart of the disformal transformation \eqref{DT} which is not a coordinate transformation. Indeed, the transformation matrix itself depends on $V$.

In order to find the transformation matrix $J^a{}_b$, we express metric in the seed frame $g$ and vector field in terms of the seed frame tetrad components \eqref{T}. The disformal transformation Eq. \eqref{DT} then gives
\begin{eqnarray}
\eta_{ab} {\tilde \theta}^a{}_\mu {\tilde \theta}^b{}_\nu = 
\Big[ A \eta_{ab} + B V_a V_b \Big] { \theta}^a{}_\mu { \theta}^b{}_\nu \,,
\end{eqnarray}
where $V_a$ are components of vector field defined as $V_\mu = V_a \theta^a{}_\mu$ in direction of the seed frame tetrad basis. Substituting from Eq. \eqref{T-map} in the above relation, we find the transformation matrix as follows
\begin{equation}\label{J}
J^a{}_b = \sqrt{A} \Big( \delta^a{}_b - \frac{\beta}{1+\beta Y} V^a V_b \Big) 
= \sqrt{A} \big( \delta_b{}^a + {\tilde \beta} {\tilde V}_b {\tilde V}^a \big) \,,
\end{equation}
where we have defined
\begin{equation}\label{beta}
{\tilde \beta} \equiv\frac{1}{\tilde Y} \bigg[\frac{1}{\sqrt{1-B{\tilde Y}}}-1\bigg] \,; \hspace{1cm}
\beta \equiv\frac{1}{Y} \bigg[\frac{\sqrt{A}}{\sqrt{A+BY}}-1\bigg] \,.
\end{equation}
In the above equation we have also defined 
\begin{equation}\label{Y-tilde}
{\tilde Y} = {\tilde V}_a {\tilde V}^a = \frac{Y}{A+BY} \,; \hspace{1cm}
{\tilde V}_a \equiv \frac{V_a}{\sqrt{A+BY}} \,,
\end{equation}
where ${\tilde V}_a = V_\mu {\tilde \theta}^\mu{}_a$ is the tetrad component of $V_\mu$ in direction of the Jordan frame tetrad basis.

Now let us look at the inverse transformation from the seed frame to the Jordan frame  which is given by
\begin{equation}\label{T-map-i}
{\theta}^a{}_\mu = T^a{}_b \, {\tilde \theta}^b{}_{\mu} \,; \hspace{1cm}
T^a{}_b=\big(J^{-1}\big)^a{}_b \,.
\end{equation}
The inverse transformation matrix can be easily obtained by substituting Eq. \eqref{J} into the relation $(J^{-1})^a{}_c J^c{}_b = \delta^a{}_b$ as follows
\begin{eqnarray}\label{J-inv}
T^a{}_b = \frac{1}{\sqrt{A}} \big( \delta^a{}_b + \beta V^a V_b \big)
= \frac{1}{\sqrt{A}} \Big( \delta^a{}_b 
- \frac{{\tilde \beta}}{1+{\tilde \beta}{\tilde Y}} {\tilde V}^a {\tilde V}_b \Big) \,.
\end{eqnarray}
Note that the invertibility of the map Eq. \eqref{T-map} is guarantied by demanding that the determinant of the transformation matrix Eq. \eqref{J}
\begin{equation}\label{J-det}
J = T^{-1} = \frac{\sqrt{-{\tilde g}}}{\sqrt{-g}}= A^{3/2} (A+BY)^{1/2} \,,
\end{equation}
where $J\equiv \mbox{det} (J^a{}_b)$ and $T \equiv \mbox{det} (T^a{}_b)$, does not vanish.
From the first equation in \eqref{g-T}, it is easy to see that the transformation between the inverse tetrad basis are given by
\begin{equation}\label{T-map-E}
{\tilde \theta}^{\mu}{}_a \equiv T^b{}_a\, {\theta}^{\mu}{}_b \,, \hspace{1cm}
{\theta}^{\mu}{}_a \equiv J^b{}_a \, {\tilde \theta}^{\mu}{}_b \,.
\end{equation}
Substituting from Eqs. \eqref{J} and \eqref{J-inv} into the above relations, we find
\begin{eqnarray}\label{T-dis}
{\tilde \theta}^\mu{}_a = \frac{1}{\sqrt{A}} \,
\Big[ {\theta}^\mu{}_a + \beta (V_{\alpha}\theta^\alpha{}_a) (g^{\mu\beta} V_\beta) \Big] \,, \hspace{1cm}
{\theta}^\mu{}_a = \sqrt{A} \, \Big[ {\tilde \theta}^\mu{}_a 
+ {\tilde \beta} \big(V_{\alpha}{\tilde \theta}^\alpha{}_a \big) 
\big({\tilde g}^{\mu\beta} V_\beta\big) \Big] \,,
\end{eqnarray}
where we have used the fact that ${\tilde V}^a {\tilde \theta}^\mu{}_a = {\tilde g}^{\mu\beta}V_\beta$. For the case of $a=0,1$, the above results are in agreement with the results obtained in \cite{BenAchour:2020fgy}. 

If the 1-form $V_\mu$ in disformal transformation \eqref{DT} lives in the subspace ${\cal S}$, {\it i.e.} belongs to the cotangent space $T^\ast({\cal S})$, then we have $V_\alpha {\tilde l}^\alpha = 0 = V_\alpha {\tilde k}^\alpha$ which also leads $V_\alpha { l}^\alpha = 0 = V_\alpha {k}^\alpha$. In this case, the null directions will not change after performing disformal transformation. However, if $V_\mu$ does not belong to $T^\ast({\cal S})$, we have $V_\alpha {\tilde l}^\alpha \neq 0 \neq V_\alpha {\tilde k}^\alpha$ in general and the null direction change as it is clear from the Eq. \eqref{T-dis}.
The tetrad components of the vector field $V_a$ are given by
\begin{eqnarray}\label{V-a-l}
V_a = V_\alpha \theta^\alpha{}_a 
= \big( V_\alpha l^\alpha, V_\alpha k^\alpha, V_\alpha m^\alpha, V_\alpha \bar{m}^\alpha \big)\,,
\end{eqnarray}
and also
\begin{eqnarray}\label{V-a-u}
V^a = \eta^{ab} V_b
= \big( - V_\alpha k^\alpha, - V_\alpha l^\alpha, V_\alpha \bar{m}^\alpha, V_\alpha m^\alpha \big)\,,
\end{eqnarray}
Substituting the above expressions into Eqs. \eqref{J} and \eqref{J-inv}, we find explicit form of the components of the transformation matrices which we need to study null geodesics in the next section.

\subsection{Disformal kinematics of a null congrurence}

Without loss of generality, we consider geodesic congruences for the null direction ${\tilde l}^\mu$ for which ${\tilde k}^\mu$ is the transverse null direction. The deviation of a general vector $W^\mu$ from the null geodesic of ${\tilde l}^\mu$ is given by
\begin{equation}\label{geod-W}
{\tilde l}^\nu \tilde{\nabla}_\nu U^\mu = {\tilde \Theta}^\mu{}_\nu U^\nu\,; \hspace{1cm}
{\tilde \Theta}_{\mu\nu} \equiv \tilde{\nabla}_\alpha \tilde{l}_\beta 
{\tilde q}^\alpha{}_\mu {\tilde q}^\beta{}_\nu \,,
\end{equation}
where $\Theta$ is the second fundamental form which clearly lives in ${\cal S}$. It can be decomposed as
\begin{equation}\label{Theta}
{\tilde \Theta}_{\mu\nu} = {\tilde\nabla}_\mu {\tilde l}_\nu 
+ {\tilde l}_\mu {\tilde k}^\alpha {\tilde\nabla}_\alpha {\tilde l}_\nu 
- {\tilde\omega}_\mu {\tilde l}_\nu \,,
\end{equation}
where $\omega_\mu$ are the components of rotation 1-form given by
\begin{equation}\label{omega}
{\tilde \omega}_\mu \equiv - {\tilde k}^\alpha {\tilde \nabla}_\mu {\tilde l}_\alpha 
- \big( {\tilde k}^\alpha {\tilde k}^\beta {\tilde\nabla}_\alpha {\tilde l}_\beta \big) {\tilde l}_\mu \,.
\end{equation}
Now, we decompose the second fundamental form as usual
\begin{equation}\label{Thete-dec}
\Theta_{\mu\nu} = \frac{1}{2} {\tilde\Theta} {\tilde q}_{\mu\nu} 
+ {\tilde \sigma}_{\mu\nu} + {\tilde \omega}_{\mu\nu} \,,
\end{equation}
where the  expansion ${\tilde \theta}$, shear ${\tilde\sigma}$, and vorticity ${\tilde \omega}_{\mu\nu}$ are defined as
\begin{eqnarray}\label{exp-shear-vort}
{\tilde \Theta} = {\tilde q}^{\alpha\beta} {\tilde \Theta}_{\alpha\beta} 
= {\tilde\nabla}_\alpha {\tilde l}^\alpha \,, 
\hspace{1cm}
{\tilde \sigma}_{\mu\nu} = {\tilde \Theta}_{(\mu\nu)} - \frac{1}{2} {\tilde\theta}{\tilde q}_{\mu\nu} \,, 
\hspace{1cm}
{\tilde \omega}_{\mu\nu} = {\tilde \Theta}_{[\mu\nu]} \,,
\end{eqnarray}
in which $t_{(\mu\nu)}=(1/2)(t_{\mu\nu}+t_{\nu\mu})$ and $t_{[\mu\nu]}=(1/2)(t_{\mu\nu}-t_{\nu\mu})$ denote symmetric and anti-symmetric parts respectively.

Let us focus on the expansion which plays a key role in identifying possible trapping horizons in a given geometry. To rewrite the expansion, shear, and vorticity that we defined in Eqs. \eqref{exp-shear-vort}, we also need the relation between covariant derivatives in two frames for a general vector $U^\mu$ which is given by \cite{Wald:1984rg}
\begin{equation}\label{CovD}
{\tilde \nabla}_\nu U^\mu = {\nabla}_\nu U^\mu + D^\mu{}_{\nu\rho} U^\rho \,;\hspace{1cm}
D^\mu{}_{\nu\rho} \equiv \frac{1}{2} {\tilde g}^{\mu\sigma}
\big( {\nabla}_{\rho}{\tilde g}_{\sigma\nu} + {\nabla}_{\nu}{\tilde g}_{\sigma\rho} 
- {\nabla}_{\sigma}{\tilde g}_{\rho\nu} \big) \,.
\end{equation}
Substituting from Eq. \eqref{T-dis} into the definition \eqref{exp-shear-vort}, and using Eqs. \eqref{J-inv}, \eqref{V-a-l}, and \eqref{CovD}, we find
\begin{eqnarray}\label{expansion}
{\tilde \Theta} &=& \nabla_\mu \big( T^a{}_0 \theta^\mu{}_a \big) 
+ D^\alpha{}_{\alpha\mu} \big( T^a{}_0 \theta^\mu{}_a \big) 
\nonumber \\
&=& \frac{1}{\sqrt{A}} \bigg\{ \Theta - l^\mu \nabla_\mu \ln\big( \sqrt{A} T \big) 
- \sqrt{A} T\, \nabla_\mu \bigg[ \frac{\beta (V_\alpha l^\alpha) (g^{\mu\beta}V_\beta)}{\sqrt{A} T} \bigg] \bigg\} \,,
\end{eqnarray}
where  $\Theta \equiv {\nabla}_\mu {l}^\mu$ are the corresponding expansions in the seed frame and we have used the formula $D^\alpha{}_{\alpha\mu}= \nabla_\mu \ln{J}=-\nabla_\mu \ln{T}$ in which $J$ and $T$ are defined in Eq. \eqref{J-det}.

The part proportional to $l^\mu$ in the r.h.s. can be removed through the redefinition of the affine parameter while the part proportional to the disformal vector $(g^{\mu\alpha}V_\alpha)$ cannot be removed by any redefinition of the affine parameter. The latter part vanishes if the disformal vector be orthogonal to the null directions as $V_\alpha l^\alpha=0$. In this case casual structure does not change as it is clear from Eq. \eqref{T-dis}. But in general, one can use the above formula to track the existence of horizons.

\section{Decomposition of the Weyl tensor}

In the following, we shall follow the approach based on the PNDs. Having determined in the previous section how the null directions transform under a disformal map, this approach will allow us to provide formula encoding how the Petrov type of a given disformally constructed solution is related to the Petrov type of the corresponding seed solution.

Consider a seed solution $(\M, g)$ in the seed frame and its disformal transformed solution $(\M, \tilde{g})$ in the Jordan frame.  By definition, the Weyl tensor of the seed geometry is given by
\begin{equation}\label{Weyl-tilde}
C^\rho{}_{\beta\mu\nu} = \frac{1}{2} 
\big(  R^\rho{}_{\beta\mu\nu} + R^\rho{}_{\nu}  g_{\beta\mu} 
+  R_{\beta\mu} \delta^\rho{}_{\nu} \big) 
+ \frac{1}{6}  R {\delta}^\rho{}_{\mu}  g_{\beta\nu} \, - \, [\mu \leftrightarrow \nu] \,.
\end{equation}
As a first step, let us derive the relation between  $\tilde{C}_{\alpha\beta\mu\nu}$ and $C_{\alpha\beta\mu\nu}$ and split it in a suitable form. Under the disformal map~\eqref{DT}, the covariant derivative $\nabla$ of a 1-form $W=W_{\mu}{d}x^{\mu}$ induces an extra term as~\cite{Wald:1984rg,Lobo:2017bfh}
\begin{equation}\label{CovD}
{\tilde \nabla}_\nu W_\mu =
{\nabla}_\nu W_\mu - D^\rho{}_{\mu\nu} W_\rho \;, \qquad \text{with} \qquad 
D^\rho{}_{\mu\nu} \equiv \frac{1}{2} {\tilde g}^{\rho\sigma}
\big( {\nabla}_{\mu}{\tilde g}_{\sigma\nu} + {\nabla}_{\nu}{\tilde g}_{\sigma\mu} 
- {\nabla}_{\sigma}{\tilde g}_{\mu\nu} \big) \,.
\end{equation}
Note that the geometrical object $D^\rho{}_{\mu\nu}$ connects the covariant derivatives in the Jordan and seed frames to each other. Indeed, it is the difference between the Christoffel symbols in the Jordan and seed frames and, therefore, it is a tensor. Using the transformation of the covariant derivative, the disformal transformation of the Riemann tensor can be expressed as
\begin{equation}\label{Riemann-tilde}
\tilde{R}^{\rho}{}_{\beta\mu\nu} = R^\rho{}_{\beta\mu\nu} + D^\rho{}_{\beta\mu\nu} \,,
\end{equation}
where the tensor $D^\rho{}_{\beta\mu\nu}$ reads
\begin{equation}\label{D}
D^\rho{}_{\beta\mu\nu} \equiv \nabla_\mu D^\rho{}_{\nu\beta} - \nabla_\nu D^\rho{}_{\mu\beta}
+ D^\rho{}_{\mu\sigma} D^\sigma{}_{\nu\beta} - D^\rho{}_{\nu\sigma} D^\sigma{}_{\mu\beta} \,.
\end{equation}
Contracting the l.h.s.\ of the Eq.~\eqref{Riemann-tilde}, we can easily obtain relations between the Ricci tensor and Ricci scalar in the Jordan and seed frames which read
\begin{equation}\label{Ricci-tilde}
\tilde{R}_{\mu\nu} = R_{\mu\nu}+ D_{\mu\nu} \,, \hspace{1cm}
{\tilde R} = \frac{1}{A} (R+D) - \frac{B}{A} \frac{(R_{\alpha\beta} + D_{\alpha\beta}) V^\alpha V^\beta}{A+BY} \,, 
\end{equation}
where $D_{\mu\nu} \equiv D^\rho{}_{\mu\rho\nu}$ and $D \equiv g^{\alpha\beta}D_{\alpha\beta}$. With these different pieces, it is straightforward to build up the Weyl tensor of the new disformed solution $(\M, \tilde{g})$ which can be decomposed as the sum of three different pieces given by
\begin{equation}\label{Weyl-JE-Sym0}
\tilde{C}_{\alpha\beta\mu\nu} = A \Big( C_{\alpha\beta\mu\nu} + D^T_{\alpha\beta\mu\nu}
+ \frac{B}{A} Z_{\alpha\beta\mu\nu} \Big) \,.
\end{equation}
The tensor $D^T_{\alpha\beta\mu\nu} = g_{\alpha \rho}D^{T\,\rho}{}_{\beta\mu\nu}$, where $D^{T\,\rho}{}_{\beta\mu\nu}$ is the Weyl part of the tensor \eqref{D} given by 
\begin{equation}\label{D-Weyl}
D^{T\,\rho}{}_{\beta\mu\nu} \equiv \frac{1}{2} 
\big( {D}^\rho{}_{\beta\mu\nu} + {D}^\rho{}_{\nu} {g}_{\beta\mu} 
+ { D}_{\beta\mu} \delta^\rho{}_{\nu} \big) 
+ \frac{1}{6} { D} {\delta}^\rho{}_{\mu} { g}_{\beta\nu} \, - \, [\mu \leftrightarrow \nu] \,.
\end{equation}
The tensor $Z_{\alpha\beta\mu\nu}$ is defined as
\begin{equation}\label{Z}
Z_{\alpha\beta\mu\nu} \equiv E_{\alpha\beta\mu\nu} + V_\alpha
\Big( C_{\rho\beta\mu\nu} + D^T_{\rho\beta\mu\nu} 
+ \frac{B}{A} E_{\rho\beta\mu\nu} \Big) V^\rho \,,
\end{equation} 
with
\begin{eqnarray}\label{E}
E^\rho{}_{\beta\mu\nu} &\equiv& \frac{1}{2} \bigg\{
(R^\rho{}_{\nu} + D^\rho{}_{\nu}) V_\beta V_\mu
- \frac{(R_{\sigma\nu}+D_{\sigma\nu})V^\sigma V^\rho}{A+BY} 
( A g_{\beta\mu} + B V_\beta V_\mu ) - \, [\mu \leftrightarrow \nu] \, 
\bigg\} 
\\ \nonumber
&+&\frac{1}{6} \bigg\{
 (R+D) \delta^\rho{}_\mu V_\beta V_\nu 
- \frac{(R_{\sigma\alpha}+D_{\sigma\alpha})V^\sigma V^\alpha}{A+BY} \delta^\rho{}_\mu
( A g_{\beta\nu} + B V_\beta V_\nu ) - \, [\mu \leftrightarrow \nu] \, 
\bigg\} \,,
\end{eqnarray}
which satisfies $g^{\alpha\beta}E_{\alpha\mu\beta\nu} = 0$ and $g^{\mu\nu}E_{\alpha\mu\beta\nu} \neq 0$. 

Let us now comment on the symmetries of the new terms. They will be crucial when decomposing the disformed Weyl tensor in term of the Weyl-type and Ricci-type scalars. Contrary to the two first pieces entering in the expression (\ref{Weyl-JE-Sym0}), the last tensor $Z_{\alpha\beta\mu\nu}$ is not traceless so that we can define
\begin{eqnarray}\label{Z-ab}
Z_{\mu\nu} &\equiv& g^{\alpha\beta} Z_{\alpha\mu\beta\nu} 
= \Big( C_{\alpha\mu\beta\nu} + D^T_{\alpha\mu\beta\nu} 
+ \frac{B}{A} E_{\alpha\mu\beta\nu} \Big) V^\alpha V^\beta \,, \nonumber \\
Z &\equiv& g^{\alpha\beta}Z_{\alpha\beta} 
= \frac{BY}{A+BY} (R_{\alpha\beta} + D_{\alpha\beta}) V^\alpha V^\beta \,.
\end{eqnarray} 
Since the tensor $Z_{\alpha\beta\mu\nu}$ shares the same properties as the Riemann tensor, this suggests to decompose it as
\begin{eqnarray}\label{Z-decom}
 Z_{\alpha\beta\mu\nu}& =& Z^T_{\alpha\beta\mu\nu}  +Z^S_{\alpha\beta\mu\nu} \,,
 \\ \label{Z-S}
Z^S_{\alpha\beta\mu\nu} & \equiv&
-\frac{1}{2} \big(Z_{\alpha\nu} g_{\beta\mu} + Z_{\beta\mu} g_{\alpha\nu}
- Z_{\alpha\mu} g_{\beta\nu} - Z_{\beta\nu} g_{\alpha\mu} \big) 
- \frac{1}{6} Z (g_{\alpha\mu} g_{\beta\nu} - g_{\alpha\nu} g_{\beta\mu}) \,,
\end{eqnarray}
where  $Z^T_{\alpha\beta\mu\nu} $ is the trace-free Weyl part, i.e.\ satisfying $g^{\alpha\beta}Z^T_{\alpha\mu\beta\nu}= 0$, while the trace of the second part reads $g^{\alpha\beta}Z^S_{\alpha\mu\beta\nu} = Z_{\mu\nu}$, and $g^{\alpha\beta}g^{\mu\nu}Z^S_{\alpha\mu\beta\nu}=Z$. Using this splitting, we then define a new tensor $B_{\alpha\beta\mu\nu}$ which captures the trace-free contribution to the two last pieces of (\ref{Weyl-JE-Sym0}) such that
\begin{eqnarray}\label{B}
&&B_{\alpha\beta\mu\nu} \equiv D^T_{\alpha\beta\mu\nu}
+ \frac{B}{A} Z^T_{\alpha\beta\mu\nu} \,.
\end{eqnarray}
By construction, this tensor has the same properties as the Weyl tensor and vanishes when $B=0$, namely for a pure conformal transformation. At the end of the day, the Weyl tensor in the Jordan and seed frames are related by
\begin{eqnarray}\label{Weyl-JE-Sym}
\tilde{C}_{\alpha\beta\mu\nu} = A \Big( C_{\alpha\beta\mu\nu} + B_{\alpha\beta\mu\nu} 
+ \frac{B}{A} Z^S_{\alpha\beta\mu\nu} \Big) \,.
\end{eqnarray}
This last expression allows one to capture the main differences between conformal and disformal transformations. It is interesting to notice that the Weyl tensor $\tilde{C}_{\alpha\beta\mu\nu}$ is no longer trace-free with respect to the untilde inverse metric $g^{\alpha\mu}$ when implementing a disformal transformation, i.e.\ when $B\neq0$, unlike the case with $B=0$.\footnote{The Weyl tensor with one upper index, which is invariant under conformal transformation can be found as $\tilde{C}^\rho{}_{\beta\mu\nu} = C^\rho{}_{\beta\mu\nu} + D^{T\,\rho}{}_{\beta\mu\nu} 
+ \frac{B}{A} E^{\rho}{}_{\beta\mu\nu} \,.$} The Weyl tensor inherits a non-traceless part with respect to the seed frame metric $g_{\mu\nu}$ given by $Z^S_{\alpha\beta\mu\nu}$. This difference from the conformal case shows up since the metrics in the seed and Jordan frames are no longer proportional to each other in the case of disformal transformation. This point will turn out to be important when decomposing this tensor on a local rest frame to build the disformed Weyl scalars. Before presenting their construction, let us nevertheless point that the Weyl tensor $\tilde{C}_{\alpha\beta\mu\nu}$ satisfies obviously the standard symmetry properties. In particular,  $\tilde{C}_{\alpha\beta\mu\nu}$ is traceless with respect to the disformal metric $\tilde{g}_{\alpha\beta}$, as should be. 

\subsection{Disformed Weyl scalars}

Following the standard approach to the Petrov classification, which amounts at identifying the number of independent PNDs of a given spacetime geometry, as well as their multiplicity, we now turn to the decomposition of the Weyl tensor. The Weyl scalars are defined by~\cite{Chandrasekhar:1985kt,Stephani:2003tm}
\begin{eqnarray}\label{WeylS-tilde-0}
\tilde{\bf \Psi}_0 & =& \tilde{C}_{\alpha\beta\mu\nu} \tilde{l}^{\alpha} {\tilde m}^{\beta} \tilde{l}^{\mu} {\tilde m}^{\nu} \,,\\
\tilde{\bf \Psi}_1 & =& \tilde{C}_{\alpha\beta\mu\nu} {\tilde l}^{\alpha} {\tilde k}^{\beta} {\tilde l}^{\mu} {\tilde m}^{\nu}  \,, \\ 
\tilde{\bf \Psi}_2 & =& {\tilde C}_{\alpha\beta\mu\nu} {\tilde l}^{\alpha} {\tilde m}^{\beta} \tilde{\bar m}^{\mu} {\tilde k}^{\nu}  \,, \\ 
\tilde{\bf \Psi}_3 & =& {\tilde C}_{\alpha\beta\mu\nu} {\tilde k}^{\alpha} {\tilde l}^{\beta}  {\tilde k}^{\mu} \tilde{\bar m}^{\nu} \,, \\ 
\tilde{\bf \Psi}_4 & =& {\tilde C}_{\alpha\beta\mu\nu} {\tilde k}^{\alpha} \tilde{\bar m}^{\beta} {\tilde k}^{\mu} \tilde{\bar m}^{\nu} \,. \label{WeylS-tilde-4}
\end{eqnarray}
Our goal now is to express these Weyl scalars $\tilde{\bf \Psi}_I$, associated to the disformed metric $\tilde{g}_{\mu\nu}$, in terms of the Weyl scalars ${\bf \Psi}_I$ associated to the seed metric $g_{\mu\nu}$. This will allow us to provide closed expressions which capture how the Petrov type of a given spacetime geometry changes under a disformal map. To do so, we rewrite the Weyl scalars in Jordan frame~\eqref{WeylS-tilde-0}-\eqref{WeylS-tilde-4}  in terms of the null tetrad in the Jordan frame. Let us first introduce the notation
\begin{eqnarray}\label{WeylS-tilde-T}
\tilde{\bf \Psi}_0 = \tilde{C}_{0202} \,, \hspace{.5cm}
\tilde{\bf \Psi}_1 = \tilde{C}_{0102} \,, \hspace{.5cm}
\tilde{\bf \Psi}_2 = {\tilde C}_{0231} \, \hspace{.5cm}
\tilde{\bf \Psi}_3 = {\tilde C}_{1013} \,, \hspace{.5cm}
\tilde{\bf \Psi}_4 = {\tilde C}_{1313}  \,,
\end{eqnarray}
where 
\begin{equation}\label{C-tilde}
\tilde{C}_{abcd} \equiv \tilde{C}_{\alpha\beta\mu\nu} 
\tilde{\theta}^{\alpha}{}_a\tilde{\theta}^{\beta}{}_b \tilde{\theta}^{\mu}{}_c \tilde{\theta}^{\nu}{}_d \,,
\end{equation}
are tetrad components of the Weyl tensor in Jordan frame with respect to the tetrads $\tilde{\theta}^{\nu}{}_a$. As it is clear from the above relations, the Weyl complex scalars are some special tetrad components of the Weyl tensor $C_{abcd}$ which describe ten independent components of the Weyl tensor. The other tetrad components of the Weyl tensor are not independent (see Eq.~\eqref{Weyl-components} in the appendix \ref{app-tetrad-rep}). Substituting Eq.~\eqref{Weyl-JE-Sym} in Eq.~\eqref{C-tilde}, we find
\begin{eqnarray}\label{C-tilde-trans0}
{\tilde C}_{efgh} = \frac{A}{4} \Big( C_{abcd} + B_{abcd} + \frac{B}{A} Z^S_{abcd} \Big) 
T^{ab}{}_{ef}T^{cd}{}_{gh} \,,
\end{eqnarray}
in which $C_{abcd} $, $B_{abcd} $, $Z^S_{abcd}$ are defined in the same manner as Eq.~\eqref{C-tilde} correspondingly. We have introduced the compact notation
\begin{eqnarray}\label{Tabcd}
T^{ab}{}_{ef} \equiv T^a{}_e T^b{}_f - T^a{}_f T^b{}_e
= \frac{1}{A} \big( \delta^a{}_e \delta^b{}_f - \delta^a{}_f \delta^b{}_e \big) 
- \frac{\beta}{A} V^{ab}{}_{ef} \,,
\end{eqnarray}
where we have used Eq.~\eqref{J-inv} in the second step. The last term is defined as
\begin{eqnarray}\label{Qabcd}
V^{ab}{}_{ef} \equiv
\delta^a{}_e V^b V_f + \delta^b{}_f V^a V_e - \delta^a{}_f V^b V_e - \delta^b{}_e V^a V_f \,.
\end{eqnarray}
Note that $T^{ab}{}_{ef}$ and $V^{ab}{}_{ef}$ are totally antisymmetric in both upper and lower pairs of indices. Moreover, $T^{ab}{}_{ef}$ is linear in $\beta$ as the quadratic parts cancel each other. Substituting Eq.~\eqref{Tabcd} in Eq.~\eqref{C-tilde-trans0} we find
\begin{eqnarray}\label{C-tilde-trans-prim}
{\tilde C}_{efgh} &=& \frac{1}{A} \Big( C_{efgh} + B_{efgh} + \frac{B}{A} Z^S_{efgh} \Big) \nonumber \\ 
&-& \frac{\beta}{2A} \Big\{ \Big( C_{efab} + B_{efab} + \frac{B}{A} Z^S_{efab} \Big) V^{ab}{}_{gh}
+ \Big( C_{ghab} + B_{ghab} + \frac{B}{A} Z^S_{ghab} \Big) V^{ab}{}_{ef} \Big\}
\nonumber \\
&+&  \frac{\beta^2}{4A} \Big( C_{abcd} + B_{abcd} + \frac{B}{A} Z^S_{abcd} \Big)
V^{ab}{}_{ef} V^{cd}{}_{gh} \,.
\end{eqnarray}
Now, the question is how many scalar quantities do we need to fully encodes the information encapsulated in the r.h.s.\ of the above equation. In the seed frame, the symmetry of the Weyl tensor allows one to use only the standard five complex Weyl scalars ${\bf \Psi}_I$ to express all components of ${C}_{abcd}$ as it is done in Eq.~\eqref{WeylS-T-a} in the appendix \ref{app-tetrad-rep}. By construction, the tensor $B_{\alpha\beta\mu\nu}$ has the same symmetry as the Weyl tensor and we only need five more complex Weyl-type scalar, which we denote ${\mathbf \Delta}_I$ ($I=0,\cdots,4$), in order to decompose it. They read
\begin{eqnarray}\label{Delta}
{\mathbf \Delta}_0 = B_{0202} \,, \hspace{.5cm}
{\mathbf \Delta}_1 = B_{0102} \,, \hspace{.5cm}
{\mathbf \Delta}_2 = B_{0231} \,, \hspace{.5cm}
{\mathbf \Delta}_3 = B_{1013} \,, \hspace{.5cm}
{\mathbf \Delta}_4 = B_{1313} \,.
\end{eqnarray}
The decomposition of the remaining trace part $Z^S_{abcd}$ is more subtle. Indeed, as pointed in the previous section, it does not share the same symmetry as the Weyl tensor and one therefore needs additional scalars to fully capture the information encoded in $Z^S_{abcd}$. From the definition Eq.~\eqref{Z-S} we find
\begin{eqnarray}\label{Zij}
Z^S_{abcd} =  
-\frac{1}{2} \big(Z_{ad} \eta_{bc} + Z_{bc} \eta_{ad}
- Z_{ac} \eta_{bd} - Z_{bd} \eta_{ac} \big) 
- \frac{1}{6} Z (\eta_{ac} \eta_{bd} - \eta_{ad} \eta_{bc}) \,,
\end{eqnarray}
which shows that $Z^S_{abcd}$ is completely expressed in terms of its Ricci part $Z_{ab}$. Therefore, it can be completely expressed in terms of the four real and three complex scalars given by
\begin{eqnarray}\label{Pi}
&&{\bf{\Pi}}_{00} = \frac{1}{2} Z_{00} \,, \hspace{.5cm}
{\bf{\Pi}}_{11} = \frac{1}{4} (Z_{01}+Z_{23}) \,, \hspace{.5cm}
{\bf{\Pi}}_{22} = \frac{1}{2} Z_{11} \,, \hspace{.5cm}
 {\bf{\Lambda^S}} = - \frac{1}{24} Z = \frac{1}{12} (Z_{01} - Z_{23}) \,, \nonumber \\
&&{\bf{\Pi}}_{01} = \frac{1}{2} Z_{02} \,, \hspace{1.5cm}
{\bf{\Pi}}_{02} = \frac{1}{2} Z_{22} \,, \hspace{1.5cm}
{\bf{\Pi}}_{12} = \frac{1}{2} Z_{12} \,.
\end{eqnarray}
These additional Ricci-type scalars, i.e.\ ${\bf \Delta}_I$ and ${\bf \Pi}_{IJ}$ and ${\bf \Lambda}^S$, allows one to write close relations between the Weyl scalars ${\bf \tilde{\Psi}}_I$  and  ${\bf \Psi}_I$ which can be compactly written as 
\begin{eqnarray}\label{C-tilde-trans}
\tilde{\bf \Psi}_I = \tilde{\bf \Psi}_I({\bf \Psi}_I,{\bf \Delta}_I,{\bf \Pi}_{IJ})  \,.
\end{eqnarray}
As a concrete example, the first disformed Weyl scalar ${\bf \tilde{\Psi}}_0$ decomposes as
\begin{eqnarray}\label{Psi0-mail}
A \tilde{\boldsymbol \Psi}_0 &=&  \gamma^2 ( {\boldsymbol \Psi}_0 + {\boldsymbol \Delta}_0 )
+ \beta^2 \big[ ( {\boldsymbol \Psi}^*_0 + {\boldsymbol \Delta}^*_0 ) (V_2V_2)^2 
+ ( {\boldsymbol \Psi}^*_4 + {\boldsymbol \Delta}^*_4 ) (V^1V^1)^2 \big]
\nonumber \\ 
&+& 2 \beta^2 V^1 V_2 \big[ 2 ( {\boldsymbol \Psi}^*_1 + {\boldsymbol \Delta}^*_1 ) V_2V_2 
+ 3 ( {\boldsymbol \Psi}^*_2 + {\boldsymbol \Delta}^*_2 ) V^1V_2 
+ 2 ( {\boldsymbol \Psi}^*_3 + {\boldsymbol \Delta}^*_3 ) V^1V^1 \big] 
\nonumber \\ 
&-& 2 \frac{B}{A} \beta \gamma \big[ 2 {\boldsymbol \Pi}_{01} V^1V_2 
+ {\boldsymbol \Pi}_{02} V^1 V^1 + {\boldsymbol \Pi}_{00} V_2 V_2 \big]
\,.
\end{eqnarray}
The explicit form of disformed Weyl scalars being quite involved, we provide their decomposition in the Appendix~\ref{app-tetrad-rep} which are given by the relations~\eqref{Psi0},~\eqref{Psi1},~\eqref{Psi2},~\eqref{Psi3}, and~\eqref{Psi4}. This provides the main result of this work.

Having the decomposition of all Weyl scalars ~\eqref{C-tilde-trans} in hand, we can study the Petrov classification after disformal transformation. The classification is performed using the following Lorentz-invariant quantities
\begin{eqnarray}\label{Petrov-quantities}
&&{\tilde I} \equiv \tilde{\bf \Psi}_0 \tilde{\bf \Psi}_4 - 4 \tilde{\bf \Psi}_1 \tilde{\bf \Psi}_3 
+ 3 \tilde{\bf \Psi}_2^2 \,, \hspace{1cm}
{\tilde J} \equiv 
\begin{vmatrix}
\tilde{\bf \Psi}_4 & \tilde{\bf \Psi}_3 & \tilde{\bf \Psi}_2 \\ 
\tilde{\bf \Psi}_3 & \tilde{\bf \Psi}_2 & \tilde{\bf \Psi}_1 \\ 
\tilde{\bf \Psi}_2 & \tilde{\bf \Psi}_1 & \tilde{\bf \Psi}_0
\end{vmatrix} \,, \hspace{1cm} 
{\tilde D} \equiv {\tilde I}^3 - 27 {\tilde J}^2 \,,
\\ \nonumber
&& {\tilde K} \equiv \tilde{\bf \Psi}_4^2 \tilde{\bf \Psi}_1 
- 3 \tilde{\bf \Psi}_4 \tilde{\bf \Psi}_3 \tilde{\bf \Psi}_2 + 2 \tilde{\bf \Psi}_3^3 \,, \hspace{1cm}
{\tilde L} \equiv \tilde{\bf \Psi}_4 \tilde{\bf \Psi}_2 - \tilde{\bf \Psi}_3^2 \,, \hspace{1cm}
{\tilde N} \equiv 12 {\tilde L}^2 - \tilde{\bf \Psi}_4^2 {\tilde I} \,.
\end{eqnarray} 
Depending on the values of these scalars, one can directly determine in a coordinate-independent and Lorentz-invariant manner the Petrov type of a given geometry. The classification is summarized in Table \ref{tab1}. Physically, the different types can be understood as follows in terms of their PNDs. A given geometry possesses four distinct PNDs at each point when it is type I, three PNDs with one repeated for type II, one triply repeated for type III, two PNDs both repeated for type D and finally only one PND but quadruply repeated for type N.
\begin{table}
	\centering
	\begin{tabular}{ |p{1cm}|p{7cm}|  }
		\hline
		\multicolumn{2}{|c|}{\bf Petrov Classification} \\
		\hline 
		\hfil Type & \hfil Conditions \\
		\hline
		\hfil O & $\tilde{\bf \Psi}_0=\tilde{\bf \Psi}_1=\tilde{\bf \Psi}_2=\tilde{\bf \Psi}_3=\tilde{\bf \Psi}_4=0$ \\
		\hfil I & $\tilde{ D}\neq0$ \\
		\hfil II & $\tilde{ D} =0$, ${\tilde I}\neq0$, ${\tilde J}\neq0$, ${\tilde K}\neq0$, ${\tilde N}\neq0$ \\
		\hfil III & ${\tilde D} =0$, ${\tilde I} = {\tilde J} =0$, ${\tilde K}\neq0$, ${\tilde L}\neq0$ \\
		\hfil N & ${\tilde D} =0$, ${\tilde I}={\tilde J}={\tilde K}={\tilde L}=0$ \\
		\hfil D & ${\tilde D} =0$, ${\tilde I}\neq0$, ${\tilde J}\neq0$, ${\tilde K}={\tilde N}=0$\\
		\hline
	\end{tabular}
	\newline
	\caption{The left column shows the Petrov types and the right column shows the corresponding desired conditions. Quantities ${\tilde I}, {\tilde J}, {\tilde D}, {\tilde K}, {\tilde L}$, and ${\tilde N}$ are defined in Eq.~\eqref{Petrov-quantities} in terms of the Weyl scalars $\tilde{\bf \Psi}_I$.}\label{tab1}
\end{table}

The results that we found in this section show that Petrov type of a given spacetime geometry can change after performing a disformal transformation. We will explicitly confirm this fact in the next section by applying our general formalism to some concrete examples. Moreover, having disformed Weyl scalars~\eqref{C-tilde-trans} in hand, we can look for conditions under which Petrov types are invariant. Therefore, one can extract general results from our analysis which are useful especially when constructing a given exact solution through a disformal transformation.

Before closing this subsection, let us consider the case of conformal map by means of our formalism. In this case, $B=0$ such that $B_{abcd} = 0$. From the results~\eqref{Psi0}-\eqref{Psi4}, it follows that $\tilde{\bf \Psi}_I = A^{-1} {\bf \Psi}_I$. From the definitions~\eqref{Petrov-quantities}, we find ${\tilde I} = A^{-2} I$, ${\tilde J} = A^{-3} J$, ${\tilde D} = A^{-6} D$, ${\tilde K} = A^{-3} K$, ${\tilde L} = A^{-2} L$, and ${\tilde N} = A^{-4} N$ where all quantities without tilde correspond to the definitions~\eqref{Petrov-quantities} replacing $\tilde{\bf \Psi}_I$ with ${\bf \Psi}$. Therefore, the Petrov type of a spacetime geometry will not change under a conformal transformation~\cite{Ajith:2020ydz}, as expected.

\subsection{Simplified disformal transformation}\label{subsec-pure-disformal}

\label{secgencond}

In this subsection, we use the results obtained in this section and analyze them in a simplified setting. As the effect of conformal transformation is already well understood, we focus on the pure disformal case with $A=1$ and constant small disformal factor $B=B_0\ll1$. Indeed, considering only constant value of the conformal factor $A=A_0$, without loss of generality $A_0=1$ can be realized through redefinition of metric. The disformal transformation~\eqref{DT} then takes the form
\begin{equation}\label{DT-S}
{\tilde g}_{\mu\nu} = {g}_{\mu\nu} + B_0\, V_\mu V_\nu \,,
\end{equation}
The above transformation captures all essential features of the disformal map while the analysis becomes significantly simpler. Moreover, an interesting application of the disformal mapping is to consider a transformation for which $ B_0\ll1$. It corresponds to small deviations form GR and provide thus an interesting regime relevant for the comparison with observations. Starting from this setup, we shall now provide simplified expressions for the relations between the Weyl scalars before and after the disformal transformations up to first order in $B_0$.

For the choice $A=1$ and $B=B_0\ll1$, Eq.~\eqref{beta} implies that $\beta \approx \frac{B_0}{2}$. The null tetrad in Jordan frame, given by Eq.~\eqref{null}, simplifies to
\begin{eqnarray}\label{kl-disform-exp}
&&{\tilde l}^\mu \approx 
{l}^\mu{} + \frac{B_0}{2} V^1 V^\mu \,, 
\hspace{1.5cm}
{\tilde k}^\mu \approx 
{k}^\mu{} - \frac{B_0}{2} V_1 V^{\mu}\,,
\nonumber \\
&&{\tilde m}^\mu \approx
{m}^\mu{} - \frac{B_0}{2} V_2 V^\mu \,, 
\hspace{1cm}
{\tilde {\bar m}}^\mu \approx
{{\bar m}}^\mu{} - \frac{B_0}{2} V^2 V^\mu \,.
\end{eqnarray}
Using these expressions, and expanding for $B=B_0\ll1$, the results~\eqref{Psi0},~\eqref{Psi1},~\eqref{Psi2},~\eqref{Psi3}, and~\eqref{Psi4} simplify to
\begin{eqnarray}\label{Psi0-ex}
\tilde{{\boldsymbol \Psi}}_0 & \approx &
{\boldsymbol \Psi}_0 \big( 1 - B_0 ( V_1V^1+V_2V^2 ) \big) +  {\boldsymbol \Delta}_0 \,,
\\[7pt]
\label{Psi1-ex}
\tilde{\boldsymbol \Psi}_1 & \approx &
{\boldsymbol \Psi}_1 \Big( 1 - \frac{B_0}{2} ( V_1V^1+V_2V^2 ) \Big) + {\boldsymbol \Delta}_1
\nonumber \\
&-& \frac{B_0}{2} \Big[ {\boldsymbol \Psi}_0 V_1 V^2
+ {\boldsymbol \Psi}^*_1 V_2 V_2
+ (2 {\boldsymbol \Psi}^*_2 - {\boldsymbol \Psi}_2 ) V^1 V_2 
+ {\boldsymbol \Psi}^*_3 V^1 V^1 
- 2 {\boldsymbol \Pi}_{01} \Big] \,,
\end{eqnarray}

\begin{eqnarray}\label{Psi2-ex}
\tilde{\boldsymbol \Psi}_2 & \approx &
{\boldsymbol \Psi}_2 \big( 1 - B_0 ( V_1V^1+V_2V^2 ) \big) + {\boldsymbol \Delta}_2
- 2B_0 {\boldsymbol \Lambda}^S \,,
\\[7pt]
\label{Psi3-ex}
\tilde{\boldsymbol \Psi}_3 & \approx &
{\boldsymbol \Psi}_3 \Big( 1 - \frac{B_0}{2} ( V_1V^1+V_2V^2 ) \Big) + {\boldsymbol \Delta}_3 
\nonumber \\ 
&-& \frac{B_0}{2} \Big[ {\boldsymbol \Psi}^*_1 V_1 V_1 
- (2 {\boldsymbol \Psi}^*_2 - {\boldsymbol \Psi}_2 ) V_1 V^2 - {\boldsymbol \Psi}_4 V^1 V_2
+ {\boldsymbol \Psi}^*_3 V^2 V^2 - 2 {\boldsymbol \Pi}^*_{12} \Big] \,,
\\[7pt]
\label{Psi4-ex}
\tilde{\boldsymbol \Psi}_4 & \approx &
{\boldsymbol \Psi}_4 \Big( 1 - B_0 ( V_1V^1+V_2V^2 ) \Big) + {\boldsymbol \Delta}_4 \,.
\end{eqnarray}
From the above results, it is more clear that Petrov types generally change under disformal transformation. For instance, suppose that we have a type O solution in Jordan frame so that all $\tilde{\boldsymbol \Psi}_I = 0$. We thus have five equations for five complex variables ${\boldsymbol \Psi}_I$ in the seed frame which give non-vanishing solutions for ${\boldsymbol \Psi}_I$ in general. Therefore a type O metric in Jordan frame will be no longer type O in the seed frame in general.

Having presented the simplified expressions for the disformed Weyl scalars for the disformal map Eq.~\eqref{DT-S}, we can now apply our general results to investigate concrete known exact solutions. This is the subject of the next section.

\section{Computing the Petrov type of disformed geometries}

In this section, we apply our general setup to several relevant solutions of the modified gravity theories to illustrate how the Petrov type of a given seed solution changes after performing a disformal transformation. We consider three different seed configurations: i) a FLRW cosmology, ii) a stealth spherically symmetric black hole and finally iii) the case of a stealth Kerr geometry. For the sake of simplicity, we perform all analysis for a pure disformal map with $B_0\ll 1$ presented in subsection \ref{subsec-pure-disformal}.

\subsection{FLRW}

Let us first consider a simple example of the FLRW cosmology. The line element for the seed solution is given by the spatially curved FLRW background in spherical coordinates $(t,r,\theta,\varphi)$ given by
\begin{equation}
\label{ds2-FLRW}
ds^2_{\rm FLRW} = - dt^2 + a(t)^2 \bigg[ 
\frac{dr^2}{1-kr^2} + r^2 d\Omega^2 
\bigg] \,.
\end{equation}
In the line element Eq.~\eqref{ds2-FLRW}, $a(t)$ is the scalar factor, $d\Omega^2 = d\theta^2 + \sin^2\theta d\varphi^2$ is the metric of the unit sphere, and $k$ is the normalized constant curvature of the spatial metric which takes the values $k=0,1,-1$ for flat, spherical, and hyperbolic 3-dimensional spatial sections respectively. Consider the following PNDs \cite{GomezLopez:2017kcw}
\begin{align}\label{null-vectors-FLRW}
l^\mu & = \frac{1}{\sqrt{2}} \Big( 1 , - \frac{\sqrt{1-k r^2}}{a},0 , 0 \Big) ,\\
\hspace{.3cm}
k^\mu & = \frac{1}{\sqrt{2}} \Big( 1 ,  \frac{\sqrt{1-k r^2}}{a}, 0 , 0 \Big) , \label{null-vectors-FLRW-k}\\
\hspace{.3cm}
m^\mu & = \frac{1}{\sqrt{2}r} \Big( 0, 0 , - \frac{1}{a}, \frac{i}{\sin\theta} \Big) ,\label{null-vectors-FLRW-m}
\end{align}
which satisfy the desired conditions~\eqref{null} as well as the vanishing of other contractions. Using the above PNDs, it is easy to confirm that ${\mathbf \Psi}_I=0$ for all $I$ and the seed FLRW solution~\eqref{ds2-FLRW} is type O as it is well known. Indeed, this is clear since the metric~\eqref{ds2-FLRW} is conformally flat and therefore the corresponding Weyl tensor vanishes.  

In order to study the effect of a disformal transformation on such cosmological background, one needs to preserve homogeneity and isotropy. The natural choice for the disformal vector is therefore $V_\mu= \partial_\mu \phi$ in which $\phi$ is a scalar field with time-dependent homogeneous vev $\phi(t)$. The vector field then takes the simple form
\begin{equation}\label{V-FLRW}
V_\mu = (\dot{\phi},0,0,0) \,,
\end{equation}
where a dot denotes derivative with respect to the time $t$. The disformed metric takes the following form
\begin{align}\label{ds2-FLRW-dis}
{\widetilde ds}^2_{\rm{FLRW}} & = { ds}^2_{\rm{FLRW}} + B_0 \dot{\phi}^2 dt^2 \\
& = - (1-B_0 \dot{\phi}^2) dt^2 + a(t)^2 \bigg[ 
\frac{dr^2}{1-kr^2} + r^2 d\Omega^2 
\bigg] \,.
\end{align}
The effects of disformal transformation in this case turns out to be quite trivial as we can define a new time coordinate $t'=\int (1-B_0 \dot{\phi}^2)^{1/2} dt$ in terms of which the metric takes the standard FLRW form. One can find the new PNDs and Weyl scalars after disformal transformation by substituting~\eqref{null-vectors-FLRW}-\eqref{V-FLRW} in~\eqref{kl-disform-exp} and~\eqref{Psi0-ex}-\eqref{Psi4-ex} respectively. However, as we have shown above, they will not change after redefining the time coordinate. Therefore, the Petrov type of the metric~\eqref{ds2-FLRW} will not change and remains type O.

\subsection{Disformed stealth Schwarzschild}

We now turn to the case where the seed solution is a spherically symmetric, static geometry of type D. A vast landscape of such exact black hole solutions has been constructed for DHOST theories~\cite{Kobayashi:2014eva, Babichev:2017guv, BenAchour:2018dap, Motohashi:2018wdq, Motohashi:2019sen, Takahashi:2020hso, Charmousis:2019vnf}. 

\subsubsection{Standard kinematics}

Consider thus a seed solution whose line element is given by
\begin{equation}\label{ds2-SS}
ds^2_{\rm SS} = - f(r) dt^2+ \frac{dr^2}{f(r)} + r^2 d\Omega^2 \,,
\end{equation}
which includes for example the form of the metric of a stealth Schwarzschild black hole. Notice that this is not the most general case, yet it provides a relevant example of the backgrounds used in modified gravity as seed.

 For such geometry, a set of PNDs is given by
\begin{eqnarray}\label{null-vectors-SS}
l^\mu & = \frac{1}{\sqrt{2}} \Big( \frac{1}{\sqrt{f}} , - \sqrt{f}, 0 , 0 \Big) , \\
\hspace{.3cm}
k^\mu & = \frac{1}{\sqrt{2}} \Big( \frac{1}{\sqrt{f}} , \sqrt{f}, 0 , 0 \Big) , \\
\hspace{.3cm}
m^\mu & = \frac{1}{\sqrt{2}r} \Big( 0 , 0 , - 1, \frac{i}{\sin\theta} \Big) ,
\end{eqnarray}
and one can check that the only non-zero Weyl scalar is
\begin{equation}\label{Weyl-SS}
{\boldsymbol \Psi}_2 = \frac{1}{6r^2} \Big( -1 + f - r f' + \frac{r^2}{2} f'' \Big) \,.
\end{equation}
The other nonzero scalars constructed by the tetrad components of the Ricci tensor which are defined in Eq.~\eqref{RicciS-T-a} are also given by
\begin{eqnarray}\label{RicciS-SS}
{\boldsymbol \Phi}_{11} = \frac{1}{4r^2} \Big( 1 - f + \frac{r^2}{2} f'' \Big) \,,
\hspace{1cm}
{\boldsymbol \Lambda} = \frac{1}{12r^2} \Big( -1 + f + 2 r f' + \frac{r^2}{2} f'' \Big) \,.
\end{eqnarray}
Therefore, using the definitions~\eqref{Petrov-quantities} and the Table \ref{tab1}, one finds as expected that the seed metric~\eqref{ds2-SS} is of Petrov type D. We can now investigate how the Petrov type changes under a disformal mapping depending on the profile of the additional field $V$.

\subsubsection{Disformed kinematics}

In order not to spoil the spherical symmetry of the seed metric~\eqref{ds2-SS}, we consider a general vector field with components depending only on the time $t$ and radial $r$ coordinates.\footnote{We have so far considered the case of a static spherically symmetric case. As for the general spherically symmetric solution (aimed to describe, e.g.\ a collapse), which can be written as $ds^2=g_{\mu\nu}\,dx^\mu dx^\nu=-f_1(t,r)^2\,dt^2+dr^2/f_2(t,r)^2+r^2\,d\Omega^2$, and for its disformed case, that is $\tilde g_{\mu\nu}=A(t,r)\,g_{\mu\nu}+B(t,r)V_\mu V_\nu $, where $V_\mu\,dx^\mu = V_t(t,r)\,dt+ V_r(t,r)\,dr$, it can be shown, by direct calculations, that they are both of Petrov type D.} Without loss of generality, the vector profile reads
\begin{equation}\label{SS-V-mu}
V_\mu = \big( V_t, V_r, 0, 0 \big) \,, \hspace{1cm} 
V^\mu = \Big( -\frac{V_t}{f}, f V_r, 0, 0 \Big) \,,
\end{equation}
where $V_t:= V_t(t,r)$ and $V_r:=V_r(t,r)$. The tetrad components of the vector field defined in Eq.~\eqref{Y-tilde} are given by 
\begin{align}\label{SS-V-a}
V_a & = \frac{1}{\sqrt{2f}} \big( V_t-f V_r , V_t + f V_r, 0, 0 \big) \,, \hspace{1cm} \\
V^a & = - \frac{1}{\sqrt{2f}} \big( V_t+ f V_r, V_t - f V_r, 0, 0 \big) \,.
\end{align}
We see that $V_3=V^2=0$ and $V^3=V_2=0$ which is equivalent to the orthogonality condition. Thus, the disformal vector $V^\mu$, given by Eq.~\eqref{SS-V-mu}, lives on the tangent space $T_p({\cal M}/{\cal S})$ and is orthogonal to the 2-dimensional surface ${\cal S}$. 

Applying the above results to the seed metric~\eqref{ds2-SS}, we find that the spherically symmetric disformed metric takes the form
\begin{equation}\label{ds2-SS-dis}
{\widetilde ds}^2_{\rm{SS}} = {ds}^2_{\rm{SS}} + B_0 \big( V_t dt + V_r dr \big)^2 \,.
\end{equation}
Starting from the PNDs in seed frame Eq.~\eqref{null-vectors-SS}, we can easily find the disformed PNDs which read
\begin{align}\label{null-vectors-SS-dis}
{\tilde l}^\mu & = l^\mu - \frac{B_0}{2} 
\Big( \frac{V_t - f V_r }{\sqrt{2f}} \Big) V^\mu \,, \hspace{1cm}\\
{\tilde k}^\mu & = k^\mu - \frac{B_0}{2} 
\Big( \frac{V_t + f V_r }{\sqrt{2f}} \Big) V^\mu \,, \label{null-vectors-SS-dis-k}
\\
\hspace{1cm}
{\tilde m}^\mu  &= m^\mu \,,\label{null-vectors-SS-dis-m}
\end{align}
where the explicit forms of undisformed PNDs and the disformal vector $V^\mu$ are given by Eqs.~\eqref{null-vectors-SS} and~\eqref{SS-V-mu} respectively.
Now, let us consider the Weyl-type and Ricci-type scalars to investigate the fate of the Petrov type. From definitions~\eqref{Delta} and~\eqref{Pi}, we find the following nonzero scalars
\begin{eqnarray}\label{Dis-Scalars-SS}
&&{\boldsymbol \Delta}_2 =
\frac{B_0}{12 r^2}
\Bigg\{
\frac{V_t^2}{f} - V_t^2 - f V_r^2 
- 2 r V_r \dot{V}_t
- f^2 V_r \left(V_r-2 r V_r'\right)
- 2 r V_t \big(r V_t''-r\dot{V}_r'+\dot{V}_r-V_t' \big)
\nonumber \\ \nonumber
&& \hspace{2.3cm}
- 2 r^2 \left(V_r (\ddot{V}_r-\dot{V}_t' )-\dot{V}_r V_t'-\dot{V}_t
V_r'+\dot{V}_r^2+ V_t'^2\right)
+ \frac{r^2 f''}{2f}
\left(V_t^2-f^2 V_r^2\right)
\\ \nonumber
&& \hspace{2.3cm} - \frac{r^2 f'^2}{2f^2}
\left( V_t^2 + f^2 V_r^2\right)
+ \frac{r f'}{f} \left(f^2 V_r
\left(2 V_r-r V_r'\right)+r V_t \left(\dot{V}_r+V_t'\right)+r V_r
\dot{V}_t\right) \Bigg\} \,,
\\
&& {\boldsymbol \Pi}_{00} = \frac{{\boldsymbol \Psi}_2}{2f} \big( V_t - f V_r \big)^2 \,, 
\hspace{.5cm}
{\boldsymbol \Pi}_{11} = - \frac{{\boldsymbol \Psi}_2}{2f} \big( V_t^2 - f^2 V_r^2 \big) \,,
\hspace{.5cm}
{\boldsymbol \Pi}_{22} = \frac{{\boldsymbol \Psi}_2}{2f} \big( V_t + f V_r \big)^2 \,,
\end{eqnarray}
where dot and prime denote derivative with respect to the coordinates $t$ and $r$ respectively. In the case of spherically symmetric metric Eq.~\eqref{ds2-SS}, using Eqs.~\eqref{Weyl-SS},~\eqref{RicciS-SS}, ~\eqref{SS-V-a}, and~\eqref{Dis-Scalars-SS}, Eqs.~\eqref{Psi0-ex}-\eqref{Psi4-ex} simplify to
\begin{eqnarray}\label{Psi-SS}
\tilde{\boldsymbol \Psi}_0 = \tilde{\boldsymbol \Psi}_1 
= \tilde{\boldsymbol \Psi}_3 = \tilde{\boldsymbol \Psi}_4 = 0 \,, \hspace{1cm}
\tilde{\boldsymbol \Psi}_2 = {\boldsymbol \Psi}_2 
\Big( 1 + \frac{B_0}{2f} \big(V_t^2 - f^2 V_r^2 \big) \Big) + {\boldsymbol \Delta}_2 \,,
\end{eqnarray}
where the explicit form of ${\boldsymbol \Delta}_2$ is given by~\eqref{Dis-Scalars-SS}. Looking at the Table \ref{tab1}, the result~\eqref{Psi-SS} shows that the disformed metric~\eqref{ds2-SS-dis} is again of Petrov type D. This is consistent with the fact that the disformed metric~\eqref{ds2-SS-dis} is spherically symmetric. Indeed, it is well known that all spherically symmetric spacetimes are of Petrov type D (or O).  We now turn to the axisymmetric case.

\subsection{Disformed stealth Kerr}

While spherically symmetric solutions are of interest to understand the description of compact object, rotating configurations are the ones required to confront modified gravity theories to current astrophysical observations. In the DHOST context, a stealth Kerr-(A)dS solution dressed with a linear time-dependent scalar profile has been derived in~\cite{Charmousis:2019vnf} while general conditions for the existence of such stealth configurations have been addressed in~\cite{Takahashi:2020hso}. More recently, such seed was used to construct the first non-stealth Kerr-like black hole solution for a large family of DHOST theories~\cite{BenAchour:2020fgy,Anson:2020trg}. 

\subsubsection{Standard kinematics}
We consider the seed metric to be the stealth Kerr solution given by
\begin{equation}\label{ds2-Kerr}
ds^2_{\rm Kerr} = -\frac{\Delta}{\rho^2} (dt - a \sin^2\theta d\varphi)^2 + \frac{\rho^2}{\Delta} dr^2 + \rho^2 d\theta^2 + \frac{\sin^2\theta}{\rho^2} 
\big( a dt - (r^2+a^2) d\varphi \big)^2 \,,
\end{equation}
where $M$ and $a$ are mass and angular momentum per unit mass and we have defined
\begin{equation}
\Delta \equiv r^2 - 2 Mr + a^2 \,, \hspace{1cm} \rho^2 \equiv r^2 + a^2 \cos^2\theta \,.
\end{equation} 
Following Ref.~\cite{Chandrasekhar:1985kt}, we work with the PNDs
\begin{align}\label{null-vectors-Kerr}
l^\mu &= \frac{1}{\Delta} \big( r^2 + a^2 , \Delta,\, 0 \, ,\, a \big) \,, \\
\hspace{2cm}
k^\mu & = \frac{1}{2\rho^2} \big( r^2+a^2 , -\Delta, \, 0 \, ,\, a \big) \,, \label{null-vectors-Kerr-k} \\
m^\mu &= \frac{1}{\sqrt{2}(r+i a \cos\theta)} \big( i a \sin\theta , \, 0 \, , \, 1, 
\, i\csc\theta \big) \,\label{null-vectors-Kerr-m}.
\end{align}
The only non-zero component of the Weyl scalars is given by
\begin{equation}\label{Weyl-Kerr}
{\boldsymbol \Psi}_2 = -\frac{M}{(r-i a \cos\theta)^3} \,.
\end{equation}
such that the seed geometry~\eqref{ds2-Kerr} is of Petrov type D, as expected.\footnote{Note that we could also work with other PNDs like \cite{GomezLopez:2017kcw}
	\begin{eqnarray}\label{null-vectors-Kerr-old}
	l^\mu = \frac{1}{\sqrt{2\Delta}\rho} \Big( \rho^2 \sqrt{\Xi}, 
	- \Delta , 0 , \frac{2Mra}{\rho^2\sqrt{\Xi}} \Big) , 
	\hspace{.2cm}
	k^\mu = \frac{1}{\sqrt{2\Delta}\rho} \Big( \rho^2 \sqrt{\Xi}, 
	\Delta ,0 , \frac{2Mra}{\rho^2\sqrt{\Xi}} \Big) , 
	\hspace{.2cm}
	m^\mu = \frac{1}{\sqrt{2}\rho} \Big( 0 , 0 , -1 , \frac{ i \, \csc\theta}{\sqrt{\Xi}} \Big) .
	\end{eqnarray}
	For the above PNDs, all Weyl scalars ${\boldsymbol \Psi}_I$ do not vanish while we saw that only ${\boldsymbol \Psi}_2$ does not vanish when we used the PNDs~\eqref{null-vectors-Kerr}-\eqref{null-vectors-Kerr-m}. Of course, the final results for the Petrov type is independent of the choice of the basis. However, working with the PNDs~\eqref{null-vectors-Kerr} is much easier in practice. Especially for our purpose in this paper, as we will show, working in basis~\eqref{null-vectors-Kerr-old} we have to compute Weyl scalars up to quadratic order for the disformal parameter ${\cal O}(B_0^2)$ while we only need to perform the analysis up to linear order ${\cal O}(B_0)$ in PNDs~\eqref{null-vectors-Kerr}. This extremely simplifies our calculations.}

\subsubsection{Disformed kinematics}

For the disformal vector field, we consider the same configuration as the static spherically symmetric case such that the co-vector field $V_{\mu}\rd x^{\mu}$ has non-vanishing components only along the time and radial direction, namely
\begin{equation}\label{Kerr-V-mu}
V_\mu = \big( V_t, V_r, 0, 0 \big) \,, \hspace{1cm} 
V^\mu = \frac{\rho^2}{\Delta}
\left(-\Xi V_t , \frac{\Delta ^2}{\rho^4}  V_r, 0,-\frac{2 a M r}{\rho^4} V_t\right) 
\,,
\end{equation}
where $V_t:=V_t(t,r)$ and $V_r:=V_r(t,r)$ and we have introduced the notation
\begin{equation}
\Xi \equiv \frac{\Delta}{\rho^2} + \frac{2Mr}{\rho^2} 
\Big( \frac{\Delta}{\rho^2} + \frac{2Mr}{\rho^2}\Big) \,.
\end{equation}
The tetrad components of the vector field can be obtained by substituting Eqs.~\eqref{Kerr-V-mu} and~\eqref{null-vectors-Kerr}-\eqref{null-vectors-Kerr-m} in Eq.~\eqref{Y-tilde} as follows 
\begin{equation}\label{Kerr-V-a}
V_a = \left( \frac{r^2+a^2}{\Delta } V_t +V_r , 
\frac{\Delta}{2\rho^2} \Big( \frac{r^2+a^2}{\Delta} V_t - V_r \Big) , 
\frac{i a \sin\theta\, V_t}{\sqrt{2} (r+i a \cos\theta)} , 
-\frac{i a \sin\theta\, V_t}{\sqrt{2} (r-i a \cos\theta)} \right) \,,
\end{equation}
and $V^a$ can be obtained from the above result simply by using $V^a=\eta^{ab}V_b$. 
We see that $V_3=V^2\neq0$, and $V^3=V_2\neq0$ and, such that the disformal co-vector $V_\mu dx^{\mu}$ is not orthogonal to the 2-dimensional surface ${\cal S}$. Notice that this situation corresponds to the disformal transformation used in~\cite{BenAchour:2020fgy} to constructed the disformed Kerr black hole, where the scalar profile depends only on time and radius but not on the $\theta$ angle. See Eq~(3.14) in~\cite{BenAchour:2020fgy}.

Using the above result, the disformed Kerr metric takes the compact form
\begin{eqnarray}\label{ds2-Kerr-dis}
{\widetilde ds}^2_{\rm{Kerr}} = {ds}^2_{\rm{Kerr}} + B_0 \big( V_t dt + V_r dr \big)^2  \,,
\end{eqnarray}
and the disformed PNDs read
\begin{eqnarray}\label{null-vectors-Kerr-dis}
&&{\tilde l}^\mu = l^\mu - \frac{B_0 }{2 }
\left( \frac{r^2+ a^2 }{\Delta} V_t+V_r\right) V^\mu \,, \hspace{1cm} \\
&&{\tilde k}^\mu = k^\mu - \frac{B_0}{4} 
\left( 
\frac{r^2+ a^2 }{\Delta} V_t-V_r\right) \frac{\Delta}{\rho^2} V^\mu
\,,\label{null-vectors-Kerr-dis-k}\\
&&{\tilde m}^\mu = m^\mu - \frac{B_0 }{2\sqrt{2}}
\frac{i a \sin\theta\, V_t }{r+i a \cos\theta} V^\mu
\,,\label{null-vectors-Kerr-dis-m}
\end{eqnarray}
where the explicit forms of undisformed PNDs and disformal vector field are given by Eqs.~\eqref{null-vectors-Kerr}-\eqref{null-vectors-Kerr-m} and~\eqref{Kerr-V-mu}. Note that, contrary to the case of static spherically symmetric case presented in~\eqref{null-vectors-SS-dis} where i) $V\in {T}_p^\star({\cal M}/{\cal S})$ and ii) the PNDs ${\tilde m}^\mu$ and $\tilde{\bar m}^\mu$ did not change, in this axisymmetric case, $V\notin {T}_p^\star({\cal M}/{\cal S})$ and all PNDs change after performing disformal transformation.

In the previous example of static spherically symmetric solution, we have computed all nonzero scalars in Eqs.~\eqref{Dis-Scalars-SS} and~\eqref{Psi-SS}. Here, we only compute those scalars that we need, illustrating if needed the efficiency of our method. We first notice that working in the null basis~\eqref{null-vectors-Kerr}-\eqref{null-vectors-Kerr-m}, all ${\boldsymbol \Psi}_I$ vanish except ${\boldsymbol \Psi}_2$.
Therefore, to first order in $B_0$, we have
\begin{align}
\tilde{\boldsymbol \Psi}_2 &= {\cal O}(B_0^0)+{\cal O}(B_0)  \\
\tilde{\boldsymbol \Psi}_I & = {\cal O}(B_0) \;, \qquad \qquad  \text{for $I\neq2$}
\end{align}
Using this result, one can show that the first non-zero contribution to ${\tilde D}$ shows up at the second order ${\cal O}(B_0^2)$ as 
\begin{equation}\label{Dtilde}
\tilde{ D} = 81 {\boldsymbol \Psi}_2^2 {\boldsymbol \Delta}_0 {\boldsymbol \Delta}_4 \,,
\end{equation}
which demonstrates that ${\tilde D}\neq0$ for ${\boldsymbol \Delta}_0\neq0$ and ${\boldsymbol \Delta}_4\neq0$. 

Before concluding, we point that since ${\tilde D}={\cal O}(B_0^2)$ for ${\boldsymbol \Delta}_I={\cal O}(B_0)$, a legitimate question would be whether one can trust this analysis up to linear order ${\cal O}(B_0)$ for $\tilde{\boldsymbol \Psi}_I$? Interestingly, it turns out that one can easily show that the result~\eqref{Dtilde} holds in the null basis~\eqref{null-vectors-Kerr}-\eqref{null-vectors-Kerr-m} even if one computes all $\tilde{\boldsymbol \Psi}_I$ up to the quadratic order ${\cal O}(B_0^2)$. Had we worked with another null basis, for instance the null basis~\eqref{null-vectors-Kerr-old}, we would had to keep calculations for $\tilde{\boldsymbol \Psi}_I$ up to the second order ${\cal O}(B_0^2)$ which would make the calculations much more involved. Thus, working with the basis~\eqref{null-vectors-Kerr}-\eqref{null-vectors-Kerr-m} allows us to only compute ${\boldsymbol \Delta}_0$  and ${\boldsymbol \Delta}_4$ up to the linear order ${\cal O}(B_0)$. The explicit expressions are complicated and we only present the results for $\theta=\pi/2$ which are given by
\begin{eqnarray}\label{kerr-Delta0}
{\boldsymbol \Delta}_0 &=& \frac{B_0 a^2}{2 r^4 } \Bigg\{
V_r^2 + \frac{4 M^2 r^2}{\Delta^2} V_t^2
+ r^2 \left(\dot{V}_r^2+ V_t'{}^2-\dot{V}_r V_t'-\dot{V}_t V_r'\right) \nonumber
\\ \nonumber
&+& \left( r \big( \ddot{V}_r - \dot{V}_t' \big)
+2 \frac{a^2+r^2}{\Delta} \dot{V}_t
+ 2 \dot{V}_r + \frac{4 M}{\Delta} V_t \right) r V_r 
+ 2 \frac{a^2+r^2}{\Delta} r \dot{V}_r V_t
\\
&+&\Big( r \big(V_t''-\dot{V}_r\big)-2 V_t' \Big) r V_t
- \frac{a^2+3 r^2}{\Delta} \bigg( \dot{V}_t
- \frac{a^2+r^2}{\Delta} \dot{V}_t \bigg) r V_t \Bigg\}
\,, \\ \label{kerr-Delta4}
{\boldsymbol \Delta}_4 &=& \frac{\Delta^2}{4r^4} \left\{ {\boldsymbol \Delta}_0 
-\frac{2 B_0 a^2}{r^4}
\left(\frac{M r}{\Delta} \frac{a^2+3 r^2}{\Delta} V_t \dot{V}_t
+ \left(\frac{2 M}{\Delta} V_t+  r \dot{V}_r\right) V_r \right) \right\} \,.
\end{eqnarray}
From Table \ref{tab1}, the above results show that the Petrov type of the stealth Kerr solution~\eqref{ds2-Kerr} changes from type D to type I. This example explicitly confirms that not only PNDs but also Petrov type of a spacetime can change after performing a disformal transformation. As a result, the non-stealth Kerr-like black hole solution derived in~\cite{BenAchour:2020fgy, Anson:2020trg} for DHOST theories is found to be algebraically general of type I.

The fact that the Petrov type of black hole solutions changes in the context of modified gravity theories was already noticed in different contexts, such as dynamical Chern-Simons and Gauss-Bonnet gravity~\cite{Yagi:2012ya,Owen:2021eez}. While the result we have obtained for the stealth Kerr parallels these results (where the slowly rotating black holes solutions analyzed there go from type D to type I too), our analysis turns out to be conceptually more subtle. Indeed, because disformal transformation reduces to field redefinition in the absence of matter fields coupled to the gravitational sector, it might seem awkward to find different Petrov type for the stealth Kerr and its disformed version. However, gravity is generated and is probed by the matter fields. Therefore, the key point is that determining the Petrov type of a given solution requires the introduction of a frame in which matter fields (including the standard model of particle physics) are minimally coupled to the metric, if one would like the classification based on the Petrov type to be physically relevant. In this frame free-falling observers follow geodesics and electromagnetic waves propagate along lightcones, while in other frames they do not in general. In this situation the Petrov classification can characterize useful physical properties of the solution measured by such observers and electromagnetic waves if and only if the classification is associated with the Weyl scalars computed in the frame where matter fields are minimally coupled to the metric. Thus, when we deal with the disformal transformation as a field redefinition which includes metric, we have to determine to which frame the matter fields are minimally coupled. Depending on this choice, gravity shows different properties. As we emphasized in the first section, the Petrov classification is performed by introducing a null tetrad which corresponds to the frame of an implicit observer moving on the underlying geometry. This step can be understood as introducing a disguised test matter field and specifying its coupling to the metric. For this reason,  as a gravitational properties of the system, the Petrov type is not invariant under disformal transformation and the analysis of the stealth Kerr metric and its disformed version illustrates this fact.

\def\dd{\mathrm{d}}
\def\ic{\vartheta}
\def\qq{\qquad}

\newcommand{\cI}{{\mathcal I}}
\newcommand{\id}{\mathbb{I}}

\chapter{Radiative spacetimes and disformal gravitational waves}
\label{Chapter7}

\textit{"The lowest kind of symmetry which we can associated with gravitational waves is that of a quadrupole. However, the significance of this result is substantially reduced by the non-linearity of the equations. [...] Perhaps the most important character of gravitational waves concerns just this. For a wave to be a wave in any real physical sense, it must convey energy: accordingly, an outgoing wave mush diminish the energy of the source and, therefore, its mass." \\
H. Bondi, M.G.J. van der Burg and A. W. K. Metzner, 1962}

\minitoc

In the previous chapter, we have seen that contrary to conformal transformations, disformal transformations can change the principal null directions of a spacetime geometry. Thus, depending on the frame a gravitational wave (GW) detector minimally couples to, the properties of GWs may change under a disformal transformation. In this chapter, based on \cite{BenAchour:2024tqt, BenAchour:2024zzk}, we provide \textit{necessary} and \textit{sufficient} conditions which determine whether GWs change under disformal transformations or not. Our argument is coordinate-independent and can be applied to any spacetime geometry at the fully non-linear level. As an example, we show that an exact radiative solution of massless Einstein-scalar gravity which admits only shear-free parallel transported frame is mapped to a disformed geometry which does not possess any shear-free parallel transported frame. This radiative geometry and its disformed counterpart provide a concrete example of the possibility to generate tensorial GWs from a disformal transformation at the fully non-linear level. 
This result shows that, at the nonlinear level, the scalar-tensor mixing descending from the higher-order terms in Horndeski dynamics can generate shear out of a pure scalar monopole. We further confirm this analysis by identifying the spin-0 and spin-2 polarizations in the disformed solution using the Penrose limit of our radiative solution. Finally, we compute the geodesic motion and the memory effects experienced by two null test particles with vanishing initial relative velocity after the passage of the pulse. 
This exact radiative solution offers a simple framework to witness nonlinear consequences of the scalar-tensor mixing in higher-order scalar-tensor theories.

\section{Spin coefficients formalism}

Consider a spacetime geometry with metric $g_{\mu\nu}$. To analyze the effects of a general DT on this geometry, and in particular on the properties measured by an idealized GW detector, it is convenient to use the local analysis based on the properties of geodesic null congruence (e.g a bundle of light rays) \cite{BenAchour:2021pla}. We thus introduce a set of tetrads $\theta^{\mu}{}_a$ at any point of the spacetime which allows us to project the metric $g_{\mu\nu}$ into the local rest frame of the observer as
\begin{align}\label{g-Tetrads}
&{ g}_{\mu\nu} = \eta_{ab}\,{\theta }^a{}_{\mu} { \theta}^b{}_{\nu} \,,
&&
\eta_{ab} = { g}_{\mu\nu}{ \theta}^{\mu}{}_a \, { \theta}^{\nu}{}_b \,,
\end{align}
where $a,b=0,1,2,3$ are the Lorentz indices and $\eta_{ab}$ is the Minkowski metric. ${\theta }^a{}_{\mu}$ is the inverse of $\theta^{\mu}{}_a$ such that
\begin{align}\label{Tetrads-complete}
&{ \theta}^a{}_{\mu} {\theta}^{\mu}{}_b = \delta^a{}_b \,, &&
{ \theta}^{\mu}{}_a {\tilde \theta}^a{}_{\nu} = \delta^\mu{}_\nu \,.
\end{align}
In the so-called Newman-Penrose formalism, we set a null tetrad basis which are given in term of four complex null vectors
\begin{align}\label{Tetrads}
&{ \theta}^a{}_{\mu} = \big( - { n}_\mu, - { \ell}_\mu, {\bar m}_\mu, { m}_\mu \big) \,, 
&&
{ \theta}^{\mu}{}_a = \big( { \ell}^\mu, { n}^\mu, {m}^\mu, {\bar m}^\mu \big) \,,
\end{align}
where ${ m}^\mu$ and ${ {\bar m}}^\mu$ are complex conjugate to each other, and the Minkowski metric takes the following null form
\begin{eqnarray}\label{eta}
\eta_{ab} = \eta^{ab} \doteq \begin{pmatrix}
0 & -1 & 0 & 0 \\
-1 & 0 & 0 & 0 \\
0 & 0 & 0 & 1 \\
0 & 0 & 1 & 0 
\end{pmatrix} \,.
\end{eqnarray}
 In this null basis, the metric $g_{\mu\nu}$ can be expressed as
\begin{equation}\label{g}
{ g}_{\mu\nu} = - { \ell}_\mu { n}_\nu - { n}_\mu { \ell}_\nu
+ { m}_\mu {\bar { m}}_\nu + {\bar { m}}_\mu { m}_\nu \,,
\end{equation}
where the null vectors satisfy the normalization and orthogonality conditions
\begin{align}\label{null}
&{ g}_{\mu\nu}{ \ell}^\mu{ n}^\nu = -1 \,,
&&{ g}_{\mu\nu}{ m}^\mu{\bar{ m}}^\nu = 1 \,,
\end{align}
and all other contractions between null vectors vanish.

The twelve complex Newman-Penrose spin coefficients for the null tetrad $\theta^\mu{}_a=(\ell^\mu, n^\mu, m^\mu, \bar{m}^\mu)$ are given by \cite{Chandrasekhar:1985kt}
\begin{align}\nonumber
&\kappa \equiv -m^{\alpha}D\ell_{\alpha} \,,
&&
\epsilon \equiv \frac{1}{2}(\bar{m}^{\alpha} Dm_{\alpha} - n^{\alpha}D\ell_{\alpha}) \,,
&&
\pi \equiv \bar{m}^{\alpha}Dn_{\alpha} \,,
\\ \nonumber
&\sigma \equiv -m^{\alpha}\delta\ell_{\alpha} \,,
&&
\beta \equiv \frac{1}{2}(\bar{m}^{\alpha}\delta{m}_{\alpha}
- n^{\alpha}\delta\ell_{\alpha}) \,,
&&
\mu \equiv \bar{m}^{\alpha}\delta {n}_{\alpha} \,,
\\ \nonumber
&\rho \equiv -m^{\alpha}\bar{\delta}\ell_{\alpha} \,,
&&
\alpha \equiv \frac{1}{2}(\bar{m}^{\alpha}\bar{\delta}m_{\alpha} - n^{\alpha}\bar{\delta}\ell_{\alpha}) \,,
&&
\lambda \equiv \bar{m}^{\alpha}\bar{\delta}n_{\alpha} \,,
\\
&\tau \equiv -m^{\alpha}\Delta\ell_{\alpha} \,,
&&
\gamma \equiv \frac{1}{2}(\bar{m}^{\alpha} \Delta m_{\alpha}
-n^{\alpha}\Delta\ell_{\alpha}) \,,
&&
\nu \equiv \bar{m}^{\alpha}\Delta n_{\alpha} \,, \nonumber
\end{align}
where the directional derivatives along the null vectors are defined as
\begin{align}\nonumber
&D = \ell^\alpha \nabla_\alpha \,,
&&
\Delta = n^\alpha \nabla_\alpha \,,
&&
\delta = m^\alpha \nabla_\alpha \,,
&&
\bar{\delta} = \bar{m}^\alpha \nabla_\alpha \,.
\end{align}
The usual kinematical quantities like the expansion $\Theta$, shear $\sigma_s$, and twist (vorticity) $\omega$ can be deduced from the spin coefficients. For the null vector $\ell^\mu$, they are defined as
\begin{align}
&\Theta \equiv - {\rm Re}\left[\rho\right] \,, 
&&\sigma_s \equiv |\sigma| \,,
&&\omega \equiv - {\rm Im}\left[\rho\right] \,.
\end{align}
We are now ready to discuss the generation of tensorial gravitational wave from a disformal transformation at the fully non-linear level.

\section{Disformal gravitational waves}

In this section, we describe how a disformal transformation impacts the properties of a parallel transported frame associated to a freely falling photon. This exercise shows that depending on the seed solution, one can indeed generate tensorial gravitational wave from a disformal transformation at the fully non-linear level. This section provides a general coordinate-free argument which can be applied to any radiative geometry. The next part of the chapter is devoted to the analysis of a concrete example illustrating the generation of disformal gravitational waves. 
This section is based in \cite{BenAchour:2024tqt}.

\subsection{Parallel transport}

We consider $\ell$ to be geodesic, i.e.
\begin{align}\label{geodesic-l}
D\ell^{\mu} = 0 \,,
\end{align}
where $D \equiv \ell^{\alpha} \nabla_{\alpha}$ is the covariant derivative along $\ell$. The properties of the null congruence are characterized by the twelve complex Newman-Penrose spin coefficients
$\kappa, \epsilon, \pi, \alpha, \beta, \rho, \sigma, \lambda, \nu, \tau, \mu, \gamma$ defined above. They describe how the bundle of light rays expand, shear and twist along each of the null directions ${ \theta}^{\mu}{}_a = \big( { \ell}^\mu, { n}^\mu, {m}^\mu, {\bar m}^\mu \big)$. The vector $\ell$ being geodesic \eqref{geodesic-l}, a PTF can be constructed by further demanding that
\begin{align}\label{PTF}
&D n^\mu = 0 \,, 
&&D m^\mu = 0 \,, 
&&D \bar{m}^\mu = 0 \,,
\end{align}
which in terms of the spin coefficients takes the form
\begin{align}\label{PT-conditions}
&\kappa = -m^{\alpha}D\ell_{\alpha} = 0 \,, 
&& \epsilon = \frac{1}{2}(\bar{m}^{\alpha} Dm_{\alpha} - n^{\alpha}D\ell_{\alpha}) = 0 \,, 
&&\pi = \bar{m}^{\alpha}Dn_{\alpha} = 0 \,.
\end{align}
The tetrads are defined up to Lorentz transformations and the  six degrees of freedom in the Lorentz transformations make it possible to always achieve the above six conditions.

With the PTF \eqref{PT-conditions} at hand, one effectively realizes a set of null Fermi coordinates \cite{Guedens:2012sz}. The remaining spin coefficients associated to this PTF provide the physical quantities characterizing the bundle of light rays that a freely falling observer will measure.

\subsection{Disformal map,  Lorentz transformations and matter coupling}

Now, let us investigate how null tetrads change under a DT
\begin{align}\label{Tetrads-dis}
{ \theta}^a{}_{\mu} = \big( -n_\mu , - { \ell}_\mu, {\bar m}^\mu, { m}^\mu \big)
&\xrightarrow{\mbox{DT}} 
{\tilde \theta}^a{}_{\mu}
= \big(- {\tilde n}_\mu,- {\tilde \ell}_\mu, \tilde{{\bar m}}^\mu, \tilde{ m}^\mu \big) \,.
\end{align}
For the sake of simplicity, we restrict our analysis to a pure constant DT with 
\begin{align}\label{DT}
&C(\phi, X) =1 \,,
&&B(\phi, X) = B_0 \,,
\end{align}
where $B_0$ is a constant. The equivalence principle guaranties that there should exist a set of disformed null tetrads ${\tilde \theta}^{\mu}{}_a$ such that
\begin{align}\label{g-T-Tetrads}
&{\tilde g}_{\mu\nu} = \eta_{ab}\,{\tilde \theta }^a{}_{\mu} {\tilde \theta}^b{}_{\nu} \,,
&&
\eta_{ab} = {\tilde g}_{\mu\nu}{\tilde \theta}^{\mu}{}_a \, {\tilde \theta}^{\nu}{}_b \,.
\end{align}
It can be easily shown that under the DT~\eqref{DT}, the relation between the disformed tetrads and original tetrads is given by  \cite{BenAchour:2021pla}
\begin{align}\label{Tetrads-disform}
{ \theta}^a{}_\mu&\xrightarrow{\mbox{DT}} {\tilde \theta}^a{}_\mu
= J^a{}_b \, { \theta}^b{}_\mu \,,
\end{align}
where
\begin{align}\label{J-def}
&J^a{}_b = \delta^a{}_b + {\cal B}\, \phi^a \phi_b \,;
&&
{\cal B}(X) \equiv \frac{1}{X}\big[\sqrt{1+B_0 X}-1\big] \,,
\end{align}
in which 
\begin{align}
&\phi_a \equiv \phi_{\alpha} \theta^\alpha{}_a \,,
&&\phi^a \equiv g^{\alpha\beta}\phi_{\alpha} \theta^a{}_\beta \,,
\end{align}
such that $\phi_\mu = \phi_a \theta^a{}_\mu$ and $\phi^\mu = \phi^a \theta^\mu{}_a$. Substituting \eqref{J-def} in \eqref{Tetrads-disform} we find
\begin{align}\label{Tetrads-disform-explicit}
{\tilde \theta}^a{}_\mu
= { \theta}^a{}_\mu + {\cal B}\, \phi^a \phi_{\mu} \,.
\end{align}
The vector $\phi_\mu$ can be expressed in terms of the null basis as
\begin{align}
\phi_\mu = - \phi_n \ell_\mu - \phi_\ell n_\mu 
+ \phi_{\bar m} m_\mu + \phi_m {\bar m}_\mu \,.
\end{align}
Substituting this in \eqref{Tetrads-disform-explicit}, we find the explicit expression of the disformed null vectors in terms of the seed null vectors.

Equipped with the disformed null vectors, it is straightforward to compute the disformed spin coefficients. In particular, we find
\begin{align}\label{PTF-dis}
\begin{split}
\kappa &\xrightarrow{\mbox{DT}} \tilde{\kappa} = \kappa + \kappa_{\rm DT} \,,
\\
\epsilon &\xrightarrow{\mbox{DT}} \tilde{\epsilon} = \epsilon + \epsilon_{\rm DT} \,,  
\\
\pi &\xrightarrow{\mbox{DT}} \tilde{\pi} = \pi + \pi_{\rm DT} \,,
\end{split} 
\end{align}
such that $\kappa_{\rm DT}=\epsilon_{\rm DT}=\pi_{\rm DT}=0$ for $B_0=0$. To first order  in $B_0$, one has
\begin{align}\nonumber
\kappa_{\rm DT} 
&= \tilde{\kappa} - \kappa =  -\tilde{m}^{\mu}\tilde{D}\tilde{\ell}_{\mu} + {m}^{\mu}{D}{\ell}_{\mu} = B_0 \kappa_{\rm DT}^{(1)} + \cdots \,, 
\\
\nonumber
\epsilon_{\rm DT} &= \tilde{\epsilon} - \epsilon 
= \frac{1}{2} (\bar{\tilde{m}}^{\mu} \tilde{D}\tilde{m}_{\mu} - \tilde{n}^{\mu}\tilde{D}\tilde{\ell}_{\mu}) 
- \frac{1}{2} (\bar{{m}}^{\mu} {D}{m}_{\mu} - {n}^{\mu}{D}{\ell}_{\mu}) 
= \epsilon + B_0 \epsilon_{\rm DT}^{(1)} + \cdots \,, \hspace{-.1cm}
\\ \nonumber
\pi_{\rm DT} &= \tilde{\pi} - \pi 
= \bar{\tilde{m}}^{\mu}\tilde{D}\tilde{n}_{\mu} 
- \bar{{m}}^{\mu}{D}{n}_{\mu} 
= \pi + B_0 \pi_{\rm DT}^{(1)} + \cdots \,,
\end{align}
where
\begin{align}\nonumber
\kappa_{\rm DT}^{(1)} &\equiv 
\kappa \phi _n \phi _\ell 
+\frac{1}{2} (2\bar{\epsilon } -\rho) \phi _m \phi _\ell 
+\frac{1}{2} \left(\tau -\bar{\pi }\right) \phi_\ell^2
- {\rm Re}\left[\bar{\kappa } \phi_m \right] \phi_m 
+ \frac{1}{2} [\phi_\ell,\phi_m]_D \,,
\\
\epsilon_{\rm DT}^{(1)} &\equiv
\frac{1}{2} \big[ \left(\bar{\epsilon }+2 \epsilon \right) \phi _n
-\alpha  \phi_m -\left(\bar{\pi }+\beta \right) \phi_{\bar{m}} \big] \phi _\ell 
- \frac{1}{2}\bar{\kappa }  \phi _m\phi _n
\nonumber \\ \nonumber
&+{\rm Re}\left[\gamma\right] \phi _\ell^2 
- i {\rm Im}\left[\epsilon\right] |\phi_{m}|^2 
+ \frac{1}{4} \big( [\phi_\ell,\phi_n]_D
- [\phi_m,\phi_{\bar{m}}]_D \big)
\,,
\\ \nonumber
\pi_{\rm DT}^{(1)} &\equiv 
-\frac{1}{2} \left(\mu  \phi_{\bar{m}}+\lambda  \phi _m-2 \pi  \phi_n\right) \phi _\ell
+  \bar{\epsilon } \phi _n \phi _{\bar{m}} 
-\frac{1}{2} \bar{\kappa }
\phi _n^2 
+\frac{1}{2} \nu  \phi _\ell^2
- {\rm Re}\left[\pi  \phi _m\right] 
- \frac{1}{2} [\phi_n,\phi_{\bar{m}}]_D
\,, \hspace{2.7cm}
\end{align}
and
\begin{align}
[\phi_\ell,\phi_m]_D\equiv\phi_\ell D\phi_m - \phi_m D \phi_\ell \,.
\end{align}
In the last steps, for the matter of presentation, we have expanded the results for $B_0\ll1$ such that $\cdots$ denotes the terms that are quadratic and higher in $B_0$. We emphasize that one can compute all $\kappa_{\rm DT}$, $\epsilon_{\rm DT}$, $\pi_{\rm DT}$ at the fully non-linear order in $B_0$.

We have shown that starting from PTF \eqref{PT-conditions} and performing the DT, we end up with \eqref{PT-conditions-DT} which is not PTF. We, however, still have the freedom to perform LTs and bring \eqref{PT-conditions-DT} to PTF. We thus perform LTs on the disformed null vectors $\theta^\mu{}_a=(\tilde{\ell}^\mu, \tilde{n}^\mu, \tilde{m}^\mu, \bar{\tilde{m}}^\mu)$. The general LTs of the null tetrad basis are given by \cite{Chandrasekhar:1985kt}
\begin{align}\label{LTs}
\begin{split}
\tilde{\ell}^\mu &\xrightarrow{\mbox{LT}} \tilde{\ell}'^\mu = \frac{1}{A} \tilde{\ell}^\mu + A |{\rm b}|^2 \tilde{n}^\mu 
+ 2 {\rm Re}\big[ e^{i \theta } \bar{\rm b} \tilde{m}^\mu \big]  
\,, 
\\
\tilde{n}^\mu &\xrightarrow{\mbox{LT}} \tilde{n}'^\mu 
= \frac{1}{A} |{\rm a}|^2 \tilde{\ell}^\mu
+ A |1+{\rm a}\bar{\rm b}|^2 \tilde{n}^\mu + 2 {\rm Re}\big[ e^{i \theta } \bar{\rm a} \left( 1 + {\rm a} \bar{\rm b} \right) \tilde{m}^\mu \big] 
\,, 
\\ 
\tilde{m}^\mu &\xrightarrow{\mbox{LT}} \tilde{m}'^\mu =  \frac{1}{A} {\rm a} \tilde{\ell}^\mu
+ A {\rm b} \left(1 + {\rm a} \bar{\rm b} \right) \tilde{n}^\mu + 2 {\rm a} {\rm Re}\big[ e^{i \theta } \bar{\rm b} \tilde{m}^\mu \big] + e^{i \theta } \tilde{m}^\mu \,, 
\end{split}
\end{align}
where functions ${\rm a}$ and ${\rm b}$ are free complex functions which characterize class I and II transformations while $A$ and $\theta$ are real functions that characterize class III transformation \cite{Chandrasekhar:1985kt}. These free functions correspond to the six free parameters of the local Lorentz group.

After performing general LTs on the disformed null basis $\theta^\mu{}_a=(\tilde{\ell}^\mu, \tilde{n}^\mu, \tilde{m}^\mu, \bar{\tilde{m}}^\mu)$, the disformed spin coefficients \eqref{PTF-dis} transform as 
\begin{align}\label{SC-LT}
\begin{split}
\kappa &\xrightarrow{\mbox{DT}} \tilde{\kappa} 
\xrightarrow{\mbox{LT}} \tilde{\kappa}' 
= \kappa + \kappa_{\rm DT} + \kappa_{\rm LT}(A,\theta,{\rm a},{\rm b})
\,, 
\\
\epsilon &\xrightarrow{\mbox{DT}} \tilde{\epsilon}
\xrightarrow{\mbox{LT}} \tilde{\epsilon}' 
= \epsilon + \epsilon_{\rm DT} + \epsilon_{\rm LT}(A,\theta,{\rm a},{\rm b})
\,,
\\
\pi &\xrightarrow{\mbox{DT}} \tilde{\pi}
\xrightarrow{\mbox{LT}} \tilde{\pi}' 
= \pi + \pi_{\rm DT} + \pi_{\rm LT}(A,\theta,{\rm a},{\rm b})
\,.
\end{split}
\end{align} 
where the explicit forms of $\kappa_{\rm LT}, \epsilon_{\rm LT}, \pi_{\rm LT}$ to first order in $B_0$ can be obtained as follows.
Considering parametrization 
\begin{align}\nonumber
&{\rm a} =B_0 {\rm a}^{(1)} + \cdots \,, 
&&{\rm b}=B_0 {\rm b}^{(1)}+ \cdots \,, 
&&A = 1 +B_0 A^{(1)}+ \cdots \,,
&&\theta=B_0\theta^{(1)}+ \cdots \,,
\end{align}
such that the LTs reduce to the unity map for $B_0=0$. Then \eqref{LTs} simplify to
\begin{align}\label{LTs-1stOrder}
\begin{split}
\tilde{\ell}'^\mu &= \tilde{\ell}^\mu - B_0 \big( A^{(1)} {\ell}^\mu - 2 {\rm Re}\big[{\rm b}^{(1)}\bar{m}^\mu\big] \big) + \cdots 
\,, 
\\
\tilde{n}'^\mu 
& = \tilde{n}^\mu + B_0 \big(A^{(1)} n^\mu
+ 2 {\rm Re}\big[{\rm a}^{(1)}\bar{m}^\mu\big]\big) + \cdots 
\,, 
\\ 
\tilde{m}'^\mu 
& = \tilde{m}^\mu + B_0 \big({\rm a}^{(1)} \ell^{\mu} + {\rm b}^{(1)} n^{\mu} + i \theta^{(1)} m^\mu\big) + \cdots 
\,.
\end{split}
\end{align} 
It is then straightforward to find
\begin{align}
\kappa_{\rm LT} &= \tilde{\kappa}' - \tilde{\kappa}
= B_0 \kappa_{\rm LT}^{(1)} + \cdots 
\,, 
\nonumber \\ \nonumber 
\epsilon_{\rm LT} &= \tilde{\epsilon} - \tilde{\epsilon} = B_0 \epsilon_{\rm LT}^{(1)} + \cdots  
\,, 
\\ \nonumber
\pi_{\rm LT} &= \tilde{\pi}' - \tilde{\pi} 
= B_0 \pi_{\rm LT}^{(1)} + \cdots 
\,, 
\end{align} 
where  
\begin{align*}\nonumber
\kappa_{\rm LT}^{(1)} &\equiv 
\big(i \theta^{(1)} -2 A^{(1)} \big) \kappa
+ {\rm b}^{(1)} ( \rho +2 \epsilon )
+ \bar{\rm b}^{(1)} \sigma
- Db^{(1)}
\,,
\\
\pi_{\rm LT}^{(1)} &\equiv
2 \bar{\rm a}^{(1)} \epsilon + \bar{\rm b}^{(1)}  \mu + {\rm b}^{(1)} \lambda - i \theta^{(1)}  \pi + D\bar{\rm a}^{(1)} 
\,,\\
\epsilon_{\rm LT}^{(1)} &\equiv \!\begin{aligned}[t]
&\bar{\rm a}^{(1)} \kappa + \bar{\rm b}^{(1)} \beta - A^{(1)} \epsilon + {\rm b}^{(1)} (\alpha+\pi) - \frac{1}{2} \left(D(A^{(1)} + i \theta^{(1)})\right) \,. \end{aligned}
\end{align*}

The results \eqref{PTF-dis} clearly show that, in general, PTF \eqref{PT-conditions} does not remain PTF after performing the DT since 
\begin{align}\label{PT-conditions-DT}
&\tilde{\kappa} \neq 0 \,, 
&&\tilde{\epsilon} \neq 0 \,, 
&&\tilde{\pi} \neq 0 \,,
\end{align} 
as  in general $\kappa_{\rm DT}\neq0$, $\epsilon_{\rm DT}\neq0$, $\pi_{\rm DT}\neq0$.
In particular, it shows that the DT induces a deviation w.r.t the geodesic motion (see Fig.~\ref{Fig}). Of course, one can compute them at the fully nonlinear level while they take complicated form to be presented here.

\subsection{Building the disformed parallel transported frame}

Let us now impose PTF conditions. With the Lorentz transformed null basis \eqref{SC-LT} at hand, we now construct the {\it disformed} PTF, imposing that
\begin{align}\label{PT-conditions-LT}
&\tilde{\kappa}' = 0 \,, 
&&\tilde{\epsilon}' = 0 \,, 
&&\tilde{\pi}' = 0 \,.
\end{align} 
Starting from the PTF seed \eqref{PT-conditions}, conditions \eqref{PT-conditions-LT} imply
\begin{align}\label{PT-conditions-LT-kappa-epsilon-pi}
\kappa_{\rm LT} &= - \kappa_{\rm DT} \,, &&\epsilon_{\rm LT}= - \epsilon_{\rm DT} \,, &&\pi_{\rm LT}= - \pi_{\rm DT} \,.
\end{align}
The six free functions in $a, b, A, \theta$ can be chosen to always satisfy the above conditions. The {\it necessary} condition to change the properties of GWs under the DT is that at least one of the free functions $a, b, A, \theta$ acquires a non-vanishing value after imposing \eqref{PT-conditions-LT}.
\begin{figure}
	\centering
	\includegraphics{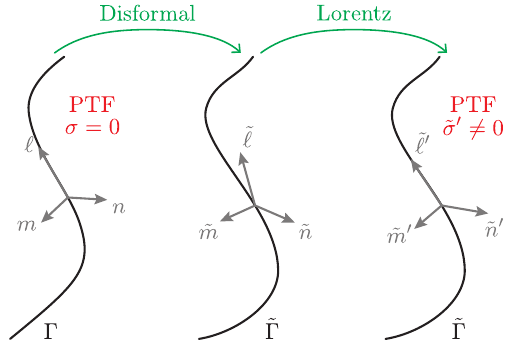}
	\caption{Transformation of a PTF of the seed under i) a disformal transformation and ii) a sequence of Lorentz transformations to build a disformed PTF. This procedure does not leave the spin coefficients invariant, e.g., shear-free PTF can turn into a PTF with non-vanishing shear.}\label{Fig}
\end{figure}

With the disformed PTF \eqref{PT-conditions-LT} at hand, one can compute the remaining disformed spin coefficients. Let us focus on the shear which transforms as
\begin{align} \label{sigma-DT-LT}
\sigma \xrightarrow{\mbox{DT}} \tilde{\sigma}
\xrightarrow{\mbox{LT}}
\tilde{\sigma}' 
& = \sigma + \sigma_{\rm DT} + \sigma_{\rm LT}
\,.
\end{align}
Having already used the six free functions in the LTs to construct the PTF~\eqref{PT-conditions-LT}, in general, one has $\sigma_{\rm DT} \neq - \sigma_{\rm LT}$ which implies that
\begin{equation}
\tilde{\sigma}' \neq \sigma \,.
\label{eq:cond-sigma-order}
\end{equation}
The same argument applies to the remaining eight spin coefficients such that, in general we have
\begin{align}\label{eq:cond-SC-order}
\tilde{\alpha}'\neq\alpha,
\quad 
\tilde{\beta}'\neq\beta,
\quad 
\tilde{\rho}'\neq\rho,
\quad
\tilde{\lambda}'\neq\lambda,
\quad 
\tilde{\nu}'\neq\nu, 
\quad
\tilde{\tau}'\neq\tau, 
\quad
\tilde{\mu}'\neq\mu,
\quad
 \tilde{\gamma}'\neq\gamma \,.
\end{align}
In this regard, the {\it sufficient} condition to change the properties of GWs under the DT is that at least one of the conditions in Eqs.~\eqref{eq:cond-sigma-order} and \eqref{eq:cond-SC-order} satisfies.

The key point we wish to emphasize here is that, in full generality, for a radiative seed, a DT will modify the properties of the GWs. In particular, even if the seed PTF is shear-free $\sigma=0$, one may find $\tilde{\sigma}'\neq0$ w.r.t the disformed PTF (see Fig.~\ref{Fig}). Thus, while a free falling (lightlike) observer w.r.t. the seed geometry can conclude on the absence of shear, i.e. $\sigma =0$, a free falling observer w.r.t. the disformed geometry can detect non-vanishing shear, i.e. $\tilde{\sigma}' \neq 0$. In the next section, we will explicitly confirm this fact by considering a concrete example.

\section{Robinson-Trautman solution with scalar hair}

In this section, we review the exact solution of the Einstein-Scalar massless system introduced in \cite{Tahamtan:2015sra} which describes a Robinson-Trautman geometry with a scalar hair.  This solution will serve as a seed to construct the non-perturbative exact solution in the Horndeski theory in the next sections. Therefore, we shall first analyze in detail the properties of this geometry and the nature of GWs it contains. Doing so, we shall confirm and improve in several ways the analysis presented initially in \cite{Tahamtan:2015sra, Tahamtan:2016fur}. See also \cite{Podolsky:2016sff} for the different Petrov types of this solution depending on the range of its parameters.
	
	Consider the Einstein-Scalar system with the action
	\be
	\label{EinS}
	S[g_{\mu\nu}, \phi] = \frac{1}{2} \int \dd^4{x} \sqrt{|g|} \left( \cR - g^{\mu\nu}\partial_{\mu} \phi \partial_{\nu} \phi \right) \,,
	\ee
	where we work in the unit $M_{\rm Pl}=(8\pi{G})^{-1/2}=1$ and metric signature $(-,+,+,+)$. Taking variation of the action w.r.t. the metric and the scalar field, the field equations take the standard form
	\begin{align}
		\cR_{\mu\nu} - \frac{1}{2} g_{\mu\nu} \cR & = \cT_{\mu\nu} \;, \\
		\label{EoM-SF}
		\Box \phi & =0 \,,
	\end{align}
	where the energy-momentum tensor is given by
	\be\label{EMT-SF}
	\cT_{\mu\nu} = \phi_{\mu} \phi_{\nu} - \frac{1}{2} g_{\mu\nu} \left(g^{\alpha\beta} \phi_{\alpha} \phi_{\beta}\right)  \,,
	\ee
	with $\phi_\mu \equiv \nabla_\mu\phi$.
	As shown in \cite{Tahamtan:2015sra}, the above field equations admit the following radiative exact solution given by
	\begin{align}\label{BG-RT}
		\rd s^2 & =  - \frac{ r \partial_u U + K(x,y) }{U(u)}  \rd u^2 - 2 \rd u \rd r + \frac{r^2 U^2(u)- C^2_0}{U(u)P^2(x,y)} (\rd x^2 + \rd y^2)  \,,\\
		\label{BG-SF}
		\phi (u, r) & = \frac{1}{\sqrt{2}} \log{\left[ \frac{r U(u) - C_0}{r U(u)+ C_0}\right]} \,;
		\hspace{1.5cm}
		U(u) = \gamma e^{\ic^2 u^2 + \eta u} \,,
	\end{align}
	where $(\ic, \eta, C_0, \gamma)$ are constants of integrations and $P(x,y)$ and $K(x,y)$ should satisfy
	\begin{align}
	\label{eq:ang-K}
	K(x,y) = \Delta \log{P}(x,y) \,, \qquad \text{and} \qquad \Delta K(x,y)  = 4 C^2_0 \ic^2 \equiv \alpha^2 \,,
	\end{align}
	with
	\be
	\Delta \equiv P^2 (\partial^2_{xx} + \partial^2_{yy}) \,.
	\ee
	At this level, the solution is parametrized by the set of four independent constants $(\ic, \eta, C_0, \gamma)$. The scalar charge is encodes in $C_0$, i.e. when $C_0=0$, the scalar field vanishes.\footnote{More precisely, the equation of motion for the scalar field \eqref{EoM-SF} can be written as $\nabla_\mu J^\mu = 0$, where $J^\mu\equiv\phi^\mu$ is the Noether current associated with the shift-symmetry. It can be easily seen that $J^\mu=0$ for $C_0=0$.} The parameter $\eta$ only affects the position of the maximum of the function $U(u)$ and it can be reabsorb by a suitable rescaling of the $u$-coordinate. Therefore, without loss of generality, we set $\eta=0$. For now on, we shall consider the case where $\gamma>0$, $\ic >0$ and $C_0 > 0$.\footnote{Note that we have excluded the case $C_0=0$. In this case, the energy-momentum tensor \eqref{EMT-SF} vanishes for the background configurations \eqref{BG-RT} and \eqref{BG-SF} and one finds the stealth Robinson-Trautman geometry: the GR Robinson-Trautman solution with a non-propagating scalar field. Stealth solutions are usually strongly coupled and we exclude this case from our consideration. It is worth mentioning that the higher-order operators may naturally resolve this issue in the context of the so-called Scordatura mechanism \cite{Motohashi:2019ymr, Gorji:2020bfl,Gorji:2021isn,DeFelice:2022xvq}.} The range of the coordinates will be discussed in the next subsection when investigating the singularities.

	It is convenient to switch from the coordinates $(u,r, x,y)$ to a new set of coordinates $(w, \rho, x,y)$ given by
	\begin{align}
		\dd{w} = \frac{\rd u}{\sqrt{U}}\,, \qquad \text{and} \qquad \rho = r \sqrt{U}  \,.
	\end{align}
	The first relation cannot be integrated into a simple analytic form. However, if one uses the error function $\text{erf}$ defined by $\text{erf}(x) = (2/\sqrt{\pi}) \int_0^{x} \exp[-t^2] \dd{t}$ and its inverse $\text{erf}^{-1}$, then one has
	\begin{equation}
		w = w_0 \;\text{erf}\left(\frac{u \ic}{\sqrt{2}}\right) \,,
	\end{equation}
	which implies
	\be
	u = \frac{\sqrt{2}}{\ic} \text{erf}^{-1}\left(\frac{w}{w_0}\right) \,, \qquad \text{and} \qquad  w_0 = \frac{\sqrt{\pi}}{\ic\sqrt{2\gamma}} \,.
	\ee
	In these coordinates, null past infinity is cast to $w = -w_0$ while null future infinity is cast to $w = +w_0$. 
	Furthermore, the line element and the scalar profile take the simpler form
	\begin{align}
		\label{met}
		\rd s^2 & = - K(x,y) \rd w^2 - 2 \rd w \rd \rho + \frac{\rho^2 - \chi^2(w)}{P^2(x,y)}(\rd x^2 + \rd y^2) \,, \\
		\label{scalprof}
		\phi(w,\rho) & = \frac{1}{\sqrt{2}} \log{\left[ \frac{\rho - \chi(w)}{\rho+ \chi(w)}\right]} \,;
		\hspace{1cm} \chi (w)= \frac{C_0}{\sqrt{U(w)}} \,.
	\end{align}
	The function $\chi(w)$ is depicted in Fig.~\ref{fig:chi}. Explicitly, one has $\chi(w\rightarrow \pm w_0) \rightarrow0$ and $\chi(0) =C_0/\sqrt{\gamma}$ such that $\chi$ remains a bounded function taking values in the range $[0, C_0/\sqrt{\gamma}]$. Finally, the kinetic energy of the scalar field reads
	\begin{align}
		\label{X}
		X \equiv g^{\alpha\beta} \phi_{\alpha} \phi_{\beta}   =  \frac{2\chi}{(\rho^2-\chi^2)^2} \left( \chi K(x,y) +2 \rho \chi'\right) \,,
	\end{align}
	where a prime denotes a derivative w.r.t. the $w$-coordinate.
	We stress that a second branch of solution can be obtained where the function $\chi(w)$ diverges in the regimes $w\rightarrow \pm w_0$ and vanishes at $w=0$. However, we discard this branch because one can show that the kinetic energy of the scalar field diverges in this case, leading to a pathological behavior. 
	\begin{figure}[!htb]
		\centering
		\includegraphics{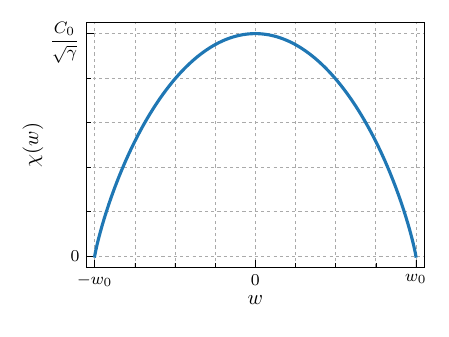}
		\caption{The function $\chi(w)$ is defined in the interval $[-w_0, w_0]$ and takes values in the interval $[0, C_0/\sqrt{\gamma}]$.}
		\label{fig:chi}
	\end{figure}
\subsection{Behavior of the solution}
	
	The effect of the scalar field can be understood by looking at the equation (\ref{eq:ang-K}) satisfied by $K(x,y)$. For $\alpha=0$, equation (\ref{eq:ang-K}) has the solution 
	\be
	\label{P0}
	P_0 = \frac{1 + x^2 + y^2}{2} \,, \qquad K_0=1 \,.
	\ee
	which is nothing but the geometry of a two-sphere in Cartesian coordinates. Thus, the deviation from the spherical symmetry is controlled by the parameter $\alpha$. In our setup, the spherical symmetric solution $\alpha = 2\ic C_0 = 0$ can be achieved in two different ways. The first possibility is when $\ic =0$ but $C_0 \neq 0$ such that the geometry reduces to a spherically symmetric time-dependent geometry with a scalar hair. The second possibility is $C_0=0$ when the geometry reduces to Minkowski spacetime $\lim_{C_0 \to 0} \rd s^2 = - \rd w^2 + 2 \rd w \rd u + \rho^2 \rd \Omega^2$ in agreement with the Birkoff's theorem. In particular, this underlies the fact that there is no mass contribution in this solution, and that the only source of curvature is the scalar monopole. However, as we have already mentioned, the case $C_0=0$ provides a stealth solution which can be problematic. We thus focus on the case $\alpha\neq0$ with $\ic >0$ and $C_0 >0$. In this case, the two-dimensional space $\cB$ spanned by $(\partial_x,\partial_y)$ has a non-homogeneous time-dependent two-dimensional curvature. As we shall see, it corresponds to a GW pulse induced by the scalar field which we shall analyze in the following. 
	
	A first analysis of the singularities and the apparent horizons with $\alpha\neq 0$ has been presented in \cite{Tahamtan:2015sra}. However, most of the expressions provided by the authors remained implicit as they did not provide an explicit solution for the profile of the function $K(x,y)$. Here, we provide an explicit solution for 
	$K(x,y)$ by solving numerically the equation (\ref{eq:ang-K}). The details of the numerical solution is summarized in the appendix of \cite{BenAchour:2024zzk} and the profile of both $K(x,y)$ and $P(x,y)$ are depicted\footnote{For clarity, we use in the plots the spherical coordinates $(\theta, \varphi)$ on the sphere instead of the coordinates $(x,y)$} in Fig.~\ref{fig:K}. Notice that while $P(x,y)$ is ${\cal C}^2$, the function $K(x,y)$ is only ${\cal C}^0$ but not ${\cal C}^1$, since it admits a cusp at the equator, i.e. at $\theta =\pi/2$. Notice however that since the metric and the kinetic energy of the scalar field $X$ do not depend on the derivatives of $K(x,y)$, both remain continuous at the equator. It follows that the two-dimensional space $\cB$ spanned by $\partial_x$ and $\partial_y$ has a positive curvature peaked at the equator.   In the following, we use this numerical solution for the function $K(x,y)$ to discuss the behavior of Tahamtan and Svitec's solution, thus improving on the analysis reported in \cite{Tahamtan:2015sra, Tahamtan:2016fur}.
	\begin{figure}[!htb]
		\centering
		\includegraphics{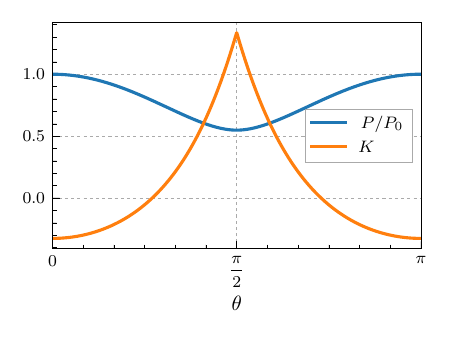}
		\caption{Plots of $P/P_0$ and $K$ obtained numerically as functions of the azimuthal angle $\theta$.}
		\label{fig:K}
	\end{figure}
	
	Now, we can analyze the singularities and the apparent horizons in this geometry. The curvature invariants are given by  
	\be
	{\cal R}  = X \,, \qquad \text{and} \qquad  \cK = {\cal R}_{\alpha\beta\mu\nu} {\cal R}^{\alpha\beta\mu\nu}= 3 {\cal R}_{\mu\nu} {\cal R}^{\mu\nu} = 3 {\cal R}^2 \,.
	\ee
	Thus the expression (\ref{X}) for $X$ reveals a singular hypersurface located at $
	\rho_s = \chi(w)$. It follows that the singularity is located on the hypersurface $\rho_s=0$ at $w= -w_0$. This hypersurface expands up to $\rho_s=C_0/\sqrt{\gamma}$ before contracting again to reach its initial position $\rho_s=0$ at $w= w_0$.
	Notice that the presence of this singularity restricts the range of the $\rho$-coordinate to $\rho > \chi(w)$ while $w\in [- w_0, + w_0]$. Moreover, notice that the curvature invariants do not involve the derivatives of $K(x,y)$. Therefore, they remain continuous at the equator just as the kinetic energy of the scalar field $X$ and the metric.
	
	To locate the possible apparent horizons, we introduce the null tetrad
	\begin{align}
		\ell^{\mu}\partial_{\mu} & = \partial_\rho \,, \\
		n^{\mu} \partial_{\mu} & = \partial_w + \frac{P(x,y)^2}{\rho^2 - \chi^2} (M_x \partial_x + M_y \partial_y)  
		- \frac{1}{2} \left[ K(x,y) - \frac{P(x,y)^2}{\rho^2 - \chi^2} \left(M_x^2 + M_y^2\right)\right] \partial_\rho \,, \\
		m^{\mu} \partial_{\mu} & = \frac{P(x,y)}{\sqrt{2(\rho^2 - \chi^2)}} \left[ (M_x + i M_y)\partial_\rho + \partial_x + i \partial_y \right] \,.
	\end{align}
	which is orthogonal to the surface $\rho = M(x, y)$, i.e. such that $g^{\mu\nu} \ell_\mu \partial_\nu M = 0$ and which satisfies the standard orthogonality relations
	\be
	\ell^{\mu} n_{\mu} = -1 \,, \qquad m^{\mu} \bar{m}_{\mu} = 1 \,, 
	\ee
	while $\ell^\mu m_\mu = n^\mu m_\mu = 0$.
	The null vectors $\ell$ and $n$ correspond respectively to out-going and in-going null rays.
	The expansions of these vectors are given by
	\begin{align}
		\Theta_{\ell} &
		= \frac{\rho}{\rho^2 - \chi^2}\,,\\
		\Theta_n &
		= \frac{ \Delta_S M-\rho K(x,y) - 2\chi(w)\chi'(w)}{2(\rho^2 - \chi^2)}  - \frac{\rho \|\nabla_S M \|^2}{2(\rho^2 - \chi^2)^2} \,.
	\end{align}
	Thus $\Theta_{\ell}$ cannot vanish. However, the expansion for the ingoing vector $n^{\mu}\partial_{\mu}$ vanishes, i.e. $\Theta_n = 0$, on the null hypersurface $\rho=M(x, y)$ when the function $M(x,y)$ satisfies the following equation
	\begin{align}
		\label{hor}
		\Delta M -M(x, y) K(x,y)  - 2\chi(w)\chi'(w)  -  \frac{\rho \|\nabla_S M\|^2}{M(x, y)^2 - \chi(w)^2}  =0 \,.
	\end{align}
	Provided this equation is satisfied, the hypersurface $\rho =M(x,y)$ corresponds to a null apparent trapping horizon. Thus, this exact solution corresponds to a singular radiative geometry with a propagating apparent horizon. The exact motion of this null hypersurface is in general too complicated to be integrated, even numerically. We stress that this challenge in investigating the motion of the apparent horizon is a general difficulty one encounters in this type of radiative solutions. See \cite{Chow:1995va, deOliveira:2009mc} for related investigations for Robinson-Trautman geometries. We shall also recover this limitation when investigating the disformed solution. Now, we would like to gain intuition on the GWs propagating in this geometry. 
	
	\subsection{Breathing wave pulse}
	
	We now show that this solution can be understood as a scalar pulse carrying finite energy.   
	
	\subsubsection{Asymptotic regimes}
	To understand the waveform propagating to $\cI^{+}$, we shall describe the different regimes of interest.
	\begin{itemize}
		\item In the remote past and far future ($w \rightarrow \pm w_0$), one has $\chi(w) \rightarrow 0$, such that the kinetic term \eqref{X} vanishes:  $X=0$. Moreover, the metric and the scalar field behave as follow:
		\begin{align}
			\label{asymmet}
			&
				\lim_{w \to \pm w_0}  \rd s^2 = - K(x,y) \rd w^2 - 2 \rd w \rd \rho + \frac{\rho^2 }{P^2(x,y)}(\rd x^2 + \rd y^2) \,,\\
			&\lim_{w \to \pm w_0}  \phi =  \frac{\sqrt{2} \chi(w)}{\rho}  \rightarrow 0 \,.
		\end{align}
		Thus, except at $\rho=0$ where the singularity lies, the scalar field and its kinetic energy vanish everywhere in these two asymptotic regimes. The metric is static and non-spherically symmetric since the term $P(x,y)$ encodes the deformation w.r.t. the unit two-sphere (corresponding to $P = P_0$). \\
		\item At the maximum of the pulse, $w=0$, the kinetic term is given by
		\be
		X = \frac{2 \gamma C^2_0 }{(\gamma \rho^2 - C^2_0)^2} K(x,y) \,,
		\ee
		which is positive, while the metric and the scalar field become
		\begin{align}
			\label{asymscal}
			&\lim_{w \to 0}  \rd s^2 = - K(x,y) \rd w^2 - 2 \rd w \rd \rho + \frac{\gamma \rho^2- C^2_0 }{\gamma P^2(x,y)}(\rd x^2 + \rd y^2) \,,  \\
			&\lim_{w \to 0}  \phi = \frac{1}{\sqrt{2}} \log{\left[ \frac{\sqrt{\gamma}\rho - C_0}{\sqrt{\gamma}\rho+ C_0}\right]} \,.
		\end{align}
		Thus, the scalar pulse which is built along time acts as a Gaussian scaling of the two-dimensional space $\cB$ spanned by $(\partial_x, \partial_y)$. After increasing, the rescaling tends to zero, giving back the initial form of the metric. 
	\end{itemize}
	To see this, it is useful to compute the curvature $\cK$ of $\cB$ whose induced metric is
	\be
	\rd s^2_{\cB} =  \frac{\rho^2 - \chi^2}{P^2(x,y)}(\rd x^2 + \rd y^2) \,.
	\ee
	Its two-dimensional curvature takes the form
	\be
	\cK(w,x,y) = \frac{1}{C^2_0} \chi^2(w) K(x,y) \,,
	\ee
	which shows the rescaling of $K(x,y)$ by the function $\chi(w)$. Indeed, at $w\rightarrow \pm w_0$, one has $\cK =0$ while at $w=0$, the curvature is $\cK(x,y) = K(x,y)/\gamma$.
	Notice that such a GW does not provide any deformation in the direction orthogonal to the direction of propagation. The only deformation takes place along the direction of propagation, providing an explicit example of a breathing GW induced by the scalar monopole (\ref{scalprof}). A careful analysis of the different polarizations a gravitational wave can carry can be found in \cite{Eardley:1973br,Eardley:1973zuo}.
	
	\subsubsection{Optical scalars}
	
	To further understand the nature of the nonlinear GW, it is useful to consider its effect on a congruence of null geodesics. To that end, let us introduce the simplified tetrad:
	\begin{align}
		\label{seednullvec1}
		\ell^{\mu}\partial_{\mu} & = \partial_\rho \,, \\
		n^{\mu} \partial_{\mu} & = \partial_w - \frac{1}{2} K(x,y) \partial_\rho \,, \\
		\label{seednullvec3}
		m^{\mu} \partial_{\mu} & = \frac{P(x,y)}{\sqrt{2(\rho^2 - \chi^2)}} \left( \partial_x + i \partial_y \right) \,.
	\end{align}
	The null vectors $\ell$ and $n$ correspond respectively to out-going and in-going null rays. The vector $\ell$ is geodesic, i.e. $\ell^{\alpha}\nabla_{\alpha} \ell^{\mu} =0$. The associated geodesic motion can be easily integrated for the associated lightlike observer. Introducing an affine parameter $\lambda$, the solution to the geodesic equation with $\ell^{\mu} \partial_{\mu}$ being the tangent vector is given by
	\begin{align}
		\label{eq:geodesic-nodisf}
		\rho(\lambda) = \lambda + \rho_0 \,, \qquad  w(\lambda) = w_0 \,, \qquad  x(\lambda) = x_0 \,, \qquad   y(\lambda) =y_0 \,,
	\end{align}
	where $(w_0, \rho_0, x_0, y_0)$ are the initial conditions. Thus the geodesic describes a photon traveling in the radial direction without any motion on the two-sphere.
	As expected, the Sachs optical parameters $( \Theta, \omega, \sigma)$ describing the expansion $\Theta$, twist $\omega$ and shear $\sigma$ of this null congruence, are given by
	\begin{align}
		\Theta (\rho, w) = - \frac{\rho}{\rho^2 - \chi^2} \,, \quad \omega = 0 \,, \qq{and} \;\; \;\;\;\; \sigma  = 0 \,.
	\end{align}
	This shows that the only effect induced by the GW consists in an expansion of the congruence, i.e. confirming the purely breathing nature of the wave. This exact solution thus belongs to the Robinson-Trautman family of geometries admitting a twist-free, shear-free but expanding null congruence. See \cite{Podolsky:2016sff} for details on the properties of Robinson-Trautman geometries. 
	
	\subsubsection{Petrov type}
	
	We can compute the Weyl scalars to identify the Petrov type of this geometry. One can show that the speciality index is given by
	\be
	S =1 \,,
	\ee 
	demonstrating that the geometry is algebraically special \cite{BenAchour:2024zzk}. Furthermore, one can consider the quantities $(L,M,  I)$ defined in the previous chapter and show that $N \neq 0$ and $M\neq 0$ when $K(x,y)$ is not constant, while $N=M=0$ when $K(x,y) =1$. This demonstrates that this spacetime is of Petrov type II when $K(x,y)$ is not a constant, while it is of type D when $K(x,y) =1$, i.e. in the spherically symmetric case. See \cite{Bini:2021aze} for details on the Petrov classification. This is consistent with previous analysis reported in \cite{Podolsky:2016sff}. It is instructive to provide the expressions of the non-vanishing Weyl scalars for our choice of null directions. They read
\begin{align}\label{Psis-seed}
\Psi_2 = \frac{1}{6} X \,, 
\qquad
\Psi_3  = -\frac{i \rho P \partial_z K}{2
	\left(\rho^2-\chi^2\right)^{3/2}} \,,
\qquad
\Psi_4 =-\frac{P^2}{2(\rho^2-\chi^2)}
\left(  \partial_z^2 K + 2 \partial_z K \partial_z \ln{P}
\right) \,,
\end{align}
where we have presented the results in terms of the complex coordinate $z=(x+iy)/\sqrt{2}$. 

Notice that $\Psi_2$ is proportional to the kinetic energy of the scalar field $X$. Since it provides the key quantity for the matter sector, we plot its behavior in Fig.~\ref{pt_Xorig}. Its behavior confirms the interpretation of the solution as a scalar energy pulse. Notice also that while $X$ is positive almost everywhere, it admits a negative value near the two poles at $w=0$. This implies that while the gradient of the scalar field is spacelike almost everywhere (for our choice of sign in the definition of $X$), it becomes time-like at the maximum of the pulse in a small region near the two poles. Whether the motion of apparent horizon located at $\Theta_n =0$ is responsible for this behavior is an open question as we were not able to integrate its time-development encoded in \eqref{hor}. Answering this question would require a complete investigation of the motion of the horizon which goes beyond the scope of this work.
	\begin{figure}[!htb]
		\centering
		\includegraphics{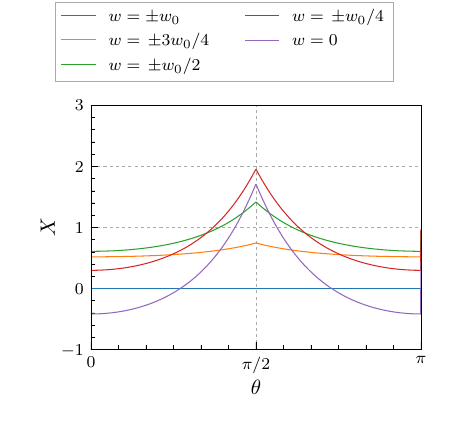}
		\caption{The kinetic energy $X$ of the scalar field is plotted for $\alpha = \alpha_0$, $C_0 = 1$, $\gamma = 1$ and $\rho = 1.5$.}
		\label{pt_Xorig}
	\end{figure}
	
	Let us emphasize that the representation of the waveform is made possible here by having first solved numerically the profile for the function $K(x,y)$, a step which was missing in \cite{Tahamtan:2015sra, Tahamtan:2016fur}. In that sense, the representation of the scalar pulse described above provides a first result of this work. Equipped with this exact solution, we can now present the new exact radiative solution in Horndeski gravity and compare its nonlinear phenomenology with its GR counterpart.

\section{Non-linear gravitational waves in DHOST gravity}

	In order to construct this exact solution in Horndeski gravity, we shall use the disformal solution-generating method \cite{BenAchour:2020wiw}. While the method applies to any scalar-tensor theory, it will be enough to consider the Einstein-Scalar action $S[g_{\mu\nu}, \phi]$ of \eqref{EinS} as a seed theory for our purposes. Performing a disformal mapping on the metric $g_{\mu\nu}$ given by
	\be
	\label{dis}
	\left( g_{\mu\nu}, \phi \right) \rightarrow \left( \tilde{g}_{\mu\nu} = g_{\mu\nu} + B_0 \phi_{\mu} \phi_{\nu}, \phi \right) \,,
	\ee
	where $B_0$ is a constant parametrizing the deviation w.r.t. the seed theory, the Einstein-Scalar action $S[g_{\mu\nu}, \phi]$ given by \eqref{EinS} gets corrected by higher-order scalar-tensor contributions giving rise to a new scalar-tensor theory $\tilde{S}[\tilde{g}_{\mu\nu}, \phi]$. This new theory belongs to the shift-symmetric Horndeski family. See \cite{Kobayashi:2019hrl} for a review on the phenomenology of the Horndeski family of theories.
	Under this transformation, the scalar kinetic term $X = g^{\mu\nu} \phi_{\mu} \phi_{\nu}$ is modified. The relations between the new kinetic term $\tilde{X}= \tilde{g}^{\mu\nu} \phi_{\mu} \phi_{\nu}$ and the seed one $X$ are given by
	\be
	\tilde{X} = \frac{X}{1+B_0 X}\,, \qquad X = \frac{\tilde{X}}{1-B_0 \tilde{X}} \,.
	\ee
	
	\subsection{Exact radiative solution}\label{C}
	
	As emphasized previously, the scalar profile remains the same such that
	\begin{align}
		\label{scalproff}
		\phi(w,\rho) = \frac{1}{\sqrt{2}} \log{\left[ \frac{\rho - \chi(w)}{\rho+ \chi(w)}\right]} \,;
		\qquad
			\chi(w) =  \frac{C_0}{\sqrt{\gamma}} \exp\left[\left(\text{erf}^{-1}\left[w/w_0\right]\right)^2\right]  \,.
	\end{align}
	Notice that since the scalar field profile does not depend on the angular coordinates $(x,y)$, it can be again considered as a pure scalar monopole.
	The new exact solution is given by
	\begin{align}
		\label{metdis}
		\rd s^2 & = - K(x,y)  \rd w^2 - 2 \rd w \rd \rho + \frac{\rho^2 - \chi^2(w)}{P^2(x,y)}(\rd x^2 + \rd y^2)   \\
		& \;\;\; + B_0 \left[ \phi^2_w  \;  \rd w^2 + 2 \phi_w \; \phi_{\rho} \;\rd w \rd \rho + \phi^2_{\rho} \; \rd \rho^2 \right] \,.
	\end{align}
	The first line corresponds to the original metric and the functions $K(x,y)$ and $P(x,y)$ remain unchanged, i.e. they are given by
	\begin{align}
		K(x,y) = \Delta \log{P}(x,y) \,, 
		\qquad
		\Delta K(x,y) = 4 C^2_0 \ic^2 = \alpha^2 \,,
	\end{align}
	where, similar to the seed metric, $C_0$ (or $\alpha$ for $\ic\neq0$) characterizes the scalar charge for the disformed metric. We recall that the functions $\chi(w)$ and $K(x,y)$ are depicted in Fig.~\ref{fig:chi}  and Fig.~\ref{fig:K} respectively. The second line in \eqref{metdis}, that is proportional to the parameter $B_0$, encodes the new effects induced by the higher-order terms.
	These new terms are constructed from the components of the scalar gradient which read explicitly
	\be
	\label{phiw}
	\phi_{\rho} = \frac{2\chi(w)}{\rho^2- \chi^2}\qquad  \text{and} \qquad  \phi_w = \frac{- 2 \rho \chi'(w)}{\rho^2- \chi^2} \,.
	\ee
	It follows that the kinetic energy of the scalar field in this new geometry can be written as
	\be
	\tilde{X}= \frac{\phi_{\rho}(K \phi_{\rho} - 2 \phi_{w} + 2 B_0 \phi_{\rho} \phi^2_{w})}{1 + B_0 \phi_{\rho}(K \phi_{\rho} + 2 \phi_{w})} \,,
	\ee
	which is depicted in Fig.~\ref{plot_X}. One can notice that it has the same qualitative behavior as for the seed solution.
	\begin{figure}[!htb]
		\centering
		\includegraphics{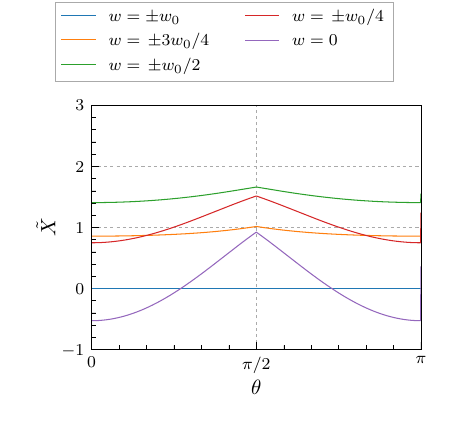}
		\caption{The kinetic energy $\tilde{X}$ of the scalar field after disformal transformation, for $\alpha = \alpha_0$, $C_0 = 1$, $\gamma = 1$ and $\rho = 1.5$.}
		\label{plot_X}
	\end{figure}
	We emphasize that this geometry is by construction a non-perturbative and exact solution of a subset of shift symmetric Horndeski theories. Now, our goal is to analyze the phenomenology of this exact solution. 
	
	Computing the Kretschmann invariant, one finds that the GR singularity at $\rho = \chi(w)$ is preserved  while additional singularities show up on the hypersurface defined by
	\be\label{singularity}
	\rho^4 - 2 \left( \rho^2 - B_0 K(x,y)\right) \chi^2 + \chi^4 + 4 B_0 \rho \chi \chi' =0 \,.
	\ee
	These new singularities can be analyzed as follows. When $w\rightarrow \pm w_0$, we have $\chi=0$ and $\chi'$ constant. Thus, in these regimes, the only solution is $\rho_s = \chi(\pm w_0)=0$ and there are no new singularities.  When the pulse reaches its maximum, at $w=0$, one has $\chi(0) = C_0/\sqrt{\gamma}$ and $\chi'(0) =0$ and the new singularity, if it exists, is located at
	\begin{align}
		\rho^2_{\ast} = \frac{C_0^2}{\gamma} \left(1 \pm \sqrt{\frac{-2B_0 K(x,y) \gamma}{C_0^2}}\right) \,.
	\end{align}
	Therefore, the disformal transformation generates a new singularity but this additional singular hypersurface remains bounded to $\rho_{\ast}$ when the pulse reaches its maximum. One can also identify the equation for the presence of an apparent trapping horizon but the explicit expression is not particularly illuminating and we shall not reproduce it here. In the end, regarding the locus of the singularities and the possible apparent horizons, the new solution behaves qualitatively the same as the seed one, demonstrating that the disformal transformation has only minimal impact on these properties. We now turn to the main focus, namely the analysis of the waveform in this new geometry.
	
	\subsection{Interpreting the GW}\label{sub-D}
	
	\subsubsection{Asymptotic regimes}
	
	To understand the time-development of the solution, we first investigate the different regimes.
	\begin{itemize}
		\item In the remote past and far future ($w\rightarrow \pm w_0$), the kinetic energy is again vanishing, i.e.  $\tilde{X}=0$. The only component of the scalar gradient which survives is $\phi_{w}$ (given by (\ref{phiw})) such that the metric and the scalar field reduce to
		\begin{align}
			& \lim_{w \to \pm w_0}  \rd s^2  = - \left[K(x,y) - \frac{Q^2}{\rho^2}\right] \rd w^2 - 2 \rd w \rd \rho  + \frac{\rho^2 }{P^2(x,y)}(\rd x^2 + \rd y^2) \,, \\
			& \lim_{w \to \pm w_0}  \phi =  \frac{\sqrt{2} \chi(w)}{\rho}  \rightarrow 0 \,.
		\end{align}
		In both asymptotic regimes, the metric is again static and non-spherically symmetric. However, there is a qualitative difference with the GR solution which shows up in the component $g_{ww}$. This term inherits a new contribution where
		\be
		\label{Q}
		Q \equiv 2 \sqrt{B_0} \lim_{w \to \pm w_0} \chi'(w) \,.
		\ee
		Thus, while the scalar field vanishes everywhere except at the singularity $\rho=0$, its $w$-derivative does not and contributes to the metric. The specific form of this new contribution suggests that it might be considered as an effective electric charge in this regime.
		
		\item At the maximum of the pulse ($w = 0$), the kinetic energy reaches a non-vanishing positive value given by
		\be
		{\tilde X} = \frac{2 \gamma  C^2_0 K(x,y)}{(\gamma\rho^2 - C_0^2)^2 + 2 B_0 \gamma C_0^2 K(x,y) } \,.
		\ee
		The metric and the scalar profile become
		\begin{align}
			& \lim_{w \to 0}  \rd s^2  = - K(x,y) \rd w^2 - 2 \rd w \rd \rho + \frac{4B_0 \gamma C^2_0}{ (\gamma \rho^2-C_0^2)^2} \rd \rho^2  + \frac{\rho^2 }{P^2(x,y)}(\rd x^2 + \rd y^2) \,,\\
			& \lim_{w \to 0}  \phi =  \frac{1}{\sqrt{2}} \log{\left[ \frac{\sqrt{\gamma}\rho - C_0}{\sqrt{\gamma}\rho+ C_0}\right]} \,.
		\end{align}
		Thus when the pulse reaches its maximum, the contribution in $g_{ww}$ present in the asymptotic regimes vanishes since $Q$ is proportional to $\chi'$. On the other hand, the metric inherits another contribution in $g_{\rho\rho}$.
	\end{itemize}
	Now, we can investigate the GWs propagating in this geometry.

	\subsubsection{Null tetrad and Weyl scalars}
	
	To do so, we proceed as follows. We first introduce four null vectors
	\begin{align}
	E^{\mu}_A = (\tilde{\ell}^\mu, \tilde{n}^\mu, \tilde{m}^\mu, \tilde{\bar{m}}^\mu) \,,
	\end{align} 
	with the standard orthogonality relations
	\begin{align}
		\label{nullvec}
		& \tilde{\ell}^{\mu} \tilde{n}_{\mu} = -1 \,, \quad \tilde{m}^{\mu} \tilde{\bar{m}}_{\mu} = 1 \,, \qquad  \tilde{\ell}^\mu \tilde{m}_\mu = \tilde{n}^\mu \tilde{m}_\mu = 0 \,.
	\end{align}
	The choice of null vector basis is not unique since it is defined up to Lorentz transformations. We choose $E^{\mu}_A$ to be the associated null tetrad
	constructed such that $\tilde{\ell}^{\mu} = E^{\mu}_{U} \rd U$ corresponds to the tangent vector of a null geodesic $\gamma$ with affine parameter $U$. The remaining vectors are selected such that $E^{\mu}_A$ corresponds to a parallel transported frame (PTF), i.e. such that
	\begin{align}
	\tilde{\ell}^{\mu} \tilde{\nabla}_{\mu} E^{\nu}_{A} =0 \,,
	\end{align}
	where the $\tilde{\nabla}_\mu$ is compatible with the disformed metric $\tilde{g}_{\mu\nu}$. In terms of the spin coefficients that are defined above, it translates into the conditions: $\tilde{\kappa} = \tilde{\epsilon} = \tilde{\pi} =0$. Physically, it corresponds to the projector onto the family of local inertial frames of an observer following the null geodesic $\gamma$. While we could have worked with a general null tetrad in this section, constructing this PTF will reveal useful later on when investigating the Penrose limit and the memory effect. As we are going to see, the main novelty is that, contrary to the GR one, a PTF cannot be shear-less in the disformed geometry. The detailed derivation of the null tetrad (\ref{nullvec}) and its spin coefficients was given at the beginning of this chapter.
	
	To simplify the different expressions, we shall assume that $B_0 \ll 1$ and expand w.r.t. the disformal parameter. This expansion enables us to obtain tractable expressions for the main quantities of interest. Yet, the reader should keep in mind that the solution described and analyzed here remains \text{exact} and \text{non-perturbative}.
	Having clarified that point, we now present the analysis up to quadratic order in $B_0$, i.e. neglecting terms of order $\cO(B^3_0)$. Schematically, the components of the PTF can be written as
	\begin{align}\label{PTF}
		E^{\mu}_A \partial_{\mu} = \left(^{(0)}\!E^{\mu}_A + B_0 \; ^{(1)}\!E^{\mu}_A + B^2_0 \; ^{(2)}\!E^{\mu}_A  \right) \partial_{\mu} \,.
	\end{align}
	The components $^{(i)}\!E^{\mu}_A$ can be solved exactly order by order in $B_0$ but their expressions are in general involved. 
	
	We now can proceed to the nonlinear analysis of the new radiative solution. First, computing the speciality index of this geometry, one can show that 
	\be
	S = 1 + B^2_0\, G(w,\rho, x,y) + \cO(B^3_0) \,,
	\ee
	where $G$ is a complicated function whose precise form is not relevant. This means that the new solution is not algebraically special, i.e. the geometry is of Petrov type I. Physically, it implies that the Weyl tensor admits four linearly independent principal null directions. While the expressions of the Weyl scalars depend on the choice of frame $E^{\mu}_A$, it will be useful to provide the expression of $\tilde{\Psi}_0$ in the following. For our specific choice of PTF, all the Weyl scalars are non-vanishing and take rather complicated expressions except for the scalar $\tilde{\Psi}_0$ which takes a rather simple form given by
	\begin{align}
		\label{psi0}
		\tilde{\Psi}_0 & = B^2_0 \frac{\chi^4 P \left( P \partial_{\bar{z}\bar{z}} K + 2 \partial_{\bar{z}} K \partial_{\bar{z}} P\right)}{(\rho^2-\chi^2)^5}  + \cO(B^3_0) \,.
	\end{align}
The behavior of $\tilde{\Psi}_0$ at a fixed value of $\rho$ is represented in Fig.~\ref{plot_psi0}. As we shall see, some of the key quantities can be expressed in terms of this scalar $\tilde{\Psi}_0$. 
	
	\begin{figure}[!htb]
		\centering
		\includegraphics{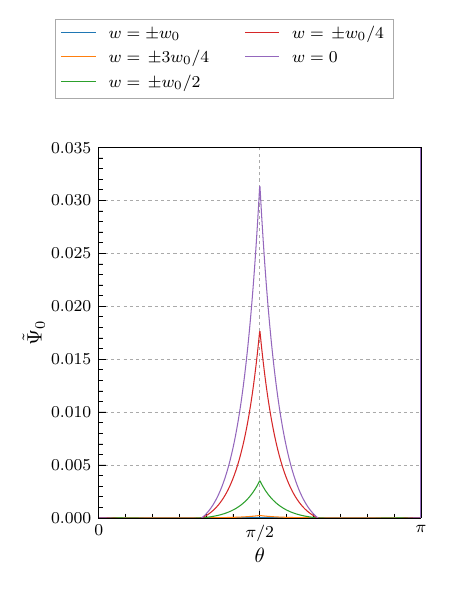}
		\caption{The Weyl scalar $\tilde{\Psi}_0$, for $\alpha = \alpha_0$, $C_0 = 1$, $\gamma = 1$ and $\rho = 1.5$.}
		\label{plot_psi0}
	\end{figure}
	
	\subsubsection{Optical scalars}
	
	To further understand qualitatively the nature of the GWs propagating in this geometry, we compute the Sachs optical scalars $(\tilde{\Theta}, \tilde{\omega}, \tilde{\sigma})$ associated to our null tetrad. 
	Neglecting terms of order $\cO(B^3_0)$, one can write the expansion scalar $\tilde{\Theta}$ as
	\be
	\tilde{\Theta} = \Theta_0 + B_0 \Theta_1 + B^2_0 \Theta_2 + \cO(B^3_0) \,,
	\ee
	and show that
	\begin{align}
		\Theta_0 & = - \frac{\rho}{\rho^2 - \chi^2} \,,\\
		\Theta_1 & = \frac{1}{4 \chi^2 (\rho^2 - \chi^2)^3} \left[ 4 \rho K \chi^4 + 2\chi\chi' \left( \rho^4  + \rho^2 \chi^2 + 2 \chi^4 \right) + \rho \chi' \log{\left( \frac{\rho - \chi(w)}{\rho+ \chi(w)}\right)} \left( \rho^4 - 2 \rho^2 \chi^2 + \chi^4  \right) \right] \,,
	\end{align}
	while the expression for $\Theta_2$ is too complicated to be reproduced here. 
	Additionally, one finds a non-vanishing shear given by
	\begin{align}
		\label{shear}
		\tilde{\sigma} & = B^2_0 \sigma_2 + \cO(B^3_0) \,,
	\end{align}
	where we have explicitly
	\begin{align}
		\sigma_2 & = - \frac{(\rho^2-\chi^2)}{96\chi^7} \left[ 30 \rho^5 \chi - 80 \rho^3 \chi^3 + 66 \rho \chi^5 + 15 (\rho^2-\chi^2)^3 \log{\left( \frac{\rho - \chi(w)}{\rho+ \chi(w)}\right)} \right]  \Psi_0 \,.
	\end{align}
	This analysis reveals that, additionally to the expansion $\tilde{\Theta}$ induced by the scalar wave, a non-vanishing shear $\tilde{\sigma}$ also shows up at the quadratic order in the disformal parameter $B_0$. 
	
	Since the shear is not Lorentz-invariant, it is natural to wonder whether it might not be a gauge artifact of the chosen frame. To answer to this question, one has to pick up a PT frame, i.e. a frame in which the vector $\tilde{\ell}^{\mu} \partial_{\mu}$ is geodesic and in which the remaining three null vectors are parallel transported along $\tilde{\ell}$. At the level of the spin coefficients, this property translates into $\tilde{\kappa} = \text{Re}(\tilde{\epsilon}) = \tilde{\pi} = 0$. This ensures that one is working in the local inertial frame associated to the null geodesic. Starting from a general frame,  one has to use a suitable sequence of Lorentz transformations to construct this PT frame. Proceeding to this construction, one can show that
	\be
	\tilde{\kappa} = \text{Re}(\tilde{\epsilon}) = \tilde{\pi} =0 \,, \qquad \text{implies that} \qquad \tilde{\sigma} \neq 0 \,.
	\ee
	Therefore, one cannot have a shear-free PT frame in this new exact radiative solution. This conclusion is Lorentz-invariant, in the sense that once the PT frame has been constructed, one can show that the residual freedom to perform Lorentz transformation while remaining PT cannot be used to set the shear $\tilde{\sigma}$ to zero. However, this residual gauge freedom can be used to set the twist to zero at second order in $B_0$. It follows that there are no shear-less PT frame in the disformed solution, an effect which descends directly from the higher-order term of the Horndeski theory. 
	
	Thus, the solution also contains a tensorial mode. While small, this effect is qualitatively completely new w.r.t. to the GR solution and can be interpreted as emerging from  the nonlinear mixing between the scalar and tensor sectors in Horndeski theory. Notice that the expression of the shear (\ref{shear}) in terms of $\tilde{\Psi}_0$ is specific to our choice of frame. While one can always rotate the frame to set $\tilde{\Psi}_0 =0$, the shear will remains non-zero and will be expressed in term of other Weyl scalars in this new frame. In the end, the presence of this non-vanishing shear $\tilde{\sigma}$ shows that the higher-order terms in the Horndeski dynamics lead to a rather surprising new phenomenology. Even in the presence of a pure scalar monopole source, the mixing between scalar and tensors sectors allows for the generation of shear. This is the main result of the present analysis.
	
	While our solution is fully nonlinear and non-perturbative in the scalar field profile, it is worth to look at the perturbative description of our solution. Since the key quantity describing the scalar pulse is $\chi(w)$, it is natural to consider the linear regime where $\phi(\rho, w)=-\chi(w)/\sqrt{2}\rho$ for $\chi(w)\ll{1}$. Then, we see that $\tilde{\Psi}_0 \simeq 0$ at the linear regime and it only acquires non-vanishing value at the order ${\cal O}(\chi^4)$. Therefore, the nonlinear effects that we have found would be lost if one linearizes the system from the beginning. For example, the shear $\tilde{\sigma}$ is proportional to $\tilde{\Psi}_0$ and is absent at the linear level. As we shall see, this can also be seen from the polarizations of the gravitational plane wave derived in the next section. Indeed, the two tensorial polarization $H_{+, \times}$ being also given by the real and imaginary part of $\tilde{\Psi}_0$, the same conclusion applies. Finally, let us point that in this linear regime, the effective electric charge (\ref{Q}) also vanishes, since it is quadratic in $\chi'$. Thus, we conclude that the two main effects induced by the scalar monopole have a fully nonlinear origin.

	Now we can describe the full process developing in the range $w \in [-w_0, + w_0]$. We start in the remote past at $w= - w_0$ from a static and non-spherically symmetric geometry. While the scalar profile vanishes there, its gradient does not which ultimately generates and effective electric charge $Q$. Then, a scalar pulse is ignited which induces both a breathing as well as a transverse GWs which propagate to $\cI^{+}$. Along the process, this nonlinear superposition of these GWs is described by an algebraically general type I geometry.
	
	To the best of our knowledge, this non-perturbative radiative solution provides a first example of a new phenomenology emerging in the fully nonlinear regime of a higher-order scalar-tensor theory. The possibility to generate a tensorial wave from a purely scalar monopole configuration is intriguing and has to be understood as a special effect inherent to i) the higher-order terms in the Horndeski dynamics and ii) to the fully nonlinear regime. We shall come back on this last point in the end. For the moment, we would like to confirm our analysis by inspecting the different polarizations of the GWs propagating in the new solution. 
	
		\section{Polarizations and memory effect}
	\label{D}
	
	\subsection{Polarizations from the Penrose limit}
	
	In the linear theory, where GWs are described by a small perturbation $h_{\mu\nu}$ around some background $g_{\mu\nu}$, one can read off the different polarizations of the wave from the structure of $h_{\mu\nu}$. However, for our non-perturbative solution, such an approach is no longer applicable. Instead, one can use the so-called Penrose limit to analyze the different GW polarizations present in a radiative spacetime.
	
	For a given spacetime $g_{\mu\nu}$, the Penrose limit consists in encoding the leading tidal effects experienced by a photon around its worldline $\gamma$ within a pp-wave \cite{Penrose}. The profile of the pp-wave is directly related to the leading geodesic deviation effect in the direction transverse to the geodesic motion. This procedure is remarkable in that it keeps track of the non-perturbative character of the initial metric. Starting from a non-perturbative radiative spacetime, it allows one to identify in a simple way the different polarizations propagating in this geometry by simply reading the profile off the associated pp-wave. We refer to \cite{Blau:2003dz,Blau:2006ar} for a detailed and pedagogical exposition of the Penrose limit.
	
	To proceed, one first constructs null Fermi coordinates adapted to a given null geodesic. The change of coordinates is in the form of a Taylor expansion in the spacelike hypersurface orthogonal to the $4$-velocity of the geodesic $\bar{\gamma}$. To that end, we consider the PTF $E^{\mu}_A$ introduced in the previous section and denote the affine parameter of this geodesic $W$ such that $\ell^{\mu} = E^{\mu}_{W} \rd W$.  Following the construction reviewed in \cite{Blau:2006ar}, we denote the null Fermi coordinates $X^{A} =(W, V, X^i)$ with $i\in\{1,2\}$. In terms of the initial coordinates $x^{\mu} = (w, x^a)$ with $x^a = (\rho, x,y)$ the coordinates in the hypersurface orthogonal to the geodesic motion, the change of coordinate between the initial and Fermi coordinates is given by
	\begin{align}
		\label{Fermi}
		X^{A} & = E^{A}_{a} x^a+ E^{A}_{\mu} \bar{\Gamma}^{\mu}{}_{ab}  x^a x^b + \cO((x^a)^3) \,,
	\end{align}
	where we restrict the change of coordinates to the quadratic terms in $x^a$. Here, the overbar means that the quantity is evaluated on the lightlike geodesic of reference $\bar{\gamma}$.  Performing the change of coordinates, the metric becomes
	\begin{align}
		\rd s^2 & = 2 \rd W \rd V + \delta_{AB} \rd X^A \rd X^B - \bar{R}_{W A W B} (W)X^A X^B \rd W^2  \\
		& \;\;\;  - \frac{4}{3} \bar{R}_{W ABC}(W) X^A X^C\rd W \rd X^B  - \frac{1}{3} \bar{R}_{ABCD} (W)X^A X^{C} \rd X^B \rd X^D  \\
		\label{fermi}
		& \;\;\; + \cO(X^3) \,,
	\end{align}
	where higher and higher-order in the expansion provide additional information on the gravitational field around the geodesic $\bar{\gamma}$. This expansion can be organized as follows. Considering the conformal transformation of the coordinates together with a rescaling of the metric $g_{AB} \rightarrow \lambda^{-2} g_{AB}$, it can be shown that the Weyl scalars $\Psi_i$ scale as $\cO(\lambda^{4-i})$ for $i\in\{0,...,4\}$. The Penrose limit amounts at approximating the geometry around the null geodesic by a Petrov type N geometry, i.e. keeping only the contribution from $\Psi_4$ at leading order in the $\lambda$ expansion \cite{Blau:2006ar, Kunze:2004qd}. This selects the first two lines in \eqref{fermi} given by
	\begin{align}
		\label{pp}
		\rd s^2 & = 2\rd W \rd V  + \delta_{AB} \rd X^A \rd X^B - H_{AB} (W) X^A X^B \rd W^2 \,,
	\end{align} 
	with the wave profile given by
	\be
	\label{prof}
	H_{AB}(W) \equiv \bar{R}_{WAWB}(W) = \bar{R}_{\mu\nu\rho\sigma} E^{\mu}_{W} E^{\nu}_A E^{\rho}_{W} E^{\sigma}_B \,.
	\ee
	This metric corresponds to a pp-wave in the standard Brinkmann coordinates and the matrix $H_{AB}$ is the key quantity encoding the polarizations of the nonlinear gravitational plane wave. In full generality, the wave profile can be decomposed as
	\begin{align}
		\label{profile}
		H_{AB} X^A X^B & = H_{\circ} (X^2 + Y^2) + \left[ H_{+} (X^2 - Y^2) +  H_{\times} X Y  \right] \,,
	\end{align} 
	where we have denoted the spatial coordinates $X^A=(X,Y)$. The quantity $H_{\circ}$ encodes the trace contribution, associated with the breathing degree of freedom, while $H_{+, \times}$ encode the two traceless tensorial polarizations of the wave. Let us emphasize that this process keeps the nonlinear nature of the original metric (\ref{metdis}) intact, in the sense that we never impose that $H_{AB}$ be small. We refer to \cite{Blau:2006ar} for further details.
	
	Thus, at leading order, once we have identified a PTF, constructing the pp-wave geometry associated to our radiative geometry reduces to computing the projection \eqref{prof}. 
	Proceeding in that way, we find that the plane wave describing the Penrose limit of our spacetime for the chosen geodesic $\bar{\gamma}$ contains three polarizations. The two tensorial polarizations can be expressed in terms of the non-vanishing Weyl scalar $\tilde{\Psi}_0$ of the original metric as
	\begin{align}
		\label{tens}
		H_{+} = \frac{1}{2} \text{Re}(\bar{\tilde{\Psi}}_0) \,, 
		\qq{and}
		H_{\times} = \frac{1}{2} \text{Im}(\bar{\tilde{\Psi}}_0) \,, 
	\end{align}
	where $\bar{\tilde{\Psi}}_0$ means that $\tilde{\Psi}_0$ is evaluated on the reference geodesic $\bar{\gamma}$. 
	
	The scalar or trace polarization $H_{\circ}$ has a complicated expression and we do not provide it. The Penrose limit confirms that at the fully non-perturbative level, the new solution contains both scalar and tensorial GWs. Finally, we stress that while the scalar sector $H_{\circ}$ receives contributions from all orders in $B_0$, the tensorial modes are triggered at quadratic order in $B_0$, following the pattern found for the Weyl scalar $\tilde{\Psi}_0$ and for the shear $\tilde{\sigma}$. 
	
	Now, we turn to the analysis of the geodesic motion of test particles and the associated memory effect induced by the pulse.

	\subsection{Geodesic motion and memory effect}
	
	As a first step, we consider the geodesic equation in the exact solution (\ref{metdis}), i.e.
	\be
	\tilde{\ell}^{\alpha} \tilde{\nabla}_{\alpha} \tilde{\ell}^{\mu} =0 \,, \qquad \text{with} \qquad \tilde{\ell}^{\mu} = \frac{\rd x^{\mu}}{\rd U} \,.
	\ee
	Since we already have the explicit expression for the vector field parallel to the geodesic, $\tilde{\ell}^\mu$, we simply need to solve the equations
	\begin{align}
		&\frac{\rd U} {\rd w} = \tilde{\ell}^w \,,  &&\frac{\rd U} {\rd \rho} = \tilde{\ell}^\rho \,, 
		&& \frac{\rd U} {\rd x} = \tilde{\ell}^x \,, && \frac{\rd U} {\rd U} = \tilde{\ell}^y \,,
		\label{eq:geodesic-l}
	\end{align}
	with initial conditions 
	\begin{equation}
		w(0) = w_0 \,,\quad \rho(0) = \rho_0 \,,\quad x(0) = x_0 \,, \qquad \text{and} \qquad  y_0 = y_0 \,.
	\end{equation}
	This can easily be done numerically, and the results of the integration of equations~\eqref{eq:geodesic-l} can be found in \cite{BenAchour:2024zzk}. The motion is modified by a small displacement on the hypersurface with the orthogonal vector $\tilde{\ell}^{\mu}$. However, it would be misleading to interpret this constant shift as a displacement memory effect as the analysis of the memories has to be done in the local inertial frame of the test particle, i.e. by constructing suitable Fermi coordinates and measuring the spatial distance in this specific frame.

	In order to discuss the memory effect induced by the scalar pulse, we consider instead the Penrose limit of our exact solution around this geodesic $\gamma$, i.e.
	\begin{align}
		\label{ppp}
		\rd s^2 & = 2\rd W \rd V  + \delta_{AB} \rd X^A \rd X^B - H_{AB} (W) X^A X^B \rd W^2 \,,
	\end{align} 
	This geometry (\ref{ppp}) encodes the leading tidal effects experienced by the lightlike observer with worldline $\gamma$. The coordinates (\ref{Fermi}) are the Fermi null coordinates, allowing one to analyze the leading memory effects induced by the GWs in the local inertial frame of the lightlike observer. Starting from this pp-wave geometry, we now solve the geodesic motion and compute the relative distance and relative velocity between two nearby photons. Notice that the Euclidean spatial distance between these two test particles can make sense only in this Fermi frame. We refer the reader to \cite{Zhang:2018srn, Flanagan:2019ezo, BenAchour:2024ucn} for different investigations of velocity memory effects in pp-wave geometries and \cite{Divakarla:2021xrd} for a derivation in the context of the wave form sourced by the coalescence of a binary black hole merger.
	
	To proceed, we first evaluate the pulse shapes $H_\circ(W)$, $H_\times(W)$ and $H_+(W)$ seen by a photon traveling along this geodesic. This is shown in Fig.~\ref{fig:pulse_plots}. One can observe from these plots that the amplitude of the scalar wave $H_\circ$ is much bigger than the amplitudes of the tensorial waves $H_\times$ and $H_+$ generated from the scalar monopole.
	
	\begin{figure}[!htb]
		\centering
		\includegraphics{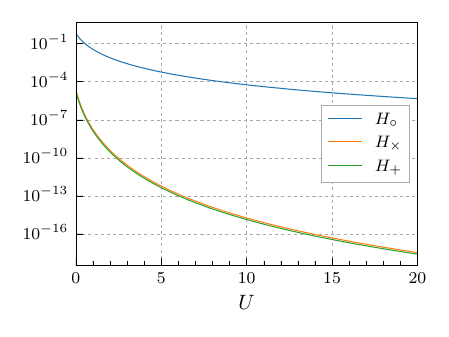}
		\caption{Plot of the amplitudes $H_\circ$, $H_\times$ and $H_+$ as functions of $U$. They encode the tensorial and breathing modes of the pp-wave seen by a photon following the null geodesic described above. One can notice that the amplitude of the tensorial waves is much smaller than that of the scalar waves.}
		\label{fig:pulse_plots}
	\end{figure}
	
	With these forms for the components of $H_{AB}$, we now compute the distance between two photons traveling initially close to the geodesic $\gamma$. Indeed, the geodesic equations are given by
	\begin{align}
		\ddot{W} & = 0 \,, \\
		\ddot{V}  & = H_{\circ} ( \dot{X}^2 + \dot{Y}^2 ) + H_{+} ( \dot{X}^2 - \dot{Y}^2 ) + 2 H_{\times} \dot{X} \dot{Y} \,,\\
		\ddot{X} & = H_\circ X + H_+ X + H_\times Y \,,\\
		\ddot{Y} & = H_\circ Y - H_+ Y + H_\times X \,.
	\end{align}
	where a dot refer to a derivative w.r.t. the affine parameter $U$. It is direct to see that $W = c U + W_0$. Choosing without loss of generality $W_0 =0$ and $c = 1$, we identify the affine parameter with the coordinate $W$. The three remaining equations can be solved numerically. Let us now focus on the geodesic motion restricted to the two-dimensional plane $(X,Y)$. In the case of a pp-wave, the geodesic equation for $(X,Y)$ coincides with the geodesic deviation equation. Thus, by solving for the motion of $(X,Y)$, one can compute the evolution of the distance $\sqrt{X^2 + Y^2}$ between a photon following exactly $\bar{\gamma}$ (having $X = 0$ and $Y = 0$ for any $U$) and a photon starting at given initial values $X_0$ and $Y_0$ for $X$ and $Y$ (respectively). The evolution of the distance $\sqrt{X^2 + Y^2}$ is given on Fig.~\ref{fig:memory-effect}.
	
	\begin{figure}[!htb]
		\centering
		\includegraphics{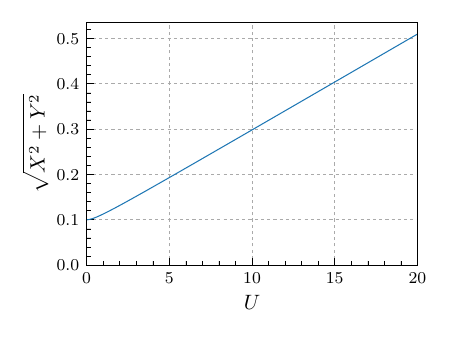}
		\caption{Plot of the distance $\sqrt{X^2 + Y^2}$ between a photon following the geodesic $\bar{\gamma}$ and a photon following the same geodesic with a small initial deviation $X = X_0$, $Y = Y_0$. The computations are done with $X_0 = Y_0 = 0.1/\sqrt{2}$.}
		\label{fig:memory-effect}
	\end{figure}
	
	One observes that the relative distance between the two null test particles grows linearly in the $W$-coordinate. It follows that for two photons with a vanishing initial relative velocity, the scalar pulse induces a constant shift in their relative velocity, an effect known as the velocity memory effect.  One can show that the effect of the transverse wave is much smaller than the breathing component, as expected. As a consequence, the correction to the velocity memory effect induced by the pure breathing wave is negligible. Notice that this memory effect has been computed using a truncation of the full geometry. Nevertheless, since this truncation contains the leading tidal effects around any observer describing the geometry around its worldline using the adapted Fermi coordinates, this approach captures the relevant contribution from our radiative solution which is encoded in the explicit expressions of the functions $(H_{\circ}, H_{+}, H_{\times})$.





\appendix



\chapter{Appendix: Disformal transformation and the Weyl scalars}
\label{app-tetrad-rep}



\setcounter{equation}{0}
\renewcommand{\theequation}{A\arabic{equation}}

\section{Riemann tensor in Einstein frame}
In this appendix we define appropriate real and complex scalars which present all tetrad components of the Riemann tensor. The Riemann tensor $R_{\alpha\beta\mu\nu}$ can be decomposed into the trace and trace-free parts according to the well-known Weyl decomposition
\begin{equation}\label{Riemann}
{R}_{\alpha\beta\mu\nu} =
{C}_{\alpha\beta\mu\nu} - 
\frac{1}{2} \big( {R}_{\alpha\nu} { g}_{\beta\mu} + { R}_{\beta\mu} g_{\alpha\nu} 
- {R}_{\alpha\mu} { g}_{\beta\nu} - { R}_{\beta\nu} g_{\alpha\mu} \big) 
- \frac{1}{6} { R} \big( g_{\alpha\mu} { g}_{\beta\nu} - g_{\alpha\nu} { g}_{\beta\mu} \big) \,,
\end{equation}
where $C_{\alpha\beta\mu\nu}$ is the Weyl part which is trace-free with respect to any index, $R_{\mu\nu} = g^{\alpha\beta}R_{\alpha\mu\beta\nu}$ is the Ricci tensor, and $R = g^{\alpha\beta} R_{\alpha\beta}$ is the Ricci scalar. We note that the above geometric decomposition is valid for any tensor with the same symmetries as the Riemann tensor. In tetrad representation, the tetrad components of the Riemann tensor are given by
\begin{equation}\label{Riemann-T}
{R}_{abcd} =
{C}_{abcd} - 
\frac{1}{2} \big( {R}_{ad} { \eta}_{bc} + { R}_{bc} \eta_{ad} 
- {R}_{ac} { \eta}_{bd} - { R}_{bd} \eta_{ac} 
\big) 
- \frac{1}{6} { R} \big( \eta_{ac} { \eta}_{bd} - \eta_{ad} { \eta}_{bc} \big) \,,
\end{equation}
where $R_{abcd} = R_{\alpha\beta\mu\nu} \theta^\alpha{}_a \theta^\beta{}_b \theta^\mu{}_c \theta^\nu{}_d$ and so on. $\theta^{\mu}{}_a$ are any tetrad basis. The Riemann tensor $R_{abcd}$ has twenty tetrad components and, in the above decomposition, ten components are encoded in the Weyl part $C_{abcd}$ and the remaining ten components are encoded in the Ricci part $R_{ab}$. 

In the null tetrad basis defined by Eq.~\eqref{T}, we can express ten components of the Weyl part by means of the five complex Weyl scalars ${\boldsymbol \Psi}_I$ as follows
\begin{eqnarray}\label{WeylS-T-a}
{\boldsymbol \Psi}_0 = C_{0202} \,, \hspace{.5cm}
{\boldsymbol \Psi}_1 = C_{0102}  \,, \hspace{.5cm}
{\boldsymbol \Psi}_2 = C_{0231} \,, \hspace{.5cm}
{\boldsymbol \Psi}_3 = C_{1013} \,, \hspace{.5cm}
{\boldsymbol \Psi}_4 = C_{1313} \,.
\end{eqnarray}
As it is clear from relations~\eqref{WeylS-T-a}, the Weyl scalars $\Psi_I$ are some special tetrad components of the Weyl tensor $C_{abcd}$ which describe ten independent components of the Weyl tensor. The other tetrad components of the Weyl tensor are not independent of the Weyl scalars and their complex conjugates and we have~\cite{Chandrasekhar:1985kt}\footnote{Note that all the scalar quantities that we define in this paper are different than those defined in~\cite{Chandrasekhar:1985kt} only by a minus sign and we use the different metric signature with the most positive signs.}
\begin{eqnarray}\label{Weyl-components}
&& C_{0203} = C_{1213} = C_{0221} = C_{0331} = 0 \,,
\\ \nonumber 
&&C_{0120} = C_{0223}\,, \hspace{1cm} 
C_{0130} = C_{0332}\,, \hspace{1cm} 
C_{0121} = C_{1232}\,, \hspace{1cm} 
C_{0131} = C_{1323}\,, 
\\ \nonumber 
&&C_{0101} = C_{2323}\,, \hspace{1cm} 
C_{0231} = \frac{1}{2} ( C_{0101} - C_{0123} ) = \frac{1}{2} ( C_{2323} - C_{0123} ) \,,
\\ \nonumber 
&& C_{0101} = {\boldsymbol \Psi}^\ast_2 + {\boldsymbol \Psi}_2 \,, \hspace{1cm} 
C_{0123} = {\boldsymbol \Psi}^\ast_2 - {\boldsymbol \Psi}_2 \,.
\end{eqnarray}
Other components can be easily obtained by using the symmetries of the Weyl tensor and also taking complex conjugate. For example, $C_{0312} = C^*_{0213} = - C^*_{0231}=-{\boldsymbol \Psi}^*_2$. Note that taking complex conjugate indices $0$ and $1$ does not change while indices $2$ and $3$ exchange with each other. The reason is that indices $0$ and $1$ correspond to the real null tetrad components $l^\mu$ and $k^\mu$ while indices $2$ and $3$ correspond to the complex null tetrad components $m^\mu$ and ${\bar m}^\mu$.

Also ten components of the Ricci part can be expressed in terms of four real scalars ${\boldsymbol \Phi}_{00}$, ${\boldsymbol \Phi}_{11}$, ${\boldsymbol \Phi}_{22}$, ${\boldsymbol \Lambda}$, and three complex scalars ${\boldsymbol \Phi}_{01}$, ${\boldsymbol \Phi}_{02}$, ${\boldsymbol \Phi}_{12}$ which are defined as
\begin{eqnarray}\label{RicciS-T-a}
&&{\boldsymbol \Phi}_{00} = \frac{1}{2} R_{00} \,, \hspace{.5cm}
{\boldsymbol \Phi}_{11} = \frac{1}{4} (R_{01}+R_{23}) \,, \hspace{.5cm}
{\boldsymbol \Phi}_{22} = \frac{1}{2} R_{11} \,, \hspace{.5cm}
{\boldsymbol \Lambda} = - \frac{1}{24} R = \frac{1}{12} (R_{01} - R_{23}) \,, \nonumber \\
&&{\boldsymbol \Phi}_{01} = \frac{1}{2} R_{02} \,, \hspace{1.5cm}
{\boldsymbol \Phi}_{02} = \frac{1}{2} R_{22} \,, \hspace{1.5cm}
{\boldsymbol \Phi}_{12} = \frac{1}{2} R_{12} \,.
\end{eqnarray}
We can express all components of the Ricci tensor in terms of the above scalar quantities and their complex conjugates. For example, $R_{01} = 2 ({\boldsymbol \Phi}_{11}+3{\boldsymbol \Lambda})$, $R_{23} = 2 ({\boldsymbol \Phi}_{11}-3{\boldsymbol \Lambda})$, $R_{03} = 2 {\boldsymbol \Phi}^\ast_{01}$, $R_{13}= 2 {\boldsymbol \Phi}^\ast_{12}$, and $R_{33} = 2 {\boldsymbol \Phi}^\ast_{02}$.
Here we considered the Riemann tensor in the Einstein frame while the results is true for the Riemann tensor in the Jordan frame ${\tilde R}_{abcd}$ and also any other tensor with the same symmetry group as the Riemann tensor.

\section{Weyl tensor in Jordan frame}

Let us use the results of the previous subsection and express the Weyl tensor in Jordan frame ${\tilde C}_{abcd}$ given by Eq.~\eqref{C-tilde-trans0} in terms of the seed frame quantities after disformal transformation (which corresponds usually to the Einstein frame). For the l.h.s.\ of Eq.~\eqref{C-tilde-trans0}, as we mentioned above, we can simply express all components of ${\tilde C}_{abcd}$ in terms of the five complex Weyl scalars in the Jordan frame similar to Eq.~\eqref{WeylS-T-a} as follows
\begin{eqnarray}\label{WeylS-tilde-T-a}
\tilde{\boldsymbol \Psi}_0 = { \tilde C}_{0202} \,, \hspace{.5cm}
\tilde{\boldsymbol \Psi}_1 = { \tilde C}_{0102}  \,, \hspace{.5cm}
\tilde{\boldsymbol \Psi}_2 = { \tilde C}_{0231} \,, \hspace{.5cm}
\tilde{\boldsymbol \Psi}_3 = { \tilde C}_{1013} \,, \hspace{.5cm}
\tilde{\boldsymbol \Psi}_4 = { \tilde C}_{1313} \,,
\end{eqnarray}
and the other components can be found with relations similar to~\eqref{Weyl-components}.

For the r.h.s.\ of Eq.~\eqref{C-tilde-trans0}, ${ C}_{abcd}$ can be expressed in terms of the five complex Weyl scalars~\eqref{WeylS-T-a}. Since the tensor $B_{\alpha\beta\mu\nu}$ has the same symmetry as the Weyl tensor, $B_{abcd}$ can be expressed in terms of five complex Weyl-type scalars as follows
\begin{eqnarray}\label{Delta-a}
{\boldsymbol \Delta}_0 = B_{0202} \,, \hspace{.5cm}
{\boldsymbol \Delta}_1 = B_{0102} \,, \hspace{.5cm}
{\boldsymbol \Delta}_2 = B_{0231} \,, \hspace{.5cm}
{\boldsymbol \Delta}_3 = B_{1013} \,, \hspace{.5cm}
{\boldsymbol \Delta}_4 = B_{1313} \,,
\end{eqnarray}
where the other components can be obtained through the same relations as~\eqref{Weyl-components} for $B_{abcd}$. The remaining part is $Z^S_{abcd}$ which from Eq.~\eqref{Z-S} we find
\begin{eqnarray}\label{Zij-a}
Z^S_{abcd} =  
-\frac{1}{2} \big(Z_{ad} \eta_{bc} + Z_{bc} \eta_{ad}
- Z_{ac} \eta_{bd} - Z_{bd} \eta_{ac} \big) 
- \frac{1}{6} Z (\eta_{ac} \eta_{bd} - \eta_{ad} \eta_{bc}) \,.
\end{eqnarray}
Hence $Z^S_{abcd}$ is completely expressed in terms of the Ricci part $Z_{ab}$ and, therefore, it can be completely expressed in terms of the four real and three complex scalars similar to relations~\eqref{RicciS-T-a}. We then define 
\begin{eqnarray}\label{Pi-a}
&&{\boldsymbol \Pi}_{00} = \frac{1}{2} Z_{00} \,, \hspace{.5cm}
{\boldsymbol \Pi}_{11} = \frac{1}{4} (Z_{01}+Z_{23}) \,, \hspace{.5cm}
{\boldsymbol \Pi}_{22} = \frac{1}{2} Z_{11} \,, \hspace{.5cm}
{\boldsymbol \Lambda}^S = - \frac{1}{24} Z = \frac{1}{12} (Z_{01} - Z_{23}) \,, \nonumber \\
&&{\boldsymbol \Pi}_{01} = \frac{1}{2} Z_{02} \,, \hspace{1.5cm}
{\boldsymbol \Pi}_{02} = \frac{1}{2} Z_{22} \,, \hspace{1.5cm}
{\boldsymbol \Pi}_{12} = \frac{1}{2} Z_{12} \,.
\end{eqnarray}
We can express all tetrad components of $Z^S_{abcd}$ in terms of the above defined scalars. Some of them are  as follows
\begin{eqnarray}\label{Z-P}
&&Z^S_{0101} = 2 ({\boldsymbol \Lambda}^S+{\boldsymbol \Pi}_{11}) \,, \hspace{.5cm}
Z^S_{0102} = {\boldsymbol \Pi}_{01} \,, \hspace{.5cm}
Z^S_{0112} = - {\boldsymbol \Pi}_{12} \,, \hspace{.5cm}
Z^S_{0123} = 0 \,,
\nonumber \\ 
&&Z^S_{0203} = {\boldsymbol \Pi}_{00} \,, \hspace{.5cm}
Z^S_{0212} = - {\boldsymbol \Pi}_{02} \,, \hspace{.5cm}
Z^S_{0213} = 2 {\boldsymbol \Lambda}^S \,, \hspace{.5cm}
Z^S_{0223} = {\boldsymbol \Pi}_{01} \,,
\nonumber \\
&&Z^S_{1212} = 0 \,, \hspace{.5cm}
Z^S_{1213} = {\boldsymbol \Pi}_{22} \,, \hspace{.5cm}
Z^S_{1223} = {\boldsymbol \Pi}_{12} \,, \hspace{.5cm}
Z^S_{2323} =  2 ({\boldsymbol \Lambda}^S-{\boldsymbol \Pi}_{11}) \,.
\end{eqnarray}
Now, from Eq.~\eqref{C-tilde-trans} we can express $\tilde{\boldsymbol \Psi}_I$ in terms of ${\boldsymbol \Psi}_I$, ${\boldsymbol \Delta}_I$, ${\boldsymbol \Pi}_{IJ}$, and ${\boldsymbol \Lambda}^S$ as follows
\begin{eqnarray}\label{Psi0}
A \tilde{\boldsymbol \Psi}_0 &=&  \gamma^2 ( {\boldsymbol \Psi}_0 + {\boldsymbol \Delta}_0 )
+ \beta^2 \big[ ( {\boldsymbol \Psi}^*_0 + {\boldsymbol \Delta}^*_0 ) (V_2V_2)^2 
+ ( {\boldsymbol \Psi}^*_4 + {\boldsymbol \Delta}^*_4 ) (V^1V^1)^2 \big]
\nonumber \\ 
&+& 2 \beta^2 V^1 V_2 \big[ 2 ( {\boldsymbol \Psi}^*_1 + {\boldsymbol \Delta}^*_1 ) V_2V_2 
+ 3 ( {\boldsymbol \Psi}^*_2 + {\boldsymbol \Delta}^*_2 ) V^1V_2 
+ 2 ( {\boldsymbol \Psi}^*_3 + {\boldsymbol \Delta}^*_3 ) V^1V^1 \big] 
\nonumber \\ 
&-& 2 \frac{B}{A} \beta \gamma \big[ 2 {\boldsymbol \Pi}_{01} V^1V_2 
+ {\boldsymbol \Pi}_{02} V^1 V^1 + {\boldsymbol \Pi}_{00} V_2 V_2 \big]
\,,
\end{eqnarray}

\begin{eqnarray}\label{Psi1}
A \tilde{\boldsymbol \Psi}_1 &= &
\gamma \big[ ({\boldsymbol \Psi} _1+{\boldsymbol \Delta} _1) \left( 1 - 2 \beta V^1 V_1 \right)
- \beta  V^2 V_1 ({\boldsymbol \Psi} _0+{\boldsymbol \Delta} _0)
+ \beta  V^1 V_2 ({\boldsymbol \Psi} _2+{\boldsymbol \Delta} _2) \big]
\nonumber \\ 
&-& \beta  \big(
V_2 V_2 \left( {\boldsymbol \Psi}^*_1+ {\boldsymbol \Delta}^*_1\right)
+ 2 V^1 V_2 \left({\boldsymbol \Psi}^*_2 + {\boldsymbol \Delta}^*_2\right)
+ V^1 V^1 \left({\boldsymbol \Psi}^*_3+{\boldsymbol \Delta}^*_3\right)
\big)
\nonumber \\ 
&+& \beta^2 \big[ 
V_1 V_2 V_2
\big(
V_2 ( {\boldsymbol \Psi}^*_0 + {\boldsymbol \Delta}^*_0 ) 
+ 4 V^1 \left({\boldsymbol \Psi}^*_1+{\boldsymbol \Delta}^*_1\right)t
\big)
- V^2 V^1V^1V^1 \left({\boldsymbol \Psi}^*_4+{\boldsymbol \Delta}^*_4\right) 
\nonumber \\ 
&&
+ 2 V^1V^1 \left({\boldsymbol \Psi}^*_3+{\boldsymbol \Delta}^*_3 \right) \left(V^1 V_1-V^2 V_2\right)
+V^1 V_2 \left({\boldsymbol \Psi}^*_2+{\boldsymbol \Delta}^*_2\right) \left(5 V^1 V_1- V^2 V_2\right) 
\big]
\nonumber \\ 
&+& \frac{B}{A} \Big\{
{\boldsymbol \Pi}_{01} \big[ 
\gamma \left( 1 - 2 \beta V^1 V_1 \right) + 2 \beta^2 V_1V^1 V_2 V^2
\big]
\nonumber \\ 
&-& \beta
\big[ 2 V^1 V_2 \left({\boldsymbol \Pi} _{11}+ 2 {\boldsymbol \Lambda}^S\right)
+\Pi_{12} V^1 V^1 - {\boldsymbol \Pi}_{02} V^2 V^1 
+ V_2 \left( {\boldsymbol \Pi}^*_{01} V_2 + {\boldsymbol \Pi} _{00} V_1\right)\big]
 \\ \nonumber 
&+& \beta^2 \big[ 
2 V_1 V_2 \big(2 V^1 V^1 \left({\boldsymbol \Pi}_{11}+{\boldsymbol \Lambda}^S\right)
+{\boldsymbol \Pi}^*_{01} V^1 V_2 + {\boldsymbol \Pi}_{00} V^2V_2 \big)
+ 2 {\boldsymbol \Pi} _{12} V_1 V^1V^1V^1
 \\ \nonumber 
&&
-  V^1 V_2 \left( {\boldsymbol \Pi}_{00} V_1 V^1 + {\boldsymbol \Pi}_{02} V^2 V^2 
+ {\boldsymbol \Pi}_{22} V^1 V^1 
+2 {\boldsymbol \Pi}^*_{12} V^1V_2
+{\boldsymbol \Pi}^*_{02} V_2 V_2\right) - 4 V^2 V_2 {\boldsymbol \Lambda}^S
\big]
\Big\} \,,
\end{eqnarray}

\begin{eqnarray}
\label{Psi2}
A \tilde{\boldsymbol \Psi}_2 &=& \gamma^2 ( {\boldsymbol \Psi}_2 + {\boldsymbol \Delta}_2 )
+ \beta^2 \big[ ( {\boldsymbol \Psi}^*_0 + {\boldsymbol \Delta}^*_0) (V_1 V_2)^2 
+ ( {\boldsymbol \Psi}^*_4 + {\boldsymbol \Delta}^*_4 ) (V^1 V^2)^2 \big]
\nonumber \\ 
&+& \beta^2 \big[ 
2 (V_1V^1-V_2V^2) \big( ( {\boldsymbol \Psi}^*_1 + {\boldsymbol \Delta}^*_1 ) V_1 V_2 
- ( {\boldsymbol \Psi}^*_3 + {\boldsymbol \Delta}^*_3 ) V^1 V^2 \big)
\nonumber \\ 
&&+ ( {\boldsymbol \Psi}^*_2 + {\boldsymbol \Delta}^*_2 ) 
\big( (V_1V^1-V_2 V^2)^2 - 2 V_1 V^1 V_2 V^2 \big)
\big] 
- 2 \frac{B}{A} {\boldsymbol \Lambda}^S \big[ 1 + 2 \beta \gamma (1-\gamma) \big]
\\ \nonumber
&-&
\frac{B}{A} \beta \gamma \big[ {\boldsymbol \Pi}_{00} V_1 V_1 - 2 {\boldsymbol \Pi}_{01} V_1 V^2 
+ {\boldsymbol \Pi}_{02} V^2 V^2 
+ 2 {\boldsymbol \Pi}^*_{12} V^1 V_2 
+ {\boldsymbol \Pi}^*_{02} V_2 V_2 + {\boldsymbol \Pi}_{22} V^1 V^1 \big] \,,
\end{eqnarray}

\begin{eqnarray}\label{Psi3}
A \tilde{\boldsymbol \Psi}_3 & = & \gamma \big[
({\boldsymbol \Psi}_3 + {\boldsymbol \Delta}_3) \left(1-2 \beta V^1 V_1\right) 
- \beta ({\boldsymbol \Psi}_2+ {\boldsymbol \Delta}_2) V^2 V_1 
+ \beta ({\boldsymbol \Psi}_4+{\boldsymbol \Delta}_4) V^1 V_2 \big]
\nonumber \\ 
&-& \beta \big[ 
V_1 V_1 \left( {\boldsymbol \Psi}^*_1 + {\boldsymbol \Delta}^*_1\right)
- 2 V^2 V_1 \left( {\boldsymbol \Psi}^*_2 + {\boldsymbol \Delta}^*_2\right)
+ V^2 V^2 \left( {\boldsymbol \Psi}^*_3 + {\boldsymbol \Delta}^*_3\right)
\big]
\nonumber \\ 
& + & \beta^2 \big[ 
2 ( {\boldsymbol \Psi}^*_1 + {\boldsymbol \Delta}^*_1) V_1^2 \left(V^1 V_1 - V^2 V_2\right)
+ V_1 V_1 V_1 V_2 \left( {\boldsymbol \Psi}^*_0+{\boldsymbol \Delta}^*_0\right)
- V^1 V^2 V^2V^2 \left( {\boldsymbol \Psi}^*_4+{\boldsymbol \Delta}^*_4\right)
\nonumber \\ 
&&
+ 4 V^1 V^2 V^2 V_1 \left( {\boldsymbol \Psi}^*_3+{\boldsymbol \Delta}^*_3\right)
+ ( {\boldsymbol \Psi}^*_2 + {\boldsymbol \Delta}^*_2) V^2 V_1 \left(V^2 V_2-5 V^1 V_1\right)
\big]
\nonumber \\ 
&+& \frac{B}{A} \Big\{ 
{\boldsymbol \Pi}^*{}_{12} \big[ 
\gamma \left( 1 - 2 \beta V^1 V_1 \right) + 2 \beta^2 V_1V^1 V_2 V^2
\big]
\nonumber \\ 
&-& \beta \big[
\left( {\boldsymbol \Pi}_{12} V^2 - {\boldsymbol \Pi}_{22} V^1 \right) V^2
-2 V^2 V_1 \left( {\boldsymbol \Pi} _{11}+2 {\boldsymbol \Lambda}^S\right)
+ {\boldsymbol \Pi}^*_{01} V_1 V_1 + {\boldsymbol \Pi}^*_{02} V_2 V_1
\big]
\nonumber \\ 
&+& \beta^2 \big[
V^2 V_1 \left( {\boldsymbol \Pi}_{02} V^2V^2 + 2 {\boldsymbol \Pi} _{12} V^1 V^2
- {\boldsymbol \Pi}_{22} V^1V^1 + {\boldsymbol \Pi}^*_{02} V_2 V_2 
- 4 V^2 V_2 {\boldsymbol \Lambda}^S \right)
- 2 {\boldsymbol \Pi}_{22} V^1 V^2V^2 V_2
\nonumber \\ 
&&+ V_1 V_1 \big( \left(2 {\boldsymbol \Pi}^*_{01} V^1 + {\boldsymbol \Pi}_{00} V^2\right)  V_1 
- 2
\left(2 V^1 \left( {\boldsymbol \Pi} _{11} + {\boldsymbol \Lambda}^S\right) 
+ {\boldsymbol \Pi}_{01} V^2 \right) V^2 \big)
\big]
\Big\} \,,
\end{eqnarray}

\begin{eqnarray}\label{Psi4}
A \tilde{\boldsymbol \Psi}_4 &=& \gamma^2 ( {\boldsymbol \Psi}_4 + {\boldsymbol \Delta}_4 ) 
+ \beta^2 \big[ ( {\boldsymbol \Psi}^*_4 + {\boldsymbol \Delta}^*_4 ) (V^2V^2)^2 
+ ( {\boldsymbol \Psi}^*_0 + {\boldsymbol \Delta}^*_0 ) (V_1V_1)^2 \big]
\nonumber \\ 
&-& 2 \beta^2 V_1 V^2 \big[ 
2 ( {\boldsymbol \Psi}^*_1 + {\boldsymbol \Delta}^*_1 ) V_1 V_1
-3 ( {\boldsymbol \Psi}^*_2 + {\boldsymbol \Delta}^*_2 ) V_1 V^2
+ 2 ( {\boldsymbol \Psi}^*_3 + {\boldsymbol \Delta}^*_3 ) V^2 V^2 \big] 
\nonumber \\ 
&+& 2 \frac{B}{A} \beta \gamma
\big[  2 {\boldsymbol \Pi}^*_{12} V_1 V^2 - {\boldsymbol \Pi}^*_{02} V_1 V_1 
- {\boldsymbol \Pi}_{22} V^2 V^2 \big]
\,,
\end{eqnarray}
where we have defined parameter
\begin{equation}\label{gamma}
\gamma \equiv 1 - \beta ( V_1V^1+V_2V^2 ) \,.
\end{equation}
The above parameter goes to unity for $B=0$ as $\beta=0$ in this case which can be seen from Eq.~\eqref{beta}.

\bibliographystyle{ThesisStyleWithEtAl}
\bibliography{Thesis}

\end{document}